\documentclass[a4paper,11pt]{article}
\pdfoutput=1 %

\usepackage{geometry}

\usepackage{jheppub} %

\usepackage[utf8]{inputenc}
\usepackage[T1]{fontenc} %

\usepackage{multirow}
\usepackage{physics}
\usepackage{slashed}
\usepackage{xifthen}
\usepackage{amsmath}
\usepackage{changepage}
\usepackage{caption}
\usepackage{listings}
\usepackage{dsfont}
\usepackage{pdflscape}
\usepackage{graphics}
\usepackage{placeins}
\usepackage{xfrac}
\usepackage[dvipsnames]{xcolor}
\usepackage{standalone}
\usepackage{enumitem}
\usepackage{slashed}
\usepackage{subcaption}

\usepackage{mathtools}
\makeatletter
\newcommand{\vasterino}{\bBigg@{3}}
\newcommand{\vast}{\bBigg@{5}}
\newcommand{\vastino}{\bBigg@{6}}
\newcommand{\Vast}{\bBigg@{7}}

\makeatother

\usepackage{tikz}
\usepackage[compat=1.1.0]{tikz-feynman}
\tikzfeynmanset{warn luatex=false}
\tikzfeynmanset{/tikzfeynman/warn luatex = false}

\tikzset{main node/.style={circle,fill=blue!20,draw,pattern=crosshatch dots,pattern color=black!30!white,minimum size=1.2cm,inner sep=0pt},
            }
            
            \tikzset{main node mini/.style={circle,fill=blue!20,draw,pattern=crosshatch dots,pattern color=black!30!white,minimum size=0.6cm,inner sep=0pt},
            }
            
\tikzset{main node2/.style={circle,fill=blue!40,draw,pattern=crosshatch dots,pattern color=black!30!white,minimum size=3cm,inner sep=0pt},
            }
            
\tikzset{main node3/.style={circle,fill=blue!40,draw,pattern=crosshatch dots,pattern color=black!30!white,minimum size=2.3cm,inner sep=0pt},
            }
            
\tikzset{main node4/.style={circle,fill=black,draw,minimum size=0.2cm,inner sep=0pt},
            }
            
\tikzset{main node5/.style={circle,fill=black,draw,minimum size=0.1cm,inner sep=0pt},
            }

            \tikzfeynmanset{
  fermion1/.style={
    /tikz/postaction={
      /tikz/decoration={
        markings,
        mark=at position 0.5 with {
          \node[
            transform shape,
            xshift=-0.5mm,
            fill,
            inner sep=0.7pt,
            draw=none,
            dart
          ] { };
        },
      },
      /tikz/decorate=true,
    },
  },
  ghost1/.style={
  dashed,
    /tikz/postaction={
      /tikz/decoration={
        markings,
        mark=at position 0.5 with {
          \node[
            transform shape,
            xshift=-0.5mm,
            fill,
            inner sep=0.7pt,
            draw=none,
            dart
          ] { };
        },
      },
      /tikz/decorate=true,
    },
  },
  fermion2/.style={
    /tikz/postaction={
      /tikz/decoration={
        markings,
        mark=at position 0.5 with {
          \arrow{[length=4pt, width=3pt]};
        },
      },
      /tikz/decorate=true,
    },
  },
}

\def \feynmantopsep{1.4mm}

\tikzfeynmanset{
   every top@@/.style={
     /tikz/preaction={draw=white,line width=\feynmantopsep} 
   },
   top/.style={
     /tikzfeynman/every top@@,
   },
}

\usepackage{tabularx}
\usepackage{xspace}
\usepackage{floatrow}
\usepackage{array}
\usepackage{rotating}
\usepackage{etex,etoolbox}
\usepackage{physics}
\usepackage{booktabs,arydshln}
\makeatletter
\def\adl@drawiv#1#2#3{%
	\hskip.5\tabcolsep
	\xleaders#3{#2.5\@tempdimb #1{1}#2.5\@tempdimb}%
	#2\z@ plus1fil minus1fil\relax
	\hskip.5\tabcolsep}
\newcommand{\cdashlinelr}[1]{%
	\noalign{\vskip\aboverulesep
		\global\let\@dashdrawstore\adl@draw
		\global\let\adl@draw\adl@drawiv}
	\cdashline{#1}
	\noalign{\global\let\adl@draw\@dashdrawstore
		\vskip\belowrulesep}}
\makeatother

\usepackage{graphicx}%

\makeatother

\usepackage{pdftexcmds}
\makeatletter

\newcommand{\stringcases}[3]{%
  \romannumeral
    \str@case{#1}#2{#1}{#3}\q@stop
}
\newcommand{\str@case}[3]{%
  \ifnum\pdf@strcmp{\unexpanded{#1}}{\unexpanded{#2}}=\z@
    \expandafter\@firstoftwo
  \else
    \expandafter\@secondoftwo
  \fi
    {\str@case@end{#3}}
    {\str@case{#1}}%
}
\newcommand{\str@case@end}{}
\long\def\str@case@end#1#2\q@stop{\z@#1}

\newcommand{\percent}{}
\newcommand{\minus}{\texttt{-}}
\newcommand*{\getRes}[1]{
  \stringcases
    {#1}
    {
        {ajjLOSGQG000mplcty}{$\mathtt{1}$}
        {ajjLOSGQG000}{$\phantom{\minus}\mathtt{5.031049\phantom{}} \cdot \mathtt{10}^{\minus\mathtt{01}}$}
        {ajjLOSGQG000delta}{$\phantom{\mathtt{0}}\phantom{\minus}\mathtt{0.0018}\percent$}
        {ajjLOTotal}{$\phantom{\minus}\mathtt{5.031049\phantom{}} \cdot \mathtt{10}^{\minus\mathtt{01}}$}
        {ajjLOTotaldelta}{$\phantom{\mathtt{0}}\phantom{\minus}\mathtt{0.0018}\percent$}
        {ajjNLOSGQG000mplcty}{$\mathtt{1}$}
        {ajjNLOSGQG000}{$\phantom{\minus}\mathtt{5.03926\phantom{0}} \cdot \mathtt{10}^{\minus\mathtt{02}}$}
        {ajjNLOSGQG000delta}{$\phantom{\mathtt{0}}\phantom{\minus}\mathtt{0.0075}\percent$}
        {ajjNLOSGQG001mplcty}{$\mathtt{2}$}
        {ajjNLOSGQG001}{$\minus\mathtt{3.14956\phantom{0}} \cdot \mathtt{10}^{\minus\mathtt{02}}$}
        {ajjNLOSGQG001delta}{$\phantom{\mathtt{0}}\phantom{\minus}\mathtt{0.018}\percent\phantom{\mathtt{0}}$}
        {ajjNLOTotal}{$\phantom{\minus}\mathtt{1.88970\phantom{0}} \cdot \mathtt{10}^{\minus\mathtt{02}}$}
        {ajjNLOTotaldelta}{$\phantom{\mathtt{0}}\phantom{\minus}\mathtt{0.036}\percent\phantom{\mathtt{0}}$}
        {ajjNLOTarget}{$\phantom{\minus}\mathtt{1.889690\phantom{}} \cdot \mathtt{10}^{\minus\mathtt{02}}$}
        {ajjNLOTargetdelta}{$\phantom{\mathtt{0}}\phantom{\minus}\mathtt{0.00053}\percent$}
        {ajjNNLOnf1SGQG015mplcty}{$\mathtt{1}$}
        {ajjNNLOnf1SGQG015}{$\minus\mathtt{4.66342\phantom{0}} \cdot \mathtt{10}^{\minus\mathtt{04}}$}
        {ajjNNLOnf1SGQG015delta}{$\phantom{\mathtt{0}}\phantom{\minus}\mathtt{0.019}\percent\phantom{\mathtt{0}}$}
        {ajjNNLOnf1SGQG025mplcty}{$\mathtt{2}$}
        {ajjNNLOnf1SGQG025}{$\phantom{\minus}\mathtt{3.8448\phantom{00}} \cdot \mathtt{10}^{\minus\mathtt{04}}$}
        {ajjNNLOnf1SGQG025delta}{$\phantom{\mathtt{0}}\phantom{\minus}\mathtt{0.036}\percent\phantom{\mathtt{0}}$}
        {ajjNNLOnf1Total}{$\minus\mathtt{8.186\phantom{000}} \cdot \mathtt{10}^{\minus\mathtt{05}}$}
        {ajjNNLOnf1Totaldelta}{$\phantom{\mathtt{0}}\phantom{\minus}\mathtt{0.20}\percent\phantom{\mathtt{00}}$}
        {ajjNNLOnf1Target}{$\minus\mathtt{8.1834\phantom{00}} \cdot \mathtt{10}^{\minus\mathtt{05}}$}
        {ajjNNLOnf1Targetdelta}{$\phantom{\mathtt{0}}\phantom{\minus}\mathtt{0.036}\percent\phantom{\mathtt{0}}$}
        {ajjNNLOSGQG023mplcty}{$\mathtt{2}$}
        {ajjNNLOSGQG023}{$\minus\mathtt{2.30886\phantom{0}} \cdot \mathtt{10}^{\minus\mathtt{03}}$}
        {ajjNNLOSGQG023delta}{$\phantom{\mathtt{0}}\phantom{\minus}\mathtt{0.017}\percent\phantom{\mathtt{0}}$}
        {ajjNNLOSGQG006mplcty}{$\mathtt{2}$}
        {ajjNNLOSGQG006}{$\phantom{\minus}\mathtt{6.42018\phantom{0}} \cdot \mathtt{10}^{\minus\mathtt{03}}$}
        {ajjNNLOSGQG006delta}{$\phantom{\mathtt{0}}\phantom{\minus}\mathtt{0.0055}\percent$}
        {ajjNNLOSGQG016mplcty}{$\mathtt{2}$}
        {ajjNNLOSGQG016}{$\minus\mathtt{6.91254\phantom{0}} \cdot \mathtt{10}^{\minus\mathtt{03}}$}
        {ajjNNLOSGQG016delta}{$\phantom{\mathtt{0}}\phantom{\minus}\mathtt{0.0046}\percent$}
        {ajjNNLOSGQG013mplcty}{$\mathtt{1}$}
        {ajjNNLOSGQG013}{$\phantom{\minus}\mathtt{3.20278\phantom{0}} \cdot \mathtt{10}^{\minus\mathtt{03}}$}
        {ajjNNLOSGQG013delta}{$\phantom{\mathtt{0}}\phantom{\minus}\mathtt{0.0084}\percent$}
        {ajjNNLOSGQG003mplcty}{$\mathtt{1}$}
        {ajjNNLOSGQG003}{$\phantom{\minus}\mathtt{1.68148\phantom{0}} \cdot \mathtt{10}^{\minus\mathtt{03}}$}
        {ajjNNLOSGQG003delta}{$\phantom{\mathtt{0}}\phantom{\minus}\mathtt{0.013}\percent\phantom{\mathtt{0}}$}
        {ajjNNLOSGQG022mplcty}{$\mathtt{2}$}
        {ajjNNLOSGQG022}{$\phantom{\minus}\mathtt{6.6698\phantom{00}} \cdot \mathtt{10}^{\minus\mathtt{04}}$}
        {ajjNNLOSGQG022delta}{$\phantom{\mathtt{0}}\phantom{\minus}\mathtt{0.027}\percent\phantom{\mathtt{0}}$}
        {ajjNNLOSGQG011mplcty}{$\mathtt{2}$}
        {ajjNNLOSGQG011}{$\minus\mathtt{1.30381\phantom{0}} \cdot \mathtt{10}^{\minus\mathtt{03}}$}
        {ajjNNLOSGQG011delta}{$\phantom{\mathtt{0}}\phantom{\minus}\mathtt{0.013}\percent\phantom{\mathtt{0}}$}
        {ajjNNLOSGQG009mplcty}{$\mathtt{2}$}
        {ajjNNLOSGQG009}{$\minus\mathtt{1.30395\phantom{0}} \cdot \mathtt{10}^{\minus\mathtt{03}}$}
        {ajjNNLOSGQG009delta}{$\phantom{\mathtt{0}}\phantom{\minus}\mathtt{0.013}\percent\phantom{\mathtt{0}}$}
        {ajjNNLOSGQG024mplcty}{$\mathtt{2}$}
        {ajjNNLOSGQG024}{$\minus\mathtt{1.6661\phantom{00}} \cdot \mathtt{10}^{\minus\mathtt{04}}$}
        {ajjNNLOSGQG024delta}{$\phantom{\mathtt{0}}\phantom{\minus}\mathtt{0.064}\percent\phantom{\mathtt{0}}$}
        {ajjNNLOSGQG020mplcty}{$\mathtt{2}$}
        {ajjNNLOSGQG020}{$\phantom{\minus}\mathtt{6.64155\phantom{0}} \cdot \mathtt{10}^{\minus\mathtt{04}}$}
        {ajjNNLOSGQG020delta}{$\phantom{\mathtt{0}}\phantom{\minus}\mathtt{0.012}\percent\phantom{\mathtt{0}}$}
        {ajjNNLOSGQG005mplcty}{$\mathtt{2}$}
        {ajjNNLOSGQG005}{$\phantom{\minus}\mathtt{2.34300\phantom{0}} \cdot \mathtt{10}^{\minus\mathtt{04}}$}
        {ajjNNLOSGQG005delta}{$\phantom{\mathtt{0}}\phantom{\minus}\mathtt{0.031}\percent\phantom{\mathtt{0}}$}
        {ajjNNLOSGQG001mplcty}{$\mathtt{1}$}
        {ajjNNLOSGQG001}{$\phantom{\minus}\mathtt{4.11063\phantom{0}} \cdot \mathtt{10}^{\minus\mathtt{04}}$}
        {ajjNNLOSGQG001delta}{$\phantom{\mathtt{0}}\phantom{\minus}\mathtt{0.017}\percent\phantom{\mathtt{0}}$}
        {ajjNNLOSGQG014mplcty}{$\mathtt{1}$}
        {ajjNNLOSGQG014}{$\phantom{\minus}\mathtt{2.41514\phantom{0}} \cdot \mathtt{10}^{\minus\mathtt{04}}$}
        {ajjNNLOSGQG014delta}{$\phantom{\mathtt{0}}\phantom{\minus}\mathtt{0.026}\percent\phantom{\mathtt{0}}$}
        {ajjNNLOSGQG017mplcty}{$\mathtt{2}$}
        {ajjNNLOSGQG017}{$\phantom{\minus}\mathtt{5.8386\phantom{00}} \cdot \mathtt{10}^{\minus\mathtt{05}}$}
        {ajjNNLOSGQG017delta}{$\phantom{\mathtt{0}}\phantom{\minus}\mathtt{0.088}\percent\phantom{\mathtt{0}}$}
        {ajjNNLOSGQG000mplcty}{$\mathtt{1}$}
        {ajjNNLOSGQG000}{$\minus\mathtt{1.75957\phantom{0}} \cdot \mathtt{10}^{\minus\mathtt{04}}$}
        {ajjNNLOSGQG000delta}{$\phantom{\mathtt{0}}\phantom{\minus}\mathtt{0.022}\percent\phantom{\mathtt{0}}$}
        {ajjNNLOTotal}{$\phantom{\minus}\mathtt{1.40910\phantom{0}} \cdot \mathtt{10}^{\minus\mathtt{03}}$}
        {ajjNNLOTotaldelta}{$\phantom{\mathtt{0}}\phantom{\minus}\mathtt{0.056}\percent\phantom{\mathtt{0}}$}
        {ajjNNLOTarget}{$\phantom{\minus}\mathtt{1.40941\phantom{0}} \cdot \mathtt{10}^{\minus\mathtt{03}}$}
        {ajjNNLOTargetdelta}{$\phantom{\mathtt{0}}\minus\mathtt{0.022}\percent\phantom{\mathtt{0}}$}
        {attxLOSGQG000mplcty}{$\mathtt{1}$}
        {attxLOSGQG000}{$\phantom{\minus}\mathtt{2.876302\phantom{}} \cdot \mathtt{10}^{+\mathtt{00}}$}
        {attxLOSGQG000delta}{$\phantom{\mathtt{0}}\phantom{\minus}\mathtt{0.00049}\percent$}
        {attxLOTotal}{$\phantom{\minus}\mathtt{2.876302\phantom{}} \cdot \mathtt{10}^{+\mathtt{00}}$}
        {attxLOTotaldelta}{$\phantom{\mathtt{0}}\phantom{\minus}\mathtt{0.00049}\percent$}
        {attxNLOSGQG000mplcty}{$\mathtt{1}$}
        {attxNLOSGQG000}{$\phantom{\minus}\mathtt{3.48276\phantom{0}} \cdot \mathtt{10}^{\minus\mathtt{01}}$}
        {attxNLOSGQG000delta}{$\phantom{\mathtt{0}}\phantom{\minus}\mathtt{0.023}\percent\phantom{\mathtt{0}}$}
        {attxNLOSGQG001mplcty}{$\mathtt{2}$}
        {attxNLOSGQG001}{$\minus\mathtt{1.46756\phantom{0}} \cdot \mathtt{10}^{\minus\mathtt{01}}$}
        {attxNLOSGQG001delta}{$\phantom{\mathtt{0}}\phantom{\minus}\mathtt{0.059}\percent\phantom{\mathtt{0}}$}
        {attxNLOTotal}{$\phantom{\minus}\mathtt{2.0152\phantom{00}} \cdot \mathtt{10}^{\minus\mathtt{01}}$}
        {attxNLOTotaldelta}{$\phantom{\mathtt{0}}\phantom{\minus}\mathtt{0.059}\percent\phantom{\mathtt{0}}$}
        {attxNLOOSSGQG000mplcty}{$\mathtt{1}$}
        {attxNLOOSSGQG000}{$\phantom{\minus}\mathtt{3.48276\phantom{0}} \cdot \mathtt{10}^{\minus\mathtt{01}}$}
        {attxNLOOSSGQG000delta}{$\phantom{\mathtt{0}}\phantom{\minus}\mathtt{0.023}\percent\phantom{\mathtt{0}}$}
        {attxNLOOSSGQG001mplcty}{$\mathtt{2}$}
        {attxNLOOSSGQG001}{$\minus\mathtt{5.8856\phantom{00}} \cdot \mathtt{10}^{\minus\mathtt{02}}$}
        {attxNLOOSSGQG001delta}{$\phantom{\mathtt{0}}\phantom{\minus}\mathtt{0.13}\percent\phantom{\mathtt{00}}$}
        {attxNLOOSTotal}{$\phantom{\minus}\mathtt{2.8942\phantom{00}} \cdot \mathtt{10}^{\minus\mathtt{01}}$}
        {attxNLOOSTotaldelta}{$\phantom{\mathtt{0}}\phantom{\minus}\mathtt{0.038}\percent\phantom{\mathtt{0}}$}
        {attxNLOOSTarget}{$\phantom{\minus}\mathtt{2.8952\phantom{00}} \cdot \mathtt{10}^{\minus\mathtt{01}}$}
        {attxNLOOSTargetdelta}{$\phantom{\mathtt{0}}\minus\mathtt{0.036}\percent\phantom{\mathtt{0}}$}
        {attxNNLOheavyfermionloopSGQG016mplcty}{$\mathtt{1}$}
        {attxNNLOheavyfermionloopSGQG016}{$\minus\mathtt{2.1076\phantom{00}} \cdot \mathtt{10}^{\minus\mathtt{03}}$}
        {attxNNLOheavyfermionloopSGQG016delta}{$\phantom{\mathtt{0}}\phantom{\minus}\mathtt{0.085}\percent\phantom{\mathtt{0}}$}
        {attxNNLOheavyfermionloopSGQG027mplcty}{$\mathtt{2}$}
        {attxNNLOheavyfermionloopSGQG027}{$\phantom{\minus}\mathtt{5.135\phantom{000}} \cdot \mathtt{10}^{\minus\mathtt{04}}$}
        {attxNNLOheavyfermionloopSGQG027delta}{$\phantom{\mathtt{0}}\phantom{\minus}\mathtt{0.39}\percent\phantom{\mathtt{00}}$}
        {attxNNLOheavyfermionloopTotal}{$\minus\mathtt{1.5941\phantom{00}} \cdot \mathtt{10}^{\minus\mathtt{03}}$}
        {attxNNLOheavyfermionloopTotaldelta}{$\phantom{\mathtt{0}}\phantom{\minus}\mathtt{0.17}\percent\phantom{\mathtt{00}}$}
        {attxNNLOnf1SGQG015mplcty}{$\mathtt{1}$}
        {attxNNLOnf1SGQG015}{$\minus\mathtt{3.2806\phantom{00}} \cdot \mathtt{10}^{\minus\mathtt{03}}$}
        {attxNNLOnf1SGQG015delta}{$\phantom{\mathtt{0}}\phantom{\minus}\mathtt{0.066}\percent\phantom{\mathtt{0}}$}
        {attxNNLOnf1SGQG026mplcty}{$\mathtt{2}$}
        {attxNNLOnf1SGQG026}{$\phantom{\minus}\mathtt{1.2309\phantom{00}} \cdot \mathtt{10}^{\minus\mathtt{03}}$}
        {attxNNLOnf1SGQG026delta}{$\phantom{\mathtt{0}}\phantom{\minus}\mathtt{0.23}\percent\phantom{\mathtt{00}}$}
        {attxNNLOnf1Total}{$\minus\mathtt{2.0497\phantom{00}} \cdot \mathtt{10}^{\minus\mathtt{03}}$}
        {attxNNLOnf1Totaldelta}{$\phantom{\mathtt{0}}\phantom{\minus}\mathtt{0.17}\percent\phantom{\mathtt{00}}$}
        {attxNNLOSGQG006mplcty}{$\mathtt{2}$}
        {attxNNLOSGQG006}{$\phantom{\minus}\mathtt{4.87662\phantom{0}} \cdot \mathtt{10}^{\minus\mathtt{02}}$}
        {attxNNLOSGQG006delta}{$\phantom{\mathtt{0}}\phantom{\minus}\mathtt{0.018}\percent\phantom{\mathtt{0}}$}
        {attxNNLOSGQG024mplcty}{$\mathtt{2}$}
        {attxNNLOSGQG024}{$\minus\mathtt{5.4290\phantom{00}} \cdot \mathtt{10}^{\minus\mathtt{03}}$}
        {attxNNLOSGQG024delta}{$\phantom{\mathtt{0}}\phantom{\minus}\mathtt{0.15}\percent\phantom{\mathtt{00}}$}
        {attxNNLOSGQG017mplcty}{$\mathtt{2}$}
        {attxNNLOSGQG017}{$\minus\mathtt{3.09477\phantom{0}} \cdot \mathtt{10}^{\minus\mathtt{02}}$}
        {attxNNLOSGQG017delta}{$\phantom{\mathtt{0}}\phantom{\minus}\mathtt{0.024}\percent\phantom{\mathtt{0}}$}
        {attxNNLOSGQG023mplcty}{$\mathtt{2}$}
        {attxNNLOSGQG023}{$\phantom{\minus}\mathtt{8.7680\phantom{00}} \cdot \mathtt{10}^{\minus\mathtt{03}}$}
        {attxNNLOSGQG023delta}{$\phantom{\mathtt{0}}\phantom{\minus}\mathtt{0.074}\percent\phantom{\mathtt{0}}$}
        {attxNNLOSGQG013mplcty}{$\mathtt{1}$}
        {attxNNLOSGQG013}{$\phantom{\minus}\mathtt{2.30085\phantom{0}} \cdot \mathtt{10}^{\minus\mathtt{02}}$}
        {attxNNLOSGQG013delta}{$\phantom{\mathtt{0}}\phantom{\minus}\mathtt{0.028}\percent\phantom{\mathtt{0}}$}
        {attxNNLOSGQG003mplcty}{$\mathtt{1}$}
        {attxNNLOSGQG003}{$\phantom{\minus}\mathtt{1.33171\phantom{0}} \cdot \mathtt{10}^{\minus\mathtt{02}}$}
        {attxNNLOSGQG003delta}{$\phantom{\mathtt{0}}\phantom{\minus}\mathtt{0.045}\percent\phantom{\mathtt{0}}$}
        {attxNNLOSGQG011mplcty}{$\mathtt{2}$}
        {attxNNLOSGQG011}{$\minus\mathtt{3.2639\phantom{00}} \cdot \mathtt{10}^{\minus\mathtt{03}}$}
        {attxNNLOSGQG011delta}{$\phantom{\mathtt{0}}\phantom{\minus}\mathtt{0.16}\percent\phantom{\mathtt{00}}$}
        {attxNNLOSGQG009mplcty}{$\mathtt{2}$}
        {attxNNLOSGQG009}{$\minus\mathtt{3.2606\phantom{00}} \cdot \mathtt{10}^{\minus\mathtt{03}}$}
        {attxNNLOSGQG009delta}{$\phantom{\mathtt{0}}\phantom{\minus}\mathtt{0.16}\percent\phantom{\mathtt{00}}$}
        {attxNNLOSGQG021mplcty}{$\mathtt{2}$}
        {attxNNLOSGQG021}{$\minus\mathtt{3.9986\phantom{00}} \cdot \mathtt{10}^{\minus\mathtt{03}}$}
        {attxNNLOSGQG021delta}{$\phantom{\mathtt{0}}\phantom{\minus}\mathtt{0.078}\percent\phantom{\mathtt{0}}$}
        {attxNNLOSGQG001mplcty}{$\mathtt{1}$}
        {attxNNLOSGQG001}{$\phantom{\minus}\mathtt{7.6693\phantom{00}} \cdot \mathtt{10}^{\minus\mathtt{03}}$}
        {attxNNLOSGQG001delta}{$\phantom{\mathtt{0}}\phantom{\minus}\mathtt{0.035}\percent\phantom{\mathtt{0}}$}
        {attxNNLOSGQG025mplcty}{$\mathtt{2}$}
        {attxNNLOSGQG025}{$\minus\mathtt{5.135\phantom{000}} \cdot \mathtt{10}^{\minus\mathtt{04}}$}
        {attxNNLOSGQG025delta}{$\phantom{\mathtt{0}}\phantom{\minus}\mathtt{0.40}\percent\phantom{\mathtt{00}}$}
        {attxNNLOSGQG018mplcty}{$\mathtt{2}$}
        {attxNNLOSGQG018}{$\phantom{\minus}\mathtt{4.257\phantom{000}} \cdot \mathtt{10}^{\minus\mathtt{04}}$}
        {attxNNLOSGQG018delta}{$\phantom{\mathtt{0}}\phantom{\minus}\mathtt{0.34}\percent\phantom{\mathtt{00}}$}
        {attxNNLOSGQG014mplcty}{$\mathtt{1}$}
        {attxNNLOSGQG014}{$\phantom{\minus}\mathtt{1.7054\phantom{00}} \cdot \mathtt{10}^{\minus\mathtt{03}}$}
        {attxNNLOSGQG014delta}{$\phantom{\mathtt{0}}\phantom{\minus}\mathtt{0.080}\percent\phantom{\mathtt{0}}$}
        {attxNNLOSGQG005mplcty}{$\mathtt{2}$}
        {attxNNLOSGQG005}{$\phantom{\minus}\mathtt{1.0696\phantom{00}} \cdot \mathtt{10}^{\minus\mathtt{03}}$}
        {attxNNLOSGQG005delta}{$\phantom{\mathtt{0}}\phantom{\minus}\mathtt{0.12}\percent\phantom{\mathtt{00}}$}
        {attxNNLOSGQG000mplcty}{$\mathtt{1}$}
        {attxNNLOSGQG000}{$\minus\mathtt{7.9301\phantom{00}} \cdot \mathtt{10}^{\minus\mathtt{04}}$}
        {attxNNLOSGQG000delta}{$\phantom{\mathtt{0}}\phantom{\minus}\mathtt{0.071}\percent\phantom{\mathtt{0}}$}
        {attxNNLOTotal}{$\phantom{\minus}\mathtt{5.6524\phantom{00}} \cdot \mathtt{10}^{\minus\mathtt{02}}$}
        {attxNNLOTotaldelta}{$\phantom{\mathtt{0}}\phantom{\minus}\mathtt{0.036}\percent\phantom{\mathtt{0}}$}
        {attxOSbench1LOSGQG000mplcty}{$\mathtt{1}$}
        {attxOSbench1LOSGQG000}{$\phantom{\minus}\mathtt{1.387586\phantom{}} \cdot \mathtt{10}^{+\mathtt{00}}$}
        {attxOSbench1LOSGQG000delta}{$\phantom{\mathtt{0}}\phantom{\minus}\mathtt{0.0011}\percent$}
        {attxOSbench1LOTotal}{$\phantom{\minus}\mathtt{1.387586\phantom{}} \cdot \mathtt{10}^{+\mathtt{00}}$}
        {attxOSbench1LOTotaldelta}{$\phantom{\mathtt{0}}\phantom{\minus}\mathtt{0.0011}\percent$}
        {attxOSbench1NLOSGQG000mplcty}{$\mathtt{1}$}
        {attxOSbench1NLOSGQG000}{$\phantom{\minus}\mathtt{2.52705\phantom{0}} \cdot \mathtt{10}^{\minus\mathtt{01}}$}
        {attxOSbench1NLOSGQG000delta}{$\phantom{\mathtt{0}}\phantom{\minus}\mathtt{0.034}\percent\phantom{\mathtt{0}}$}
        {attxOSbench1NLOSGQG001mplcty}{$\mathtt{2}$}
        {attxOSbench1NLOSGQG001}{$\phantom{\minus}\mathtt{1.80050\phantom{0}} \cdot \mathtt{10}^{\minus\mathtt{01}}$}
        {attxOSbench1NLOSGQG001delta}{$\phantom{\mathtt{0}}\phantom{\minus}\mathtt{0.049}\percent\phantom{\mathtt{0}}$}
        {attxOSbench1NLOTotal}{$\phantom{\minus}\mathtt{4.3276\phantom{00}} \cdot \mathtt{10}^{\minus\mathtt{01}}$}
        {attxOSbench1NLOTotaldelta}{$\phantom{\mathtt{0}}\phantom{\minus}\mathtt{0.028}\percent\phantom{\mathtt{0}}$}
        {attxOSbench1NLOTarget}{$\phantom{\minus}\mathtt{4.32831\phantom{0}} \cdot \mathtt{10}^{\minus\mathtt{01}}$}
        {attxOSbench1NLOTargetdelta}{$\phantom{\mathtt{0}}\minus\mathtt{0.018}\percent\phantom{\mathtt{0}}$}
        {attxOSbench1NNLOheavyfermionloopSGQG016mplcty}{$\mathtt{1}$}
        {attxOSbench1NNLOheavyfermionloopSGQG016}{$\phantom{\minus}\mathtt{6.990\phantom{000}} \cdot \mathtt{10}^{\minus\mathtt{04}}$}
        {attxOSbench1NNLOheavyfermionloopSGQG016delta}{$\phantom{\mathtt{0}}\phantom{\minus}\mathtt{0.21}\percent\phantom{\mathtt{00}}$}
        {attxOSbench1NNLOheavyfermionloopSGQG027mplcty}{$\mathtt{2}$}
        {attxOSbench1NNLOheavyfermionloopSGQG027}{$\minus\mathtt{1.39\phantom{0000}} \cdot \mathtt{10}^{\minus\mathtt{05}}$}
        {attxOSbench1NNLOheavyfermionloopSGQG027delta}{$\phantom{\minus}\mathtt{17}\percent\phantom{\mathtt{.0000}}$}
        {attxOSbench1NNLOheavyfermionloopTotal}{$\phantom{\minus}\mathtt{6.851\phantom{000}} \cdot \mathtt{10}^{\minus\mathtt{04}}$}
        {attxOSbench1NNLOheavyfermionloopTotaldelta}{$\phantom{\mathtt{0}}\phantom{\minus}\mathtt{0.40}\percent\phantom{\mathtt{00}}$}
        {attxOSbench1NNLOnf1SGQG015mplcty}{$\mathtt{1}$}
        {attxOSbench1NNLOnf1SGQG015}{$\minus\mathtt{1.0022\phantom{00}} \cdot \mathtt{10}^{\minus\mathtt{03}}$}
        {attxOSbench1NNLOnf1SGQG015delta}{$\phantom{\mathtt{0}}\phantom{\minus}\mathtt{0.17}\percent\phantom{\mathtt{00}}$}
        {attxOSbench1NNLOnf1SGQG026mplcty}{$\mathtt{2}$}
        {attxOSbench1NNLOnf1SGQG026}{$\minus\mathtt{4.6982\phantom{00}} \cdot \mathtt{10}^{\minus\mathtt{03}}$}
        {attxOSbench1NNLOnf1SGQG026delta}{$\phantom{\mathtt{0}}\phantom{\minus}\mathtt{0.081}\percent\phantom{\mathtt{0}}$}
        {attxOSbench1NNLOnf1Total}{$\minus\mathtt{5.7004\phantom{00}} \cdot \mathtt{10}^{\minus\mathtt{03}}$}
        {attxOSbench1NNLOnf1Totaldelta}{$\phantom{\mathtt{0}}\phantom{\minus}\mathtt{0.073}\percent\phantom{\mathtt{0}}$}
        {attxOSbench1NNLOnf1Target}{$\minus\mathtt{5.6982\phantom{00}} \cdot \mathtt{10}^{\minus\mathtt{03}}$}
        {attxOSbench1NNLOnf1Targetdelta}{$\phantom{\mathtt{0}}\phantom{\minus}\mathtt{0.038}\percent\phantom{\mathtt{0}}$}
        {attxOSbench1NNLOSGQG006mplcty}{$\mathtt{2}$}
        {attxOSbench1NNLOSGQG006}{$\phantom{\minus}\mathtt{5.6351\phantom{00}} \cdot \mathtt{10}^{\minus\mathtt{03}}$}
        {attxOSbench1NNLOSGQG006delta}{$\phantom{\mathtt{0}}\phantom{\minus}\mathtt{0.14}\percent\phantom{\mathtt{00}}$}
        {attxOSbench1NNLOSGQG024mplcty}{$\mathtt{2}$}
        {attxOSbench1NNLOSGQG024}{$\phantom{\minus}\mathtt{3.8886\phantom{00}} \cdot \mathtt{10}^{\minus\mathtt{02}}$}
        {attxOSbench1NNLOSGQG024delta}{$\phantom{\mathtt{0}}\phantom{\minus}\mathtt{0.031}\percent\phantom{\mathtt{0}}$}
        {attxOSbench1NNLOSGQG017mplcty}{$\mathtt{2}$}
        {attxOSbench1NNLOSGQG017}{$\phantom{\minus}\mathtt{1.76075\phantom{0}} \cdot \mathtt{10}^{\minus\mathtt{02}}$}
        {attxOSbench1NNLOSGQG017delta}{$\phantom{\mathtt{0}}\phantom{\minus}\mathtt{0.055}\percent\phantom{\mathtt{0}}$}
        {attxOSbench1NNLOSGQG023mplcty}{$\mathtt{2}$}
        {attxOSbench1NNLOSGQG023}{$\phantom{\minus}\mathtt{5.1058\phantom{00}} \cdot \mathtt{10}^{\minus\mathtt{03}}$}
        {attxOSbench1NNLOSGQG023delta}{$\phantom{\mathtt{0}}\phantom{\minus}\mathtt{0.15}\percent\phantom{\mathtt{00}}$}
        {attxOSbench1NNLOSGQG013mplcty}{$\mathtt{1}$}
        {attxOSbench1NNLOSGQG013}{$\phantom{\minus}\mathtt{8.8163\phantom{00}} \cdot \mathtt{10}^{\minus\mathtt{03}}$}
        {attxOSbench1NNLOSGQG013delta}{$\phantom{\mathtt{0}}\phantom{\minus}\mathtt{0.078}\percent\phantom{\mathtt{0}}$}
        {attxOSbench1NNLOSGQG003mplcty}{$\mathtt{1}$}
        {attxOSbench1NNLOSGQG003}{$\phantom{\minus}\mathtt{9.200\phantom{000}} \cdot \mathtt{10}^{\minus\mathtt{04}}$}
        {attxOSbench1NNLOSGQG003delta}{$\phantom{\mathtt{0}}\phantom{\minus}\mathtt{0.79}\percent\phantom{\mathtt{00}}$}
        {attxOSbench1NNLOSGQG011mplcty}{$\mathtt{2}$}
        {attxOSbench1NNLOSGQG011}{$\phantom{\minus}\mathtt{6.7284\phantom{00}} \cdot \mathtt{10}^{\minus\mathtt{03}}$}
        {attxOSbench1NNLOSGQG011delta}{$\phantom{\mathtt{0}}\phantom{\minus}\mathtt{0.10}\percent\phantom{\mathtt{00}}$}
        {attxOSbench1NNLOSGQG009mplcty}{$\mathtt{2}$}
        {attxOSbench1NNLOSGQG009}{$\phantom{\minus}\mathtt{6.7300\phantom{00}} \cdot \mathtt{10}^{\minus\mathtt{03}}$}
        {attxOSbench1NNLOSGQG009delta}{$\phantom{\mathtt{0}}\phantom{\minus}\mathtt{0.10}\percent\phantom{\mathtt{00}}$}
        {attxOSbench1NNLOSGQG021mplcty}{$\mathtt{2}$}
        {attxOSbench1NNLOSGQG021}{$\phantom{\minus}\mathtt{3.7418\phantom{00}} \cdot \mathtt{10}^{\minus\mathtt{03}}$}
        {attxOSbench1NNLOSGQG021delta}{$\phantom{\mathtt{0}}\phantom{\minus}\mathtt{0.14}\percent\phantom{\mathtt{00}}$}
        {attxOSbench1NNLOSGQG001mplcty}{$\mathtt{1}$}
        {attxOSbench1NNLOSGQG001}{$\phantom{\minus}\mathtt{3.5114\phantom{00}} \cdot \mathtt{10}^{\minus\mathtt{03}}$}
        {attxOSbench1NNLOSGQG001delta}{$\phantom{\mathtt{0}}\phantom{\minus}\mathtt{0.12}\percent\phantom{\mathtt{00}}$}
        {attxOSbench1NNLOSGQG025mplcty}{$\mathtt{2}$}
        {attxOSbench1NNLOSGQG025}{$\phantom{\minus}\mathtt{2.3361\phantom{00}} \cdot \mathtt{10}^{\minus\mathtt{03}}$}
        {attxOSbench1NNLOSGQG025delta}{$\phantom{\mathtt{0}}\phantom{\minus}\mathtt{0.12}\percent\phantom{\mathtt{00}}$}
        {attxOSbench1NNLOSGQG018mplcty}{$\mathtt{2}$}
        {attxOSbench1NNLOSGQG018}{$\phantom{\minus}\mathtt{2.0126\phantom{00}} \cdot \mathtt{10}^{\minus\mathtt{03}}$}
        {attxOSbench1NNLOSGQG018delta}{$\phantom{\mathtt{0}}\phantom{\minus}\mathtt{0.089}\percent\phantom{\mathtt{0}}$}
        {attxOSbench1NNLOSGQG014mplcty}{$\mathtt{1}$}
        {attxOSbench1NNLOSGQG014}{$\phantom{\minus}\mathtt{8.222\phantom{000}} \cdot \mathtt{10}^{\minus\mathtt{04}}$}
        {attxOSbench1NNLOSGQG014delta}{$\phantom{\mathtt{0}}\phantom{\minus}\mathtt{0.19}\percent\phantom{\mathtt{00}}$}
        {attxOSbench1NNLOSGQG005mplcty}{$\mathtt{2}$}
        {attxOSbench1NNLOSGQG005}{$\phantom{\minus}\mathtt{2.0845\phantom{00}} \cdot \mathtt{10}^{\minus\mathtt{03}}$}
        {attxOSbench1NNLOSGQG005delta}{$\phantom{\mathtt{0}}\phantom{\minus}\mathtt{0.083}\percent\phantom{\mathtt{0}}$}
        {attxOSbench1NNLOSGQG000mplcty}{$\mathtt{1}$}
        {attxOSbench1NNLOSGQG000}{$\minus\mathtt{7.242\phantom{000}} \cdot \mathtt{10}^{\minus\mathtt{04}}$}
        {attxOSbench1NNLOSGQG000delta}{$\phantom{\mathtt{0}}\phantom{\minus}\mathtt{0.14}\percent\phantom{\mathtt{00}}$}
        {attxOSbench1NNLOTotal}{$\phantom{\minus}\mathtt{1.04214\phantom{0}} \cdot \mathtt{10}^{\minus\mathtt{01}}$}
        {attxOSbench1NNLOTotaldelta}{$\phantom{\mathtt{0}}\phantom{\minus}\mathtt{0.024}\percent\phantom{\mathtt{0}}$}
        {attxOSbench1NNLOTarget}{$\phantom{\minus}\mathtt{1.0386\phantom{00}} \cdot \mathtt{10}^{\minus\mathtt{01}}$}
        {attxOSbench1NNLOTargetdelta}{$\phantom{\mathtt{0}}\phantom{\minus}\mathtt{0.34}\percent\phantom{\mathtt{00}}$}
        {attxOSbench2LOSGQG000mplcty}{$\mathtt{1}$}
        {attxOSbench2LOSGQG000}{$\phantom{\minus}\mathtt{7.312112\phantom{}} \cdot \mathtt{10}^{\minus\mathtt{01}}$}
        {attxOSbench2LOSGQG000delta}{$\phantom{\mathtt{0}}\phantom{\minus}\mathtt{0.0013}\percent$}
        {attxOSbench2LOTotal}{$\phantom{\minus}\mathtt{7.312112\phantom{}} \cdot \mathtt{10}^{\minus\mathtt{01}}$}
        {attxOSbench2LOTotaldelta}{$\phantom{\mathtt{0}}\phantom{\minus}\mathtt{0.0013}\percent$}
        {attxOSbench2NLOSGQG000mplcty}{$\mathtt{1}$}
        {attxOSbench2NLOSGQG000}{$\phantom{\minus}\mathtt{3.44269\phantom{0}} \cdot \mathtt{10}^{\minus\mathtt{01}}$}
        {attxOSbench2NLOSGQG000delta}{$\phantom{\mathtt{0}}\phantom{\minus}\mathtt{0.018}\percent\phantom{\mathtt{0}}$}
        {attxOSbench2NLOSGQG001mplcty}{$\mathtt{2}$}
        {attxOSbench2NLOSGQG001}{$\phantom{\minus}\mathtt{1.82308\phantom{0}} \cdot \mathtt{10}^{\minus\mathtt{01}}$}
        {attxOSbench2NLOSGQG001delta}{$\phantom{\mathtt{0}}\phantom{\minus}\mathtt{0.047}\percent\phantom{\mathtt{0}}$}
        {attxOSbench2NLOTotal}{$\phantom{\minus}\mathtt{5.2658\phantom{00}} \cdot \mathtt{10}^{\minus\mathtt{01}}$}
        {attxOSbench2NLOTotaldelta}{$\phantom{\mathtt{0}}\phantom{\minus}\mathtt{0.020}\percent\phantom{\mathtt{0}}$}
        {attxOSbench2NLOTarget}{$\phantom{\minus}\mathtt{5.2670\phantom{00}} \cdot \mathtt{10}^{\minus\mathtt{01}}$}
        {attxOSbench2NLOTargetdelta}{$\phantom{\mathtt{0}}\minus\mathtt{0.023}\percent\phantom{\mathtt{0}}$}
        {attxOSbench2NNLOheavyfermionloopSGQG016mplcty}{$\mathtt{1}$}
        {attxOSbench2NNLOheavyfermionloopSGQG016}{$\phantom{\minus}\mathtt{3.489\phantom{000}} \cdot \mathtt{10}^{\minus\mathtt{04}}$}
        {attxOSbench2NNLOheavyfermionloopSGQG016delta}{$\phantom{\mathtt{0}}\phantom{\minus}\mathtt{0.53}\percent\phantom{\mathtt{00}}$}
        {attxOSbench2NNLOheavyfermionloopSGQG027mplcty}{$\mathtt{2}$}
        {attxOSbench2NNLOheavyfermionloopSGQG027}{$\minus\mathtt{1.84\phantom{0000}} \cdot \mathtt{10}^{\minus\mathtt{06}}$}
        {attxOSbench2NNLOheavyfermionloopSGQG027delta}{$\phantom{\minus}\mathtt{160}\percent\phantom{\mathtt{.000}}$}
        {attxOSbench2NNLOheavyfermionloopTotal}{$\phantom{\minus}\mathtt{3.471\phantom{000}} \cdot \mathtt{10}^{\minus\mathtt{04}}$}
        {attxOSbench2NNLOheavyfermionloopTotaldelta}{$\phantom{\mathtt{0}}\phantom{\minus}\mathtt{0.0}\percent\phantom{\mathtt{000}}$}
        {attxOSbench2NNLOnf1SGQG015mplcty}{$\mathtt{1}$}
        {attxOSbench2NNLOnf1SGQG015}{$\minus\mathtt{3.6012\phantom{00}} \cdot \mathtt{10}^{\minus\mathtt{03}}$}
        {attxOSbench2NNLOnf1SGQG015delta}{$\phantom{\mathtt{0}}\phantom{\minus}\mathtt{0.094}\percent\phantom{\mathtt{0}}$}
        {attxOSbench2NNLOnf1SGQG026mplcty}{$\mathtt{2}$}
        {attxOSbench2NNLOnf1SGQG026}{$\minus\mathtt{7.6585\phantom{00}} \cdot \mathtt{10}^{\minus\mathtt{03}}$}
        {attxOSbench2NNLOnf1SGQG026delta}{$\phantom{\mathtt{0}}\phantom{\minus}\mathtt{0.068}\percent\phantom{\mathtt{0}}$}
        {attxOSbench2NNLOnf1Total}{$\minus\mathtt{1.12597\phantom{0}} \cdot \mathtt{10}^{\minus\mathtt{02}}$}
        {attxOSbench2NNLOnf1Totaldelta}{$\phantom{\mathtt{0}}\phantom{\minus}\mathtt{0.055}\percent\phantom{\mathtt{0}}$}
        {attxOSbench2NNLOnf1Target}{$\minus\mathtt{1.1275\phantom{00}} \cdot \mathtt{10}^{\minus\mathtt{02}}$}
        {attxOSbench2NNLOnf1Targetdelta}{$\phantom{\mathtt{0}}\minus\mathtt{0.14}\percent\phantom{\mathtt{00}}$}
        {attxOSbench2NNLOSGQG006mplcty}{$\mathtt{2}$}
        {attxOSbench2NNLOSGQG006}{$\phantom{\minus}\mathtt{2.9987\phantom{00}} \cdot \mathtt{10}^{\minus\mathtt{02}}$}
        {attxOSbench2NNLOSGQG006delta}{$\phantom{\mathtt{0}}\phantom{\minus}\mathtt{0.034}\percent\phantom{\mathtt{0}}$}
        {attxOSbench2NNLOSGQG024mplcty}{$\mathtt{2}$}
        {attxOSbench2NNLOSGQG024}{$\phantom{\minus}\mathtt{6.1123\phantom{00}} \cdot \mathtt{10}^{\minus\mathtt{02}}$}
        {attxOSbench2NNLOSGQG024delta}{$\phantom{\mathtt{0}}\phantom{\minus}\mathtt{0.028}\percent\phantom{\mathtt{0}}$}
        {attxOSbench2NNLOSGQG017mplcty}{$\mathtt{2}$}
        {attxOSbench2NNLOSGQG017}{$\phantom{\minus}\mathtt{2.0873\phantom{00}} \cdot \mathtt{10}^{\minus\mathtt{02}}$}
        {attxOSbench2NNLOSGQG017delta}{$\phantom{\mathtt{0}}\phantom{\minus}\mathtt{0.062}\percent\phantom{\mathtt{0}}$}
        {attxOSbench2NNLOSGQG023mplcty}{$\mathtt{2}$}
        {attxOSbench2NNLOSGQG023}{$\phantom{\minus}\mathtt{1.4989\phantom{00}} \cdot \mathtt{10}^{\minus\mathtt{02}}$}
        {attxOSbench2NNLOSGQG023delta}{$\phantom{\mathtt{0}}\phantom{\minus}\mathtt{0.078}\percent\phantom{\mathtt{0}}$}
        {attxOSbench2NNLOSGQG013mplcty}{$\mathtt{1}$}
        {attxOSbench2NNLOSGQG013}{$\phantom{\minus}\mathtt{2.7887\phantom{00}} \cdot \mathtt{10}^{\minus\mathtt{02}}$}
        {attxOSbench2NNLOSGQG013delta}{$\phantom{\mathtt{0}}\phantom{\minus}\mathtt{0.041}\percent\phantom{\mathtt{0}}$}
        {attxOSbench2NNLOSGQG003mplcty}{$\mathtt{1}$}
        {attxOSbench2NNLOSGQG003}{$\phantom{\minus}\mathtt{1.4006\phantom{00}} \cdot \mathtt{10}^{\minus\mathtt{02}}$}
        {attxOSbench2NNLOSGQG003delta}{$\phantom{\mathtt{0}}\phantom{\minus}\mathtt{0.11}\percent\phantom{\mathtt{00}}$}
        {attxOSbench2NNLOSGQG011mplcty}{$\mathtt{2}$}
        {attxOSbench2NNLOSGQG011}{$\phantom{\minus}\mathtt{3.4020\phantom{00}} \cdot \mathtt{10}^{\minus\mathtt{02}}$}
        {attxOSbench2NNLOSGQG011delta}{$\phantom{\mathtt{0}}\phantom{\minus}\mathtt{0.038}\percent\phantom{\mathtt{0}}$}
        {attxOSbench2NNLOSGQG009mplcty}{$\mathtt{2}$}
        {attxOSbench2NNLOSGQG009}{$\phantom{\minus}\mathtt{3.4045\phantom{00}} \cdot \mathtt{10}^{\minus\mathtt{02}}$}
        {attxOSbench2NNLOSGQG009delta}{$\phantom{\mathtt{0}}\phantom{\minus}\mathtt{0.036}\percent\phantom{\mathtt{0}}$}
        {attxOSbench2NNLOSGQG021mplcty}{$\mathtt{2}$}
        {attxOSbench2NNLOSGQG021}{$\phantom{\minus}\mathtt{1.56073\phantom{0}} \cdot \mathtt{10}^{\minus\mathtt{02}}$}
        {attxOSbench2NNLOSGQG021delta}{$\phantom{\mathtt{0}}\phantom{\minus}\mathtt{0.063}\percent\phantom{\mathtt{0}}$}
        {attxOSbench2NNLOSGQG001mplcty}{$\mathtt{1}$}
        {attxOSbench2NNLOSGQG001}{$\phantom{\minus}\mathtt{9.8039\phantom{00}} \cdot \mathtt{10}^{\minus\mathtt{03}}$}
        {attxOSbench2NNLOSGQG001delta}{$\phantom{\mathtt{0}}\phantom{\minus}\mathtt{0.095}\percent\phantom{\mathtt{0}}$}
        {attxOSbench2NNLOSGQG025mplcty}{$\mathtt{2}$}
        {attxOSbench2NNLOSGQG025}{$\phantom{\minus}\mathtt{3.4813\phantom{00}} \cdot \mathtt{10}^{\minus\mathtt{03}}$}
        {attxOSbench2NNLOSGQG025delta}{$\phantom{\mathtt{0}}\phantom{\minus}\mathtt{0.11}\percent\phantom{\mathtt{00}}$}
        {attxOSbench2NNLOSGQG018mplcty}{$\mathtt{2}$}
        {attxOSbench2NNLOSGQG018}{$\phantom{\minus}\mathtt{3.3269\phantom{00}} \cdot \mathtt{10}^{\minus\mathtt{03}}$}
        {attxOSbench2NNLOSGQG018delta}{$\phantom{\mathtt{0}}\phantom{\minus}\mathtt{0.076}\percent\phantom{\mathtt{0}}$}
        {attxOSbench2NNLOSGQG014mplcty}{$\mathtt{1}$}
        {attxOSbench2NNLOSGQG014}{$\phantom{\minus}\mathtt{1.9795\phantom{00}} \cdot \mathtt{10}^{\minus\mathtt{03}}$}
        {attxOSbench2NNLOSGQG014delta}{$\phantom{\mathtt{0}}\phantom{\minus}\mathtt{0.13}\percent\phantom{\mathtt{00}}$}
        {attxOSbench2NNLOSGQG005mplcty}{$\mathtt{2}$}
        {attxOSbench2NNLOSGQG005}{$\phantom{\minus}\mathtt{2.9860\phantom{00}} \cdot \mathtt{10}^{\minus\mathtt{03}}$}
        {attxOSbench2NNLOSGQG005delta}{$\phantom{\mathtt{0}}\phantom{\minus}\mathtt{0.090}\percent\phantom{\mathtt{0}}$}
        {attxOSbench2NNLOSGQG000mplcty}{$\mathtt{1}$}
        {attxOSbench2NNLOSGQG000}{$\minus\mathtt{1.6027\phantom{00}} \cdot \mathtt{10}^{\minus\mathtt{03}}$}
        {attxOSbench2NNLOSGQG000delta}{$\phantom{\mathtt{0}}\phantom{\minus}\mathtt{0.11}\percent\phantom{\mathtt{00}}$}
        {attxOSbench2NNLOTotal}{$\phantom{\minus}\mathtt{2.72512\phantom{0}} \cdot \mathtt{10}^{\minus\mathtt{01}}$}
        {attxOSbench2NNLOTotaldelta}{$\phantom{\mathtt{0}}\phantom{\minus}\mathtt{0.015}\percent\phantom{\mathtt{0}}$}
        {attxOSbench2NNLOTarget}{$\phantom{\minus}\mathtt{2.7329\phantom{00}} \cdot \mathtt{10}^{\minus\mathtt{01}}$}
        {attxOSbench2NNLOTargetdelta}{$\phantom{\mathtt{0}}\minus\mathtt{0.28}\percent\phantom{\mathtt{00}}$}
        {attxOSbenchHELOSGQG000mplcty}{$\mathtt{1}$}
        {attxOSbenchHELOSGQG000}{$\phantom{\minus}\mathtt{1.509262\phantom{}} \cdot \mathtt{10}^{+\mathtt{01}}$}
        {attxOSbenchHELOSGQG000delta}{$\phantom{\mathtt{0}}\phantom{\minus}\mathtt{0.000064}\percent$}
        {attxOSbenchHELOTotal}{$\phantom{\minus}\mathtt{1.509262\phantom{}} \cdot \mathtt{10}^{+\mathtt{01}}$}
        {attxOSbenchHELOTotaldelta}{$\phantom{\mathtt{0}}\phantom{\minus}\mathtt{0.000064}\percent$}
        {attxOSbenchHENLOSGQG000mplcty}{$\mathtt{1}$}
        {attxOSbenchHENLOSGQG000}{$\minus\mathtt{6.3725\phantom{00}} \cdot \mathtt{10}^{\minus\mathtt{01}}$}
        {attxOSbenchHENLOSGQG000delta}{$\phantom{\mathtt{0}}\phantom{\minus}\mathtt{0.071}\percent\phantom{\mathtt{0}}$}
        {attxOSbenchHENLOSGQG001mplcty}{$\mathtt{2}$}
        {attxOSbenchHENLOSGQG001}{$\phantom{\minus}\mathtt{1.22702\phantom{0}} \cdot \mathtt{10}^{+\mathtt{00}}$}
        {attxOSbenchHENLOSGQG001delta}{$\phantom{\mathtt{0}}\phantom{\minus}\mathtt{0.039}\percent\phantom{\mathtt{0}}$}
        {attxOSbenchHENLOTotal}{$\phantom{\minus}\mathtt{5.8977\phantom{00}} \cdot \mathtt{10}^{\minus\mathtt{01}}$}
        {attxOSbenchHENLOTotaldelta}{$\phantom{\mathtt{0}}\phantom{\minus}\mathtt{0.11}\percent\phantom{\mathtt{00}}$}
        {attxOSbenchHENLOTarget}{$\phantom{\minus}\mathtt{5.9047\phantom{00}} \cdot \mathtt{10}^{\minus\mathtt{01}}$}
        {attxOSbenchHENLOTargetdelta}{$\phantom{\mathtt{0}}\minus\mathtt{0.12}\percent\phantom{\mathtt{00}}$}
        {attxOSbenchHENNLOheavyfermionloopSGQG016mplcty}{$\mathtt{1}$}
        {attxOSbenchHENNLOheavyfermionloopSGQG016}{$\phantom{\minus}\mathtt{2.8484\phantom{00}} \cdot \mathtt{10}^{\minus\mathtt{02}}$}
        {attxOSbenchHENNLOheavyfermionloopSGQG016delta}{$\phantom{\mathtt{0}}\phantom{\minus}\mathtt{0.061}\percent\phantom{\mathtt{0}}$}
        {attxOSbenchHENNLOheavyfermionloopSGQG027mplcty}{$\mathtt{2}$}
        {attxOSbenchHENNLOheavyfermionloopSGQG027}{$\minus\mathtt{2.3718\phantom{00}} \cdot \mathtt{10}^{\minus\mathtt{02}}$}
        {attxOSbenchHENNLOheavyfermionloopSGQG027delta}{$\phantom{\mathtt{0}}\phantom{\minus}\mathtt{0.11}\percent\phantom{\mathtt{00}}$}
        {attxOSbenchHENNLOheavyfermionloopTotal}{$\phantom{\minus}\mathtt{4.766\phantom{000}} \cdot \mathtt{10}^{\minus\mathtt{03}}$}
        {attxOSbenchHENNLOheavyfermionloopTotaldelta}{$\phantom{\mathtt{0}}\phantom{\minus}\mathtt{0.66}\percent\phantom{\mathtt{00}}$}
        {attxOSbenchHENNLOnf1SGQG015mplcty}{$\mathtt{1}$}
        {attxOSbenchHENNLOnf1SGQG015}{$\phantom{\minus}\mathtt{2.6658\phantom{00}} \cdot \mathtt{10}^{\minus\mathtt{02}}$}
        {attxOSbenchHENNLOnf1SGQG015delta}{$\phantom{\mathtt{0}}\phantom{\minus}\mathtt{0.059}\percent\phantom{\mathtt{0}}$}
        {attxOSbenchHENNLOnf1SGQG026mplcty}{$\mathtt{2}$}
        {attxOSbenchHENNLOnf1SGQG026}{$\minus\mathtt{8.388\phantom{000}} \cdot \mathtt{10}^{\minus\mathtt{03}}$}
        {attxOSbenchHENNLOnf1SGQG026delta}{$\phantom{\mathtt{0}}\phantom{\minus}\mathtt{0.30}\percent\phantom{\mathtt{00}}$}
        {attxOSbenchHENNLOnf1Total}{$\phantom{\minus}\mathtt{1.8270\phantom{00}} \cdot \mathtt{10}^{\minus\mathtt{02}}$}
        {attxOSbenchHENNLOnf1Totaldelta}{$\phantom{\mathtt{0}}\phantom{\minus}\mathtt{0.16}\percent\phantom{\mathtt{00}}$}
        {attxOSbenchHENNLOnf1Target}{$\phantom{\minus}\mathtt{1.8296\phantom{00}} \cdot \mathtt{10}^{\minus\mathtt{02}}$}
        {attxOSbenchHENNLOnf1Targetdelta}{$\phantom{\mathtt{0}}\minus\mathtt{0.15}\percent\phantom{\mathtt{00}}$}
        {attxOSbenchHENNLOSGQG006mplcty}{$\mathtt{2}$}
        {attxOSbenchHENNLOSGQG006}{$\minus\mathtt{3.52337\phantom{0}} \cdot \mathtt{10}^{\minus\mathtt{01}}$}
        {attxOSbenchHENNLOSGQG006delta}{$\phantom{\mathtt{0}}\phantom{\minus}\mathtt{0.027}\percent\phantom{\mathtt{0}}$}
        {attxOSbenchHENNLOSGQG024mplcty}{$\mathtt{2}$}
        {attxOSbenchHENNLOSGQG024}{$\phantom{\minus}\mathtt{6.3163\phantom{00}} \cdot \mathtt{10}^{\minus\mathtt{02}}$}
        {attxOSbenchHENNLOSGQG024delta}{$\phantom{\mathtt{0}}\phantom{\minus}\mathtt{0.11}\percent\phantom{\mathtt{00}}$}
        {attxOSbenchHENNLOSGQG017mplcty}{$\mathtt{2}$}
        {attxOSbenchHENNLOSGQG017}{$\phantom{\minus}\mathtt{5.6646\phantom{00}} \cdot \mathtt{10}^{\minus\mathtt{02}}$}
        {attxOSbenchHENNLOSGQG017delta}{$\phantom{\mathtt{0}}\phantom{\minus}\mathtt{0.14}\percent\phantom{\mathtt{00}}$}
        {attxOSbenchHENNLOSGQG023mplcty}{$\mathtt{2}$}
        {attxOSbenchHENNLOSGQG023}{$\phantom{\minus}\mathtt{1.1244\phantom{00}} \cdot \mathtt{10}^{\minus\mathtt{02}}$}
        {attxOSbenchHENNLOSGQG023delta}{$\phantom{\mathtt{0}}\phantom{\minus}\mathtt{0.51}\percent\phantom{\mathtt{00}}$}
        {attxOSbenchHENNLOSGQG013mplcty}{$\mathtt{1}$}
        {attxOSbenchHENNLOSGQG013}{$\minus\mathtt{1.83770\phantom{0}} \cdot \mathtt{10}^{\minus\mathtt{01}}$}
        {attxOSbenchHENNLOSGQG013delta}{$\phantom{\mathtt{0}}\phantom{\minus}\mathtt{0.023}\percent\phantom{\mathtt{0}}$}
        {attxOSbenchHENNLOSGQG003mplcty}{$\mathtt{1}$}
        {attxOSbenchHENNLOSGQG003}{$\minus\mathtt{7.9531\phantom{00}} \cdot \mathtt{10}^{\minus\mathtt{02}}$}
        {attxOSbenchHENNLOSGQG003delta}{$\phantom{\mathtt{0}}\phantom{\minus}\mathtt{0.054}\percent\phantom{\mathtt{0}}$}
        {attxOSbenchHENNLOSGQG011mplcty}{$\mathtt{2}$}
        {attxOSbenchHENNLOSGQG011}{$\phantom{\minus}\mathtt{5.2105\phantom{00}} \cdot \mathtt{10}^{\minus\mathtt{02}}$}
        {attxOSbenchHENNLOSGQG011delta}{$\phantom{\mathtt{0}}\phantom{\minus}\mathtt{0.094}\percent\phantom{\mathtt{0}}$}
        {attxOSbenchHENNLOSGQG009mplcty}{$\mathtt{2}$}
        {attxOSbenchHENNLOSGQG009}{$\phantom{\minus}\mathtt{5.2171\phantom{00}} \cdot \mathtt{10}^{\minus\mathtt{02}}$}
        {attxOSbenchHENNLOSGQG009delta}{$\phantom{\mathtt{0}}\phantom{\minus}\mathtt{0.094}\percent\phantom{\mathtt{0}}$}
        {attxOSbenchHENNLOSGQG021mplcty}{$\mathtt{2}$}
        {attxOSbenchHENNLOSGQG021}{$\phantom{\minus}\mathtt{3.4996\phantom{00}} \cdot \mathtt{10}^{\minus\mathtt{02}}$}
        {attxOSbenchHENNLOSGQG021delta}{$\phantom{\mathtt{0}}\phantom{\minus}\mathtt{0.11}\percent\phantom{\mathtt{00}}$}
        {attxOSbenchHENNLOSGQG001mplcty}{$\mathtt{1}$}
        {attxOSbenchHENNLOSGQG001}{$\phantom{\minus}\mathtt{2.8263\phantom{00}} \cdot \mathtt{10}^{\minus\mathtt{02}}$}
        {attxOSbenchHENNLOSGQG001delta}{$\phantom{\mathtt{0}}\phantom{\minus}\mathtt{0.10}\percent\phantom{\mathtt{00}}$}
        {attxOSbenchHENNLOSGQG025mplcty}{$\mathtt{2}$}
        {attxOSbenchHENNLOSGQG025}{$\phantom{\minus}\mathtt{2.520\phantom{000}} \cdot \mathtt{10}^{\minus\mathtt{03}}$}
        {attxOSbenchHENNLOSGQG025delta}{$\phantom{\mathtt{0}}\phantom{\minus}\mathtt{0.73}\percent\phantom{\mathtt{00}}$}
        {attxOSbenchHENNLOSGQG018mplcty}{$\mathtt{2}$}
        {attxOSbenchHENNLOSGQG018}{$\phantom{\minus}\mathtt{9.106\phantom{000}} \cdot \mathtt{10}^{\minus\mathtt{03}}$}
        {attxOSbenchHENNLOSGQG018delta}{$\phantom{\mathtt{0}}\phantom{\minus}\mathtt{0.19}\percent\phantom{\mathtt{00}}$}
        {attxOSbenchHENNLOSGQG014mplcty}{$\mathtt{1}$}
        {attxOSbenchHENNLOSGQG014}{$\minus\mathtt{7.994\phantom{000}} \cdot \mathtt{10}^{\minus\mathtt{03}}$}
        {attxOSbenchHENNLOSGQG014delta}{$\phantom{\mathtt{0}}\phantom{\minus}\mathtt{0.13}\percent\phantom{\mathtt{00}}$}
        {attxOSbenchHENNLOSGQG005mplcty}{$\mathtt{2}$}
        {attxOSbenchHENNLOSGQG005}{$\phantom{\minus}\mathtt{2.5486\phantom{00}} \cdot \mathtt{10}^{\minus\mathtt{02}}$}
        {attxOSbenchHENNLOSGQG005delta}{$\phantom{\mathtt{0}}\phantom{\minus}\mathtt{0.060}\percent\phantom{\mathtt{0}}$}
        {attxOSbenchHENNLOSGQG000mplcty}{$\mathtt{1}$}
        {attxOSbenchHENNLOSGQG000}{$\minus\mathtt{1.96633\phantom{0}} \cdot \mathtt{10}^{\minus\mathtt{02}}$}
        {attxOSbenchHENNLOSGQG000delta}{$\phantom{\mathtt{0}}\phantom{\minus}\mathtt{0.044}\percent\phantom{\mathtt{0}}$}
        {attxOSbenchHENNLOTotal}{$\minus\mathtt{3.0760\phantom{00}} \cdot \mathtt{10}^{\minus\mathtt{01}}$}
        {attxOSbenchHENNLOTotaldelta}{$\phantom{\mathtt{0}}\phantom{\minus}\mathtt{0.061}\percent\phantom{\mathtt{0}}$}
        {attxOSbenchHENNLOTarget}{$\minus\mathtt{3.0818\phantom{00}} \cdot \mathtt{10}^{\minus\mathtt{01}}$}
        {attxOSbenchHENNLOTargetdelta}{$\phantom{\mathtt{0}}\minus\mathtt{0.19}\percent\phantom{\mathtt{00}}$}
        {epemjjjLOSGQG000mplcty}{$\mathtt{1}$}
        {epemjjjLOSGQG000}{$\phantom{\minus}\mathtt{1.01001\phantom{0}} \cdot \mathtt{10}^{\minus\mathtt{02}}$}
        {epemjjjLOSGQG000delta}{$\phantom{\mathtt{0}}\phantom{\minus}\mathtt{0.048}\percent\phantom{\mathtt{0}}$}
        {epemjjjLOSGQG002mplcty}{$\mathtt{2}$}
        {epemjjjLOSGQG002}{$\phantom{\minus}\mathtt{1.84151\phantom{0}} \cdot \mathtt{10}^{\minus\mathtt{03}}$}
        {epemjjjLOSGQG002delta}{$\phantom{\mathtt{0}}\phantom{\minus}\mathtt{0.0077}\percent$}
        {epemjjjLOTotal}{$\phantom{\minus}\mathtt{1.19416\phantom{0}} \cdot \mathtt{10}^{\minus\mathtt{02}}$}
        {epemjjjLOTotaldelta}{$\phantom{\mathtt{0}}\phantom{\minus}\mathtt{0.040}\percent\phantom{\mathtt{0}}$}
        {epemjjjLOTarget}{$\phantom{\minus}\mathtt{1.1955\phantom{00}} \cdot \mathtt{10}^{\minus\mathtt{02}}$}
        {epemjjjLOTargetdelta}{$\phantom{\mathtt{0}}\minus\mathtt{0.11}\percent\phantom{\mathtt{00}}$}
        {epemjjjNLOnf1ReSGQG015mplcty}{$\mathtt{1}$}
        {epemjjjNLOnf1ReSGQG015}{$\phantom{\minus}\mathtt{2.77\phantom{0000}} \cdot \mathtt{10}^{\minus\mathtt{05}}$}
        {epemjjjNLOnf1ReSGQG015delta}{$\phantom{\minus}\mathtt{14}\percent\phantom{\mathtt{.0000}}$}
        {epemjjjNLOnf1ReSGQG036mplcty}{$\mathtt{2}$}
        {epemjjjNLOnf1ReSGQG036}{$\minus\mathtt{1.912\phantom{000}} \cdot \mathtt{10}^{\minus\mathtt{04}}$}
        {epemjjjNLOnf1ReSGQG036delta}{$\phantom{\mathtt{0}}\phantom{\minus}\mathtt{3.3}\percent\phantom{\mathtt{000}}$}
        {epemjjjNLOnf1ReTotal}{$\minus\mathtt{1.635\phantom{000}} \cdot \mathtt{10}^{\minus\mathtt{04}}$}
        {epemjjjNLOnf1ReTotaldelta}{$\phantom{\mathtt{0}}\phantom{\minus}\mathtt{5.5}\percent\phantom{\mathtt{000}}$}
        {epemjjjNLOnf1ReTarget}{$\minus\mathtt{1.51\phantom{0000}} \cdot \mathtt{10}^{\minus\mathtt{04}}$}
        {epemjjjNLOnf1ReTargetdelta}{$\phantom{\mathtt{0}}\phantom{\minus}\mathtt{3.3}\percent\phantom{\mathtt{000}}$}
        {epemjjjNLOnf1ImSGQG015mplcty}{$\mathtt{1}$}
        {epemjjjNLOnf1ImSGQG015}{$\phantom{\minus}\mathtt{3\phantom{}} \cdot \mathtt{10}^{\minus\mathtt{26}}$}
        {epemjjjNLOnf1ImSGQG036mplcty}{$\mathtt{2}$}
        {epemjjjNLOnf1ImSGQG036}{$\phantom{\minus}\mathtt{6\phantom{}} \cdot \mathtt{10}^{\minus\mathtt{18}}$}
        {epemjjjNLOnf1ImTotal}{$\phantom{\minus}\mathtt{6\phantom{}} \cdot \mathtt{10}^{\minus\mathtt{18}}$}
        {epemjjjNLOnf1ImTarget}{$\phantom{\minus}\mathtt{0\phantom{}} \cdot \mathtt{10}^{+\mathtt{00}}$}
        {epemjjjNLOReSGQG006mplcty}{$\mathtt{2}$}
        {epemjjjNLOReSGQG006}{$\phantom{\minus}\mathtt{1.272\phantom{000}} \cdot \mathtt{10}^{\minus\mathtt{03}}$}
        {epemjjjNLOReSGQG006delta}{$\phantom{\mathtt{0}}\phantom{\minus}\mathtt{6.6}\percent\phantom{\mathtt{000}}$}
        {epemjjjNLOReSGQG034mplcty}{$\mathtt{2}$}
        {epemjjjNLOReSGQG034}{$\phantom{\minus}\mathtt{1.579\phantom{000}} \cdot \mathtt{10}^{\minus\mathtt{03}}$}
        {epemjjjNLOReSGQG034delta}{$\phantom{\mathtt{0}}\phantom{\minus}\mathtt{2.2}\percent\phantom{\mathtt{000}}$}
        {epemjjjNLOReSGQG013mplcty}{$\mathtt{1}$}
        {epemjjjNLOReSGQG013}{$\minus\mathtt{2.79\phantom{0000}} \cdot \mathtt{10}^{\minus\mathtt{04}}$}
        {epemjjjNLOReSGQG013delta}{$\phantom{\mathtt{0}}\phantom{\minus}\mathtt{5.5}\percent\phantom{\mathtt{000}}$}
        {epemjjjNLOReSGQG035mplcty}{$\mathtt{2}$}
        {epemjjjNLOReSGQG035}{$\phantom{\minus}\mathtt{5.53\phantom{0000}} \cdot \mathtt{10}^{\minus\mathtt{05}}$}
        {epemjjjNLOReSGQG035delta}{$\phantom{\mathtt{0}}\phantom{\minus}\mathtt{9.9}\percent\phantom{\mathtt{000}}$}
        {epemjjjNLOReSGQG000mplcty}{$\mathtt{1}$}
        {epemjjjNLOReSGQG000}{$\minus\mathtt{2.778\phantom{000}} \cdot \mathtt{10}^{\minus\mathtt{04}}$}
        {epemjjjNLOReSGQG000delta}{$\phantom{\mathtt{0}}\phantom{\minus}\mathtt{0.86}\percent\phantom{\mathtt{00}}$}
        {epemjjjNLOReSGQG014mplcty}{$\mathtt{1}$}
        {epemjjjNLOReSGQG014}{$\phantom{\minus}\mathtt{3.88\phantom{0000}} \cdot \mathtt{10}^{\minus\mathtt{05}}$}
        {epemjjjNLOReSGQG014delta}{$\phantom{\mathtt{0}}\phantom{\minus}\mathtt{4.4}\percent\phantom{\mathtt{000}}$}
        {epemjjjNLOReSGQG003mplcty}{$\mathtt{1}$}
        {epemjjjNLOReSGQG003}{$\minus\mathtt{3.8878\phantom{00}} \cdot \mathtt{10}^{\minus\mathtt{03}}$}
        {epemjjjNLOReSGQG003delta}{$\phantom{\mathtt{0}}\phantom{\minus}\mathtt{0.037}\percent\phantom{\mathtt{0}}$}
        {epemjjjNLOReSGQG011mplcty}{$\mathtt{2}$}
        {epemjjjNLOReSGQG011}{$\minus\mathtt{1.073\phantom{000}} \cdot \mathtt{10}^{\minus\mathtt{04}}$}
        {epemjjjNLOReSGQG011delta}{$\phantom{\mathtt{0}}\phantom{\minus}\mathtt{3.3}\percent\phantom{\mathtt{000}}$}
        {epemjjjNLOReSGQG009mplcty}{$\mathtt{2}$}
        {epemjjjNLOReSGQG009}{$\minus\mathtt{1.080\phantom{000}} \cdot \mathtt{10}^{\minus\mathtt{04}}$}
        {epemjjjNLOReSGQG009delta}{$\phantom{\mathtt{0}}\phantom{\minus}\mathtt{2.2}\percent\phantom{\mathtt{000}}$}
        {epemjjjNLOReSGQG027mplcty}{$\mathtt{2}$}
        {epemjjjNLOReSGQG027}{$\phantom{\minus}\mathtt{1.04\phantom{0000}} \cdot \mathtt{10}^{\minus\mathtt{05}}$}
        {epemjjjNLOReSGQG027delta}{$\phantom{\minus}\mathtt{11}\percent\phantom{\mathtt{.0000}}$}
        {epemjjjNLOReSGQG031mplcty}{$\mathtt{2}$}
        {epemjjjNLOReSGQG031}{$\phantom{\minus}\mathtt{5.729\phantom{000}} \cdot \mathtt{10}^{\minus\mathtt{05}}$}
        {epemjjjNLOReSGQG031delta}{$\phantom{\mathtt{0}}\phantom{\minus}\mathtt{3.3}\percent\phantom{\mathtt{000}}$}
        {epemjjjNLOReSGQG033mplcty}{$\mathtt{2}$}
        {epemjjjNLOReSGQG033}{$\phantom{\minus}\mathtt{9.19\phantom{0000}} \cdot \mathtt{10}^{\minus\mathtt{06}}$}
        {epemjjjNLOReSGQG033delta}{$\phantom{\mathtt{0}}\phantom{\minus}\mathtt{0.0}\percent\phantom{\mathtt{000}}$}
        {epemjjjNLOReSGQG005mplcty}{$\mathtt{2}$}
        {epemjjjNLOReSGQG005}{$\phantom{\minus}\mathtt{3.872\phantom{000}} \cdot \mathtt{10}^{\minus\mathtt{05}}$}
        {epemjjjNLOReSGQG005delta}{$\phantom{\mathtt{0}}\phantom{\minus}\mathtt{0.68}\percent\phantom{\mathtt{00}}$}
        {epemjjjNLOReSGQG028mplcty}{$\mathtt{2}$}
        {epemjjjNLOReSGQG028}{$\phantom{\minus}\mathtt{8.40\phantom{0000}} \cdot \mathtt{10}^{\minus\mathtt{06}}$}
        {epemjjjNLOReSGQG028delta}{$\phantom{\mathtt{0}}\phantom{\minus}\mathtt{9.9}\percent\phantom{\mathtt{000}}$}
        {epemjjjNLOReSGQG001mplcty}{$\mathtt{1}$}
        {epemjjjNLOReSGQG001}{$\phantom{\minus}\mathtt{2.3495\phantom{00}} \cdot \mathtt{10}^{\minus\mathtt{05}}$}
        {epemjjjNLOReSGQG001delta}{$\phantom{\mathtt{0}}\phantom{\minus}\mathtt{0.22}\percent\phantom{\mathtt{00}}$}
        {epemjjjNLOReTotal}{$\minus\mathtt{1.567\phantom{000}} \cdot \mathtt{10}^{\minus\mathtt{03}}$}
        {epemjjjNLOReTotaldelta}{$\phantom{\mathtt{0}}\phantom{\minus}\mathtt{3.3}\percent\phantom{\mathtt{000}}$}
        {epemjjjNLOReTarget}{$\minus\mathtt{1.515\phantom{000}} \cdot \mathtt{10}^{\minus\mathtt{03}}$}
        {epemjjjNLOReTargetdelta}{$\phantom{\mathtt{0}}\phantom{\minus}\mathtt{4.4}\percent\phantom{\mathtt{000}}$}
        {epemjjjNLOImSGQG006mplcty}{$\mathtt{2}$}
        {epemjjjNLOImSGQG006}{$\phantom{\minus}\mathtt{2\phantom{}} \cdot \mathtt{10}^{\minus\mathtt{05}}$}
        {epemjjjNLOImSGQG034mplcty}{$\mathtt{2}$}
        {epemjjjNLOImSGQG034}{$\phantom{\minus}\mathtt{7\phantom{}} \cdot \mathtt{10}^{\minus\mathtt{22}}$}
        {epemjjjNLOImSGQG013mplcty}{$\mathtt{1}$}
        {epemjjjNLOImSGQG013}{$\minus\mathtt{1\phantom{}} \cdot \mathtt{10}^{\minus\mathtt{12}}$}
        {epemjjjNLOImSGQG035mplcty}{$\mathtt{2}$}
        {epemjjjNLOImSGQG035}{$\phantom{\minus}\mathtt{7\phantom{}} \cdot \mathtt{10}^{\minus\mathtt{14}}$}
        {epemjjjNLOImSGQG000mplcty}{$\mathtt{1}$}
        {epemjjjNLOImSGQG000}{$\minus\mathtt{3\phantom{}} \cdot \mathtt{10}^{\minus\mathtt{06}}$}
        {epemjjjNLOImSGQG014mplcty}{$\mathtt{1}$}
        {epemjjjNLOImSGQG014}{$\minus\mathtt{1\phantom{}} \cdot \mathtt{10}^{\minus\mathtt{38}}$}
        {epemjjjNLOImSGQG003mplcty}{$\mathtt{1}$}
        {epemjjjNLOImSGQG003}{$\minus\mathtt{2\phantom{}} \cdot \mathtt{10}^{\minus\mathtt{07}}$}
        {epemjjjNLOImSGQG011mplcty}{$\mathtt{2}$}
        {epemjjjNLOImSGQG011}{$\minus\mathtt{7.2755\phantom{}} \cdot \mathtt{10}^{\minus\mathtt{04}}$}
        {epemjjjNLOImSGQG011delta}{$\phantom{\mathtt{0}}\phantom{\minus}\mathtt{0.18}\percent\phantom{\mathtt{00}}$}
        {epemjjjNLOImSGQG009mplcty}{$\mathtt{2}$}
        {epemjjjNLOImSGQG009}{$\phantom{\minus}\mathtt{7.2625\phantom{}} \cdot \mathtt{10}^{\minus\mathtt{04}}$}
        {epemjjjNLOImSGQG009delta}{$\phantom{\mathtt{0}}\phantom{\minus}\mathtt{0.18}\percent\phantom{\mathtt{00}}$}
        {epemjjjNLOImSGQG027mplcty}{$\mathtt{2}$}
        {epemjjjNLOImSGQG027}{$\phantom{\minus}\mathtt{2\phantom{}} \cdot \mathtt{10}^{\minus\mathtt{06}}$}
        {epemjjjNLOImSGQG031mplcty}{$\mathtt{2}$}
        {epemjjjNLOImSGQG031}{$\minus\mathtt{6\phantom{}} \cdot \mathtt{10}^{\minus\mathtt{07}}$}
        {epemjjjNLOImSGQG033mplcty}{$\mathtt{2}$}
        {epemjjjNLOImSGQG033}{$\phantom{\minus}\mathtt{9\phantom{}} \cdot \mathtt{10}^{\minus\mathtt{06}}$}
        {epemjjjNLOImSGQG005mplcty}{$\mathtt{2}$}
        {epemjjjNLOImSGQG005}{$\phantom{\minus}\mathtt{2\phantom{}} \cdot \mathtt{10}^{\minus\mathtt{07}}$}
        {epemjjjNLOImSGQG028mplcty}{$\mathtt{2}$}
        {epemjjjNLOImSGQG028}{$\phantom{\minus}\mathtt{4\phantom{}} \cdot \mathtt{10}^{\minus\mathtt{08}}$}
        {epemjjjNLOImSGQG001mplcty}{$\mathtt{1}$}
        {epemjjjNLOImSGQG001}{$\phantom{\minus}\mathtt{5\phantom{}} \cdot \mathtt{10}^{\minus\mathtt{16}}$}
        {epemjjjNLOImTotal}{$\phantom{\minus}\mathtt{2\phantom{}} \cdot \mathtt{10}^{\minus\mathtt{05}}$}
        {epemjjjNLOImTarget}{$\phantom{\minus}\mathtt{0\phantom{}} \cdot \mathtt{10}^{+\mathtt{00}}$}
        {epemttxhLOinclSGQG000mplcty}{$\mathtt{1}$}
        {epemttxhLOinclSGQG000}{$\phantom{\minus}\mathtt{6.17234\phantom{0}} \cdot \mathtt{10}^{\minus\mathtt{04}}$}
        {epemttxhLOinclSGQG000delta}{$\phantom{\mathtt{0}}\phantom{\minus}\mathtt{0.0097}\percent$}
        {epemttxhLOinclSGQG002mplcty}{$\mathtt{2}$}
        {epemttxhLOinclSGQG002}{$\phantom{\minus}\mathtt{9.4622\phantom{00}} \cdot \mathtt{10}^{\minus\mathtt{04}}$}
        {epemttxhLOinclSGQG002delta}{$\phantom{\mathtt{0}}\phantom{\minus}\mathtt{0.044}\percent\phantom{\mathtt{0}}$}
        {epemttxhLOinclTotal}{$\phantom{\minus}\mathtt{1.56345\phantom{0}} \cdot \mathtt{10}^{\minus\mathtt{03}}$}
        {epemttxhLOinclTotaldelta}{$\phantom{\mathtt{0}}\phantom{\minus}\mathtt{0.027}\percent\phantom{\mathtt{0}}$}
        {epemttxhLOinclTarget}{$\phantom{\minus}\mathtt{1.56331\phantom{0}} \cdot \mathtt{10}^{\minus\mathtt{03}}$}
        {epemttxhLOinclTargetdelta}{$\phantom{\mathtt{0}}\phantom{\minus}\mathtt{0.0092}\percent$}
        {epemttxhNLOinclF000001mplcty}{$\mathtt{2}$}
        {epemttxhNLOinclF000001}{$\minus\mathtt{2.504\phantom{000}} \cdot \mathtt{10}^{\minus\mathtt{06}}$}
        {epemttxhNLOinclF000001delta}{$\phantom{\mathtt{0}}\phantom{\minus}\mathtt{5.5}\percent\phantom{\mathtt{000}}$}
        {epemttxhNLOinclF000002mplcty}{$\mathtt{2}$}
        {epemttxhNLOinclF000002}{$\phantom{\minus}\mathtt{3.1908\phantom{00}} \cdot \mathtt{10}^{\minus\mathtt{05}}$}
        {epemttxhNLOinclF000002delta}{$\phantom{\mathtt{0}}\phantom{\minus}\mathtt{0.11}\percent\phantom{\mathtt{00}}$}
        {epemttxhNLOinclF000003mplcty}{$\mathtt{2}$}
        {epemttxhNLOinclF000003}{$\phantom{\minus}\mathtt{1.3472\phantom{00}} \cdot \mathtt{10}^{\minus\mathtt{04}}$}
        {epemttxhNLOinclF000003delta}{$\phantom{\mathtt{0}}\phantom{\minus}\mathtt{0.72}\percent\phantom{\mathtt{00}}$}
        {epemttxhNLOinclF000004mplcty}{$\mathtt{1}$}
        {epemttxhNLOinclF000004}{$\minus\mathtt{1.47653\phantom{0}} \cdot \mathtt{10}^{\minus\mathtt{04}}$}
        {epemttxhNLOinclF000004delta}{$\phantom{\mathtt{0}}\phantom{\minus}\mathtt{0.061}\percent\phantom{\mathtt{0}}$}
        {epemttxhNLOinclF000005mplcty}{$\mathtt{2}$}
        {epemttxhNLOinclF000005}{$\minus\mathtt{3.2336\phantom{00}} \cdot \mathtt{10}^{\minus\mathtt{06}}$}
        {epemttxhNLOinclF000005delta}{$\phantom{\mathtt{0}}\phantom{\minus}\mathtt{0.16}\percent\phantom{\mathtt{00}}$}
        {epemttxhNLOinclF000006mplcty}{$\mathtt{2}$}
        {epemttxhNLOinclF000006}{$\phantom{\minus}\mathtt{5.5828\phantom{00}} \cdot \mathtt{10}^{\minus\mathtt{05}}$}
        {epemttxhNLOinclF000006delta}{$\phantom{\mathtt{0}}\phantom{\minus}\mathtt{0.049}\percent\phantom{\mathtt{0}}$}
        {epemttxhNLOinclF000007mplcty}{$\mathtt{2}$}
        {epemttxhNLOinclF000007}{$\minus\mathtt{1.18480\phantom{0}} \cdot \mathtt{10}^{\minus\mathtt{04}}$}
        {epemttxhNLOinclF000007delta}{$\phantom{\mathtt{0}}\phantom{\minus}\mathtt{0.042}\percent\phantom{\mathtt{0}}$}
        {epemttxhNLOinclF000008mplcty}{$\mathtt{2}$}
        {epemttxhNLOinclF000008}{$\minus\mathtt{2.4706\phantom{00}} \cdot \mathtt{10}^{\minus\mathtt{04}}$}
        {epemttxhNLOinclF000008delta}{$\phantom{\mathtt{0}}\phantom{\minus}\mathtt{0.058}\percent\phantom{\mathtt{0}}$}
        {epemttxhNLOinclF000009mplcty}{$\mathtt{2}$}
        {epemttxhNLOinclF000009}{$\phantom{\minus}\mathtt{4.7475\phantom{00}} \cdot \mathtt{10}^{\minus\mathtt{05}}$}
        {epemttxhNLOinclF000009delta}{$\phantom{\mathtt{0}}\phantom{\minus}\mathtt{0.097}\percent\phantom{\mathtt{0}}$}
        {epemttxhNLOinclF000010mplcty}{$\mathtt{2}$}
        {epemttxhNLOinclF000010}{$\phantom{\minus}\mathtt{2.0875\phantom{00}} \cdot \mathtt{10}^{\minus\mathtt{04}}$}
        {epemttxhNLOinclF000010delta}{$\phantom{\mathtt{0}}\phantom{\minus}\mathtt{0.064}\percent\phantom{\mathtt{0}}$}
        {epemttxhNLOinclF000011mplcty}{$\mathtt{2}$}
        {epemttxhNLOinclF000011}{$\minus\mathtt{9.1283\phantom{00}} \cdot \mathtt{10}^{\minus\mathtt{05}}$}
        {epemttxhNLOinclF000011delta}{$\phantom{\mathtt{0}}\phantom{\minus}\mathtt{0.041}\percent\phantom{\mathtt{0}}$}
        {epemttxhNLOinclTotal}{$\minus\mathtt{1.315\phantom{000}} \cdot \mathtt{10}^{\minus\mathtt{04}}$}
        {epemttxhNLOinclTotaldelta}{$\phantom{\mathtt{0}}\phantom{\minus}\mathtt{0.76}\percent\phantom{\mathtt{00}}$}
        {epemttxhNLOinclTarget}{$\minus\mathtt{1.3205\phantom{00}} \cdot \mathtt{10}^{\minus\mathtt{04}}$}
        {epemttxhNLOinclTargetdelta}{$\phantom{\mathtt{0}}\minus\mathtt{0.38}\percent\phantom{\mathtt{00}}$}
        {epemttxhNLOinclImALLGRAPHSmplcty}{$\mathtt{1}$}
        {epemttxhNLOinclImALLGRAPHS}{$\minus\mathtt{2\phantom{}} \cdot \mathtt{10}^{\minus\mathtt{06}}$}
        {epemttxhNLOinclImTotal}{$\minus\mathtt{2\phantom{}} \cdot \mathtt{10}^{\minus\mathtt{06}}$}
        {epemttxhNLOinclImTarget}{$\phantom{\minus}\mathtt{0\phantom{}} \cdot \mathtt{10}^{+\mathtt{00}}$}
        {epemttxhLOsinclSGQG000mplcty}{$\mathtt{1}$}
        {epemttxhLOsinclSGQG000}{$\phantom{\minus}\mathtt{1.39990\phantom{0}} \cdot \mathtt{10}^{\minus\mathtt{04}}$}
        {epemttxhLOsinclSGQG000delta}{$\phantom{\mathtt{0}}\phantom{\minus}\mathtt{0.014}\percent\phantom{\mathtt{0}}$}
        {epemttxhLOsinclSGQG002mplcty}{$\mathtt{2}$}
        {epemttxhLOsinclSGQG002}{$\phantom{\minus}\mathtt{2.20996\phantom{0}} \cdot \mathtt{10}^{\minus\mathtt{04}}$}
        {epemttxhLOsinclSGQG002delta}{$\phantom{\mathtt{0}}\phantom{\minus}\mathtt{0.024}\percent\phantom{\mathtt{0}}$}
        {epemttxhLOsinclTotal}{$\phantom{\minus}\mathtt{3.60986\phantom{0}} \cdot \mathtt{10}^{\minus\mathtt{04}}$}
        {epemttxhLOsinclTotaldelta}{$\phantom{\mathtt{0}}\phantom{\minus}\mathtt{0.015}\percent\phantom{\mathtt{0}}$}
        {epemttxhLOsinclTarget}{$\phantom{\minus}\mathtt{3.6070\phantom{00}} \cdot \mathtt{10}^{\minus\mathtt{04}}$}
        {epemttxhLOsinclTargetdelta}{$\phantom{\mathtt{0}}\phantom{\minus}\mathtt{0.079}\percent\phantom{\mathtt{0}}$}
        {epemttxhNLOsinclF000001mplcty}{$\mathtt{2}$}
        {epemttxhNLOsinclF000001}{$\phantom{\minus}\mathtt{3.57\phantom{0000}} \cdot \mathtt{10}^{\minus\mathtt{07}}$}
        {epemttxhNLOsinclF000001delta}{$\phantom{\mathtt{0}}\phantom{\minus}\mathtt{6.6}\percent\phantom{\mathtt{000}}$}
        {epemttxhNLOsinclF000002mplcty}{$\mathtt{2}$}
        {epemttxhNLOsinclF000002}{$\phantom{\minus}\mathtt{9.125\phantom{000}} \cdot \mathtt{10}^{\minus\mathtt{06}}$}
        {epemttxhNLOsinclF000002delta}{$\phantom{\mathtt{0}}\phantom{\minus}\mathtt{0.16}\percent\phantom{\mathtt{00}}$}
        {epemttxhNLOsinclF000003mplcty}{$\mathtt{2}$}
        {epemttxhNLOsinclF000003}{$\phantom{\minus}\mathtt{2.598\phantom{000}} \cdot \mathtt{10}^{\minus\mathtt{05}}$}
        {epemttxhNLOsinclF000003delta}{$\phantom{\mathtt{0}}\phantom{\minus}\mathtt{0.65}\percent\phantom{\mathtt{00}}$}
        {epemttxhNLOsinclF000004mplcty}{$\mathtt{1}$}
        {epemttxhNLOsinclF000004}{$\minus\mathtt{3.2604\phantom{00}} \cdot \mathtt{10}^{\minus\mathtt{05}}$}
        {epemttxhNLOsinclF000004delta}{$\phantom{\mathtt{0}}\phantom{\minus}\mathtt{0.042}\percent\phantom{\mathtt{0}}$}
        {epemttxhNLOsinclF000005mplcty}{$\mathtt{2}$}
        {epemttxhNLOsinclF000005}{$\minus\mathtt{3.4399\phantom{00}} \cdot \mathtt{10}^{\minus\mathtt{07}}$}
        {epemttxhNLOsinclF000005delta}{$\phantom{\mathtt{0}}\phantom{\minus}\mathtt{0.25}\percent\phantom{\mathtt{00}}$}
        {epemttxhNLOsinclF000006mplcty}{$\mathtt{2}$}
        {epemttxhNLOsinclF000006}{$\phantom{\minus}\mathtt{1.5349\phantom{00}} \cdot \mathtt{10}^{\minus\mathtt{05}}$}
        {epemttxhNLOsinclF000006delta}{$\phantom{\mathtt{0}}\phantom{\minus}\mathtt{0.071}\percent\phantom{\mathtt{0}}$}
        {epemttxhNLOsinclF000007mplcty}{$\mathtt{2}$}
        {epemttxhNLOsinclF000007}{$\minus\mathtt{2.5882\phantom{00}} \cdot \mathtt{10}^{\minus\mathtt{05}}$}
        {epemttxhNLOsinclF000007delta}{$\phantom{\mathtt{0}}\phantom{\minus}\mathtt{0.052}\percent\phantom{\mathtt{0}}$}
        {epemttxhNLOsinclF000008mplcty}{$\mathtt{2}$}
        {epemttxhNLOsinclF000008}{$\minus\mathtt{6.2182\phantom{00}} \cdot \mathtt{10}^{\minus\mathtt{05}}$}
        {epemttxhNLOsinclF000008delta}{$\phantom{\mathtt{0}}\phantom{\minus}\mathtt{0.066}\percent\phantom{\mathtt{0}}$}
        {epemttxhNLOsinclF000009mplcty}{$\mathtt{2}$}
        {epemttxhNLOsinclF000009}{$\phantom{\minus}\mathtt{1.4429\phantom{00}} \cdot \mathtt{10}^{\minus\mathtt{05}}$}
        {epemttxhNLOsinclF000009delta}{$\phantom{\mathtt{0}}\phantom{\minus}\mathtt{0.091}\percent\phantom{\mathtt{0}}$}
        {epemttxhNLOsinclF000010mplcty}{$\mathtt{2}$}
        {epemttxhNLOsinclF000010}{$\phantom{\minus}\mathtt{4.4976\phantom{00}} \cdot \mathtt{10}^{\minus\mathtt{05}}$}
        {epemttxhNLOsinclF000010delta}{$\phantom{\mathtt{0}}\phantom{\minus}\mathtt{0.084}\percent\phantom{\mathtt{0}}$}
        {epemttxhNLOsinclF000011mplcty}{$\mathtt{2}$}
        {epemttxhNLOsinclF000011}{$\minus\mathtt{1.9691\phantom{00}} \cdot \mathtt{10}^{\minus\mathtt{05}}$}
        {epemttxhNLOsinclF000011delta}{$\phantom{\mathtt{0}}\phantom{\minus}\mathtt{0.052}\percent\phantom{\mathtt{0}}$}
        {epemttxhNLOsinclTotal}{$\minus\mathtt{3.048\phantom{000}} \cdot \mathtt{10}^{\minus\mathtt{05}}$}
        {epemttxhNLOsinclTotaldelta}{$\phantom{\mathtt{0}}\phantom{\minus}\mathtt{0.59}\percent\phantom{\mathtt{00}}$}
        {epemttxhNLOsinclTarget}{$\minus\mathtt{3.0570\phantom{00}} \cdot \mathtt{10}^{\minus\mathtt{05}}$}
        {epemttxhNLOsinclTargetdelta}{$\phantom{\mathtt{0}}\minus\mathtt{0.28}\percent\phantom{\mathtt{00}}$}
        {epemttxhNLOsinclImALLGRAPHSmplcty}{$\mathtt{1}$}
        {epemttxhNLOsinclImALLGRAPHS}{$\phantom{\minus}\mathtt{9\phantom{}} \cdot \mathtt{10}^{\minus\mathtt{08}}$}
        {epemttxhNLOsinclImTotal}{$\phantom{\minus}\mathtt{9\phantom{}} \cdot \mathtt{10}^{\minus\mathtt{08}}$}
        {epemttxhNLOsinclImTarget}{$\phantom{\minus}\mathtt{0\phantom{}} \cdot \mathtt{10}^{+\mathtt{00}}$}
    }
    {[N/A]}
}

\makeatother

\usepackage{glossaries}
\usepackage{glossaries-extra}

\setabbreviationstyle[acronym]{long-short}

\makeatletter
\newcommand*{\glsplainhyperlink}[2]{%
  \colorlet{currenttext}{.}%
  \colorlet{currentlink}{\@linkcolor}%
  \hypersetup{linkcolor=currenttext}%
  \hyperlink{#1}{#2}%
  \hypersetup{linkcolor=currentlink}%
}
\let\@glslink\glsplainhyperlink
\makeatother

\newacronym{SM}{SM}{Standard Model}
\newacronym{OS}{OS}{On-Shell}
\newacronym{dod}{dod}{superficial degree of divergence}
\newacronym{IBP}{IBP}{integration-by-parts}
\newacronym{LTD}{LTD}{Loop-Tree Duality}
\newacronym{cLTD}{cLTD}{Manifestly Causal Loop-Tree Duality}
\newacronym{TOPT}{TOPT}{Time-Ordered Perturbation Theory}
\newacronym{LU}{LU}{Local Unitarity}
\newacronym{QCD}{QCD}{Quantum Chromodynamics}
\newacronym{FSR}{FSR}{Final-State Radiation}
\newacronym{ISR}{ISR}{Initial-State Radiation}
\newacronym{IR}{IR}{InfraRed}
\newacronym{UV}{UV}{UltraViolet}
\newacronym{KLN}{KLN}{Bloch-Nordsieck/Kinoshita–Lee–Nauenberg}
\newacronym{ODE}{ODE}{Ordinary Differential Equation}
\newacronym{LSZ}{LSZ}{Lehmann-Symanzik-Zimmermann}
\newacronym{LEP}{LEP}{Large Electron-Positron Collider}
\newacronym{FT}{FT}{Feynman-Tree}
\newacronym{EW}{weak}{weak}
\newacronym{1PI}{1PI}{One-Particle-Irreducible}
\newacronym{LO}{LO}{Leading-Order}
\newacronym{QFT}{QFT}{Quantum Field Theory}
\newacronym{pQFT}{pQFT}{perturbative Quantum Field Theory}
\newacronym{NLO}{NLO}{Next-to-Leading-Order}
\newacronym{NNLO}{NNLO}{Next-to-Next-to-Leading-Order}
\newacronym{N3LO}{N3LO}{Next-to-Next-to-Next-to-Leading-Order}
\newacronym{PDF}{PDF}{Parton Distribution Function}
\newacronym{BRST}{BRST}{Becchi-Rouet-Stora-Tyutin}
\newacronym{BPHZ}{BPHZ}{Bogoliubov-Parasiuk-Hepp-Zimmermann}
\newacronym{MG5aMC}{\texttt{MG5aMC}}{\texttt{MadGraph5\_aMC@NLO}}

\newcommand{\glss}[1]{
  \stringcases
    {#1}
    {
       {PDF}{\unskip\ignorespaces PDF\unskip\ignorespaces}
    }
    {\unskip\ignorespaces\gls{#1}\unskip\ignorespaces}
    \xspace
}

\glssetcategoryattribute{acronym}{nohyperfirst}{true}

\newcommand*{\halfway}{0.5*\pgfdecoratedpathlength+.5*5pt}
\tikzset{->-/.style={decoration={markings, mark=at position #1 with {\arrow{latex}}},postaction={decorate}},
	->-/.default=\halfway}
	
\makeatletter
\newcommand{\subalign}[1]{%
  \vcenter{%
    \Let@ \restore@math@cr \default@tag
    \baselineskip\fontdimen10 \scriptfont\tw@
    \advance\baselineskip\fontdimen12 \scriptfont\tw@
    \lineskip\thr@@\fontdimen8 \scriptfont\thr@@
    \lineskiplimit\lineskip
    \ialign{\hfil$\m@th\scriptstyle##$&$\m@th\scriptstyle{}##$\hfil\crcr
      #1\crcr
    }%
  }%
}
\makeatother

\allowdisplaybreaks

\definecolor{mmblue}{rgb}{0.37, 0.51, 0.71}
\definecolor{mmyellow}{rgb}{0.88, 0.61, 0.14}
\definecolor{mmgreen}{rgb}{0.56, 0.69, 0.19}
\definecolor{mmred}{rgb}{0.92, 0.39, 0.21}
    
\title{Local Unitarity: cutting raised propagators and localising renormalisation}

\author[a]{Zeno Capatti,}
\author[b]{Valentin Hirschi,}
\author[a]{and Ben Ruijl}

\affiliation[a]{ETH Z\"urich,\\
R\"amistrasse 101, %
8092 Z\"urich, Switzerland}

\affiliation[b]{CERN,\\
Espl. des Particules 1, 1211 Meyrin, Switzerland
}

\emailAdd{zeno.ca@gmail.com}
\emailAdd{valentin.hirschi@gmail.com}
\emailAdd{benruyl@gmail.com}

\abstract{The Local Unitarity (LU) representation of differential cross-sections locally realises the cancellations of infrared singularities predicted by the Kinoshita-Lee-Nauenberg theorem. In this work we solve the two remaining challenges to enable practical higher-loop computations within the LU formalism. The first concerns the generalisation of the LU representation to graphs with raised propagators. The solution to this problem results in a generalisation of distributional Cutkosky rules. The second concerns the regularisation of ultraviolet and spurious soft singularities, solved using a fully automated and local renormalisation procedure based on Bogoliubov's $R$-operation. We detail an all-order construction for the hybrid $\overline{\text{MS}}$ and On-Shell scheme whose only analytic input is single-scale vacuum diagrams.
 Using this novel technology, we provide (semi-)inclusive results for two multi-leg processes at NLO, study limits of individual supergraphs up to N3LO and present the first physical NNLO cross-sections computed fully numerically in momentum-space, namely for the processes $\gamma^* \rightarrow j j$ and $\gamma^* \rightarrow t \bar{t}$.
}

\begin{document} 
\maketitle
\flushbottom

\section{Introduction}

The lack of clear evidence for physics beyond the \glss{SM} within our current collider observations pushes High-Energy Physics into the high-intensity frontier and ushers in the Precision Era. In the context of hadronic collisions, one obstacle to precise comparisons between theory and experiment, is the limited accuracy of theoretical predictions.
This limitation is for a large part driven by our (in)ability to compute higher-order corrections in the perturbative expansion of \glss{QFT} coupling constants, which is made especially severe by the strength of \glss{QCD} interactions.

This problem is well-studied, and several decades of efforts from the theoretical community have managed to postpone the point at which theoretical uncertainties of fixed-order origin become the main limitation to our ability to interpret collider data.
Historically, and especially during the \glss{LEP} era, these efforts have mostly pursued a fully analytical approach to the computation of (semi-)inclusive cross-sections. With the advent of parton showers and the need for supporting ever more complicated differential observables, phase-space integrals quickly became predominantly computed numerically, using slicing and/or subtraction approaches for regularising \glss{IR} singularities~\cite{Czakon:2014oma,Campbell:2022gdq,Asteriadis:2019dte,Gehrmann-DeRidder:2005btv,Currie:2013vh,Somogyi:2006cz,Somogyi:2009ri,Caola:2017dug,Catani:2007vq,Boughezal:2015dva,TorresBobadilla:2020ekr}.
Then, at \glss{NLO}, the reduction of one-loop amplitudes also transitioned to numerical methods as soon as stable and fast numerical reduction algorithms became available~\cite{Ossola:2007ax,Denner:2016kdg,Peraro:2014cba,Hirschi:2016mdz}.
Nowadays, traditional multi-loop computations welcome increasingly more numerical aspects, for example for the reduction through numerical reconstruction over finite fields~\cite{vonManteuffel:2014ixa, Peraro:2019svx, Klappert:2019emp, Klappert:2020nbg, Heller:2021qkz}, or solutions of systems of differential equations~\cite{Caffo:1998du,Czakon:2020vql} through series expansions~\cite{Hidding:2020ytt,Moriello:2019yhu,Liu:2022chg}, or also by direct numerical computation of amplitudes in momentum space~\cite{Anastasiou:2018rib,Anastasiou:2020sdt,PhysRevD.79.033005,2013,Buchta:2015wna,Capatti:2020ytd,2020}.

This slow-moving transition to numerics still maintains the historical divide of the task of computing phase-space and loop integrals, with the important consequence that \glss{IR} singularities must be regularised separately within each of these two classes of integrals. 
One notable exception is the method of reverse unitarity~\cite{Anastasiou:2002yz,Anastasiou:2002wq,Anastasiou:2015vya,Duhr:2019kwi} since it turns phase-space integrals into loop ones in order to compute both together analytically. Although applications of reverse unitarity provided important \glss{N3LO} fixed-order cross-sections for $2\rightarrow 1$ processes~\cite{Anastasiou:2016cez,Mistlberger:2018etf,Chen:2019lzz,Duhr:2020sdp,Duhr:2021vwj,Chen:2021isd}, it is as of now not possible to generalise this approach to higher multiplicity processes and arbitrary differential observables.

In general, the separate treatment of loop and phase-space integrals cannot take advantage of the inherent simplicity of the \glss{IR} cancellation pattern featured in the proof of the \glss{KLN} theorem~\cite{Kinoshita:1962ur,Lee:1964is,Bloch:1937pw}.
Motivated by this realisation, we focused on establishing a fundamentally different framework for the computation of differential cross-sections whereby \glss{IR} singularities cancel locally, without any subtraction procedure. We refer to the resulting formulation of the differential cross-section as its \glss{LU} expression and we recently published its detailed construction in ref.~\cite{2021}, as well as the proof that it is free of any final-state \glss{IR} singularities at any perturbative order and for any observable. We also point the reader to the proceedings of ref.~\cite{Capatti:2021bsm} for a more concise summary of \glss{LU}.
The objective of this construction is two-fold. First, we aim at discovering new theoretical aspects of perturbative expansions at fixed order in \glss{QFT}, and with a particular attention towards new techniques for matching them to a (parton shower) resummation.
Second, we want to apply the \glss{LU} construction in order to build a competitive tool for the fully numerical computation of predictions of arbitrary collider observables.

In this work we address two key remaining challenges for practical higher-loop computations using \glss{LU}. The first challenge is the identification of a universal treatment of raised propagators introduced from self-energy insertions that is more convenient than the procedure introduced in ref.~\cite{2021}.  The new treatment involves taking higher-order residues, which results in derivatives of amplitudes that are efficiently and automatically computed.
The second challenge is the treatment of \glss{UV} divergences. We use Bogoliubov's $R$-operation and construct local \glss{UV} subtraction terms that render any amplitude of any order \glss{UV}-finite. We then modify our subtraction terms so as to also subtract spurious soft singularities and also automatically reproduce the \glss{OS} mass renormalisation conditions. Next, we determine integrated level counterterms, consisting of only single-scale massive vacuum graphs, which are defined such that we obtain results directly renormalised in the commonly used hybrid $\overline{\text{MS}}$ and \glss{OS} scheme.

Using this novel technology, we provide (semi-)inclusive results for $e^+ e^- \rightarrow \gamma^\star \rightarrow jjj$ and $e^+ e^- \rightarrow \gamma \rightarrow t\bar{t} H$ at \glss{NLO} and the first physical cross-sections at \glss{NNLO} fully computed in momentum-space, namely for processes $\gamma^* \rightarrow d \bar{d}$ and $\gamma^* \rightarrow t \bar{t}$. We also verify the correct subtraction of \glss{IR} and \glss{UV} limits of specific individual supergraphs contributing up to \glss{N3LO}.

The outline of this work is as follows. We present the \glss{LU} construction in presence of raised propagators in sect.~\ref{sec:selfenergies_lu}. In sect.~\ref{sec:R-operation} we construct the $R$-operation for \glss{UV} subtraction in \glss{LU}. Next, we construct local counterterms for subtracting \glss{UV} and spurious \glss{IR} singularities in sect.~\ref{sec:subtraction_operators}. 
In sect.~\ref{sect:localised_renormalisation}, we construct the integrated counterterms so as to automatically produce results renormalised in the hybrid $\overline{\text{MS}}$ and \glss{OS} scheme. In sect.~\ref{sec:gauge_invariance} we show how gauge invariance is realised in \glss{LU} and in sect.~\ref{sec:results} we provide numerical results supporting the validity of our construction. Finally, we present our conclusion in sect.~\ref{sec:conclusion}.

\section{Local Unitarity in the presence of raised propagators}
\label{sec:selfenergies_lu}

Perturbative cross-sections generally include the effect of self-energy corrections to particles participating in the process. Diagrammatically, self-energy corrections amount to the insertion of a \glss{1PI} graph on an edge which, in turn, results in a doubling of the edge itself. Repeated edges correspond to (raised) propagators with powers higher than one after substituting the Feynman rules.
As the \glss{LU} representation requires taking residues stemming from propagator poles, the question then arises of what is the correct residue-taking procedure in the case of raised propagators. Ordinary Cutkosky rules, for example, cannot account for the correct contribution.
Moreover, one wonders what is the effect of taking residues of such raised propagators on the local \glss{IR} cancellations guaranteed by the \glss{LU} representation. 

The discussion of and solution to this problem, given in the proof of ref.~\cite{2021} is correct, but difficult to apply in practice and is opaque in its interpretation. The underlying idea is that the \glss{LU} representation associated with a forward-scattering diagram with raised propagators should equal the \glss{LU} representation of the same forward-scattering diagram in which fictitious momenta are introduced to eliminate raised propagators, and only at the end the limit to zero of such fictitious momenta is taken. While convenient for the purpose of the proof, taking limits numerically is notoriously cumbersome, and we seek a more direct expression of the \glss{LU} representation in the presence of raised propagators. In other words, we wish to identify the correct generalisation of the Cutkosky rule which yields the right expression when taking the cut of a repeated edge, that is:

\begin{center}
\resizebox{3.cm}{!}{%
\begin{tikzpicture}

     \node[main node4]  (1) {};
     \node[] (L1) [left = 0.1cm of 1] {};
     \node[]  (A2) [right = 1cm of 1] {};
     
     \node[]  (A3) [right = 1.25cm of 1] {};
     \node[main node4]  (A4) [right = 2.5cm of 1] {};
     
     \node[main node4]  (B1) [right = 0.6cm of 1] {};
     \node[main node4]  (B2) [left = 0.6cm of A4] {};
     
     \node[]  (M2) [left = 0.95cm of A4]{$...$};
     
     \node[] (X1) [above right = 0.5cm and 1.13cm of 1] {};
     \node[] (LABEL) [right = 0.001cm of X1] {$\color{red}\boldsymbol{\pm}$};
    \node[] (X2) [below right = 0.5cm and 1.13cm of 1] {};
    
    \draw[thick, red] (X1) to[out= -60, in=120] (X2);

\begin{feynman}
\draw[thick, momentum=\(p\)] (1) to (B1);
\draw[thick, momentum=\(p\)] (B2) to (A4);

\draw[thick] (B1) to (A2);
\draw[thick] (B2) to (A3);

\end{feynman}

\end{tikzpicture}

}
\end{center}
In this section, we will show how to achieve this. Such a generalisation of the \glss{LU} representation is based on a conceptual understanding of the relationship between raised propagators and the residue formula for higher-order poles. This, in turn, provides a clear candidate for the generalisation of the Cutkosky distributional rule. 

Our discussion will leverage many of the key results presented in ref.~\cite{2021}, as well as employ a similar notation. We therefore refer the reader to that work for more details on the quantities manipulated in this section. We also refer the reader to ref.~\cite{Agarwal:2021ais} for a comprehensive review of the modern treatment of IR singularities and to ref.~\cite{Kreimer:2020mwn,Kreimer:2021jni,Berghoff:2020bug} for a recent and original approach to the topic of Cutkosky cuts and cutting rules in general.
The reader only interested in the final expression for the generalised cutting rules applicable to raised propagators can proceed to sect.~\ref{sec;generalised_cutting_rules}.

\subsection{Local Unitarity for raised propagators (residue theorem approach)}
\label{sec:raised_lu}

We now give a summary of the \glss{LU} representation.
Let us start with a three-dimensional representation $f_{\text{3d}}(G)$ of a graph $G$. In the following, we will call $G$ a supergraph, in anticipation of it being identified with the parent diagram from which interference diagrams are constructed. One may use \gls{LTD}~\cite{LTDRodrigoOrigin2008,LTDRodrigoMultiLoop2010,Capatti:2019ypt,Runkel:2019yrs}, \gls{cLTD}~\cite{Capatti:2020ytd,JesusAguilera-Verdugo:2020fsn,Aguilera-Verdugo:2020set,Berghoff:2020bug} or \gls{TOPT}~\cite{sterman_1993,Schwartz:2013pla,Mantovani_2016,bourjaily2020sequential} to obtain such a three-dimensional representation from the four-dimensional one, since the three formulations are locally equivalent.
In other words, these local representations are just different mathematical expressions of the same integrand, which is obtained from the sum over all residues arising when considering the integration over the energy components of all loop momenta. Importantly, this also implies that the physical threshold structure coincides in all three representations. 

\begin{figure}[t!]
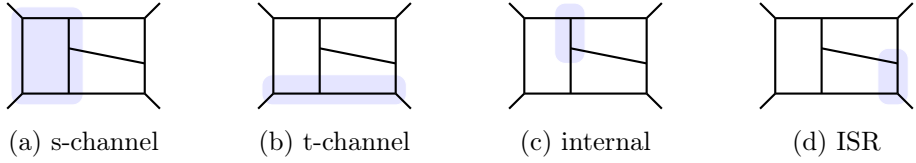

\centering
\begin{subfigure}[b]{1.3in}
\begin{center}
\input{diagrams/thresholds/s_channel}
\subcaption{s-channel}
\label{fig:vacuum_diagram}
\end{center}
\end{subfigure}%
\begin{subfigure}[b]{1.3in}
\begin{center}
\input{diagrams/thresholds/t_channel}%
\subcaption{t-channel}
\label{fig:embedding_1}
\end{center}
\end{subfigure}%
\begin{subfigure}[b]{1.3in}
\begin{center}
\input{diagrams/thresholds/internal}%
\caption{internal}
\label{fig:embedding_2}
\end{center}
\end{subfigure}%
\begin{subfigure}[b]{1.3in}
\begin{center}
\input{diagrams/thresholds/iss}%
\caption{ISR}
\label{fig:embedding_3}
\end{center}
\end{subfigure}%
\label{fig:connectedness}
\caption{Pictorial representation of thresholds for a $2\rightarrow 2$ diagram. Highlighted in blue are the vertex sets $\mathbf{s}$, corresponding to subgraphs of the original graph. The complement of each subgraph corresponds to a threshold of the same type.} 
\label{fig:embedding}
\end{figure}

Assume that $G$ is a connected supergraph, corresponding to a couplet $G=(\mathbf{v},\mathbf{e}=\mathbf{e}_{\text{ext}}\cup\mathbf{e}_{\text{int}})$, where $\mathbf{v}$ is the set of vertices of the graph and $\mathbf{e}$ is the set of edges. We assign to the external, on-shell, particles the edges in $\mathbf{e}_{\text{ext}}=\mathbf{a}_{\text{in}}\cup \mathbf{a}_{\text{out}}$. The particles in $\mathbf{a}_{\text{in}}$ are incoming and those in $\mathbf{a}_{\text{out}}$ are outgoing, and the sign of their on-shell energy is fixed accordingly. For forward-scattering diagrams, $G$ has the same incoming and outgoing particles, and each incoming particle has an outgoing partner with same momentum. We define the reduced graph $G_\mathrm{r}=(\mathbf{v},\mathbf{e},\boldsymbol{\gamma})$ to be the graph where each edge $e$ of $G$ with $\gamma_e$ occurrences is fused into one representative in $G_\mathrm{r}$ raised to power $\gamma_e$. From here on, any graph quantity will refer to the reduced graph rather than the original graph. Any threshold of the three-dimensional representation of the supergraph can be associated to a connected subgraph identified by the set of vertices $\mathbf{s}\subset \mathbf{v}$ (the subgraph is identified from the graph by the set $\mathbf{s}$ and the collection of the edges of the original graph whose vertices are in $\mathbf{s}$). Let $\mathcal{E}$ be the collection of all such connected subgraphs. The implicit equation defining the location of each of such thresholds is the support of the function $\eta_{(\mathbf{s},\alpha)}$, defined by
\begin{equation}
\label{eq:threshold_equation}
    \eta_{(\mathbf{s},\alpha)}=\sum_{e\in\delta(\mathbf{s})\cap\mathbf{a}_{\text{in}}}E_e-\sum_{e\in\delta(\mathbf{s})\cap\mathbf{a}_{\text{out}}}E_e-\alpha\sum_{e\in\delta(\mathbf{s})\setminus\mathbf{e}_{\text{ext}} }E_e=0,
\end{equation}
where $E_e=\sqrt{|\vec{p}_e|^2+m_e^2}$ is the on-shell energy associated to the particular edge $e$ carrying the spatial momentum $\vec{p}_e$, $\alpha\in\{\pm 1\}$, and $\delta(\mathbf{s})=\{e=\{v,v'\}\in\mathbf{e} \, | \, v\in\mathbf{s}, \ v'\in \mathbf{v}\setminus\mathbf{s}\}$ denotes the collection of all edges that are at the boundary of the subgraph identified by the set $\mathbf{s}\subset\mathbf{v}$. When momentum conservation conditions are assumed to hold on external particles, we have that $\eta_{(\mathbf{s},\alpha)}=-\eta_{(\mathbf{v}\setminus\mathbf{s},-\alpha)}$. Instead, $\eta_{(\mathbf{s},\alpha)}$ and $\eta_{(\mathbf{s},-\alpha)}$ correspond in general to different thresholds. In light of this definition, the thresholds can be classified into four categories based on specific properties of $\mathbf{s}$. We list below the precise definition of these categories and give diagrammatic representatives of each category in fig.~\ref{fig:embedding}: 
\begin{itemize}
    \item s-channel: if $\delta(\mathbf{s})\cap\mathbf{e}_{\text{ext}}=\mathbf{a}_{\text{in}}$ or $\delta(\mathbf{s})\cap\mathbf{e}_{\text{ext}}=\mathbf{a}_{\text{out}}$. We then write that $(\mathbf{s},1)\in\mathcal{E}^{+,\text{in}}_{\text{s-ch}}$, $(\mathbf{s},-1)\in\mathcal{E}^{-,\text{in}}_{\text{s-ch}}$, in the first case and $(\mathbf{s},1)\in\mathcal{E}^{+,\text{out}}_{\text{s-ch}}$, $(\mathbf{s},-1)\in\mathcal{E}^{-,\text{out}}_{\text{s-ch}}$ in the second. Because of the equivalence $\eta_{(\mathbf{s},\alpha)}=-\eta_{(\mathbf{v}\setminus\mathbf{s},-\alpha)}$, we set $\mathcal{E}^{+}_{\text{s-ch}}=\mathcal{E}^{+,\text{in}}_{\text{s-ch}}=\mathcal{E}^{-,\text{out}}_{\text{s-ch}}$ and $\mathcal{E}^{-}_{\text{s-ch}}=\mathcal{E}^{-,\text{in}}_{\text{s-ch}}=\mathcal{E}^{+,\text{out}}_{\text{s-ch}}$. Observe that, if the initial-state particles have energies larger than zero, any surface in $\mathcal{E}^-_{\text{s-ch}}$ is empty, i.e. eq.~\eqref{eq:threshold_equation} has no solution.
    \item t-channel: if $\delta(\mathbf{s})\cap\mathbf{a}_{\text{in}}\neq \emptyset, \, \mathbf{a}_{\text{in}}$ and $\delta(\mathbf{s})\cap\mathbf{a}_{\text{out}}\neq\emptyset, \, \mathbf{a}_{\text{out}}$. Such thresholds are generally non empty for both $\alpha=1$ and $\alpha=-1$.
    \item internal-like: if $\delta(\mathbf{s})\cap\mathbf{e}_{\text{ext}}\in\{\emptyset,\mathbf{e}_{\text{ext}}\}$. The locus of such singularities can only be a soft location, when all the particles in $\delta(\mathbf{s})$ are massless. For these thresholds, it also holds that $\eta_{(\mathbf{s},\alpha)}=-\eta_{(\mathbf{s},-\alpha)}$.
    \item ISR-like: if it is not s-channel, t-channel or internal-like.
\end{itemize}

For what concerns this paper and ref.~\cite{2021} (which focuses on \gls{FSR} singularities), a Cutkosky cut is a connected subgraph of the supergraph $\mathbf{s}$ such that either $\delta(\mathbf{s})\cap\mathbf{e}_{\text{ext}}=\mathbf{a}_{\text{in}}$ or $\delta(\mathbf{s})\cap\mathbf{e}_{\text{ext}}=\mathbf{a}_{\text{out}}$. That is, each Cutkosky cut is in one-to-one correspondence with s-channel thresholds in $\mathcal{E}^{+}_{\text{s-ch}}$. Alternatively, the Cutkosky cut can be denoted by the set $\mathbf{c}_\mathbf{s}=\delta(\mathbf{s})\setminus e_{\text{ext}}$. Generalisations of the notion of Cutkosky cuts can be used to also include t-channel or \gls{ISR} cuts~\cite{ISR_LU}.

In the following,  we will assume that $\mathcal{E}=\mathcal{E}_{\text{s-ch}}^+\cup\mathcal{E}_{\text{s-ch}}^-\cup\mathcal{E}_{\text{int}}$. For the purposes of \gls{IR} finiteness, this is equivalent to assuming that there is no t-channel or \gls{ISR}-like singularity, which is the assumption that we will consider in this paper, for simplicity. This assumption is manifestly true for two-point functions, due to $\mathbf{a}_{\text{in}}$ and $\mathbf{a}_{\text{out}}$ each having one element only.

Alternatively, one can prove local \gls{IR}-finiteness of the \gls{LU} representation in the complete phase-space minus small volumes around the location of t-channel and \gls{ISR}-like singularities: this is the strategy that was used in ref.~\cite{2021}. \glss{ISR}-like and t-channel singularities would traditionally be handled within the parton model paradigm and using a factorisation approach.

The three-dimensional representation of the supergraph can be written as
\begin{equation}
    f_{\mathrm{3d}}(G_\mathrm{r})=\frac{g(\vec{k})}{\prod_{e\in\mathbf{e}}E_e^{\beta_e} \prod_{\mathbf{s}\in\mathcal{E}_{\text{s-ch}}^+}\eta_\mathbf{s}^{\beta_\mathbf{s}}\prod_{\mathbf{s}\in\mathcal{E}_{\text{s-ch}}^-}\eta_\mathbf{s}^{\beta_\mathbf{s}}\prod_{\mathbf{s}\in\mathcal{E}_{\text{int}}}\eta_\mathbf{s}^{\beta_\mathbf{s}}},
\end{equation}
where $g$ is a polynomial in the on-shell energies and spatial momenta.
We stress that $f_{\mathrm{3d}}$ also depends on the original graph $G$ for what concerns the expression of the numerator.
We denote with $\vec{k}=(\vec{k}_1,...,\vec{k}_L)$ the collection of all spatial loop momenta of the graph.  In this way, we have extracted all the singularities of $f_{\text{3d}}(G_\mathrm{r})$ with their respective powers $\beta_e$ (for the inverse energies) and $\beta_\mathbf{s}$ (for the thresholds), which can be determined in terms of the raising powers $\boldsymbol{\gamma}$ of the propagators of $G_\mathrm{r}=(\mathbf{v},\mathbf{e},\boldsymbol{\gamma})$. The specific structure of $g$ is irrelevant for what concerns the proof of cancellation of thresholds, that is the cancellation of the enhancements associated with any of the $\eta_\mathbf{s}$, for $\mathbf{s}\in\mathcal{E}_{\text{s-ch}}^+$. On the other hand, the structure of $g$ is important for a rigorous proof of the integrability of soft singularities. For this section, we will write the function $g$ in the following manner, which is valid for any arbitrary supergraph:
\begin{equation}
    g(\vec{k})=\sum_{F\in\mathcal{F}} c_{F}(\vec{k}) \prod_{\mathbf{s}\in \mathcal{E}\setminus F}  \,  \eta_\mathbf{s}^{\beta_\mathbf{s}},
\end{equation}
where $\mathcal{F}$ is the collection of all cross-free families of connected cuts of the supergraph, with the added constraint that any subgraph in any of the families cannot be written as the union of its children (see fig.~10 of ref.~\cite{2021}). $c(\vec{k})$ is a polynomial in the on-shell energies and the spatial components of the external momenta and loop moments. This is a weak constraint on the structure of $g$, and stronger constraints may be formulated. Such constraint is relevant for what concerns the soft scaling of $f_{\text{3d}}(G_\mathrm{r})$, and in particular it can be used (as in ref.~\cite{2021}) to show that in gauge theories the \gls{LU} representation is locally \gls{IR} finite unless there are spurious soft singularities (see sect.~\ref{sect:soft_higher_order_propagators}). 

Given this, we would like to construct interference diagrams from $f_{\text{3d}}(G_\mathrm{r})$. Cutkosky's original work~\cite{Cutkosky:1960sp} clearly showed that interference diagrams correspond to thresholds of supergraphs, and we would like to replicate such result here, but at the local level. In order to do this, we introduce an auxiliary variable that allows us to clearly parametrise the distance of a point to any threshold. Such auxiliary variable corresponds to the group parameter of the causal flow introduced in sect. 3.2 of ref.~\cite{2021}. In particular, since any Cutkosky cut corresponds to an s-channel threshold, let us consider a vector field $\kappa$ such that
\begin{equation}
\label{eq:causal_prescription}
    \kappa\cdot\nabla \eta_\mathbf{s}<0, \quad \forall\vec{k}\in\partial\eta_\mathbf{s}, \quad \forall\mathbf{s}\in\mathcal{E}_{\text{s-ch}}^+,
\end{equation}
i.e., it has positive projection onto the outward-pointing normal to any s-channel threshold when evaluated on that threshold. Such constraint arises from the causal $i\varepsilon$ prescription, which forces $\text{Im}[\eta_\mathbf{s}]<0$ (see ref.~\cite{2020} for more details).  Furthermore, we shall require that 
\begin{equation}
\label{eq:branch_cut_check}
    \lim_{|q_e|\rightarrow 0}\frac{|\vec{Q}_e(\kappa)|}{E_e}<c, \quad \forall e\in\mathbf{e},
\end{equation}
where $c$ is finite and if $\vec{q}_e=\sum_{i=1}^L s_{ei} \vec{k}_i+\vec{p}$ for the chosen loop momentum basis, then $\vec{Q}_e(\vec{k})=\vec{q}_e-\vec{p}$. Such constraint is required in order to obtain an integrable behaviour close to soft singularities. We have shown in ref.~\cite{2020} that such a vector field $\kappa$ always exists for \textit{any} given graph and we described a deterministic procedure that allows one to generically construct it. In the same reference, we have also shown that for any two-point graph with a massive external four-momentum set in its rest-frame, a simple and valid choice of vector field $\kappa$ is given by
\begin{equation}
    \kappa_{\text{2p}}=-\lambda\vec{k}, \quad \lambda\in(0,1].
\end{equation}
In particular, $\kappa_{\text{2p}}$ satisfies both eq.~\eqref{eq:causal_prescription} and eq.~\eqref{eq:branch_cut_check} when the two-point graph is evaluated in the rest frame of its massive external momentum. We note that $\kappa_{\text{2p}}$ also has the clear interpretation of being the deformation vector field needed to contour-deform the thresholds in $\mathcal{E}_{\text{s-ch}}^+$.

Given the vector field $\kappa$, we consider the following Cauchy problem
\begin{equation}
\label{eq:causal_flow_ODE}
    \begin{cases}
    \partial_t \phi_t=\kappa \circ \phi_t \\
    \phi_0=\vec{k}
    \end{cases}\,.
\end{equation}
This Cauchy problem defines a flow $\phi_t$, which we call causal flow in connection with the fact that the vector field generating the flow, $\kappa$, satisfies the causal prescription given in eq.~\eqref{eq:causal_prescription}. For any two-point supergraph with a massive external momentum, the causal flow can be obtained in an analytic form. Indeed, in this case we can choose $\kappa=\kappa_{\text{2p}}$ for which the \gls{ODE} system has the following solution:
\begin{equation}
    \phi_t^{\text{2p}}=e^{-\lambda t}\vec{k}\;.
\end{equation}
The causal flow inherits a series of crucial properties from the field $\kappa$ when interpreted as a contour deformation. First, for any given $\vec{k}$, the curve $\phi_t$ satisfying eq.~\eqref{eq:causal_flow_ODE} intersects any of the thresholds $\mathbf{s}\in\mathcal{E}_{\text{s-ch}}^+$ at most once, for otherwise the vector field $\kappa$ could not flow consistently inwards of the surface defined by $\eta_\mathbf{s}=0$, as established by eq.~\eqref{eq:causal_prescription}. Second, given the curve $\phi_t$ for $t\in(-\infty,\infty)$ and the unique value $t^\star_\mathbf{s}$ for which $\eta_\mathbf{s}\circ \phi_{t^\star_\mathbf{s}}=0$, we must have $\eta_\mathbf{s}\circ\phi_t\approx (t-t^\star_\mathbf{s})$ for $t$ approaching $t^\star_\mathbf{s}$. These two key properties can be summarised as follows. Let $\vec{k}$ be fixed and let
\begin{equation}
    \mathcal{E}_{\text{s-ch},\vec{k}}=\{\mathbf{s}\in\mathcal{E}_{\text{s-ch}}^+ \, | \, \exists t^\star_\mathbf{s}\in(-\infty,\infty) \text{ with } \eta_\mathbf{s}\circ \phi_{t^\star_\mathbf{s}}=0 \}.
\end{equation}
Then the expansion of $\eta_\mathbf{s}\circ\phi_t$, for any threshold $\mathbf{s}\in\mathcal{E}_{\text{s-ch},\vec{k}}$ around a unique zero $t=t^\star_\mathbf{s}$ reads
\begin{equation}
    \eta_\mathbf{s}\circ\phi_t=(t-t^\star_\mathbf{s}) \, \partial_t (\eta_\mathbf{s}\circ\phi_t)|_{t=t^\star_\mathbf{s}} +\mathcal{O}\left((t-t^\star_\mathbf{s})^2\right)=(t-t^\star_\mathbf{s}) \, (\kappa\cdot\nabla\eta_\mathbf{s})|_{\vec{k}=\phi_{t^\star_\mathbf{s}}} + \mathcal{O}\left((t-t^\star_\mathbf{s})^2\right),
\end{equation}
where we used the chain rule and eq.~\eqref{eq:causal_flow_ODE} to obtain the second equality. In particular, we observe that $(\kappa\cdot\nabla\eta_\mathbf{s})|_{\vec{k}=\phi_{t^\star_\mathbf{s}}}$ is always guaranteed to be strictly positive in virtue of eq.~\eqref{eq:causal_prescription} and thus the expansion of $\eta_\mathbf{s}\circ\phi_t$ is always guaranteed to scale linearly in $t-t^\star_\mathbf{s}$ close to a threshold location. In other words, any threshold $\mathbf{s}$ with $\beta_{\mathbf{s}}=1$ appears as a simple pole along the causal flow lines. In tun, this means that each threshold $\eta_\mathbf{s}$ with power $\beta_\mathbf{s}$ appears as a pole of order $\beta_\mathbf{s}$ of the integrand. Another useful way to state this same idea is the following. Consider the function
\begin{equation}
    w_t(\vec{k})=\prod_{\mathbf{s}\in\mathcal{E}_{\text{s-ch},\vec{k}}} \frac{(t-t^\star_\mathbf{s})^{\beta_\mathbf{s}}}{[(\eta_\mathbf{s}\circ\phi_t)(\vec{k})]^{\beta_\mathbf{s}}}.
\end{equation}
Then $w_t$ is bounded for any $\vec{k}$ and any $t$. The introduction of $w_t$ allows us to clearly isolate the scaling behaviour of $f_{\text{3d}}(G)$ close to a threshold:
\begin{equation}
\label{eq:composition_supergraph}
    f_{\text{3d}}(G_\mathrm{r})\circ\phi_t=\frac{w_t(\vec{k})}{\prod_{\mathbf{s}\in\mathcal{E}_{\text{s-ch}}^+}(t-t^\star_\mathbf{s})^{\beta_\mathbf{s}}}f\circ\phi_t, \quad f=\frac{g}{\prod_{e\in\mathbf{e}}E_e^{\beta_e} \prod_{\mathbf{s}\in\mathcal{E}_{\text{s-ch}}^-}\eta_\mathbf{s}^{\beta_\mathbf{s}}\prod_{\mathbf{s}\in\mathcal{E}_{\text{int}}}\eta_\mathbf{s}^{\beta_\mathbf{s}}}\;,
\end{equation}
where $f$ is an integrable function (provided $g$ vanishes fast enough in the UV region). We are now ready to construct the \gls{LU} representation generalised to the case of raised propagators. From the construction above, it is clear that $\eta_\mathbf{s}$ scales like $(t-t^\star_\mathbf{s})$ in the limit of $t\rightarrow t^\star_\mathbf{s}$.
Thus, if a threshold $\eta_\mathbf{s}$ itself appears raised to a power $\beta_{\mathbf{s}}$ within $f_{\text{3d}}(G)$, that threshold corresponds to a pole in the variable $t$ of order $\beta_{\mathbf{s}}$ of the function $f_{\text{3d}}(G)\circ\phi_t$. Cutkosky cuts that intersect any propagator raised to a power larger than one then correspond to higher-order residues in the variable $t$ of the \gls{LU} representation of the supergraph.

\subsubsection{Local Unitarity representation}

We are now ready to write the \gls{LU} representation for $f_{\text{3d}}(G)$, generalised to the case of $\beta_{\mathbf{s}}\ge 1$ by a straightfroward application of the residue formula:
\begin{equation}
\label{eq:LU_formula}
    \sigma_\mathrm{d}^G(\vec{k})=\sum_{\mathbf{s}\in\mathcal{E}_{\text{s-ch},\vec{k}}} \frac{1}{(\beta_\mathbf{s}-1)!} \lim_{t\rightarrow t^\star_\mathbf{s}}\frac{\mathrm{d}^{\beta_\mathbf{s}-1}}{\mathrm{d}t^{\beta_\mathbf{s}-1}}\left[(t-t^\star_\mathbf{s})^{\beta_\mathbf{s}} \, (f_{\mathrm{3d}}(G_\mathrm{r})\circ\phi_t) \, f_\mathbf{s}(\mathcal{O}) \, \mathbb{J}[\phi_t] \, h(t)\right],
\end{equation}
for a given normalised function $h(t)$. $\mathbb{J}[\phi_t]$ is the Jacobian of the causal flow $\phi$ with respect to the variables $\vec{k}$ (for more details on $h(t)$ and $\mathbb{J}[\phi_t]$, we refer the reader to ref.~\cite{2021}). $f_\mathbf{s}(\mathcal{O})$ is a final-state density that measures the value of an observable $\mathcal{O}$. From $\sigma_\mathrm{d}$ we obtain the \gls{LU} representation for the contribution from the supergraph $G$ to any differential cross-section
\begin{equation}
    \frac{\mathrm{d}\sigma^G}{\mathrm{d}\mathcal{O}}=\int \left[\prod_{i=1}^L\frac{\mathrm{d}\vec{k}_i}{(2\pi)^3}\right] \sigma_\mathrm{d}^G(\vec{k}).
\end{equation}
We already mentioned that eq.~\eqref{eq:LU_formula} could in principle be obtained from its analogue presented in ref.~\cite{2021} by first assigning fictitious momenta to the raised propagators, then computing the \gls{LU} representation, and finally sending such fictitious momenta to zero. Let us perform this exact procedure.

The fictitious momenta that we plan to introduce can be added to propagators before the integration of the energy components of loop momenta or to the thresholds after the integration of the energies. The simplest path to the result is that of introducing them after the integration over the energy components. Let us start then with the integrand
\begin{equation}
\label{eq:raised_3d}
    f_{\mathrm{3d}}(G_\mathrm{r})=\int \left[\prod_{i=1}^L \frac{\mathrm{d} k_i^0}{(2\pi)}\right]\frac{N}{\prod_{e\in\mathbf{e}}(q_e^2-m_e^2+\mathrm{i}\epsilon)^{\gamma_e}}=\frac{f}{\prod_{\mathbf{s}\in\mathcal{E}_{\text{s-ch}}^+}\eta_\mathbf{s}^{\beta_\mathbf{s}}},
\end{equation}
 On the right-hand side, we explicitly performed the integration of energy components and made explicit the dependence on non-empty s-channel thresholds only. For two-point supergraphs, $f$ is only singular at soft points but it is integrable ($f$ is defined in eq.~\eqref{eq:composition_supergraph}).

We now introduce fictitious shifts to the surfaces $\eta_\mathbf{s}$. In particular, consider the threshold
\begin{equation}
\label{eq:shifted_thresholds}
    \eta_{\mathbf{s}i}=\eta_\mathbf{s}+p_{\mathbf{s}i}^0, \quad \quad i=1,...,\beta_\mathbf{s},
\end{equation}
where $p_{\mathbf{s}i}^0$ are real constants, and $p_{\mathbf{s}1}^0=0$. The dependence of $\eta_{\mathbf{s}i}$ on the index $i$ is fully contained in the fictitious shift $p_{\mathbf{s}i}^0$. Setting $\eta_{\mathbf{s}i}$ to zero gives the implicit equation for a level surface of $\eta_\mathbf{s}$. Then, we can write eq.~\eqref{eq:raised_3d} as a limit of a quantity with no raised propagator
\begin{equation}
\label{eq:limit_raised}
    \tilde{f}_{\mathrm{3d}}(G_\mathrm{r})=\frac{f}{\prod_{\mathbf{s}\in\mathcal{E}_{\text{s-ch}}^+} \prod_{i=1}^{\beta_\mathbf{s}}\eta_{\mathbf{s}i}}, \quad f_{\mathrm{3d}}=\lim_{\{p^0_{\mathbf{s}i}\}\rightarrow 0}\tilde{f}_{\mathrm{3d}}\;.
\end{equation}
The \gls{LU} representation associated with the function of which we are taking the limit now only involves single poles. Furthermore, if the shifts $p_{ej}^0$ are small enough, then the causal flow for the thresholds eq.~\eqref{eq:shifted_thresholds} can be set to be the same as the causal flow for the thresholds with shifts set to zero. Let $\phi_t$ be such a causal flow. We will exchange the limiting procedure associated with the Local Unitarity representation with that associated with the fictitious momenta going to zero. %

We now consider the value of $\tilde{f}_{\mathrm{3d}}$ along the flow lines identified by $\phi_t$, and cast it into a form analogous to eq.~\eqref{eq:composition_supergraph},
\begin{equation}
    \tilde{f}_{\mathrm{3d}}\circ\phi_t= \frac{(f\circ\phi_t) \, w_t}{\prod_{\mathbf{s}\in\mathcal{E}_{\text{s-ch}}^+} \prod_{i=1}^{\beta_\mathbf{s }}(t-t^\star_{\mathbf{s}i})}, \quad w_t=\prod_{\mathbf{s}\in\mathcal{E}_{\text{s-ch}}^+} \prod_{i=1}^{\beta_\mathbf{s }} \frac{(t-t^\star_{\mathbf{s}i})}{\eta_{\mathbf{s}i}}.
\end{equation}
where $t^\star_{\mathbf{s}i}$ is the solution in $t$ of the equation $\eta_{\mathbf{s}i}\circ\phi_t=0$. Because of the way we introduced the fictitious shifts, it is straightforward to determine the one-to-one relation between $t^\star_{\mathbf{s}i}$ and $p_{\mathbf{s}i}^0$,
\begin{equation}
\label{eq:rel_shift_t_star}
    p_{\mathbf{s}i}^0(t^\star_{\mathbf{s}i})=-\eta_\mathbf{s}\circ\phi_{t^\star_{\mathbf{s}i}},
\end{equation}
which is also invertible for small $p_{\mathbf{s}i}^0$. We are ready to construct the \gls{LU} representation for $\tilde{f}_{\mathrm{3d}}$. $\tilde{f}_{\mathrm{3d}}$ now only features simple poles in $t$, and we can directly use the formula from ref.~\cite{2021}. Let us focus, in particular, on the sum of all residues obtained by fixing $\mathbf{s}$ and varying $i$; from this result, after summing over all thresholds $\mathbf{s}$, we will directly obtain the desired result of eq.~\eqref{eq:LU_formula}. We have
\begin{equation}
\label{eq:divided_differences}
    \tilde{\sigma}_\mathbf{s}=\sum_{i=1}^{\beta_\mathbf{s}}\lim_{t\rightarrow t^\star_{\mathbf{s}i}}(t-t^\star_{\mathbf{s}i})\tilde{f}_{\mathrm{3d}}\circ\phi_t=\sum_{i=1}^{\beta_\mathbf{s}}\frac{y_{\mathbf{s},i}}{\prod_{\substack{j=1 \\ j\neq i}}^{\beta_\mathbf{s}} (t^\star_{\mathbf{s}i}-t^\star_{\mathbf{s}j})},
    \quad y_{\mathbf{s},i}= \lim_{t\rightarrow t^\star_{\mathbf{s}i}}\prod_{\substack{j=1 \\ j\neq i}}^{\beta_\mathbf{s}}(t-t^\star_{\mathbf{s}j})\tilde{f}_{\mathrm{3d}}.
\end{equation}
The expression on the right-hand-side of eq.~\eqref{eq:divided_differences} is a divided difference in the variables $t^\star_{\mathbf{s}1},...,t^\star_{\mathbf{s}\beta_\mathbf{s}}$ (see analogous discussion in sect. 2.2 of ref.~\cite{Capatti:2020ytd}). Finally, we take the limit of all shifts $p_{\mathbf{s}i}^0$ going to zero. Observe that eq.~\eqref{eq:rel_shift_t_star} shows that the limit of $p_{\mathbf{s}i}$ going to zero is equivalent to the limit of $t^\star_{\mathbf{s}i}$ going to $t^\star_{\mathbf{s}}$. We then have
\begin{equation}
\label{eq:proof_of_LU_formula}
    \lim_{\{p_{\mathbf{s}i}^0\}_i\rightarrow 0} \tilde{\sigma}_\mathbf{s}=\lim_{t\rightarrow t^\star_{\mathbf{s}}}\frac{1}{(\beta_\mathbf{s}-1)!}\frac{\mathrm{d}^{\beta_\mathbf{s}-1}}{\mathrm{d}t^{\beta_\mathbf{s}-1}} (t-t^\star_{\mathbf{s}})^{\beta_\mathbf{s}} f_{\text{3d}}(G_\mathrm{r}),
\end{equation}
obtained directly from the application of divided differences. Eq.~\eqref{eq:proof_of_LU_formula} finalises our argument supporting the explicit \gls{LU} expression of eq.~\eqref{eq:LU_formula}. It shows that the correct way to interpret Cutkosky cuts that go through raised propagators is that of higher-order residues in the variable that is used to approach the corresponding threshold. 

The overarching argument that directly guides the construction of eq.~\eqref{eq:LU_formula} is that interference diagrams are weighted residues of a parent diagram, the supergraph. Very much related to this principle is that of \gls{IR}-finiteness. The \gls{KLN} cancellation pattern exhibited here and in ref.~\cite{2021} is a direct consequence of the pattern of divided differences that arises from taking different residues of one single integrand. 

In particular, we stress that the local \gls{FSR}-finiteness of $\sigma_\mathrm{d}$ as defined in eq.~\eqref{eq:LU_formula} is guaranteed by the work of ref.~\cite{2021} since we constructed eq.~\eqref{eq:LU_formula} as a specific limit of the \gls{LU} formula of ref.~\cite{2021}. This implies that there is a deep relationship between higher-order poles, the derivatives that are needed to compute their residues and local IR-finiteness.

\subsection{Interference diagrams (distributional approach)}
\label{sec;generalised_cutting_rules}

Although the \gls{LU} expression of eq.~\eqref{eq:LU_formula} is compact and well-suited to prove its \gls{FSR}-finiteness, it is difficult to use it in practice because it requires explicitly taking a limit in $t$ of a function. It is much easier then, to construct the integrand \textit{after} that limit has been taken. In particular, we will show now that it is possible to obtain eq.~\eqref{eq:LU_formula} starting from interference diagrams directly, and finding the correct distributional cutting rules for raised propagators. Because we started motivating eq.~\eqref{eq:LU_formula} with the three-dimensional representation of a Feynman diagram with raised propagators, we can already anticipate that these cutting rules involve derivatives in the energy components of the loop momenta of the supergraph. It is also manifest from eq.~\eqref{eq:LU_formula} that these cutting rules should ultimately also involve derivatives in the $t$ variable.

\subsubsection{Generalised cutting rules}

We start by defining a distribution whose action on a test function gives the divided differences of the test function $f$ at given loci $x_1,...,x_n$. It is defined by recursion as
\begin{equation}
\label{eq:delta_divided_difference}
    \delta[x-x_1,...,x-x_n]=\frac{\delta[x-x_2,...,x-x_n]-\delta[x-x_1,...,x-x_{n-1}]}{x_n-x_{1}}, \quad \delta[x]=\delta(x).
\end{equation}
We can then define a new distribution $\delta^{(n)}[x]$ as the limit of that in eq.~\eqref{eq:delta_divided_difference} when $x_i=x_j$ for any $i$ and $j$. In this limit both the denominator and numerator of eq.~\eqref{eq:delta_divided_difference} vanish. \begin{equation}
\label{eq:our_distribution}
    \delta^{(n)}[x]=\lim_{\{x_i\}_{i=1}^n\rightarrow 0}\delta[x-x_1,...,x - x_n].
\end{equation}
Carrying out the distributional limit, we have that the action of $\delta^{(n+1)}[x]$ produces the $n$-th coefficient of the Taylor expansion of $f$, 
\begin{equation}
\label{eq:our_delta}
    \int \mathrm{d}x \,  \delta^{(n+1)}[x] f(x)=  \frac{1}{n!}\frac{\mathrm{d}^{n}f}{\mathrm{d}x^{n}}\Bigg|_{x=0}\;,
\end{equation}
that is, the effect of $\delta^{(n)}[x]$ is to compute the $n$-th derivative of the test function divided by the factorial of $n$.
We observe that $\delta^{(n)}[x]$ does not correspond to the usual definition of a derivative of a Dirac delta function. Instead, $\delta^{(n)}[x]$ implements divided differences of the test function $f$, in the limit of all the variables defining the divided difference converging to the same value. This limit is computed explicitly on the right-hand-side of eq.~\eqref{eq:our_delta}.  The extra combinatorial factor $1/n!$ arising from divided differences is fundamental in order to obtain the equivalence with the application of the residue theorem given in sect.~\ref{sec:raised_lu}. 

Two properties of the $\delta^{(n)}[x]$ distribution are especially important. The first one concerns the action of the derivative on $\delta^{(n)}[x]$
\begin{equation}
\label{eq:generalised_delta_composition_rule}
    \frac{\mathrm{d}}{\mathrm{d}x}\delta^{(n)}[x]=n \, \delta^{(n+1)}[x],
\end{equation}
i.e., the derivative simply raises the degree of the $\delta^{(n)}$ distribution by one. The second property is the composition rule
\begin{equation}
\label{eq:delta_composition_rule}
    \delta^{(n)}[g(x)]= \delta^{(n)}[x-g^{-1}(0)]w_n[g](x),
\end{equation}
with
\begin{equation}
    w_n[g](x)=\text{sgn}[g'(g^{-1}(0))]\lim_{x'\rightarrow x}\left[\frac{(x'-g^{-1}(0))}{g(x')}\right]^{n}
\end{equation}
assuming $g(x)$ has a unique zero located at $g^{-1}(0)$, and $w_1[g](x)$ corresponds to the composition rule of the Dirac delta distribution.
 Note that while $w_n[g](g^{-1}(0))=1/g'(g^{-1}(0))^n$, it is incorrect to directly substitute $w_n[g](x)$ with its value at $x=g^{-1}(0)$ in eq.~\eqref{eq:delta_composition_rule}, as $\delta^{(n)}$ will eventually set $x=g^{-1}(0)$ only \textit{after} taking $n-1$ derivatives of it.  Even though the function $w_n[g]$ is given by a limit, it can be explicitly written in terms of a power series in $x$ around $g^{-1}(0)$, so that its derivatives become trivial to compute (see example in eq.~\ref{eq:wn_explicit}). The expression one obtains is not compact, so we do not report it here. Alternatively, one can use a change of variables in the integration in order to show that
 \begin{equation}
     \int \mathrm{d}x \delta^{(n)}[g(x)] f(x)=\frac{1}{(n-1)!}\frac{\mathrm{d}^{n-1}}{\mathrm{d}y^{n-1} } \left[\frac{f(g^{-1}(y))}{|g'(g^{-1}(y))|}\right]_{y=0},
 \end{equation}
 where we stress again that the assignment $y=0$ takes place \emph{after} the $n-1$ derivatives have been evaluated.
 The result for $g$ having many zero can be trivially obtained by recalling that divided differences are linear. Finally, we are ready to present the modified cutting rules. A propagator raised to a power $n$ must be substituted with the distribution
\begin{equation}
\label{eq:cutkosky_rule}
\raisebox{-0.3cm}{$\underbrace{
\resizebox{3.cm}{!}{%
\begin{tikzpicture}

     \node[main node4]  (1) {};
     \node[] (L1) [left = 0.1cm of 1] {};
     \node[]  (A2) [right = 1cm of 1] {};
     
     \node[]  (A3) [right = 1.25cm of 1] {};
     \node[main node4]  (A4) [right = 2.5cm of 1] {};
     
     \node[main node4]  (B1) [right = 0.6cm of 1] {};
     \node[main node4]  (B2) [left = 0.6cm of A4] {};
     
     \node[]  (M2) [left = 0.95cm of A4]{$...$};
     
     \node[] (X1) [above right = 0.5cm and 1.13cm of 1] {};
     \node[] (LABEL) [right = 0.001cm of X1] {$\color{red}\boldsymbol{\pm}$};
    \node[] (X2) [below right = 0.5cm and 1.13cm of 1] {};
    
    \draw[thick, red] (X1) to[out= -60, in=120] (X2);

\begin{feynman}
\draw[thick, momentum=\(p\)] (1) to (B1);
\draw[thick, momentum=\(p\)] (B2) to (A4);

\draw[thick] (B1) to (A2);
\draw[thick] (B2) to (A3);

\end{feynman}

\end{tikzpicture}

}
}_{n \ \mathrm{ times}}$} \ = -2\pi\mathrm{i} \, \delta^{(n, \pm)}[p^2-m^2]= \ \mp 2\pi\mathrm{i}\frac{ \delta^{(n)}[p^0\pm E_{\vec{p}}] }{(p^0\pm E_{\vec{p}})^n},
\end{equation}
which enforces on-shellness of $p$ through the modified distribution $\delta^{(n-1)}[x]$, effectively computing the residue corresponding to the factor $(p^0\mp E_{\vec{p}})^{-n}$ obtained by factoring $(p^2-m^2)^{-n}$. The denominator then features the remaining part of the raised propagator, that is $(p^0\pm E_{\vec{p}})^{-n}$.

\subsubsection{Example of application of the generalised cutting rules}
\label{sec:generalised_cutting_rules_example}

We will now apply the modified cutting rules to an example interference diagram and show that it reproduces what is expected from the residue theorem. Let us start from the following example supergraph:

\begin{equation}
\label{eq:supergraph_ex}
I \ = \ \raisebox{-1.cm}{\resizebox{5.5cm}{!}{%
\begin{tikzpicture}[baseline=-2cm]
    
    \node[main node2]  (1) {$G_{\mathrm{L}}$};
    \node[main node2]  (3) [right = 2.4cm of 1]  {$G_{\mathrm{R}}$};

     \node[main node4]  (A1) [right = 0.6cm of 1] {};
     \node[main node4]  (A2) [right = 1.6cm of 1] {};

     \node[main node4]  (B1) [above right = -0.cm and 1.6cm of 1] {};

        \begin{feynman}
        \vertex[left=2.5cm of 1] (att1);
        \vertex[right=2.5cm of 3] (att2);
        \draw[thick, momentum=\(q\)] (att1) to (1);
        \draw[thick, momentum=\(q\)] (3) to (att2);
    	\draw[thick, momentum=\(p_2\)] (1) to[out= 30, in=150] (3);
    	\draw[thick, momentum=\(p_3\)] (1) to[out= 0, in=180] (3);
    	\draw[thick, momentum=\(p_1\)] (1) to[out= -30, in=-150] (3);
    	\end{feynman}

\end{tikzpicture}
}}
\end{equation}

The mathematical expression for the interference diagram obtained from $I$ by cutting the edges labelled $p_1$, $p_2$ and $p_3$ using the generalised cutting rules of eq.~\eqref{eq:cutkosky_rule} is:
\vspace{-0.2cm}
\begin{equation}
\label{eq:generic_raised2}
\hspace{-0.1cm}\raisebox{-0.4cm}{\resizebox{2.1cm}{!}{%
\begin{tikzpicture}[baseline=-2.5cm]
    
    \node[main node2]  (1) {};
    \node[main node2]  (3) [right = 2.4cm of 1]  {};

     \node[main node4]  (A1) [right = 0.6cm of 1] {};
     \node[main node4]  (A2) [right = 1.6cm of 1] {};

     \node[main node4]  (B1) [above right = -0.cm and 1.6cm of 1] {};
     
     \node[] (L1) [below right = -1.cm and 0.4cm of 1] {};
     \node[] (L2) [below right = 0.2cm and 1.3cm of 1] {};
     \node[] (L3) [above right = 0.2cm and 1.3cm of 1] {};

     \node[] (Cut1) [below right = 0.6cm and 2cm of 1] {};
     \node[] (Cut2) [above right = 0.6cm and 0.7cm of 1] {};

        \begin{feynman}
        \vertex[left=2.5cm of 1] (att1);
        \vertex[right=2.5cm of 3] (att2);

        \draw[line width=0.07cm] (att1) to (1);
        \draw[line width=0.07cm] (3) to (att2);
    	\draw[line width=0.07cm] (1) to[out= 30, in=150] (3);
    	\end{feynman}
    	\draw[line width=0.07cm] (1) to[out= 0, in=180] (3);
    	\draw[line width=0.07cm] (1) to[out= -30, in=-150] (3);
    	
    	\draw[line width=0.07cm,red] (Cut1) to[out= 120, in=-60] (Cut2);

\end{tikzpicture}

}}\ = \ \int \left[\prod_{i=1}^3 \mathrm{d}^4p_i\frac{\delta^{(i)}[p_i^0+E_i]}{(p_i^0-E_i)^i}\right] \delta\left(\sum_{i=1}^3p_i-q\right) G_\textrm{L} G_\textrm{R}^\dagger f_3.
\end{equation}
where $f_3$ is a test function of the momenta $p_1$, $p_2$ and $p_3$, and for simplicity we normalised the integration measure of $p_i$ by $-(2\pi \mathrm{i})$.
 Note that we aligned the labelling of the edges with the power of the propagators in order to obtain a more compact expression (i.e. for this example the propagator with momentum $p_i$ is raised to power $i$). In order to solve all the delta distributions explicitly, we must choose a particular order in which to solve them. We start with the delta function enforcing energy-momentum conservation, which we solve using the $p_3$ integration. Next, we solve the Dirac delta function enforcing on-shellness of $p_1$ using the $p_1^0$ integration, then the delta function enforcing on-shellness of $p_2$ using the $p_2^0$ integration, and finally the delta function enforcing on-shellness of $p_3$ using the causal flow, i.e. the integration over the parameter $t$. This particular ordering choice is arbitrary, and any other choice would be locally equivalent. We begin by solving the overall momentum-conservation delta function and the delta function enforcing on-shellness of $p_1$:
\begin{equation}
\hspace{-0.1cm}\raisebox{-0.4cm}{\resizebox{2.1cm}{!}{%
\begin{tikzpicture}[baseline=-2.5cm]
    
    \node[main node2]  (1) {};
    \node[main node2]  (3) [right = 2.4cm of 1]  {};

     \node[main node4]  (A1) [right = 0.6cm of 1] {};
     \node[main node4]  (A2) [right = 1.6cm of 1] {};

     \node[main node4]  (B1) [above right = -0.cm and 1.6cm of 1] {};
     
     \node[] (L1) [below right = -1.cm and 0.4cm of 1] {};
     \node[] (L2) [below right = 0.2cm and 1.3cm of 1] {};
     \node[] (L3) [above right = 0.2cm and 1.3cm of 1] {};

     \node[] (Cut1) [below right = 0.6cm and 2cm of 1] {};
     \node[] (Cut2) [above right = 0.6cm and 0.7cm of 1] {};

        \begin{feynman}
        \vertex[left=2.5cm of 1] (att1);
        \vertex[right=2.5cm of 3] (att2);

        \draw[line width=0.07cm] (att1) to (1);
        \draw[line width=0.07cm] (3) to (att2);
    	\draw[line width=0.07cm] (1) to[out= 30, in=150] (3);
    	\end{feynman}
    	\draw[line width=0.07cm] (1) to[out= 0, in=180] (3);
    	\draw[line width=0.07cm] (1) to[out= -30, in=-150] (3);
    	
    	\draw[line width=0.07cm,red] (Cut1) to[out= 120, in=-60] (Cut2);

\end{tikzpicture}

}} \ =
    \int  \mathrm{d}^3\vec{p}_1\mathrm{d}^4p_2 \frac{ \delta^{(2)}[p_2^0-E_2]\delta^{(3)}[-p_2^0+q^0-E_3-E_1] f_3}{2E_1(-p_2^0+q^0+E_3-E_1)^3(p_2^0+E_2)^2}G_\textrm{L} G_\textrm{R}^\dagger \Bigg|_{p_1^0=E_1}.
\end{equation}
The next delta function is then solved in the variable $p_2^0$. This time the generalised cutting rule yields a derivative in that variable; specifically
\begin{equation}
\hspace{-0.2cm}\raisebox{-0.4cm}{\resizebox{2.1cm}{!}{%
\begin{tikzpicture}[baseline=-2.5cm]
    
    \node[main node2]  (1) {};
    \node[main node2]  (3) [right = 2.4cm of 1]  {};

     \node[main node4]  (A1) [right = 0.6cm of 1] {};
     \node[main node4]  (A2) [right = 1.6cm of 1] {};

     \node[main node4]  (B1) [above right = -0.cm and 1.6cm of 1] {};
     
     \node[] (L1) [below right = -1.cm and 0.4cm of 1] {};
     \node[] (L2) [below right = 0.2cm and 1.3cm of 1] {};
     \node[] (L3) [above right = 0.2cm and 1.3cm of 1] {};

     \node[] (Cut1) [below right = 0.6cm and 2cm of 1] {};
     \node[] (Cut2) [above right = 0.6cm and 0.7cm of 1] {};

        \begin{feynman}
        \vertex[left=2.5cm of 1] (att1);
        \vertex[right=2.5cm of 3] (att2);

        \draw[line width=0.07cm] (att1) to (1);
        \draw[line width=0.07cm] (3) to (att2);
    	\draw[line width=0.07cm] (1) to[out= 30, in=150] (3);
    	\end{feynman}
    	\draw[line width=0.07cm] (1) to[out= 0, in=180] (3);
    	\draw[line width=0.07cm] (1) to[out= -30, in=-150] (3);
    	
    	\draw[line width=0.07cm,red] (Cut1) to[out= 120, in=-60] (Cut2);

\end{tikzpicture}

}} \ =
    \int \mathrm{d}^3\vec{p}_1\mathrm{d}^3\vec{p}_2 \frac{\mathrm{d}}{\mathrm{d}p_2^0}\left[\frac{ \delta^{(3)}[-p_2^0+q^0-E_3-E_1] \left[G_\textrm{L} G_\textrm{R}^\dagger f_3\right]_{p_1^0=E_1} }{2E_1(-p_2^0+q^0+E_3-E_1)^3(p_2^0+E_2)^2}\right]_{p_2^0=E_2}.
\end{equation}
When unfolding the action of this derivative explicitly, we obtain two terms, one corresponding to the derivative acting on the remaining generalised Dirac delta distribution, and the other resulting from the derivative acting on all other terms. The derivative of the generalised delta distribution is obtained following eq.~\eqref{eq:generalised_delta_composition_rule}:
\begin{align}
\label{eq:after_ltd}
    \raisebox{-0.4cm}{\resizebox{2.1cm}{!}{%
\begin{tikzpicture}[baseline=-2.5cm]
    
    \node[main node2]  (1) {};
    \node[main node2]  (3) [right = 2.4cm of 1]  {};

     \node[main node4]  (A1) [right = 0.6cm of 1] {};
     \node[main node4]  (A2) [right = 1.6cm of 1] {};

     \node[main node4]  (B1) [above right = -0.cm and 1.6cm of 1] {};
     
     \node[] (L1) [below right = -1.cm and 0.4cm of 1] {};
     \node[] (L2) [below right = 0.2cm and 1.3cm of 1] {};
     \node[] (L3) [above right = 0.2cm and 1.3cm of 1] {};

     \node[] (Cut1) [below right = 0.6cm and 2cm of 1] {};
     \node[] (Cut2) [above right = 0.6cm and 0.7cm of 1] {};

        \begin{feynman}
        \vertex[left=2.5cm of 1] (att1);
        \vertex[right=2.5cm of 3] (att2);

        \draw[line width=0.07cm] (att1) to (1);
        \draw[line width=0.07cm] (3) to (att2);
    	\draw[line width=0.07cm] (1) to[out= 30, in=150] (3);
    	\end{feynman}
    	\draw[line width=0.07cm] (1) to[out= 0, in=180] (3);
    	\draw[line width=0.07cm] (1) to[out= -30, in=-150] (3);
    	
    	\draw[line width=0.07cm,red] (Cut1) to[out= 120, in=-60] (Cut2);

\end{tikzpicture}

}} & \ = \ \int  \mathrm{d}^3\vec{p}_1\mathrm{d}^3\vec{p}_2  \frac{ 3 \delta^{(4)}\left[q^0-\sum_{i=1}^3 E_i\right]\left[G_\textrm{L} G_\textrm{R}^\dagger f_3\right]_{p_1^0=E_1, \, p_2^0=E_2} }{2E_1(q^0+E_3-E_2-E_1)^3(2E_2)^2} \\ 
&\hspace{-1cm}+\delta^{(3)}\left[q^0-\sum_{i=1}^3 E_i\right]\frac{\mathrm{d}}{\mathrm{d}p_2^0}\left[\frac{ \left[G_\textrm{L} G_\textrm{R}^\dagger f_3\right]_{p_1^0=E_1} }{2E_1(-p_2^0+q^0+E_3-E_1)^3(p_2^0+E_2)^2}\right]_{p_2^0=E_2}.
\end{align}
Such distinction is important because the two terms have different pole orders, and thus require a different number of derivatives when acted upon by the last generalised delta distribution.

This final Dirac delta distribution must then be solved using the causal flow. Let us introduce the one-parameter group $\phi_t$, solution of the \gls{ODE} in eq.~\eqref{eq:causal_flow_ODE}, by first introducing a resolution of the identity by rewriting $1$ as the integral of a normalised function $h(t)$. Next, we change variables from $\vec{p}_1, \ \vec{p}_2$ to $(\phi_{t})_1, \ (\phi_{t})_2$. We can then solve the remaining delta function using the variable $t$:
\begin{align}
\label{eq:cutkosky_rule_solved}
    \raisebox{-0.4cm}{\resizebox{2.1cm}{!}{%
\begin{tikzpicture}[baseline=-2.5cm]
    
    \node[main node2]  (1) {};
    \node[main node2]  (3) [right = 2.4cm of 1]  {};

     \node[main node4]  (A1) [right = 0.6cm of 1] {};
     \node[main node4]  (A2) [right = 1.6cm of 1] {};

     \node[main node4]  (B1) [above right = -0.cm and 1.6cm of 1] {};
     
     \node[] (L1) [below right = -1.cm and 0.4cm of 1] {};
     \node[] (L2) [below right = 0.2cm and 1.3cm of 1] {};
     \node[] (L3) [above right = 0.2cm and 1.3cm of 1] {};

     \node[] (Cut1) [below right = 0.6cm and 2cm of 1] {};
     \node[] (Cut2) [above right = 0.6cm and 0.7cm of 1] {};

        \begin{feynman}
        \vertex[left=2.5cm of 1] (att1);
        \vertex[right=2.5cm of 3] (att2);

        \draw[line width=0.07cm] (att1) to (1);
        \draw[line width=0.07cm] (3) to (att2);
    	\draw[line width=0.07cm] (1) to[out= 30, in=150] (3);
    	\end{feynman}
    	\draw[line width=0.07cm] (1) to[out= 0, in=180] (3);
    	\draw[line width=0.07cm] (1) to[out= -30, in=-150] (3);
    	
    	\draw[line width=0.07cm,red] (Cut1) to[out= 120, in=-60] (Cut2);

\end{tikzpicture}

}} \ = \ 
-\frac{1}{2}\int & \mathrm{d}^3\vec{p}_1\mathrm{d}^3\vec{p}_2
\frac{\mathrm{d}^3}{\mathrm{d}t^3}\left[\frac{ \left[h(t) \, w_4(t) \,  \mathbb{J}\phi_t \,  
G_\textrm{L} G_\textrm{R}^\dagger f_3\right]_{p_1^0=E_1, \, p_2^0=E_2} }{2E_1(q^0+E_3-E_2-E_1)^3(2E_2)^2}\right]_{t=t^\star} \nonumber\\ & 
\hspace{-1.2cm}+\frac{\mathrm{d}^2}{\mathrm{d}t^2}\left[\frac{\mathrm{d}}{\mathrm{d}p_2^0}\left[
\frac{\left[w_3(t) \, h(t) \,  \mathbb{J}\phi_t \, G_\textrm{L} G_\textrm{R}^\dagger f_3\right]_{p_1^0=E_1}}
{2E_1(-p_2^0+q^0+E_3-E_2)^3(p_2^0+E_2)^2}\right]_{p_2^0=E_2}\right]_{t=t^\star}.
\end{align}
where we introduced the Jacobian of the change of variables $\mathbb{J}\phi_t$ and the derivative factor $w_n(t)$ arising from solving the delta distribution in the variable $t$, which is also subject to the derivative in $t$ (see eq.~\eqref{eq:delta_composition_rule}). The explicit expression of the $w_n$ derivative factor is:
\begin{equation}
\label{eq:wn_explicit}
    w_n(t)=\left[\frac{t-t^\star}{(E_1+E_2+E_3-q^0)\circ\phi_t}\right]^n,
\end{equation}
and $t^\star$ is the unique value of $t$ such that $(E_1+E_2+E_3-q^0)\circ\phi_t=0$.
This concludes the explicit solving of all the delta distributions associated with the phase space measure of the example interference diagram introduced in eq.~\eqref{eq:generic_raised2}.

\subsection{Dual number representations and efficient computation of derivatives}
\label{sect:dual_numbers_section}

We saw in the previous sections that the application of the \glss{LU} representation of differential cross-section to supergraphs with repeated edges requires taking derivatives of amplitudes.
While the derivatives could be computed symbolically at the time of generating the code for the \glss{LU} integrands, this would come at the cost of a significant overhead of both generation and run time. A far more elegant and efficient solution is to compute these derivatives using a numerical implementation of them based on the chain-rule, which can be achieved by using multivariate dual numbers (e.g. see survey of auto-differentiation tools and method in ref.~\cite{JMLR:v18:17-468}). This procedure is both \emph{exact} and \emph{numerically stable}. Note that such dual numbers were already used in the context of ref.~\cite{2020} for efficiently computing the Jacobian of complicated contour deformations and are also commonly used in the context of machine learning when implementing the \emph{automatic differentiation} necessary for back-propagation algorithms. 
We first recall here the formal construction of dual numbers. We start by considering a nilpotent object with degree $n+1$
\begin{equation}
\label{eq:truncation}
    \varepsilon^{n+1}=0.
\end{equation}
We then consider all polynomials that can be obtained from such a nilpotent element. We let $\mathcal{D}(n)$ be the set of all such polynomials. Its explicit definition reads
\begin{equation}
\label{eq:monovoriate_dual_definition}
    \mathcal{D}(n)=\left\{\,\sum_{k=0}^n \frac{c_k}{k!}\varepsilon^k \ \Bigg| \ c_i\in\mathbb{C}, \ \forall i=0,\ldots,n \right\}.
\end{equation}
An element $\bar{x}\in\mathcal{D}(n) $ is called a dual. As we will see, duals can be used to compute up to the $n$-th derivative of $f$. In particular, the truncation rule established by eq.~\eqref{eq:truncation} allows to implements duals in $\mathcal{D}(n)$ on a computer by simply storing the $n$ coefficients $c_i$ in an array.
The action of a function $f$ originally defined on complex numbers is extended to an action on a generic dual number $\bar{x}$ through the Taylor expansion of $f$ around $\bar{x}|_{\varepsilon=0}$:
\begin{equation}
\label{eq:monovariate_dual_application}
f(\bar{x}) = \sum_{k=0}^\infty \frac{\left( \bar{x}-\bar{x}|_{\varepsilon=0} \right)^k }{k!} f^{(k)}( \bar{x}|_{\varepsilon=0} ) = \sum_{k=0}^\infty \frac{1}{k!} f^{(k)} \left( c_0 \right) \left(\sum_{i=1}^\infty \frac{c_i}{i!} \varepsilon^i\right)^k.
\end{equation}
When applying the nilpotency condition of eq.~\eqref{eq:truncation}, we can truncate the expansion explicitly as follows:
\begin{equation}
\label{eq:truncated_monovariate_dual_application}
    f(\bar{x}\in \mathcal{D}(n)) = \sum_{k=0}^n \frac{1}{k!} \left( \sum_{i=0}^k B_{n,k}(c_1,\ldots,c_{n-k+1})  f^{(k)}\left( c_0 \right) \right) \varepsilon^k\;,
\end{equation}
where $B_{n,k}$ is a Bell polynomial.
The two key properties of this definition of the application of functions to dual numbers is that it follows the composition rule and satisfies:
\begin{equation}
    f(c_0+\varepsilon) = \sum_{k=0}^n \frac{1}{k!} f^{(k)}\left( c_0 \right) \varepsilon^k\;,
\end{equation}
thus allowing one to read the $j$-th derivative of the function $f$ evaluated at $c_0$ directly from the coefficient multiplying the $j$-th power of the dual $\varepsilon$. 

So far, dual numbers do not seem especially useful, since at first glance it seems like implementing the action of an arbitrary function $f$ on duals in $\mathcal{D}(n)$ would anyway require the prior symbolic computation of all the $n$-th first derivatives of $f$ in order to construct the coefficients of eq.~\eqref{eq:truncated_monovariate_dual_application}.
However, a much more efficient implementation can be obtained by realising that most complicated functions typically implemented on a computer are composites of simple elementary ones.
One can then use eq.~\eqref{eq:truncated_monovariate_dual_application} to construct the implementation of all elementary arithmetic and intrinsic operations $f$ (e.g. $\sin{x}$, $\cos{x}$, $\sqrt{x}$, etc.) from their trivial derivatives. Since this construction satisfies the composition rule, dual numbers provide a fast, exact and automated numerical computation of up to the $n$-th derivative of any complicated function that corresponds to iterated compositions of elementary ones. 
In that case it is useful to extend eq.~\eqref{eq:monovariate_dual_application} to a function with $v>1$ arguments that are functions of $x$. Multiplication is a prime example of a $2$-ary elementary function necessary for implementing useful function compositions.
Applying a $v$-ary function to $v$ duals can be obtained using Taylor expansions again:
\begin{equation}
\label{eq:monovariate_v_ary_dual_application}
f(\bar{x}_1,\ldots,\bar{x}_v) = \sum_{k_1,\ldots,k_v=0}^\infty \left(\prod_{j=1,v} \frac{\left( \bar{x}_j-\bar{x}_j|_{\varepsilon=0} \right)^{k_j} }{k_j!} \right) f^{(k_1,\ldots,k_v)} \left( \bar{x}_1|_{\varepsilon=0},\ldots,\bar{x}_v|_{\varepsilon=0} \right),
\end{equation}
where $f^{(k_1,\ldots,k_m)}$ denotes the partial derivative with $k_i$ derivatives in the variable $x_i$. The infinite sum of eq.~\eqref{eq:monovariate_v_ary_dual_application} is truncated once the powers of dual numbers $(\bar{x}_j-\bar{x}_j|_{\varepsilon=0})^{k_j}$ are expanded and the truncation rule $\varepsilon^{n+1}=0$ is imposed. Let us stress again the two fundamental steps required to successfully apply dual numbers to the numerical computation of derivatives of a function $f$:
\begin{itemize}
    \item Hardcode the action of all elementary ($v$-ary) functions on a $v$-uplet of dual numbers, according to eq.~\eqref{eq:truncated_monovariate_dual_application} and eq.~\eqref{eq:monovariate_v_ary_dual_application}.
    \item Write $f$ as a composition of elementary functions, and evaluate $f(x+\varepsilon)$.
\end{itemize}
This straightforward two-step procedure summarises the discussion on the use of dual numbers to compute derivatives of a single-variable function. To conclude we propose a simple example of the application of this technology to the computation of first and second derivative of a function.

\vspace{0.2cm}
\begin{center}
\begin{minipage}{14cm}
\paragraph{Example:} Let $f(x)=x\sin(x)$, and write $f(x)=g_\times(x,s(x))$ with $g_\times(x,y)=xy$ and $s(x)=\sin(x)$. In order to compute the derivative of $f$, we consider the following two rules for the dual evaluations of the sine function and the product function at dual numbers $\bar{x}=c_0+c_1\varepsilon + c_2\varepsilon^2/2$ and $\bar{x}'=c_0'+c_1'\varepsilon + c_2' \varepsilon^2/2$:
\begin{align}
    s\left(\bar{x}\right)=\sin(c_0)+c_1 \cos(c_0) \varepsilon +\frac{1}{2}\left(c_2 \cos(c_0)-c_1 \sin(c_0)\right)\varepsilon^2, \\
    g_\times(\bar{x},\bar{x}')=c_0 c_0'+(c_0 c_1'+c_0' c_1)\varepsilon+\left(c_1 c_1'+\frac{c_0 c_2'}{2}+ \frac{c_0'c_2}{2}\right) \varepsilon^2.
\end{align}
This allows us to compute the full derivative of $f$ knowing the expansion of the elementary functions $s$ and $g_\times$. In particular, this yields
\begin{align}
    f(x+\varepsilon)&=g_\times(x+\varepsilon, s(x+\varepsilon))=g_\times\left(x+\varepsilon, \sin(x)+\cos(x)\varepsilon -\frac{1}{2}\sin(x)\varepsilon^2\right)\nonumber\\
    &=x\sin(x)+(\sin(x)+x\cos(x))\varepsilon+\frac{1}{2}(2\cos(x)-x\sin(x))\varepsilon^2.
\end{align}
We see that the coefficients of the power series reproduce the derivatives of $f$ divided by $1/n!$. In other words, it is the Taylor expansion of $f(x+\varepsilon)$ around $\varepsilon=0$.
\end{minipage}
\end{center}
\vspace{0.2cm}

\noindent This construction thus far only allows us to compute derivatives in a single variable, but its generalisation to what we call multivariate duals is straightforward. 
Given a vector of positive integers $\vec{n}\in\mathbb{N}^m$, we consider $m$ objects $\varepsilon_i$, $i=1,\ldots,m$ with the following truncation rules:
\begin{equation}
\label{eq:dual_multivariate_truncation}
 \varepsilon_{i}^{n_i+1}=0\;,
\end{equation}
and we will write $\vec{\varepsilon}=\{\varepsilon_1,\ldots,\varepsilon_m\}$.
These $m$-variate duals will be used to support the computation of up to $n_i$ derivative in each of the $m$ variables $x_i$. For this purpose, we must consider all multivariate polynomials in the $\varepsilon_i$ objects
\begin{equation}
\label{eq:multivariate_dual}
    \mathcal{D}(\vec{n})=\left\{\sum_{k_1=0}^{n_1} \ldots \sum_{k_m=0}^{n_m} \frac{c_{k_1\ldots k_m}}{k_1!\ldots k_m!} \prod_{i=1}^m \varepsilon_{i}^{k_i} \ \Bigg| \ c_{k_1\cdots k_m}\in\mathbb{C}, \ \forall k_i=0,\ldots,n_i, \ \forall i=1,\ldots,m\right\}.
\end{equation}
An element $\mathbf{\bar{x}}\in \mathcal{D}(\vec{n})$ is then a $m$-variate dual number. Given a $v$-ary function $f$, its action on $v$ $m$-variate dual numbers $\mathbf{\bar{x}}_i\in \mathcal{D}(\vec{n})$ is obtained in complete analogy w.r.t the monovariate case of eq.~\eqref{eq:monovariate_v_ary_dual_application}:
\begin{equation}
\label{eq:mvariate_v_ary_dual_application}
    f(\mathbf{\bar{x}}_1,\ldots,\mathbf{\bar{x}}_v) = \sum_{k_1,\ldots,k_v=0}^\infty \left(\prod_{j=1,v} \frac{\left( \mathbf{
    \bar{x}}_j-\mathbf{\bar{x}}_j|_{\vec{\varepsilon}=\vec{0}} \right)^{k_j} }{k_j!} \right) f^{(k_1,\ldots,k_v)} \left( \bar{x}_1|_{\vec{\varepsilon}=\vec{0}},\ldots,\bar{x}_v|_{\vec{\varepsilon}=\vec{0}} \right).
\end{equation}
Upon application of the truncation rules of eq.~\eqref{eq:dual_multivariate_truncation}, $f$ will thus evaluate to an element of $\mathcal{D}(\vec{n})$. With these definitions, the partial derivatives of any $m$-ary function $f$ can be easily obtained by evaluating it at $(x_1+\varepsilon_1,\ldots,x_m+\varepsilon_m)$, and then reading the coefficient of each monomial in the result. We show in appendix~\ref{sect:multivariate_dual} an example of the application of multi-variate duals to the computation of partial derivatives of a multi-variate function.

Finally, on top of the truncation rules given in eq.~\eqref{eq:dual_multivariate_truncation}, one may wish to also add the following rule for optimisation purposes:
\begin{equation}
\label{eq:extra_truncation}
    \prod_{i=1}^m \varepsilon_{i}^{k_i}=0\;\;\forall\;\{k_1,\dots,k_m\} \ \text{s.t.} \ \sum_{i=1}^m k_i > N.
\end{equation}
We call $\mathcal{D}(N;\vec{n})$ the polynomials obtained from multivariate dual numbers that satisfy both eq.~\eqref{eq:dual_multivariate_truncation} and eq.~\eqref{eq:extra_truncation}.

Note that the implementation of dual numbers supporting more than one derivative is often performed through a nested formulation, that is by writing $\bar{x} = c^\prime_n + \bar{x}' \varepsilon_x$, with $\bar{x}\in\mathcal{D}(n)$ and $\bar{x}'\in\mathcal{D}(n-1)$, instead of the flattened version of eq.~\eqref{eq:truncated_monovariate_dual_application}.
The obvious advantage of this more common formulation is that it only requires implementing the elementary operations for the single-derivative duals in $\mathcal{D}(1)$, after which all deeper derivatives can easily be obtained at an arbitrary depth from properly templated data structures.
However, this formulation is less efficient and very inconvenient for implementing the complicated truncation rules in the multivariate case for anything other than $\mathcal{D}(N\times m;N,\ldots,N)$. 
In the context of the \glss{LU} implementation in the presence of raised propagators, we can easily know a priori the derivative structure required for each term, i.e. what $N$ and $\vec{n}$ should minimally be. We therefore opted to explicitly implement the few dedicated dual structures necessary for the computation of cross-sections up to a given perturbative order, which we shall discuss next.

\subsection{Solving the distributional rules}

Up to $\text{N}^m\text{LO}$, there can be at most $m$ self-energy insertions and therefore at most $m$-raised cut propagators. We sort the propagators crossed by a Cutkosky cut $\mathbf{c}_\mathbf{s}$ of multiplicity $n=|\mathbf{c}_\mathbf{s}|$ based on their power in ascending order and write the $n$-tuplet $\vec{r}=(r_1,\ldots,r_n)$ that specifies the power of each raised cut propagator \textit{minus} one, and write $(\mathbf{0},x_1,\ldots,x_k)$ for a tuplet that can have an arbitrary number of zeros at the start and has non-zero $x_i$. At $\text{N}^m\text{LO}$ the number of powers has an upper bound, and specifically one has $\sum_{i=1}^n r_i\le m$.   The onshellness condition of each of the first $n-1$ raised cut propagators will be solved in the energy component $p_i^0$ of its independent momentum, whereas the onshellness condition of the last $n$-th propagator will be solved in the causal flow parameter $t$. Irrespective of the ordering, the derivative in $t$ always has to be performed to the order equal to the sum of all raised powers of the Cutkosky cut edges, i.e. $\sum_i r_i$.

We denote with $\mathbf{x}=(p_1^0,\ldots,p_{n-1}^0,t)$ these variables in which we will solve the on-shell conditions of the Cutkosky cuts, and thus possibly compute derivatives of the \glss{LU} integrand.
Equipped with this notation, we can now write the generic structure of the integrand for any Cutkosky cut $\mathbf{s}^{(\vec{r})}$ crossing $n$ edges, as follows:
\begin{equation}
\label{eq:general_raised_LU_expression}
    I_{\mathbf{s}^{(\vec{r})}} = \int \prod_{i=1}^{n-1}\left(\mathrm{d}p_i^0 \delta^{(r_i+1)} [p_i^0-E_i]\right) \delta^{(r_n+1)} \left[q^0-p_{1}^0-\ldots -p_{n-1}^0 - E_n \right] \mathcal{M}(\{p_i\}) \;,
\end{equation}
where $\mathcal{M}(\{p_i\})$ is a function that contains, among all other \glss{LU} factors (cutting rule factors $1/(p_i^0+E_i)^{r_i+1}$, observable, Jacobians, etc\ldots), the (c)\glss{LTD} representation of the original amplitude graphs together with their \glss{UV} counterterms obtained from applying the $R$-operator to them (as will be discussed in sect.~\ref{sec:R-operation}).
The quantity $\mathcal{M}$ also contains the observable-dependent final-state density $f(\mathcal{O})$.

One key observation is that the first $n-1$ distributions, when solved and turned into derivatives, do not act on each other since they depend on the independent variables $p_i^0$. 
This means that, for a given Cutkosky cut $\mathbf{s}^{(\vec{r})}$, we find that the necessary multivariate dual structure is simply $\mathcal{D}( \sum_{i=1}^n r_i; {r_1,\ldots,r_{n-1}, \sum_{i=1}^n r_i } )$ for the variables $(p_1^0,\ldots,p_{n-1}^0,t)$.
We stress that even though eq.~\eqref{eq:general_raised_LU_expression} contains many terms once fully expanded, a \emph{single} evaluation of $\mathcal{M}$ with dual arguments in the appropriate structure is sufficient to evaluate \emph{all} necessary derivatives, as all the different combinations of derivatives needed are present in the various dual components of that single evaluation. For completeness, we report the exhaustive list of dual structures required up to \glss{N3LO} in tab.~\eqref{tab:DualStructures}.
\begin{table}[h!]
\begin{center}
{\setlength\doublerulesep{1.5pt}   %
 \aboverulesep=0ex 
 \belowrulesep=0ex
\begin{tabular}{c|c|cc|ccc}
\toprule[1pt]
	$\vec{r}$ & $(\mathbf{0},1)$ & $(\mathbf{0},2)$ & $(\mathbf{0},1,1)$ & $(\mathbf{0},3)$ & $(\mathbf{0},1,2)$ & $(\mathbf{0},1,1,1)$
\\ \midrule[0.5pt]
    order & \glss{NLO} & \multicolumn{2}{c|}{ \glss{NNLO} } & \multicolumn{3}{c}{ \glss{N3LO} }
\\ \midrule[0.5pt]
dual & $\mathcal{D}(1;1)$ & $\mathcal{D}(2;2)$  & $\mathcal{D}(2;1,2)$ & $\mathcal{D}(3;3)$ & $\mathcal{D}(3;1,3)$ & $\mathcal{D}(3;1,1,3)$
\\
\midrule[0.3pt]\bottomrule[1pt]
\end{tabular}
}
\end{center}
\caption{\label{tab:DualStructures} Multivariate dual structures $\mathcal{D}$ for the variables $(\ldots,p_{n-1}^0,t)$ required for implementing a Cutkosky cut $\mathbf{s}^{(\vec{r})}$ of the \glss{LU} representation up to \glss{N3LO}, with a specific configuration $\vec{r}$ of the raised propagators crossed by the cut and its first order of appearance. We refer to the text for details.
}
\end{table}

The dedicated implementation of all dual structures listed in tab.~\ref{tab:DualStructures} is then sufficient for the implementation of the \glss{LU} representation of the differential cross-section of arbitrary processes up to \glss{N3LO}. Notice that for perturbative orders up to N3LO, second-order derivatives w.r.t one energy component are never needed (it only first becomes necessary at N4LO, and only for the $\vec{r}=(\mathbf{0},2,2)$ configuration).
Moreover, for QCD corrections up to \glss{N3LO} of processes with only two external colour-charged particles, at most one derivative in any energy component is needed, together with at most three in the $t$ parameter.
We stress that the use of dual numbers for computing derivatives yields no extra steps during the generation of the integrand (since it is merely a type-redefinition of the integrand arguments) and it slows down run-time evaluations by a factor roughly given by the number of terms generated by the multiplication operator of two truncated dual numbers.

\subsection{Generalised cutting rules and truncated Green's functions}
\label{sec:derivatives_OS_scheme}

We have seen that constructing a local representation of cross-sections that is \glss{FSR}-finite requires a careful treatment of raised propagators, which results in derivatives of amplitudes on both sides of the cut.
An even more noteworthy feature of eq.~\eqref{eq:LU_formula} is that the derivatives in $p^0_1,...,p^0_{n-1},t$ act on the final-state density $f(\mathcal{O})$ associated with the observable $\mathcal{O}$. Such derivatives are never be considered within the traditional approach in which external propagators, including their self-energy corrections, are truncated, taking advantage of the \glss{OS} renormalisation scheme. In the following we will clarify the interplay between the \glss{OS} scheme, raised propagators and the request of local IR finiteness. Let us discuss the \glss{OS} renormalisation of masses and fields separately:

\paragraph{\glss{OS} mass counterterm:}

Let us start by showing that we can eliminate the derivatives of amplitudes arising from raised propagators when we renormalise the masses in the \glss{OS} scheme. As an example, we will focus on self-energy corrections within a scalar theory, and consider the application of generalised cutting rules on \glss{OS} renormalised quantities:
\begin{equation}
    \frac{(\Sigma(p^2)-\delta m^{os})^n}{(p^2-m^2)^{n+1}}f(p)\rightarrow (-2\pi \mathrm{i})\frac{(\Sigma(p^2)-\delta m^{os})^n}{(p^0+E_{\vec{p}})^{n+1}}\delta^{(n+1)}[p^0-E_{\vec{p}}]f(p).
\end{equation}
where $f$ is a test function (for example, it could be the product of two amplitudes, contracted and already integrated over all common external momenta aside from $p$). The resolution of the generalised cutting rule yields
\begin{equation}
\label{eq:no_derivatives_in_OS_scheme}
    \int \mathrm{d}p^0 \frac{(\Sigma(p^2)-\delta m^{os})^n}{(p^0+E_{\vec{p}})^{n+1}}\delta^{(n+1)}[p^0-E_{\vec{p}}]f(p)=\frac{1}{n!}\frac{\partial^n}{(\partial p^0)^n}\left[\frac{(\Sigma(p^2)-\delta m^{os})^n }{(p^0+E_{\vec{p}})^{n+1}}f(p)\right]_{p^0=E_{\vec{p}}}.
\end{equation}
and using the integrated level identity $\Sigma(m^2)=\delta m^{\text{os}}$ stemming from \glss{OS} renormalisation conditions, we obtain:
\begin{equation}
\label{eq:no_derivatives_in_OS_scheme_1}
    \int \mathrm{d}p^0 \frac{(\Sigma(p^2)-\delta m^{os})^n}{(p^0+E_{\vec{p}})^{n+1}}\delta^{(n+1)}[p^0-E_{\vec{p}}]f(p)=\frac{\Sigma'(m^2)^n}{2E_{\vec{p}}}f(p^{os})
\end{equation}
This confirms that within the \glss{OS} scheme, all contributions arising from derivatives of amplitudes or observables vanish, and we are left with the \glss{OS} field renormalisation counter-term $\Sigma'(m^2)$ multiplying the truncated integrand. Importantly, we note that the last equality of eq.~\eqref{eq:no_derivatives_in_OS_scheme} was obtained by enforcing that $\Sigma(m^2)=\delta m^{\text{os}}$, which is a result that holds only at the integrated level, but not at the local level.

The fact that $\Sigma(m^2)=\delta m^{\text{os}}$ holds at the integrated level and not at the local level is precisely why generalised cutting rules are required. At the local level, the effect of raised propagators and the derivatives generated by the cutting rules is relevant, so much so that not including such derivatives would break the \glss{LU} IR cancellation pattern. In other words, because the cancellation of raised propagators given in eq.~\eqref{eq:no_derivatives_in_OS_scheme_1} does not hold locally, then the \glss{LU} construction forces to first include the full effect of the generalised cutting rules and only later renormalise masses in the \glss{OS} scheme.

\paragraph{\glss{OS} field counterterm:}
We now discuss \glss{OS} field counterterms within \glss{LU}. The \glss{OS} field counterterms contain \glss{IR} poles as well as \glss{UV} poles, which are needed to achieve local IR finiteness within \glss{LU}. Thus, performing \glss{OS} field renormalisation before the construction of the \glss{LU} renormalisation would break the \glss{KLN} cancellation mechanism upon which \glss{LU} relies. In order to preserve the \glss{KLN} cancellation pattern, one must renormalise fields in the \glss{OS} scheme (if desired) only \emph{after} the application of \glss{LU}, through couplings redefinition, which we will detail in sect.~\ref{sec:os_mass_renormalisation}. We view the separation of handling \glss{IR} and \glss{UV} singularities as theoretically more appealing than the usual dimensional-regularisation approach which blurs the distinction of these two opposite regimes by using a single regulator for both.
\\

In summary, our handling of external self-energy corrections is separated into two independent steps: a) we first construct an unrenormalised and locally IR-finite integrand and b) we locally subtract its UV divergences (see sect.~\ref{sec:local_UV_CT}), and add compensating terms (see sect.~\ref{sec:integrated_UV_CT}) to accommodate a choice of renormalisation conditions.
Within this two-step procedure, it is clear that the role of the derivatives introduced within the \glss{LU} formulation of sect.~\ref{sec:raised_lu} is only that of guaranteeing \textit{local} \glss{IR}-finiteness and they bring no overall contribution to 
the differential cross-section when working within the \glss{OS} renormalisation scheme. We also stress that the formalism of generalised cutting rules opens the possibility of constructing a \glss{LSZ} formula that allows for the renormalisation of external particles in generic schemes.

Finally, we briefly discuss the effect of taking derivatives of the observable functions. We distinguish the case of continuous observables and piece-wise constant observables.
Continuous observable densities often correspond to some variant of an event shape. In that case, the derivatives of the observable density induced by the generalised cutting rule must \emph{always} be accounted for since the terms of sub-leading order in $(t-t^\star)$ that are generated in this way are essential for guaranteeing local FSR-finiteness of the \glss{LU} representation (much like it is the case for derivatives of the $h(t)$ normalising function).
This is however not an obstacle since in practice the continuous observables typically considered are differentiable, such that computing their derivatives using the dual numbers introduced in sect.~\ref{sect:dual_numbers_section} is straightforward.
Other common observables are piece-wise constant, i.e. a histogram. In that case, the observable derivatives are zero everywhere except at the bin-boundary where the observable is technically not \glss{IR}-safe (which leads to the common misbinning feature). At the bin boundaries however, the observable derivatives are delta functions whose contributions are however typically excluded from the bin weights (and anyway zero at the integrated level when renormalizing external masses in the \glss{OS} scheme, as we already discussed).

\clearpage

\section{The $R$-operation}
\label{sec:R-operation}

We have seen that the Local Unitarity representation allows to cast the sum over all interference diagrams arising from a given supergraph as the integral of a function that is free of \glss{IR} singularities. Interference diagrams, however, also feature \glss{UV} singularities. Each interference diagram is split by the Cutkosky cut into two diagrams which may have loops and, consequently, \glss{UV} divergences. These singularities need to be regulated locally, in order for the \glss{LU} framework to be effective. Thankfully, the understanding of \glss{UV} divergences of amplitudes is far more advanced that that of \glss{IR} divergences and a fully generic framework for their treatment is known.

In particular, for renormalisable theories, \glss{UV} poles of complete amplitudes are known to be factorisable and can be removed through parameter re-definitions. Diagrammatic proofs of this fact usually follow the $R$ formalism~\cite{Bogoliubov:1957gp,Caswell:1981ek}, which establishes a recursive subtraction procedure, or the \glss{BPHZ}~\cite{Bogoliubov:1957gp,Zimmermann:1969jj,Hepp:1966eg,Herzog:2017jgk} formalism, the unfolding of the $R$ procedure in terms of a forest subtraction formula. These procedures can be applied to an individual diagram and render it \glss{UV}-finite. Although the generic framework was available since the end of the sixties and has been extended to subtract internal soft divergences in the eighties~\cite{Chetyrkin:1982nn,Chetyrkin:1984xa,Smirnov:1986me,Herzog:2017bjx,Chetyrkin:2017ppe}, there have been few automated implementations of the framework for relevant and complete physical theories. In particular, rarely is it discussed how to construct the renormalisation operator $K$, upon which the $R$-formalism critically relies. A proper definition of $K$ requires disentangling its role as a subtraction operator and as a renormalisation operator. Furthermore, if the $R$-operation is to be applied within the \glss{LU} framework, it also needs to satisfy local properties that are crucial to the realisation of both \glss{IR} and \glss{UV} cancellations.

In this section, we will start by discussing the general features of the $R$ formalism, including the definition of the objects participating in its construction, namely the renormalisation operator $K$ and the \emph{wood} of a graph. We will then discuss four important properties that we wish to impose on our construction. Finally, we will briefly comment on the interplay between the \glss{KLN} cancellation pattern that underlies the construction of the \glss{LU} formula and the \glss{UV} subtraction implemented by the $R$-operation.

\subsection{The $R$-operation master formula}
\label{sec:local_UV_CT}

The first step in constructing a \glss{UV} subtraction formula is to determine all divergent \glss{1PI} subgraphs, classified according to their \glss{dod}. The \glss{dod} of a graph is defined as the integer corresponding to the sum of all mass dimensions of vertices and edges of the graph (i.e., a quark propagator has mass dimension -1, a triple gluon vertex +1) and its integration measure. If the degree of divergence is greater or equal to zero, the graph is ultraviolet divergent, and requires subtraction. The degree of divergence of a given subgraph has the direct interpretation of being the leading power in the Taylor expansion around the infinity limit of a parameter rescaling all loop momentum components to infinity. Note that it can be that the superficial degree of divergence of a graph is higher than its actual degree of divergence.

For gauge theories, we can provide an explicit formula for the degree of divergence of a graph. For a given graph $\Gamma$ with $N_f$ external fermionic lines and $N_A$ external bosonic lines, the \glss{dod} can be shown to be
\begin{equation}
    \text{dod}(\Gamma)=4- \frac{3}{2} N_f - N_A.
\end{equation}

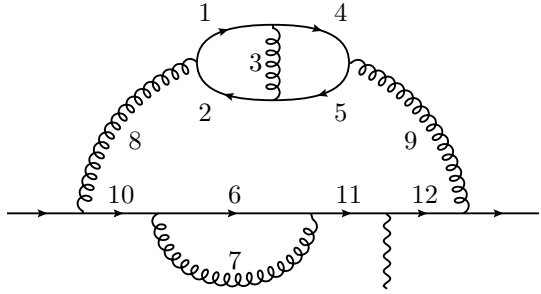
\begin{figure}
    \centering
\begin{tikzpicture}
    \tikzstyle{every node}=[font=\small]
    \begin{feynman}
      \vertex (a3);
      \vertex[right=1cm of a3] (a4);
      \vertex[right=1cm of a4] (a5);
      \vertex[right=1cm of a5] (a6);
      \vertex[right=1.5cm of a6] (bb6);
      \vertex[right=2cm of a6] (a7);
      \vertex[right=1cm of a7] (a8);
      \vertex[right=1cm of a8] (a9);
      \vertex[right=1cm of a9] (a10);

      \vertex[below=1cm of a8] (b1);

      \vertex[above=of bb6] (c1);
      \vertex[above=1cm of c1] (c3);
      \vertex at ($(c1)!0.5!(c3) - (1cm, 0)$) (c2);
      \vertex[right=2cm of c2] (c4);

      \diagram* {
        {
         (a4) --[fermion1, line width=0.25mm] (a5) -- [fermion1, edge label=${10}$, line width=0.25mm] (a6) -- [fermion1, edge label=${6}$, line width=0.25mm] (a7) -- [fermion1, edge label=${11}$, line width=0.25mm] (a8) -- [fermion1, edge label=${12}$, line width=0.25mm] (a9) --[fermion1, line width=0.25mm] (a10),
        },

        (c1) -- [fermion1, out=180,in=-90, edge label={$2$}, line width=0.25mm] (c2) -- [fermion1, out=90,in=180,edge label={$1$}, line width=0.25mm] (c3) -- [fermion1, out=0, in=90,edge label={$4$}, line width=0.25mm] (c4) -- [fermion1, out=-90,in=0,edge label={$5$}, line width=0.25mm] (c1),
        (a9) -- [gluon, bend right, edge label={$9$}, line width=0.25mm] (c4),
        (c1) -- [gluon, edge label={$3$}, line width=0.25mm] (c3),
        (c2) -- [gluon, bend right, edge label={$8$}, line width=0.25mm] (a5),
        (a6) -- [gluon, half right, edge label={$7$}, line width=0.25mm] (a7),
        (a8) -- [boson, line width=0.25mm] (b1),
      };
    \end{feynman}
  \end{tikzpicture}
  \caption{\label{fig:spinneys_examples}Example graph $\Gamma$ with nested ultraviolet divergences. Its collection of \glss{UV} divergent subgraphs is $D(\Gamma)=\{\gamma_{6 7}, \gamma_{1 2 3}, \gamma_{3 4 5}, \gamma_{1 2 4 5}, \gamma_{1 2 3 4 5}, \gamma_{1 \ldots 12}\}$, with \glss{dod}'s being $1$, $0$, $0$, $0$, $2$ and $0$ respectively.}
  \label{fig:uvexample}
\end{figure}

As an example, we list all divergent subgraphs for the Feynman diagram $\Gamma$ in fig.~\ref{fig:uvexample}. The \glss{UV} divergent subgraphs $\gamma$ are characterised by the numbers that label the edges of the subgraph. The vertices at the endpoint of the edges are understood to be part of the subgraph as well. The set of all \glss{UV} divergent subgraphs is $D(\Gamma) = \{\gamma_{6 7}, \gamma_{1 2 3}, \gamma_{3 4 5}, \gamma_{1 2 4 5}, \gamma_{1 2 3 4 5}, \gamma_{1 \ldots 12}\}$, with the \glss{dod} being $1$, $0$, $0$, $0$, $2$ and $0$ respectively.

Given the set of all divergent subgraphs $D(\Gamma)$, one can construct all families of subgraphs for which any two subgraphs are disjunct, i.e. they share no edge or vertex. We refer to these families as spinneys and the collection of all spinneys as the wood $W(D(\Gamma))=:W(\Gamma)$. The wood obtained from the divergent subgraphs for the example of fig.~\ref{fig:spinneys_examples} is:
\begin{align}
\label{eq:UV_master_equation}
\begin{split}
    W(\Gamma) =
    \{
        &\{\},
        \{\gamma_{6 7}\}, \{\gamma_{1 2 3}\}, \{\gamma_{3 4 5}\}, \{\gamma_{1 2 4 5}\}, \{\gamma_{1 2 3 4 5}\}, \{\gamma_{1\ldots12}\},\\
        &\{\gamma_{1 2 3},\gamma_{6 7}\}, \{\gamma_{3 4 5}, \gamma_{6 7}\}, \{\gamma_{1 2 4 5}, \gamma_{6 7}\}, \{\gamma_{1 2 3 4 5}, \gamma_{6 7}\}
    \} \,.
\end{split}
\end{align}
Having defined the wood of a graph, we now present the $R$-procedure~\cite{Bogoliubov:1957gp} that makes any diagram $\Gamma$ \glss{UV} finite:
\begin{equation}
\label{eq:R_formula}
    R(\Gamma) = \sum_{S \in W(\Gamma)}\Gamma \setminus S * \prod_{\gamma \in S} Z(\gamma) ,\quad Z(\gamma) = -K\left(\sum_{S \in W(\gamma)\setminus \gamma } \gamma \setminus S * \prod_{\gamma' \in S} Z(\gamma')  \right) \,.
\end{equation}
where $Z(\gamma)$ is the local \glss{UV} counterterm operator and is defined recursively through $Z(\emptyset)$ = 1. $Z(\gamma)$ is the object that captures the divergence of the entire \glss{UV} subgraph $\gamma$ going to infinity with the subdivergences of $\gamma$ subtracted. It is a polynomial in the external momenta and masses of $\gamma$.
$\Gamma\setminus S$ is the graph where the edges of the \glss{UV} divergent graphs contained in $S$ have all been shrunk to a point serving as an effective vertex of the remaining graph. Its corresponding integrand is contracted with the counterterm $Z(\gamma)$ obtained from $\gamma$. In general, both $Z(\gamma)$ and $\Gamma\setminus S$ are tensors stemming from the numerator structure of the edges and vertices contained with the graph. We leave these tensor indices implicit, but remind their presence by using the $*$ symbol to indicate the combination and contraction of indices when `multiplying' the tensorial representations of the two subgraphs. The minus sign in the definition of $Z(\Gamma)$ takes care of the proper subtraction of overlapping divergences through an inclusion-exclusion principle.
The operator $K$ is a scheme-dependent renormalization operator that is not uniquely defined, but must reproduce the divergent part of the integral expressed in some regulator. We will further constrain $K$ in sect.~\ref{sect:constraints_on_K}.

Finally, note that the $R$-operation can also be written in terms of the Zimmermann's forest formula~\cite{Zimmermann:1969jj,Hepp:1966eg}, which however displays in a less manifest way the nested structure of renormalisation and introduces some subtleties regarding the operator ordering of $K$s which are captured more clearly in the iterated definition of $Z$.

In order to properly identify the numerator specific to each subgraph and compute its approximation obtained by the application of $K$, it is important that the numerator algebra be performed after the application of the $R$-operation~\cite{Caswell:1981ek,Herzog:2017bjx}.

For the example graph in fig.~\ref{fig:uvexample}, the contribution to $R(\Gamma)$ from the spinney $\{\gamma_{1 2 3 4 5},\gamma_{6 7}\}$ yields:
\begin{align}
    \prod_{\gamma \in \{\gamma_{1 2 3 4 5},\gamma_{6 7}\}}& Z(\gamma) * \Gamma \setminus \{\gamma_{1 2 3 4 5},\gamma_{6 7}\} = \\
    &K(\gamma_{6 7}) * K\Big(\gamma_{1 2 3 4 5} - K(\gamma_{1 2 3}) * \gamma_{4 5} - K(\gamma_{3 4 5}) * \gamma_{1 2} - K(\gamma_{1 2 4 5}) * \gamma_{3} \Big) * \gamma_{8 \ldots 12} \nonumber \;.
\end{align}
As long as $K$ is a well-defined renormalisation operator, the application of $R$ on any graph is ensured to yield a quantity that is \glss{UV}-finite and that reproduces a consistent renormalisation scheme. $K$ itself can be defined at the local or integrated level and in the following we will focus on a local level construction of it that is compatible with Local Unitarity.

\subsection{Constraints on the renormalisation operator $K$}
\label{sect:constraints_on_K}

$K$, being a renormalisation operator, is constrained by definition to subtract the superficial UV divergence of a graph it acts on and at the same time to implement a valid renormalisation scheme. Because we want the $R$-operation to be compatible with Local Unitarity, we stress that $R$ acts on the Minkowski representation of a graph's integrand. In other words, given a graph's integrand $\gamma(k_1,\ldots,k_L)$ that is a rational polynomial in the loop momenta $\{k_i\}_{i=1}^L$, external momenta $\{p_i\}_{i=1}^n$ and masses $\mathbf{m}$ then
\begin{equation}
K(\gamma)=\frac{P\left(\{k_i\}_{i=1}^L,\{p_i\}_{i=1}^n,\mathbf{m}\right)}{Q\left(\{k_i\}_{i=1}^L,\{p_i\}_{i=1}^n,\mathbf{m}\right)}.
\end{equation}
where $P$ and $Q$ are polynomials. We formulate four more constraints on the action of $K$:
\begin{enumerate}
    \item \textbf{Local UV convergence}: the result of applying the $R$-operation to a graph's $G$ \textit{integrand} with $K$ being the subtraction operator yields a \textit{locally} UV finite quantity.
    We require that:
    \begin{itemize}
    \item 
    $K$ locally subtracts the superficial UV divergence of any graph it is applied to, i.e. if $\lambda k_1,\ldots,\lambda k_L$ are the loop momenta of the graph $\gamma$, then
    \begin{equation}
        \lim_{\lambda\rightarrow\infty}\lambda^{4L}|\gamma-K(\gamma)|\in\mathbb{R}^+,
    \end{equation}
    i.e. the subtracted graph is finite when all its loop momenta go to infinity.
        \item $K$ preserves nested cancellations, i.e. if $\lambda k_1,\ldots,\lambda k_{L_2}$ are the loop momenta of the graph $\gamma_2$, and $ \tilde{k}_1,\ldots,\tilde{k}_{L_1}$ are the loop momenta of the graph $\gamma_1$, then
    \begin{align}
        &\lim_{\lambda\rightarrow\infty}\lambda^{4L_2}|K(\gamma_1 * \gamma_2)-K(\gamma_1 * K(\gamma_2))|\in\mathbb{R}^+, \\
        &\lim_{\lambda\rightarrow\infty}\lambda^{4L_2}|K(\gamma_1) * \gamma_2-K(K(\gamma_1) * \gamma_2)|\in\mathbb{R}^+.
    \end{align}
    \end{itemize}
    
    \item \textbf{Spurious soft subtraction}: $R(\Gamma)$ should be locally free of soft singularities for non-exceptional external momenta (see sect.~\ref{sect:soft_higher_order_propagators}).
    \item \textbf{Minimal analytic complexity}: the construction of $K$ should only involve the analytic computation of at most single-scale vacuum diagrams, with all propagators having the same mass.
    \item \textbf{Hybrid} $\overline{\mathbf{MS}}$\textbf{+OS renormalisation}: $R(\Gamma)$ should give the renormalised expression for $\Gamma$, where masses of massive particles are renormalised in the \glss{OS} scheme and the fields and vertices are renormalised in $\overline{\text{MS}}$.
    
\end{enumerate}
In the following, we will construct a renormalisation operator that satisfies all four constraints. In order to compartmentalise the discussion, we decompose the operator $K$ as follows:
\begin{equation}
    K=\mathbf{T}+\bar{K},
\end{equation}
where $\mathbf{T}$ the \textit{local} subtraction operator and $\bar{K}[\Gamma]$ is a \glss{UV} (and \glss{IR}) finite function of the loop momenta of the graph $\Gamma$, and a polynomial in the external momenta of $\Gamma$. In sect.~\ref{sec:subtraction_operators}, we will construct the local subtraction operator $\mathbf{T}$ that manifestly satisfies the first two constraints. In sect.~\ref{sect:localised_renormalisation}, we will determine $\bar{K}$ and show that such a choice also satisfies the last two constraints.

\subsection{IR-finiteness of the UV forest}
\label{sect:UV_cts_in_LU}

We briefly discuss the interplay between subtraction of \glss{UV} divergences and the pattern of cancellations of \glss{IR} divergences predicted by the \glss{KLN} theorem, which itself underlies the local \glss{IR} finiteness of the \glss{LU} representation. Let us thus consider a fixed forward-scattering (connected) diagram $\Gamma$ and the set of all Cutkosky cuts $\mathcal{C}=\{\mathbf{c}_i\}$ that can be operated on it. Recall that a Cutkosky cut is characterised by a subset $\mathbf{c}$ of the edges of $\Gamma$ such that the deletion of the edges in $\mathbf{c}$ produces two connected diagrams, and that is minimal under this property (any subset of $\mathbf{c}$ does not satisfy such property). The \glss{KLN} cancellation mechanism ensures that the sum over all interference diagrams obtained from the Cutkosky cuts $\{\mathbf{c}_i\}$ is free of final state \glss{IR} singularities (see fig.~\ref{fig:cutkosky_sum}).

The structure described above of identifying diagrammatic contributions to transition probabilities as arising from the couplets of forward-scattering diagrams and Cutkosky cuts, should be preserved by the \glss{UV} subtraction procedure. Given $\Gamma$ and a fixed Cutkosky cut $\mathbf{c}$ on it, we denote by $\Gamma_1$ and $\Gamma_2$ the graphs obtained from the deletion of $\mathbf{c}$ from $\Gamma$. We then consider the counterterms obtained from the application of the $R$-operation on each of these two graphs; specifically, to each element in the wood $S\in W[\Gamma_i]$, corresponds a \glss{UV} counterterm. Given $\Gamma_i$ and $S$, the reduced graph is obtained by contracting to a point all of the subgraphs $\gamma\in S$ of $\Gamma_i$, denoted with $\Gamma_i^{S}$, and it is a diagrammatic representation of the \glss{UV} counterterm constructed from $S$. The contracted graph does not have \glss{IR} singularities, since the propagators of \glss{IR}-divergent graphs, and only those, will have been assigned a \glss{UV} mass.

This diagrammatic representation of the \glss{UV} counterterms of $\Gamma_1$ and $\Gamma_2$ can be included into a similar representation for the interference diagram $(\Gamma,\mathbf{c})$. Indeed, given two elements $S_1\in W[\Gamma_1]$ and $S_2\in W[\Gamma_2]$, we can consider the reduced diagram $\Gamma^{S_1 S_2}$, which is divided by $\mathbf{c}$ into the two reduced diagrams $\Gamma_1^{S_1}$ and $\Gamma_2^{S_2}$. $(\Gamma^{S_1 S_2},\mathbf{c})$ is then an interference diagram that could potentially feature \glss{IR} divergences. Thus we have to show that the $R$-operation applied to the full sum of interference diagrams arising from $G$ also yields all of the other reduced interference diagrams that can be constructed by operating Cutkosky cuts on the reduced forward-scattering diagram $\Gamma^{S_1 S_2}$.

This is straightforward to prove: for any other Cutkosky cut $\mathbf{c}'$ of $\Gamma^{S_1 S_2}$, the two graphs obtained from deletion of the edges in $\mathbf{c'}$ of the \emph{original} graph $\Gamma$ have forests with unique elements $S_1$ and $S_2$ belonging to them. Thus, the reduced interference diagram $(\Gamma^{S_1 S_2}, \mathbf{c}')$ is also produced by the subtraction procedure. For fixed $S_1$ and $S_2$, each reduced interference diagram is in a one-to-one correspondence with Cutkosky cuts (although different elements of the forest could produce the same reduced diagram). This ensures that the \glss{KLN} cancellation cancellation mechanism is preserved across \glss{UV} counterterms. In general, all discussion and results pertaining to the \glss{LU} representation of a forward-scattering diagram equally apply to the forward-scattering diagrams $\Gamma$ and to any individual term stemming from the $R$-operation and generating forward-scattering diagrams like $\Gamma^{S_1S_2}$.

We can restate this result in other words: let $(S,\mathbf{c})$ be the couplet obtained from an element of the wood $S\in W[\Gamma]$ of the forward-scattering diagram $\Gamma$ and a Cutkosky cut $\mathbf{c}$, such that any subgraph $\gamma\in S$ does not share an edge with $\mathbf{c}$. Then, $S$ does not intersect $\mathbf{c}$. Let $W_{\text{ext}}$ be the set of all such elements. If we define $C_\mathbf{c}$ as cutting only \emph{non-contracted} parts of a reduced graph or $R$ as being the \glss{UV} subtraction operator that contracts only \emph{non-cut} parts of the graph, then
\begin{equation}
    R_{\text{cut}}(\Gamma)=R \left(\sum_{\mathbf{c}\in\mathcal{C}} C_\mathbf{c} \Gamma \right) = \sum_{\mathbf{c}\in\mathcal{C}} C_\mathbf{c} R(\Gamma) = \sum_{(S,\mathbf{c})\in W_{\text{ext}}} C_\mathbf{c} \left[ \Gamma\setminus S\right] * \prod_{\gamma\in S} Z(\gamma)%
\end{equation}
is both \glss{IR} and \glss{UV} finite, where $C_\mathbf{c}$ is the cut-taking operator, that substitutes the propagators corresponding to edges in $\mathbf{c}$ with their cut versions. The diagrammatic application of both cutting rules and the $R$-operations are especially clear. In fig.~\ref{fig:R_to_the_interference} we show the result of applying the $R$-operation to interference diagrams, for an example forward-scattering diagram. We then show in fig.~\ref{fig:collecting_IR_finite} that the result of applying R on the whole sum of interference diagrams can be cast in a way that manifestly exhibits its \glss{IR} finiteness.

 \begin{figure}[t!]
 \begin{tabular}{ccccccc}
 $\sum_{\mathbf{c}}C_\mathbf{c}$\Bigg[\raisebox{0.1cm}{\resizebox{1.5cm}{!}{\input{diagrams/UV_KLN/graph1}}}\Bigg]$=$ &
 \hspace{-0.3cm}\raisebox{0.1cm}{\resizebox{1.4cm}{!}{\input{diagrams/UV_KLN/graphcut1}}}$\, +$ &
 \hspace{-0.3cm}\raisebox{0.1cm}{\resizebox{1.4cm}{!}{\input{diagrams/UV_KLN/graphcut2}}}$\, +$ &
 \hspace{-0.3cm}\raisebox{0.1cm}{\resizebox{1.4cm}{!}{\input{diagrams/UV_KLN/graphcut3}}}$\, +$ &
 \hspace{-0.3cm}\raisebox{0.1cm}{\resizebox{1.4cm}{!}{\input{diagrams/UV_KLN/graphcut4}}}$\, +$ &
 \hspace{-0.3cm}\raisebox{0.1cm}{\resizebox{1.4cm}{!}{\input{diagrams/UV_KLN/graphcut5}}}$\, +$ &
 \hspace{-0.3cm}\raisebox{0.1cm}{\resizebox{1.4cm}{!}{\input{diagrams/UV_KLN/graphcut6}}} \\
  \end{tabular}
  \caption{The sum over all Cutkosky cuts of a forward-scattering diagram is \glss{IR}-finite.}
  \label{fig:cutkosky_sum}
\end{figure}

\begin{figure}[t!]
    \begin{tabular}{cccccc}
         $(R-\mathds{1})$\Bigg[ \raisebox{0.1cm}{\hspace{-0.1cm}\resizebox{1.4cm}{!}{\input{diagrams/UV_KLN/graphcut1}}} \hspace{-0.1cm}\Bigg]=&  
         \hspace{-0.3cm}\raisebox{0.1cm}{\resizebox{1.4cm}{!}{\input{diagrams/UV_KLN/graphcutUV1}}}$+$&
         \hspace{-0.3cm}\raisebox{0.1cm}{\resizebox{1.4cm}{!}{\input{diagrams/UV_KLN/graphcutUV4}}}$+$&
         \hspace{-0.3cm}\raisebox{0.1cm}{\resizebox{1.4cm}{!}{\input{diagrams/UV_KLN/graphcutUV3}}},&
         $(R-\mathds{1})$\Bigg[\hspace{-0.1cm} \raisebox{0.1cm}{\resizebox{1.4cm}{!}{\input{diagrams/UV_KLN/graphcut2}}}\hspace{-0.1cm} \Bigg]=&
         \hspace{-0.3cm}\raisebox{0.1cm}{\resizebox{1.4cm}{!}{\input{diagrams/UV_KLN/graphcutUV2}}}
         ,\\
         &&&&&\\
         $(R-\mathds{1})$\Bigg[ \raisebox{0.1cm}{\hspace{-0.1cm}\resizebox{1.4cm}{!}{\scalebox{-1}[1]{\input{diagrams/UV_KLN/graphcut1}}}} \hspace{-0.1cm}\Bigg]=&  
         \hspace{-0.3cm}\raisebox{0.1cm}{\resizebox{1.4cm}{!}{\scalebox{-1}[1]{\input{diagrams/UV_KLN/graphcutUV1}}}}$+$&
         \hspace{-0.3cm}\raisebox{0.1cm}{\resizebox{1.4cm}{!}{\scalebox{-1}[1]{\input{diagrams/UV_KLN/graphcutUV4}}}}$+$&
         \hspace{-0.3cm}\raisebox{0.1cm}{\resizebox{1.4cm}{!}{\scalebox{-1}[1]{\input{diagrams/UV_KLN/graphcutUV3}}}},&
         $(R-\mathds{1})$\Bigg[\hspace{-0.1cm} \raisebox{0.1cm}{\resizebox{1.4cm}{!}{\scalebox{-1}[1]{\input{diagrams/UV_KLN/graphcut2}}}}\hspace{-0.1cm} \Bigg]=&
         \hspace{-0.3cm}\raisebox{0.1cm}{\scalebox{-1}[1]{\resizebox{1.4cm}{!}{\input{diagrams/UV_KLN/graphcutUV2}}}}.
    \end{tabular}

    \caption{The $R$-operation can be appropriately defined so as to act on interference diagrams.}
    \label{fig:R_to_the_interference}
\end{figure}

\begin{figure}[t!]
    \begin{tabular}{l}
      $(R-\mathds{1})\sum_{\mathbf{c}}C_\mathbf{c}$\Bigg[\raisebox{0.1cm}{\resizebox{1.5cm}{!}{\input{diagrams/UV_KLN/graph1}}}\Bigg]$=$ \Bigg[\raisebox{0.1cm}{\resizebox{1.4cm}{!}{\input{diagrams/UV_KLN/graphcutUV1}}} $+$ \raisebox{0.1cm}{\resizebox{1.4cm}{!}{\input{diagrams/UV_KLN/graphcutUV2}}}\Bigg] $+$
      \Bigg[\raisebox{0.1cm}{\resizebox{1.4cm}{!}{\scalebox{-1}[1]{\input{diagrams/UV_KLN/graphcutUV1}}}} $+$ 
      \raisebox{0.1cm}{\resizebox{1.4cm}{!}{\scalebox{-1}[1]{\input{diagrams/UV_KLN/graphcutUV2}}}}\Bigg] $+$
      \\
      \\
       \raisebox{0.1cm}{\resizebox{1.4cm}{!}{\scalebox{-1}[1]{\input{diagrams/UV_KLN/graphcutUV3}}}} $+$  
       \raisebox{0.1cm}{\resizebox{1.4cm}{!}{\scalebox{-1}[1]{\input{diagrams/UV_KLN/graphcutUV4}}}} $+$ 
       \raisebox{0.1cm}{\resizebox{1.4cm}{!}{\scalebox{1}[1]{\input{diagrams/UV_KLN/graphcutUV3}}}} $+$
       \raisebox{0.1cm}{\resizebox{1.4cm}{!}{\scalebox{1}[1]{\input{diagrams/UV_KLN/graphcutUV4}}}}
    \end{tabular}
    \caption{Applying the $R$-operation on the sum of all the interference diagrams arising from a given forward-scattering diagram yields \glss{IR}-finite contributions. In particular, each expression in square brackets of the first line is separately \glss{IR} finite, and each diagram on the second line is also \glss{IR} finite, provided that $q^2>0$.}
    \label{fig:collecting_IR_finite}
\end{figure}

\section{Subtraction operators}
\label{sec:subtraction_operators}
Having expressed the local renormalisation operator $K$ as the sum of a subtraction operator $\mathbf{T}$ and a finite quantity $\bar{K}$, and having defined the relevant constraints on $K$, we proceed to construct the subtraction operator $\mathbf{T}$. In the following, we will limit ourselves to introducing $\mathbf{T}$ and arguing that it satisfies the first two constraints of those laid out in sect.~\ref{sect:constraints_on_K}, associated with the \textit{local} UV convergence of $R(\Gamma)$ and to the absence of spurious soft enhancements due to raised massless propagators. If we wish the subtraction implemented by $K$ to be local, so that the integral corresponding to such amplitudes can be obtained by direct Monte-Carlo integration, we need both $\mathbf{T}$ and $\bar{K}$ to have a local representation. 
The construction of $\mathbf{T}$ follows four steps:
\begin{enumerate}
    \item The construction of $T$, a local subtraction operator that satisfies the first constraint of sect.~\ref{sect:constraints_on_K}, but not the second one.
    \item The construction of $\tilde{T}$, a local subtraction operator that satisfies the second constraint of sect.~\ref{sect:constraints_on_K}, but not the first one.
    \item The construction of $\hat{T}=\tilde{T}+T-T\tilde{T}$, a local subtraction operator that satisfies both constraints of sect.~\ref{sect:constraints_on_K}.
    \item The construction of $\mathbf{T}$, a refined version of $\hat{T}$, whose action on massive fermionic self-energy corrections also reproduces the \glss{OS} renormalisation of the fermion's mass. 
\end{enumerate}
Ultimately, a full understanding of the $\mathbf{T}$ operator and its properties can only be obtained after it is combined with $\bar{K}$, and $K$ is then shown to also satisfy the last two constraints laid out in sect.~\ref{sect:constraints_on_K}. This last step is performed in sect.~\ref{sect:localised_renormalisation}.

\subsection{The UV subtraction operator}

In order to construct local \glss{UV} counterterms, we consider a modified expansion of the propagators so that the denominator of each approximated propagator includes an arbitrary mass term $m_\text{UV}$ that regulates the soft behaviour of the \glss{UV} counterterms. Let $\gamma$ be a subgraph depending on the external momenta $p_1,\ldots,p_n$, loop momenta $k_1,\ldots,k_L$ and masses $\mathbf{m}=(m_1,\ldots,m_l)$. Let us then start by considering the case in which we write $K(\gamma)=T_{\text{dod}}(\gamma)$, where $T$ is the Taylor expansion operator that expands in the parameter $\lambda$ that rescales the external scales, including masses, of the graph $\gamma$. The `\glss{dod}' subscript refers to the degree of divergence of $\gamma$ which also corresponds to the order in $\lambda$ at which the Taylor series is truncated. By setting $K=T$, we can already assess the \glss{UV} convergence of the $R$-operation (Note however that we will eventually use a different version for $K$). More precisely, let $\gamma^\lambda$ be defined as
\begin{equation}
\label{eq:lambda_rescaling_definition}
    \gamma^\lambda=\frac{\mathcal{N}\left(\{\lambda p_i\}_{i=1}^n,\{\lambda m_j\}_{j=1}^l, \{k_m\}_{m=1}^L\right)}{\prod_{e\in\mathbf{e}}D_e^\lambda},
\end{equation}
where $\mathcal{N}$ is a polynomial numerator. We have $\gamma^{1}=\gamma$. We stress that the numerator of $\gamma$ is \emph{not} equivalent to the numerator of the complete supergraph $\Gamma$, and that the expansion is performed only in the numerator and denominator of the subgraph $\gamma$. $D_e^\lambda$ is the inverse propagator defined as
\begin{equation}
\label{eq:prop_expansion}
    D_e^\lambda=k_e^2 -m_{\text{UV}}^2 +2\lambda k_e \cdot p_e + \lambda^2 p_e^2 - \lambda^2 (m_e^2-m_{\text{UV}}^2),
\end{equation}
where $p_e$ is constrained to be a linear combination of the external momenta $p_1,\ldots,p_m$, $k_e$ is a linear combination of the loop momenta $k_1,\ldots,k_L$, and $m_e$ can be either zero or equal one of the masses $m_1,\ldots,m_l$. $m_{\text{UV}}$ is a \glss{UV} mass that is introduced as a regulator for the \glss{IR} behaviour of the \glss{UV} subtraction counterterm.
We define $K(\gamma)=T_{\text{dod}}(\gamma)$, with
\begin{equation}
\label{eq:expansion_rule}
    T(\gamma)=T_{\text{dod}}\left(\gamma\right)=\sum_{j=0}^{\text{dod}} \frac{1}{j!}\frac{\mathrm{d}^j}{\mathrm{d}\lambda^j}\gamma^\lambda \Bigg|_{\lambda=0}
\end{equation}
We note that with this definition, the approximation of a graph always corresponds to a polynomial in the external momenta and internal masses, with coefficients that are vacuum tensor integrands with a single scale $m^2_\text{UV}$ appearing as the mass of each denominator.
Importantly, the infrared rearrangement~\cite{Lowenstein:1974qt,Vladimirov:1979zm} operated by introducing a $m_{\text{UV}}$ does not spoil the local cancellation of the counterterm. The substitution of eq.~\eqref{eq:lambda_rescaling_definition} is \emph{not} equivalent to an expansion in the propagator mass $m_e^2$ around $m_e^2=m_\text{UV}^2$, and this is especially manifest when considering a numerator with a dependence on $m$.

For our example spinney $\{\gamma_{1 2 3 4 5},\gamma_{6 7}\}$, we obtain the following contribution to $R(\Gamma)$:
\begin{eqnarray}
\label{eq:UV_t_substituted}
    &&Z(\gamma_{1 2 3 4 5}) Z(\gamma_{6 7}) * \Gamma \setminus \{\gamma_{1 2 3 4 5},\gamma_{6 7}\} = \nonumber \\
    &&T_1(\gamma_{6 7}) T_2\Big(\gamma_{1 2 3 4 5} - T_0(\gamma_{1 2 3}) * \gamma_{4 5} - T_0(\gamma_{3 4 5}) * \gamma_{1 2} - T_0(\gamma_{1 2 4 5}) * \gamma_{3} \Big) * \gamma_{8 \ldots 12} \;.
\end{eqnarray}
The Taylor expansions $T_\text{dod}$ yield tensor graphs in terms of the polynomial in internal masses and the external momenta of the subgraph. We remind the reader that our use of the symbol $*$ indicates that indices that are in common between the subgraphs multiplied are contracted whereas those that are different combine as in a tensor product. 
Note that in the case of nested $T_\text{dod}$ approximations, the outermost approximation must expand over the $m_{\text{UV}}$ factors introduced in the numerator by the innermost Taylor expansion (see eq.~\eqref{eq:expansion_rule}).

The separation of the dependence of $D_e^\lambda$ and $\mathcal{N}$ into linear combinations of \emph{loop} momenta $k_e$ (not rescaled with $\lambda$) and linear combinations of \emph{external} momenta $p_e$ (rescaled with $\lambda$) is in principle ambiguous as it depends on a particular choice of a loop momentum basis for performing the expansion. 
The simplest way to make this identification is to perform a change of loop momentum basis on the complete graph $\Gamma$ such that in the new basis any $n$-loop UV subgraph $\gamma$ of a spinney contains $n$ basis momenta. The Taylor expansion of the subgraph is then performed in this newly constructed basis. As an example, we can perform such change of basis for the term $T_0(\gamma_{1 2 3})*\gamma_{4 5}$:
\begin{equation}
\label{eq:diagrammatic_R}
  T\left(\hspace{-0.2pt}
  \raisebox{0.1cm}{\scalebox{0.7}{
  \begin{tikzpicture}[baseline=(c2.center)]
    \tikzstyle{every node}=[font=\large]
    \begin{feynman}
      \vertex (c1);
      \vertex[above=1cm of c1] (c3);
      \vertex at ($(c1)!0.5!(c3) - (1cm, 0)$) (c2);
      \vertex[right=2cm of c2] (c4);
      \vertex[left=0.5cm of c2] (d1);
      \vertex[right=1cm of c4] (d2);
      \vertex[right=0.7cm of c3] (e1);
      \vertex[right=0.7cm of c1] (e2);
      \diagram* {
        (e2) -- [fermion1,edge label=$k+l-p$,near start, line width=0.3mm] (c1) -- [fermion1,edge label=$k-p$, line width=0.3mm] (c2) -- [fermion1,edge label=$k$, line width=0.3mm] (c3) -- [fermion1,edge label=$k+l$, line width=0.3mm] (e1),
        (c1) -- [boson, edge label'={$l$}, line width=0.3mm] (c3),
        (d1) -- [boson, edge label={$p$}, line width=0.3mm] (c2),
      };
    \end{feynman}
  \end{tikzpicture}
  }}
  \hspace{-0.2pt}\right)
  *
  \raisebox{0.1cm}{\scalebox{0.7}{
  \begin{tikzpicture}[baseline=(c2.center)]
    \tikzstyle{every node}=[font=\large]
    \begin{feynman}
      \vertex (c1);
      \vertex[above=1cm of c1] (c3);
      \vertex[crossed dot] at ($(c1)!0.5!(c3) - (1cm, 0)$) (c2) {};
      \vertex[right=1cm of c2] (c4);
      \vertex[left=0.5cm of c2] (d1);
      \vertex[right=0.5cm of c4] (d2);
      \vertex[right=0.5cm of c3] (e1);
      \vertex[right=0.5cm of c1] (e2);
      \diagram* {
        (c4) -- [fermion1,half left, edge label=$k+l-p$, line width=0.3mm] (c2) -- [fermion1,half left,edge label=$k+l$, line width=0.3mm] (c4),
        (d1) -- [boson, edge label={$p$}, line width=0.3mm] (c2),
        (d2) -- [boson, edge label'={$p$}, line width=0.3mm] (c4),
      };
    \end{feynman}
  \end{tikzpicture}
  }}
  \hspace{-0.5cm}\rightarrow
  T\left(\hspace{-0.2pt}
  \raisebox{0.1cm}{\scalebox{0.7}{
    \begin{tikzpicture}[baseline=(c2.center)]
      \tikzstyle{every node}=[font=\large]
      \begin{feynman}
        \vertex (c1);
        \vertex[above=1cm of c1] (c3);
        \vertex at ($(c1)!0.5!(c3) - (1cm, 0)$) (c2);
        \vertex[right=1cm of c2] (c4);
        \vertex[left=0.5cm of c2] (d1);
        \vertex[right=1cm of c4] (d2);
        \vertex[right=0.7cm of c3] (e1);
        \vertex[right=0.7cm of c1] (e2);
        \diagram* {
          (e2) -- [fermion1,edge label=$c_2-p$, line width=0.3mm] (c1) -- [fermion1,edge label=$c_1-p$, line width=0.3mm] (c2) -- [fermion1,edge label=$c_1$, line width=0.3mm] (c3) -- [fermion1,edge label=$c_2$, line width=0.3mm] (e1),
          (c1) -- [boson, edge label'={$c_2 - c_1$}, line width=0.3mm] (c3),
          (d1) -- [boson, edge label={$p$}, line width=0.3mm] (c2),
        };
      \end{feynman}
    \end{tikzpicture}
    }}
    \hspace{-0.2pt}\right)
    *
    \raisebox{0.1cm}{\scalebox{0.7}{
    \begin{tikzpicture}[baseline=(c2.center)]
      \tikzstyle{every node}=[font=\large]
      \begin{feynman}
        \vertex (c1);
        \vertex[above=1cm of c1] (c3);
        \vertex[crossed dot] at ($(c1)!0.5!(c3) - (1cm, 0)$) (c2) {};
        \vertex[right=1cm of c2] (c4);
        \vertex[left=0.5cm of c2] (d1);
        \vertex[right=0.5cm of c4] (d2);
        \vertex[right=0.5cm of c3] (e1);
        \vertex[right=0.5cm of c1] (e2);
        \diagram* {
          (c4) -- [fermion1,half left,edge label=$c_2 - p$, line width=0.3mm] (c2) -- [fermion1,half left,edge label=$c_2$, line width=0.3mm] (c4),
          (d1) -- [boson, edge label={$p$}, line width=0.3mm] (c2),
          (c4) -- [boson, edge label={$p$}, line width=0.3mm] (d2),
        };
      \end{feynman}
    \end{tikzpicture}
    }}
\end{equation}
where on the right we changed the loop momentum basis that is used as a reference for the Taylor expansion into the expansion basis $c_1=k, c_2=k+l$.
After performing the expansion, all terms are transformed back into the original basis in order to have local \glss{UV} cancellations between all terms of the $R$ construction. It is important that the expansion basis be chosen consistently, namely the same subgraph momenta of $\gamma$ should be in the expansion basis chosen for each term of the $R$-operation that involves the \glss{UV} subgraph $\gamma$. In the example above, the (unfolded) $R$-operation terms $T_0(\gamma_{1 2 3})*\gamma_{4 5}$ and $-T_2(T_0(\gamma_{1 2 3})*\gamma_{4 5})$ will cancel in the \glss{UV} limit of the subgraph momenta of $\gamma_{4 5}$ only upon such a consistent choice of expansion basis. In order to better illustrate this point, we give here an example of an inconsistent choice of expansion basis, where one elects the basis $c_1=k, c_2=k+l$ for the implementation of $T_0(\gamma_{1 2 3})*\gamma_{4 5}$ and $c_1'=k-p, c_2'=k+l$ for implementing $-T_2(T_0(\gamma_{1 2 3})*\gamma_{4 5})$. Then, there would be local \glss{UV} mis-cancellation when $l\rightarrow \infty$:
\begin{align}
\begin{split}
    T_0(\gamma_{1 2 3})*\gamma_{4 5} -T_2(T_0(\gamma_{1 2 3})*\gamma_{4 5}) &%
        \underset{l \rightarrow \infty}{\rightarrow} \frac{\mathcal{N}_1'(k,l,p)}{k^4 l^4} - \frac{\mathcal{N}_2'(k,l,p)}{(k-p)^4 l^4} + \ldots \;.
    \end{split}
\end{align}
A similar miscancellation occurs when selecting the basis $c_1=k,c_2=k+l-p$ for $-T_2(T_0(\gamma_{1 2 3})*\gamma_{4 5})$ and the basis $c_1'=k,c_2'=l$ for $T_2(\gamma_{12345})$. In the limit of $k\rightarrow\infty$ these two terms should cancel, but they do not since the UV propagator in $\gamma_{45}$ has a different shift. These problems can be avoided by never allowing an external momentum of $\Gamma$ to appear in the basis transformation, i.e. disallowing affine terms in the basis transformation.

Sect.~\ref{sect:UV_cts_in_LU} proves that the \glss{UV}-subtracted \glss{LU} expression is \glss{IR}-finite, but for the arguments laid out in that section to hold, it is important that the \emph{same} expansion basis momenta be chosen across all identical \glss{UV} subgraphs appearing in the woods of amplitudes arising from different the Cutkosky cuts. This is because the \glss{UV} counterterms of the amplitudes defined by these Cutkosky cuts must locally match on collinear limits so as to preserve the local cancellations of \glss{IR} singularities featured by the \glss{LU} representation. This can be achieved by electing, at the supergraph level, preferred momenta for the choice of expansion basis of each \glss{UV} subgraph that will then be the same irrespective of the Cutkosky cut considered.

\subsection{Spurious soft divergences}
\label{sect:soft_higher_order_propagators}

In sect.~\ref{sec:raised_lu}, we have argued that in order to faithfully reproduce the \glss{KLN} cancellation pattern at the local level, Cutkosky cuts crossing raised propagators must be interpreted in terms of residues of higher-order poles. The \glss{KLN} cancellation pattern then eliminates all collinear enhancements. In turn, this should make the scaling around the soft region tame enough so that also soft singularities become integrable, and can be completely eliminated (i.e. bounded) using a multi-channeling procedure. 

\begin{figure}[ht!]
\begin{center}
\resizebox{8cm}{!}{
\input{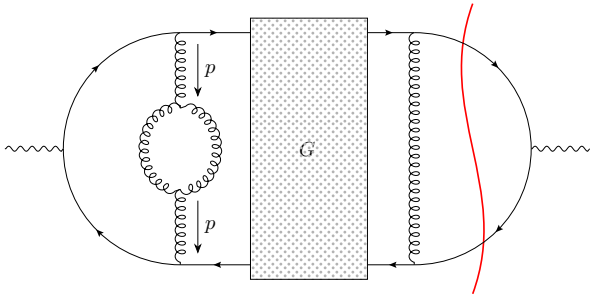}
}
\end{center}
\caption{
\label{fig:soft_IR_diag}
Example of an amplitude diagram featuring a spurious local soft singularity in the limit $p^\mu \rightarrow 0$ (i.e. spurious because it leads to no pole at the integrated level) due to the insertion of a  self-energy subgraph on an internal loop line. The cancellation of such a spurious soft singularity is completely unrelated to the \glss{KLN} cancellation pattern, and its regularisation within the \glss{LU} integrand therefore necessitates a subtraction procedure.
}
\end{figure}

This last statement, however, is not manifestly true when self-energies are inserted on propagators of massless particles. 
For example, we consider the diagram shown in fig.~\ref{fig:soft_IR_diag},
where $G$ is any graph featuring four external (anti-)quarks. In the soft limit $p^\mu \rightarrow 0$, the integrand corresponding to the amplitude of fig.~\ref{fig:soft_IR_diag} scales at best as $(p^2)^{-2}$ (excluding the scaling of the integral measure), independent of the value of the cut momenta. In general, the lower bound on the soft \glss{dod} of a massless vector propagator subject to $n$ self-energy corrections is $4-2(n+1)$ and it is $4-(n+1)$ for a massless quark. Notice that in this soft limit $p^\mu \rightarrow 0$, no particle other than the gluon itself, and in particular not any of the propagators internal to the self-energy inserted, is forced to lie on its mass-shell. These two observations establish that such soft singularities are not related to the \glss{KLN} cancellation mechanism, and are therefore expected to remain present in the \glss{LU} representation of cross-sections.

 However, these soft singularities are spurious in gauge theories. In other words, gauge symmetry guarantees that \emph{at the integrated level}, the gluonic self-energy itself scales quadratically in the soft region. This is clear given that the general tensor structure of the integrated self-energy of a massless vector reads $A(p^2) (p^2 g^{\mu \nu} + p^\mu p^\nu)$, where $A$ behaves like a constant in the soft limit. Similarly, the massless fermionic integrated self-energy scales linearly given its tensor structure of $B(p^2) \slashed{p}$, where $B$ is again a constant in the soft limit. This ensures that the \glss{IR} pole structure of integrated loop-amplitudes remains unaffected by such spurious soft singularities. However, at the local level the generic configuration of fig.~\ref{fig:soft_IR_diag} has a soft divergence.
 Such configurations can happen only at \gls{NNLO} and beyond, where the momentum of a propagator that is corrected by one or more self-energies can itself become soft.
 
In this work, we will seek to remedy these spurious soft divergences by the introduction of local soft counterterms.
As we shall see, the soft counterterms we will construct satisfy the following two key requirements: they are inherently local and, in gauge theories and when regularising spurious soft singularities stemming from the insertion of self-energy subgraphs, they directly integrate to zero.

\subsubsection{Subtraction of spurious soft singularities}

Local counterterms for the spurious soft divergence discussed in this section can be obtained by Taylor-expanding around $p \rightarrow 0$ all $n$ self-energy subgraphs correcting the raised propagator. The depth of this expansion, henceforth referred to as soft expansion depth, is chosen to be the minimal order required in order to eliminate the additional soft scaling introduced by the insertion of the self-energy subgraphs. The expansion is only performed in the external momentum $p$ of the self-energy subgraphs, and all other scales, which for a two-point-function can only be masses, are kept unexpanded. Once combined with such a soft counterterm, the soft scaling of the subtracted graph together with its two neighbouring repeated propagators is restored to that of a single propagator without self-energy insertion, thus ensuring the \gls{IR}-finiteness property of the \glss{LU} representation is recovered. Quite remarkably, for renormalisable theories the soft expansion depth in the momentum external to a self-energy is always equal to the \glss{UV} \glss{dod} of the self-energy graph minus one. For example, in \glss{QCD}, we consider the following soft expansions:
\vspace{-0.1cm}
\begin{align}
   \label{eq:Tdodm1_gluon}
    \tilde{T}_{\text{dod}-1}\Bigg(
    \raisebox{-0.4cm}{
    \resizebox{2.4cm}{!}{
    \begin{tikzpicture}
        \tikzstyle{every node}=[font=\small]
         \node[main node]  (1) {};
        \begin{feynman}
          \vertex[left=1.5cm of 1] (a4);
          \vertex[right=1.5cm of 1] (a5);
          \diagram* {
            {
             (a4) -- [gluon, momentum=\(p\)] (1) -- [gluon] (a5),
            },
          };
        \end{feynman}
      \end{tikzpicture}
      }
      }
      \,
    \Bigg)&=\raisebox{-0.3cm}{
    \resizebox{2.0cm}{!}{
    \begin{tikzpicture}
        \tikzstyle{every node}=[font=\small]
         \node[main node]  (1) {};
        \begin{feynman}
          \vertex[left=1.5cm of 1] (a4);
          \vertex[right=1.5cm of 1] (a5);
          \diagram* {
            {
             (a4) -- [gluon] (1) -- [gluon] (a5),
            },
          };
        \end{feynman}
      \end{tikzpicture}
      }
      }
      \Bigg|_{p=0}
      +p^\mu \partial_\mu \raisebox{-0.3cm}{
    \resizebox{2.0cm}{!}{
    \begin{tikzpicture}
        \tikzstyle{every node}=[font=\small]
         \node[main node]  (1) {};
        \begin{feynman}
          \vertex[left=1.5cm of 1] (a4);
          \vertex[right=1.5cm of 1] (a5);
          \diagram* {
            {
             (a4) -- [gluon] (1) -- [gluon] (a5),
            },
          };
        \end{feynman}
      \end{tikzpicture}
      }
      }
      \Bigg|_{p=0} \\
    \label{eq:Tdodm1_quark}
    \tilde{T}_{\text{dod}-1}\Bigg(
    \raisebox{-0.4cm}{
    \resizebox{2.4cm}{!}{
    \begin{tikzpicture}
        \tikzstyle{every node}=[font=\small]
         \node[main node]  (1) {};
        \begin{feynman}
          \vertex[left=1.5cm of 1] (a4);
          \vertex[right=1.5cm of 1] (a5);
          \diagram* {
            {
             (a4) -- [fermion1, momentum=\(p\)] (1) -- [fermion1] (a5),
            },
          };
        \end{feynman}
      \end{tikzpicture}
      }
      }
      \,
    \Bigg)&=\raisebox{-0.32cm}{
    \resizebox{2.0cm}{!}{
    \begin{tikzpicture}
        \tikzstyle{every node}=[font=\small]
         \node[main node]  (1) {};
        \begin{feynman}
          \vertex[left=1.5cm of 1] (a4);
          \vertex[right=1.5cm of 1] (a5);
          \diagram* {
            {
             (a4) -- [fermion1] (1) -- [fermion1] (a5),
            },
          };
        \end{feynman}
      \end{tikzpicture}
      }
      }
      \Bigg|_{p=0}\;,
 \end{align}
where $\text{dod}$ in the subscript of the soft-expansion operator $\tilde{T}$ refers to the \glss{UV} \glss{dod} of the graph in argument. As shown in eqs.~\eqref{eq:Tdodm1_gluon} and~\eqref{eq:Tdodm1_quark}, the soft expansion depth $\text{dod}-1$ is always equal to $1$ for any self-energy graph correcting a vector propagator and always $0$ when correcting a fermionic propagator.
The application of the operator $(1-\tilde{T}_{\text{dod}-1})$ to any self-energy subgraph guarantees that, together with its two adjacent propagators, its soft scaling becomes identical to that of a single soft propagator. Finally, we report the generic action of $\tilde{T}_{\text{dod}}$ on an arbitrary graph $\Gamma$ depending on external momenta $\{p_i\}_{i=1}^n$, loop momenta $\{k_j\}_{j=1}^L$, and internal masses $\mathbf{m}$, as it will be useful for future discussion. It is the multi-variate Taylor expansion of $\Gamma$ in the rescaled external momenta only
\begin{equation}
    \tilde{T}_{\text{dod}}\left(\Gamma(\{p_i\}_{i=0}^n,\{k_j\}_{j=1}^L,\mathbf{m})\right)=\sum_{i=0}^\text{dod} \frac{1}{i!}\frac{\mathrm{d}^i}{\mathrm{d}\lambda^i}\Gamma\left(\{\lambda p_i\}_{i=1}^n,\{k_j\}_{j=1}^L,\mathbf{m}\right) \Bigg|_{\lambda=0}
\end{equation}
In order to study the soft scaling of the \emph{subtracted} object (whose actual soft scaling cannot be easily related to the \glss{UV} \glss{dod} anymore), we explicitly define the soft \gls{dod} of an integrand $f(k_1,\ldots,k_m)$ of a graph $\Gamma$ in a particular soft limit $k_1,\ldots,k_m \rightarrow 0$ ($k_1,\ldots,k_m$ are independent momenta), to be the leading power of its Taylor expansion in the parameter $\lambda$ rescaling the momenta, $k_i \rightarrow 1/\lambda\; k_i$, both in the arguments of the integrand and in the integration measure for the momenta involved in the limit. In other words, for a soft limit
\begin{equation}
    \text{dod}^{\text{soft}}_{k_1,\ldots,k_m}(\Gamma)=-4m+n, \quad \text{where } \lim_{\lambda\rightarrow 0} \lambda^{n} f(\lambda k_1,\ldots,\lambda k_m)\in\mathbb{R}\setminus\{0\}.
\end{equation}

Specifically $n$ is the integer corresponding to the leading term in the expansion of the integrand in the limit $\lambda\rightarrow 0$. Defined this way, a soft \glss{dod} of 0 indicates a logarithmic divergence whereas a negative \glss{dod} indicates convergence.
We will now explore this type of subtraction when multiple self-energy corrections are present for two specific cases, before presenting the general construction. 

Our first example, shown in fig.~\ref{fig:consecutive_gluon_SEs}, features the complete soft subtraction of two consecutive self-energy corrections. The overall leading soft scaling of the original graph when the momentum $p$ of the external gluon becomes soft is $(p^2)^{-3}$ at the local level. Once subtracted, this overall leading soft scaling becomes $(p^2)^{-1}$. In other words, after soft-subtraction the propagator dressed with two successive self-energy corrections still scales in the soft limit exactly as a single propagator would.

\begin{figure}[ht!]
\captionsetup[subfigure]{labelformat=empty}
\begin{subfigure}{1.0\textwidth}
\resizebox{3cm}{!}{
    \begin{tikzpicture}
        \tikzstyle{every node}=[font=\normalsize]
        \begin{feynman}
          \vertex (a3);
          \vertex[right=0.7cm of a3] (a4);
          \vertex[right=1cm of a4] (a5);
          \vertex[right=0.7cm of a5] (a6);
          \vertex[right=1cm of a6] (a7);
          \vertex[right=0.7cm of a7] (a8);
          \diagram* {
            {
             (a3) -- [gluon, edge label=$p$, line width=0.25mm] (a4) -- [gluon, half right, line width=0.25mm] (a5) -- [gluon, line width=0.25mm] (a6) -- [gluon, half right, line width=0.25mm] (a7) -- [gluon, line width=0.25mm] (a8),
            },
            (a5) -- [gluon,half right, line width=0.25mm] (a4);
            (a7) -- [gluon,half right, line width=0.25mm] (a6);
          };
        \end{feynman}
    \end{tikzpicture}}
    \raisebox{0.3cm}{$\longrightarrow\bigg($}\hspace{-0.3cm}
       \raisebox{-0.cm}{
       \resizebox{1.8cm}{!}{
        \begin{tikzpicture}
            \tikzstyle{every node}=[font=\small]
            \begin{feynman}
              \vertex (a3);
              \vertex[right=0.7cm of a3] (a4);
              \vertex[right=1cm of a4] (a5);
              \vertex[right=0.7cm of a5] (a6);
              \diagram* {
                {
                 (a3) -- [gluon, line width=0.25mm] (a4) -- [gluon, half right, line width=0.25mm] (a5) -- [gluon, line width=0.25mm] (a6),
                },
                (a5) -- [gluon,half right, line width=0.25mm] (a4);
              };
            \end{feynman}
          \end{tikzpicture}
          }
          }\hspace{-0.2cm}
       \raisebox{0.3cm}{$-\tilde{T}_1\bigg($}\hspace{-0.3cm}
       \raisebox{-0.cm}{
       \resizebox{1.8cm}{!}{
        \begin{tikzpicture}
            \tikzstyle{every node}=[font=\small]
            \begin{feynman}
              \vertex (a3);
              \vertex[right=0.7cm of a3] (a4);
              \vertex[right=1cm of a4] (a5);
              \vertex[right=0.7cm of a5] (a6);
              \diagram* {
                {
                 (a3) -- [gluon, line width=0.25mm] (a4) -- [gluon, half right, line width=0.25mm] (a5) -- [gluon, line width=0.25mm] (a6),
                },
                (a5) -- [gluon,half right, line width=0.25mm] (a4);
              };
            \end{feynman}
          \end{tikzpicture} 
          }
          }\hspace{-0.3cm}
       \raisebox{0.3cm}{$\bigg)\bigg)
       \bigg($}\hspace{-0.3cm}
       \raisebox{-0.cm}{
       \resizebox{1.8cm}{!}{
        \begin{tikzpicture}
            \tikzstyle{every node}=[font=\small]
            \begin{feynman}
              \vertex (a3);
              \vertex[right=0.7cm of a3] (a4);
              \vertex[right=1cm of a4] (a5);
              \vertex[right=0.7cm of a5] (a6);
              \diagram* {
                {
                 (a3) -- [gluon, line width=0.25mm] (a4) -- [gluon, half right, line width=0.25mm] (a5) -- [gluon, line width=0.25mm] (a6),
                },
                (a5) -- [gluon,half right, line width=0.25mm] (a4);
              };
            \end{feynman}
          \end{tikzpicture}   
          }
          }\hspace{-0.2cm}
       \raisebox{0.3cm}{$-\tilde{T}_1\bigg($}\hspace{-0.3cm}
       \raisebox{-0.cm}{
       \resizebox{1.8cm}{!}{
        \begin{tikzpicture}
            \tikzstyle{every node}=[font=\small]
            \begin{feynman}
              \vertex (a3);
              \vertex[right=0.7cm of a3] (a4);
              \vertex[right=1cm of a4] (a5);
              \vertex[right=0.7cm of a5] (a6);
              \diagram* {
                {
                 (a3) -- [gluon, line width=0.25mm] (a4) -- [gluon, half right, line width=0.25mm] (a5) -- [gluon, line width=0.25mm] (a6),
                },
                (a5) -- [gluon,half right, line width=0.25mm] (a4);
              };
            \end{feynman}
          \end{tikzpicture}
          }
          }\hspace{-0.3cm}
       \raisebox{0.3cm}{$\bigg)\bigg)$}
\end{subfigure}
    
    \caption{\label{fig:consecutive_gluon_SEs}Soft subtraction for two consecutive gluonic self-energy corrections. The two \glss{1PI} subgraphs result in an $R$-style subtraction once the product above is expanded. In this form shown however, the correct scaling of the subtracted quantity is manifest: the subtracted consecutive self-energy scales like a single propagator would in the soft limit.}
    \label{fig:chained_bubbles}
\end{figure}

\begin{figure}[ht!]
\captionsetup[subfigure]{labelformat=empty}
\begin{subfigure}{1.0\textwidth}

\begin{tabular}{cc}

\hspace{0.7cm}\raisebox{-1.4cm}{
     \begin{tikzpicture}
            \tikzstyle{every node}=[font=\small]
            \node[main node]  (1) {$G_1$};
            
            \node[main node]  (2) [above = 0.65cm of 1] {$G_2$};
            \begin{feynman}
              \vertex[right=1.5cm of 1] (a4);
              \vertex[left=1.5cm of 1] (a6);
              \diagram* {
                {
                 (1) -- [gluon] (a4),
                 (a6) -- [gluon, momentum=\(p\)] (1),
                 (2) -- [gluon, in=50, out=-50] (1),
                 (1) -- [gluon, momentum=\(q\), in=-130, out=130] (2),
                },

              };
            \end{feynman}
          \end{tikzpicture}}
          
           &
          
            \hspace{0.6cm}$-$ \hspace{0.4cm}
          \raisebox{-0.8cm}{
          \begin{tikzpicture}
            \tikzstyle{every node}=[font=\small]
            \node[main node]  (1) {};
            
            \node[main node4]  (2) [above = 0.45cm of 1] {};
            \begin{feynman}
              \vertex[right=1.5cm of 1] (a4);
              \vertex[left=1.5cm of 1] (a6);
              \diagram* {
                {
                 (1) -- [gluon] (a4),
                 (a6) -- [gluon] (1),
                 (1) -- [gluon, in=0, out=50] (2),
                 (2) -- [gluon, in=130, out=-180] (1),
                },

              };
            \end{feynman}
          \end{tikzpicture}}
          $* \ \tilde{T}_1\Bigg($
          \raisebox{-0.5cm}{
          \begin{tikzpicture}
        \tikzstyle{every node}=[font=\small]
         \node[main node]  (1) {};
        \begin{feynman}
          \vertex[left=1.5cm of 1] (a4);
          \vertex[right=1.5cm of 1] (a5);
          \diagram* {
            {
             (a4) -- [gluon] (1) -- [gluon] (a5),
            },
          };
        \end{feynman}
      \end{tikzpicture}}
      $\Bigg)$
      \end{tabular}
      
      \end{subfigure}

\vspace{0.5cm}

\begin{subfigure}{1.0\textwidth}

\begin{tabular}{cc}

     \hspace{-0.5cm}$-\tilde{T}_1\Vast($
     \raisebox{-1.4cm}{
     \begin{tikzpicture}
            \tikzstyle{every node}=[font=\small]
            \node[main node]  (1) {};
            
            \node[main node]  (2) [above = 0.65cm of 1] {};
            \begin{feynman}
              \vertex[right=1.5cm of 1] (a4);
              \vertex[left=1.5cm of 1] (a6);
              \diagram* {
                {
                 (1) -- [gluon] (a4),
                 (a6) -- [gluon] (1),
                 (2) -- [gluon, in=50, out=-50] (1),
                 (1) -- [gluon, in=-130, out=130] (2),
                },

              };
            \end{feynman}
          \end{tikzpicture}
          }
          $\Vast)$
          
           &

          $+\tilde{T}_1\vast($
          \raisebox{-0.8cm}{\begin{tikzpicture}
            \tikzstyle{every node}=[font=\small]
            \node[main node]  (1) {};
            
            \node[main node4]  (2) [above = 0.45cm of 1] {};
            \begin{feynman}
              \vertex[right=1.5cm of 1] (a4);
              \vertex[left=1.5cm of 1] (a6);
              \diagram* {
                {
                 (1) -- [gluon] (a4),
                 (a6) -- [gluon] (1),
                 (1) -- [gluon, in=0, out=50] (2),
                 (2) -- [gluon, in=130, out=-180] (1),
                },

              };
            \end{feynman}
          \end{tikzpicture}}
          $* \ \tilde{T}_1\Bigg($
          \raisebox{-0.5cm}{
          \begin{tikzpicture}
        \tikzstyle{every node}=[font=\small]
         \node[main node]  (1) {};
        \begin{feynman}
          \vertex[left=1.5cm of 1] (a4);
          \vertex[right=1.5cm of 1] (a5);
          \diagram* {
            {
             (a4) -- [gluon] (1) -- [gluon] (a5),
            },
          };
        \end{feynman}
      \end{tikzpicture}}
      $\Bigg)\vast)$

\end{tabular}
\end{subfigure}

\caption{Soft subtraction for a gluonic self-energy nested within another gluonic self-energy. The collection of all \glss{1PI} two-point subgraphs of $\Gamma$ is $\mathcal{B}(\Gamma)=\{G_2,G_{12}\}$. The wood of this graph, that is the collection of all its spinneys, is $W[\mathcal{B}]=\{\{\},\{G_2\},\{G_{12}\}\}$.}
\label{fig:nested_soft_subtraction}
\end{figure}
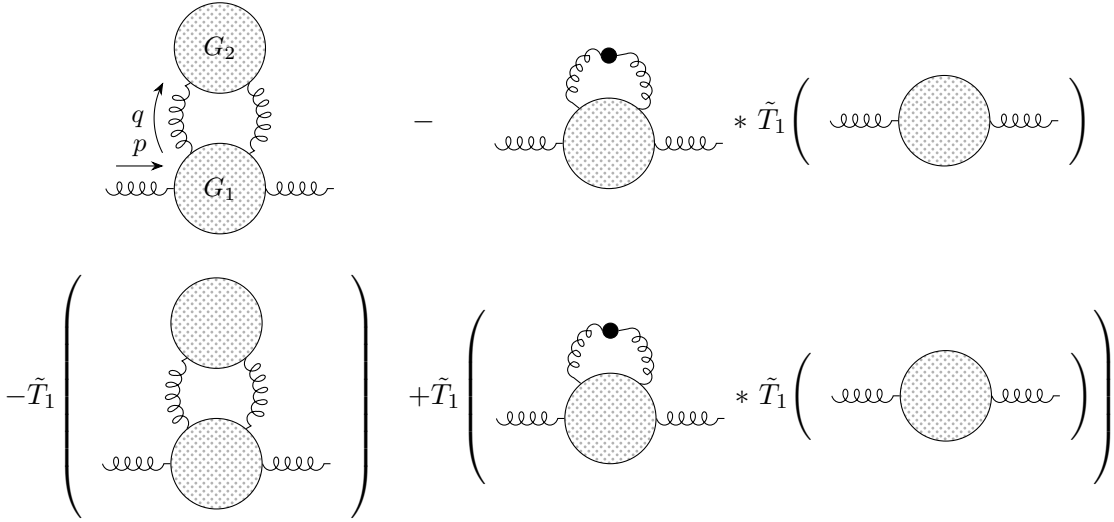

Our second example is the more complicated case involving the nesting of spurious soft divergences, as shown in fig.~\ref{fig:nested_soft_subtraction} that features the nesting of a \glss{1PI} gluonic self-energy correction within another \glss{1PI} gluonic self-energy correction. The spurious soft subtraction structure in that cases bears a striking resemblance with the structure generated by the $R$-operation for \glss{UV} divergences. The main difference being that subtraction terms $\tilde{T}_{\text{dod}-1}(\gamma)$ are now constructed for self-energy subgraphs only, and Taylor-expanded around the spurious soft limit $p^\mu=0$ only (e.g. no expansion of masses) and up to a depth given by the \glss{UV} \glss{dod} of the graph in argument \emph{minus one}.

In more details, we denote the local tensorial expression of the subraphs $G_1$, resp. $G_2$ in fig.~\ref{fig:nested_soft_subtraction} with $\Gamma_1$, resp. $\Gamma_2$, and write $\Gamma_{12}$ as the complete graph, including its two adjacent propagators $\frac{1}{(p^2)^2}$ that can become soft. We can then write the soft-subtraction of $\Gamma_{12}$ in the Feynman gauge as follows:
\begin{align}
\label{eq:subtraction_soft}
\mathcal{F}(\Gamma_{12})=\frac{(\Gamma_1-\Gamma_1\big|_{p=0}-p^\mu\partial_\mu\Gamma_1\big|_{p=0})^{\mu_1\mu_2\mu_3\mu_4}(\Gamma_2-\Gamma_2\big|_{q=0}-q^\nu \partial_\nu\Gamma_2\big|_{q=0})_{\mu_2\mu_3}}{(q^2)^2 (p^2)^2}.
\end{align}
This expression of eq.~\eqref{eq:subtraction_soft} is factorised since $q$ and $p$ are treated as independent momenta (this cannot be done in general, but it can in this case and it simplifies the discussion), however $\Gamma_1$ is a function of both $q$ and $p$. We now investigate the two spurious soft limits $q\rightarrow 0$ and $p\rightarrow 0$. We start by rescaling $p$ with $\lambda$, and investigate the limit $p\rightarrow 0$ . In the limit of $p\rightarrow 0$ with $q$ finite, the leading behaviour of $\mathcal{F}(\Gamma_{12})$ is
\begin{equation}
   \text{dod}^{\text{soft}}_p(\Gamma_{12})=-2\;.
\end{equation}
We now consider the spurious soft limit $q\rightarrow 0$ for non-zero $p$. In that limit, $\Gamma_1\big|_{p=0}$ and $p^\mu\partial_\mu\Gamma_1\big|_{p=0}$ may scale like a negative power of $q^2$, which is potentially problematic. We want to derive a bound on this power within massless gauge theories (adding masses can only improve the soft behaviour). A diagrammatic analysis shows that
\begin{equation}
    \text{dod}^{\text{soft}}_q(\Gamma_{12})= -1.
\end{equation}
We finally look at the double limit in which $p,q\rightarrow 0$. The scaling in this case equals
\begin{equation}
    \text{dod}^{\text{soft}}_{p,q}(\Gamma_{12})=-2.
\end{equation}
which leads to the conclusion that in the nested case the worst possible soft scaling (in this case the limit $q\rightarrow 0$) has a \glss{dod} equal to -1, which is still convergent.

As already mentioned, the generic form of the forest matches that obtained by unfolding the subtracted quantity obtained through the $R$-operation. Let us consider a graph $G$ and the collection $\mathcal{B}$ of \glss{1PI} two-point subgraphs of $\Gamma$. Assume, for the purpose of the correct soft power-counting, that the particles external to $\Gamma$ are all off-shell (the enhancements created by putting external particles on-shell are exactly those cancelled by the Local Unitarity procedure). Then consider the set of all spinneys created from $\mathcal{B}$, denoted as $W[\mathcal{B}]$. Finally, we set up the $R$-subtraction procedure of eq.~\eqref{eq:R_formula}, using $K=\tilde{T}_\text{dod}$ as an approximant. The resulting subtracted integrand does not have any non-integrable soft singularities.

In principle, one may perform the subtraction of spuriou soft divergences before applying the \glss{UV} $R$-operation completely independently, since the introduction of the \glss{UV} mass $m_\text{UV}$  prevents the \glss{UV} counterterms from reintroducing any \glss{IR} singularity. However, in the next section, we will study a more efficient alternative whereby both spurious soft and \glss{UV} subtractions are performed simultaneously within a single modified $R$-operation.

\subsubsection{Embedding spurious soft subtraction within the UV $R$-operation}

A more refined way to perform the subtraction of spurious soft and \glss{UV} divergences is to combine the two subtractions into one unique $R$-type of operation. 
The merging of the two different forest structures stemming from the spurious soft and \gls{UV} subtractions is a delicate process, and achieving this relies on the scaling properties of renormalisable theories, and the ensuing relationship between \glss{UV} and spurious soft power-counting. A systematic approach for constructing such combined $R$ operator is to seek for a subtraction operator that regulates \emph{both} the spurious soft and \glss{UV} behaviour of certain subgraphs. This was the original inspiration to refs.~\cite{Zimmermann:1969jj, Lowenstein}. In this section, we will consider a specific subtraction operator for which the all-order proof of ref.~\cite{Lowenstein} holds and which also coincides with the subtraction of ref.~\cite{Zimmermann:1969jj} when applied to $\phi^4$ theory. In particular, let us consider 
\begin{equation}
    \hat{T}_{\text{dod}}=T_{\text{dod}}+\tilde{T}_{\text{dod}-1}- T_{\text{dod}}\tilde{T}_{\text{dod}-1},
\end{equation}
where the subscript $\text{dod}$ in that case refers to \glss{UV} \glss{dod} of the graph that the operator $\hat{T}_{\text{dod}}$ acts on, and it will also dictate the depth of the Taylor expansion it contains. 
In short, $\hat{T}$ is a subtraction operator that includes the full \glss{UV} subtraction of the graph and the soft subtraction of the graph minus the \glss{UV} behaviour of the soft subtraction, so that it is manifestly a good \glss{UV} approximant of the graph it acts on. As an example, let us consider a gluonic 1PI graph and the action of $\hat{T}$ on it. We consider specifically a graph that features an incoming gluon momentum $p$ and an internal quark of mass $m$ (for example, the top quark). In this case
\begin{equation}
\label{eq:gluonic_soft_T}
    \hat{T}_{2}\Big(
    \raisebox{-0.3cm}{
    \resizebox{1.6cm}{!}{
    \begin{tikzpicture}
        \tikzstyle{every node}=[font=\small]
         \node (1) {};
        \begin{feynman}
          \vertex[left=0.8cm of 1] (a4);
          \vertex[left=0.4cm of 1] (b4);
          \vertex[right=0.8cm of 1] (a5);
          \vertex[right=0.4cm of 1] (b5);
          \diagram {
            {
             (a4) -- [gluon] (b4),
             (b5) -- [gluon] (a5),
             (b4) -- [fermion1, half right] (b5),
             (b5) -- [fermion1, half right] (b4),
            },
          };
        \end{feynman}
      \end{tikzpicture}
      }
      }
      \,
    \Big)
    =
    \raisebox{-0.3cm}{
    \resizebox{1.6cm}{!}{
    \begin{tikzpicture}
        \tikzstyle{every node}=[font=\small]
         \node (1) {};
        \begin{feynman}
          \vertex[left=0.8cm of 1] (a4);
          \vertex[left=0.4cm of 1] (b4);
          \vertex[right=0.8cm of 1] (a5);
          \vertex[right=0.4cm of 1] (b5);
          \diagram {
            {
             (a4) -- [gluon] (b4),
             (b5) -- [gluon] (a5),
             (b4) -- [fermion1, half right] (b5),
             (b5) -- [fermion1, half right] (b4),
            },
          };
        \end{feynman}
      \end{tikzpicture}
      }
      }\hspace{-0.1cm}\Big|_{p=0}+p^\mu\partial_\mu \raisebox{-0.3cm}{
    \resizebox{1.6cm}{!}{
    \begin{tikzpicture}
        \tikzstyle{every node}=[font=\small]
         \node (1) {};
        \begin{feynman}
          \vertex[left=0.8cm of 1] (a4);
          \vertex[left=0.4cm of 1] (b4);
          \vertex[right=0.8cm of 1] (a5);
          \vertex[right=0.4cm of 1] (b5);
          \diagram {
            {
             (a4) -- [gluon] (b4),
             (b5) -- [gluon] (a5),
             (b4) -- [fermion1, half right] (b5),
             (b5) -- [fermion1, half right] (b4),
            },
          };
        \end{feynman}
      \end{tikzpicture}
      }
      }\hspace{-0.1cm}\Big|_{p=0}
      +\frac{p^\mu p^\nu}{2} \partial_\mu\partial_\nu
      \raisebox{-0.3cm}{
    \resizebox{1.6cm}{!}{
    \begin{tikzpicture}
        \tikzstyle{every node}=[font=\small]
         \node (1) {};
        \begin{feynman}
          \vertex[left=0.8cm of 1] (a4);
          \vertex[left=0.4cm of 1] (b4);
          \vertex[right=0.8cm of 1] (a5);
          \vertex[right=0.4cm of 1] (b5);
          \diagram {
            {
             (a4) -- [gluon] (b4),
             (b5) -- [gluon] (a5),
             (b4) -- [fermion1, half right, line width=0.35mm] (b5),
             (b5) -- [fermion1, half right, line width=0.35mm] (b4),
            },
          };
        \end{feynman}
      \end{tikzpicture}
      }
      }\hspace{-0.1cm}\Big|_{\substack{p=0 \\ m = 0\\ m_{\text{UV}}}}
\end{equation}
where thicker lines indicate the introduction of a UV mass in the denominators of the Feynman rules (according to the introduction of UV masses established by eq.~\eqref{eq:lambda_rescaling_definition} and eq.~\eqref{eq:prop_expansion}) and the light line indicates that the propagator retains its original mass, i.e.
\begin{align}
\raisebox{-0.05cm}{\begin{tikzpicture}
        \tikzstyle{every node}=[font=\small]
        \begin{feynman}
        \vertex (1) {};
          \vertex[right=1.5cm of 1] (a1);
          \diagram {
             (1) -- [fermion1] (a1),
             };
        \end{feynman}
\end{tikzpicture}}&=i\frac{\slashed{p}+m}{p^2-m^2+i\epsilon} &\longrightarrow
    \raisebox{-0.05cm}{\begin{tikzpicture}
        \tikzstyle{every node}=[font=\small]
        \begin{feynman}
        \vertex (1) {};
          \vertex[right=1.5cm of 1] (a1);
          \diagram {
             (1) -- [fermion1, line width=0.35mm] (a1),
             };
        \end{feynman}
\end{tikzpicture}}&=i\frac{\slashed{p}}{p^2-m_{\text{UV}}^2+i\varepsilon}\\
\begin{tikzpicture}
        \tikzstyle{every node}=[font=\small]
        \begin{feynman}
        \vertex (1) {};
          \vertex[right=1.5cm of 1] (a1);
          \diagram {
             (1) -- [gluon] (a1),
             };
        \end{feynman}
\end{tikzpicture}&=-i \frac{g^{\mu\nu}}{p^2+i\varepsilon} &\longrightarrow
    \begin{tikzpicture}
        \tikzstyle{every node}=[font=\small]
        \begin{feynman}
        \vertex (1) {};
          \vertex[right=1.5cm of 1] (a1);
          \diagram {
             (1) -- [gluon, line width=0.35mm] (a1),
             };
        \end{feynman}
\end{tikzpicture}&=-i\frac{g^{\mu\nu}}{p^2-m_{\text{UV}}^2+i\varepsilon}
\end{align}
in the Feynman gauge (see sect.~\ref{sec:gauge_invariance} for more details on gauge invariance in unitarity methods). Eq.~\eqref{eq:gluonic_soft_T} makes it clear that the effect of $\hat{T}$ is to consider the soft approximation for the first $\text{dod}-1$ terms of expansion in the \glss{UV} limit and the \glss{UV} approximation for the term of order $\text{dod}$. Observe specifically that the first two orders in the expansion of the gluonic self-energy are evaluated at internal mass $m$ that is unchanged with respect to that of the original graph, whereas the logarithmic order is evaluated at $m=0$ and to every propagator a mass $m_{\text{UV}}$ is assigned. Had we not assigned a mass to every propagator, the last term would feature a new IR singularity in the loop momentum of the quark. On the other hand the first two terms do not require a UV mass.

We also note that for any graph that is \glss{UV} divergent at most logarithmically, we have $\hat{T}_0=T$. Thus, in \glss{QCD}, for the four-gluon vertex and the quark-gluon interaction vertices, we have $\hat{T}=T$. On the other hand, for the three-gluonic vertex the situation is different since it is linearly divergent in the \glss{UV}, and thus the action of $\hat{T}$ on it does include a soft counterterm:
\begin{equation}
\hat{T}_1 \vastino (
\raisebox{-1.3cm}{
\begin{tikzpicture}
        \tikzstyle{every node}=[font=\small]
         
        \begin{feynman}
        \vertex (1) {};
          \vertex[above right=0.6cm and 0.6cm of 1] (a1);
          
          \vertex[right=1.5cm of a1] (a2);
          \vertex[right=0.75cm of a1] (m);
          
          \vertex[above=1.3cm of m] (c1);
          \vertex[above=0.72cm of c1] (c2);

          \vertex[below right=0.5cm and 0.5cm of a2] (a3);
          
          \diagram {
             (1) -- [gluon] (a1),
             (a1) -- [gluon] (a2),
             (a2) -- [gluon] (a3),
             (c1) -- [gluon] (a1),
             (a2) -- [gluon] (c1),
             (c2) -- [gluon] (c1),
            };
        \end{feynman}
\end{tikzpicture}} \hspace{0.3cm}
\vastino )=
\raisebox{-1.3cm}{
\begin{tikzpicture}
        \tikzstyle{every node}=[font=\small]
         
        \begin{feynman}
        \vertex (1) {};
          \vertex[above right=0.6cm and 0.6cm of 1] (a1);
          
          \vertex[right=1.5cm of a1] (a2);
          \vertex[right=0.75cm of a1] (m);
          
          \vertex[above=1.3cm of m] (c1);
          \vertex[above=0.72cm of c1] (c2);

          \vertex[below right=0.5cm and 0.5cm of a2] (a3);
          
          \diagram {
             (1) -- [gluon] (a1),
             (a1) -- [gluon] (a2),
             (a2) -- [gluon] (a3),
             (c1) -- [gluon] (a1),
             (a2) -- [gluon] (c1),
             (c2) -- [gluon] (c1),
            };
        \end{feynman}
\end{tikzpicture}}\hspace{0.1cm}\vastino|_{\substack{p_1=0 \\ p_2=0 \\ p_3=0}}
+ p_i^\mu \frac{\partial}{\partial p_i^\mu} \raisebox{-1.3cm}{
\begin{tikzpicture}
        \tikzstyle{every node}=[font=\small]
         
        \begin{feynman}
        \vertex (1) {};
          \vertex[above right=0.6cm and 0.6cm of 1] (a1);
          
          \vertex[right=1.5cm of a1] (a2);
          \vertex[right=0.75cm of a1] (m);
          
          \vertex[above=1.3cm of m] (c1);
          \vertex[above=0.72cm of c1] (c2);

          \vertex[below right=0.5cm and 0.5cm of a2] (a3);
          
          \diagram {
             (1) -- [gluon] (a1),
             (a1) -- [gluon, line width=0.35mm] (a2),
             (a2) -- [gluon] (a3),
             (c1) -- [gluon, line width=0.35mm] (a1),
             (a2) -- [gluon, line width=0.35mm] (c1),
             (c2) -- [gluon] (c1),
            };
        \end{feynman}
\end{tikzpicture}}\hspace{0.1cm}\vastino|_{\substack{p_1=0 \\ p_2=0 \\ p_3=0 \\ m_{\text{UV}}}}
\end{equation}
where in the last term on the right-hand side a \glss{UV} mass has been assigned to the gluons (again, only in denominators, as established by eq.~\eqref{eq:lambda_rescaling_definition} and eq.~\eqref{eq:prop_expansion}), and summation over all indices is assumed. We observe again that due to the relation between soft and \glss{UV} power-counting in renormalisable theories, the first diagram on the rhs of the equation does not have a non-integrable singularity in the limit of the gluon in the loop becoming soft. A singularity would instead be present in the second term if no \glss{UV} mass was assigned.

Because $\hat{T}$ is a valid and consistent subtraction operator, the subtraction of \glss{UV} divergences performed with the $R$-operation and $K=\hat{T}$ is guaranteed. At the same time, the subtraction of IR divergences is guaranteed by the fact that the wood constructed using $T$ contains the soft wood constructed in the previous section.

Finally, as an example of the nested application of $\hat{T}$, we work out the $R$-operation for a gluonic self-energy nested in a fermionic self-energy, as shown in fig.~\ref{fig:soft_uv_forest}. The spinneys and the wood are by definition the same as that of the \glss{UV} $R$-operation: in this case, the two spinneys correspond with the nested gluonic self-energy and the full fermionic self-energy. The \glss{UV} subtraction operator and the soft subtraction operator only differ by the way masses are expanded (and re-arranged, for the \glss{UV} operator), and this particular way of treating masses in the merged wood is key in order to obtain the right cancellations. Specifically, observe that in the last two lines of fig.~\ref{fig:soft_uv_forest}, corresponding to the nesting of the $\hat{T}$ operation, the second term has all masses set to zero and to every propagator a mass $m_\text{UV}$ is assigned.

Finally, we discuss the case in which we want to compute QCD corrections to processes that involve the Higgs or electro-weak bosons. In this case, studying the superficial degree of divergence of all three-point graphs connecting QCD particles to electro-weak bosons, we realise that we only need to discuss the action of $\hat{T}$ in the following cases
\begin{itemize}
    \item a graph coupling two gluons or ghosts with a Higgs boson: we observe that such vertex must contain a fermionic loop from which the Higgs boson is spawned. So, even if the superficial degree of divergence yields that such vertex diverges linearly in the ultraviolet region, in practice the linearly divergent term vanishes identically once gamma traces are performed for the fermionic loop spawning the Higgs boson. That is, if $\Gamma$ is a vertex coupling two gluons or ghosts and a Higgs, then
    \begin{equation}
        \tilde{T}(\Gamma)=0.
    \end{equation}
    Thus, for all purposes, $\hat{T}=T$ on such fermionic loop corrections.
    \item a graph $\Gamma$ coupling one or two electroweak bosons with two or one gluons. In this case, the application of $\hat{T}$ works exactly the same as the gluonic three-vertex, i.e. 
    \begin{equation}
        \hat{T}(\Gamma)=\tilde{T}_0 (\Gamma) + T_1 (\Gamma) - T_1 \tilde{T}_0 (\Gamma).
    \end{equation}
    \item for all graphs coupling a quark, an anti-quark and a gauge boson or a Higgs boson, then the superficial degree of divergence yields a logarithmic ultraviolet behaviour, so that $\hat{T}=T$.
\end{itemize}

This concludes the discussion of $\hat{T}$ on all relevant graphs. $\hat{T}$ manifestly satisfies both constraints laid out it sect.~\ref{sect:constraints_on_K}.

\subsection{On-shell counterterms for massive fermions}

The arguments we presented for the soft subtraction hold for any particle content, including massive particles. Specifically, one can also use $\hat{T}$ to subtract the divergent \glss{UV} behaviour of a massive fermionic self-energy correction. However, the soft limit of a massive self-energy does not have a particular physical meaning, and specifically it does not correspond to the massive particle becoming on-shell. This implies that the integrated soft counterterms of a massive self-energy (or a vertex in which a massive particle interacts) is non-zero and the analytic integration of the corresponding integrated \glss{UV} counterterm would involve multiple scales. 
To avoid these complications, we propose yet another modification for the subtraction operator:
\begin{equation}
\label{eq:bold_T}
    T^{\text{os}}=\frac{1}{2}(T^{\text{os}+}-T T^{\text{os}+})+\frac{1}{2}(T^{\text{os}-}-TT^{\text{os}-})+T,
\end{equation}
where $T^{\text{os}\pm}$ is the on-shell subtraction operator. We will show in sect.~\ref{sect:integrated_os_cts} that computing its integrated counterterm is not needed, as it reproduces the \glss{OS} mass counterterm.

The action of $T^{\text{os}\pm}$ is only defined for a given fermionic \glss{1PI} self-energy correction $\Sigma$
\begin{equation}
    T^{\text{os}\pm}(\Sigma(p,\mathbf{m}))=(1\pm \gamma^0)\Sigma(\pm p^{os},\mathbf{m}),
\end{equation}
where $p^{os}=(m,0,0,0)$ is the at-rest on-shell momentum for massive particles, and $p^{os}=(0,0,0,0)$ for massless particles, $m$ is the mass of the fermion and $\mathbf{m}$ is any collection of masses the self-energy might depend on. In general, one can substitute $p^{os}$ with any fixed on-shell momentum and $\gamma^0$ with $\slashed{p}^{os}/m$. 

First, we simplify the action of $T^{\text{os}}$ on a self-energy correction to a massive particle by filling in all operators participating in the definition of eq.~\eqref{eq:bold_T}:
\begin{equation}
    T^{\text{os}}(\Sigma(p,\mathbf{m}))=\sum_{\sigma\in\{\pm 1\}}\frac{1+\sigma \gamma^0}{2}\Sigma( \sigma p^{os},\mathbf{m})+(p-\gamma^0 p^{os})^\mu \frac{\partial}{\partial p^\mu}\Sigma( p,\mathbf{m})\Bigg|_{\substack{\hspace{-0.1cm}p=0, \\ \mathbf{m}=0, \\ \hspace{-0.2cm}m_{\text{uv}}}},
\end{equation}
where $m_{\text{uv}}$ again indicates that every propagator of the self-energy has been given the UV mass $m_{\textrm{uv}}$. Eq.~\eqref{eq:bold_T} is manifestly a valid \glss{UV} subtraction operator. However, in order to show that it can be used in general, one has to discuss the nesting of $T^{\text{os}}$.

Let us study the action of $T^{\text{os}}$ on two relevant examples. To keep the notation compact, let us define the operator
\begin{equation}
    P_{\pm}=\frac{1}{2}(1\pm \gamma^0),
\end{equation}
which is a projector on the first (last) two spinor components.
We start with a massive fermionic self-energy correction nested in a gluonic self-energy correction (fig.~\ref{fig:os_soft_subtraction}). In this case the $R$-operation with subtraction operator ${T^{\text{os}}}$, which we will write as $R_{T^{\text{os}}}$, achieves three objectives simultaneously. First, it regulates the \glss{UV} divergent behaviour of the entire graph. Second, it subtracts the fictitious soft divergence associated with the outer gluonic correction, i.e. $R_{T^{\text{os}}} (\Gamma) \approx q^2$ as $q\rightarrow 0$. Third, as we will see in the next section, it implements on-shell mass renormalisation of the fermionic two-point function.

The second example is that of a massive fermionic self-energy nested in a massive fermionic self-energy (fig.~\ref{fig:os_os_forest}). $T^{\text{os}}$ is a valid subtraction operator, and thus $R_{T^{\text{os}}}(\Gamma)$ is \glss{UV} finite. As we will see in sect.~\ref{sect:integrated_os_cts}, $R_{T^{\text{os}}}(\Gamma)$ also contains the necessary pieces needed to perform \glss{OS} renormalisation of both self-energies, in virtue of the nested structure of $R_{T^{\text{os}}}$. %

\subsection{Summary: the \textbf{T} subtraction operator}
\label{sect:summary_bold_T}

The discussion of on-shell counterterms concludes the study of subtraction operators required for our purposes. We will now define the $\mathbf{T}$ operator, which includes the requirements of UV convergence and absence of spurious soft propagators, including the correct extension of the soft subtraction operator to self-energy corrections of massive particles. The reasons behind the construction of the $\mathbf{T}$ operator can only be fully explained after its relationship with the renormalisation operator $K$ is clarified in the next section.

In this work we limit ourselves to summarising the action of $\textbf{T}$ on all vertices and self-energies that are relevant for QCD corrections to any process. We will explicitly separate cases that could be more compactly grouped together for the sake of clarity.
\begin{equation}
    \mathbf{T}(\Gamma)=\begin{cases}
    \hat{T}(\Gamma)=T_{0}(\Gamma) \quad \text{for } \Gamma\in V_0
\\

\hat{T}(\Gamma)=\tilde{T}_{0}(\Gamma)+T_1(\Gamma)-T_1(\tilde{T}_{0}(\Gamma))  \quad \text{for } \Gamma\in V_1 \\

\hat{T}(\Gamma)=\tilde{T}_{1}(\Gamma)+T_2(\Gamma)-T_2(\tilde{T}_{1}(\Gamma)) \quad \text{for } \Gamma\in\Big\{
            \raisebox{-0.1cm}{
            \resizebox{1.6cm}{!}{
            \begin{tikzpicture}
                    \tikzstyle{every node}=[font=\small]
                     
                     \node[main node mini] (1) {};
                     
                    \begin{feynman}
                    \vertex[left=1cm of 1 ] (2);
                      \vertex[right=1cm of 1] (3);
                      \vertex[above=1cm of 1] (4);

                      \diagram {
                         (2) -- [gluon] (1),
                         (1) -- [gluon] (3),
                        };
                    \end{feynman}
            \end{tikzpicture}}},
             \raisebox{-0.1cm}{
            \resizebox{1.6cm}{!}{
            \begin{tikzpicture}
                    \tikzstyle{every node}=[font=\small]
                     
                     \node[main node mini] (1) {};
                     
                    \begin{feynman}
                    \vertex[left=1cm of 1 ] (2);
                      \vertex[right=1cm of 1] (3);
                      \vertex[above=1cm of 1] (4);

                      \diagram {
                         (2) -- [ghost1] (1),
                         (1) -- [ghost1] (3),
                        };
                    \end{feynman}
            \end{tikzpicture}}}

\Big\} \\

   T^{\text{os}}(\Gamma)=\sum_{\sigma\in\{\pm 1\}}(T^{\text{os},\sigma}-T_1T^{\text{os},\sigma})(\Gamma)+T_1(\Gamma) \quad \text{for } \Gamma\in\Big\{       \raisebox{-0.1cm}{
            \resizebox{1.6cm}{!}{
            \begin{tikzpicture}
                    \tikzstyle{every node}=[font=\small]
                     
                     \node[main node mini] (1) {};
                     
                    \begin{feynman}
                    \vertex[left=1cm of 1 ] (2);
                      \vertex[right=1cm of 1] (3);
                      \vertex[above=1cm of 1] (4);

                      \diagram {
                         (2) -- [fermion1] (1),
                         (1) -- [fermion1] (3),
                        };
                    \end{feynman}
            \end{tikzpicture}}}
\Big\}
    \end{cases}
\end{equation}
where $V_0$ and $V_1$ are the vertices with \textit{true, local} dod being equal to $0$ and $1$ respectively, that is
\begin{equation*}
\begin{aligned}
    V_0=\Bigg\{
            \raisebox{-0.4cm}{
            \resizebox{1.6cm}{!}{
            \begin{tikzpicture}
                    \tikzstyle{every node}=[font=\small]
                     
                     \node[main node mini] (1) {};
                     
                    \begin{feynman}
                    \vertex[left=1cm of 1 ] (2);
                      \vertex[right=1cm of 1] (3);
                      \vertex[above=1cm of 1] (4);

                      \diagram {
                         (2) -- [fermion1] (1),
                         (1) -- [fermion1] (3),
                         (1) -- [gluon] (4),
                        };
                    \end{feynman}
            \end{tikzpicture}}},
            \raisebox{-0.4cm}{
            \resizebox{1.6cm}{!}{
            \begin{tikzpicture}
                    \tikzstyle{every node}=[font=\small]
                     
                     \node[main node mini] (1) {};
                     
                    \begin{feynman}
                    \vertex[left=1cm of 1 ] (2);
                      \vertex[right=1cm of 1] (3);
                      \vertex[above=1cm of 1] (4);

                      \diagram {
                         (2) -- [fermion1] (1),
                         (1) -- [fermion1] (3),
                         (1) -- [boson] (4),
                        };
                    \end{feynman}
            \end{tikzpicture}}},
            \raisebox{-0.4cm}{
            \resizebox{1.2cm}{!}{
            \begin{tikzpicture}
                    \tikzstyle{every node}=[font=\small]
                     
                     \node[main node mini] (1) {};
                     
                    \begin{feynman}
                    \vertex[above left=0.7cm and 0.7cm of 1 ] (2);
                      \vertex[above right=0.7cm and 0.7cm of 1] (3);
                      \vertex[below left=0.7cm and 0.7cm of 1] (4);
                      \vertex[below right=0.7cm and 0.7cm of 1] (5);

                      \diagram {
                         (2) -- [gluon] (1),
                         (1) -- [gluon] (3),
                         (1) -- [gluon] (4),
                         (1) -- [gluon] (5),
                        };
                    \end{feynman}
            \end{tikzpicture}}}, 
            \raisebox{-0.4cm}{
            \resizebox{1.2cm}{!}{
            \begin{tikzpicture}
                    \tikzstyle{every node}=[font=\small]
                     
                     \node[main node mini] (1) {};
                     
                    \begin{feynman}
                    \vertex[above left=0.7cm and 0.7cm of 1 ] (2);
                      \vertex[above right=0.7cm and 0.7cm of 1] (3);
                      \vertex[below left=0.7cm and 0.7cm of 1] (4);
                      \vertex[below right=0.7cm and 0.7cm of 1] (5);

                      \diagram {
                         (2) -- [gluon] (1),
                         (1) -- [gluon] (3),
                         (1) -- [boson] (4),
                         (1) -- [boson] (5),
                        };
                    \end{feynman}
            \end{tikzpicture}}},
            \raisebox{-0.4cm}{
            \resizebox{1.2cm}{!}{
            \begin{tikzpicture}
                    \tikzstyle{every node}=[font=\small]
                     
                     \node[main node mini] (1) {};
                     
                    \begin{feynman}
                    \vertex[above left=0.7cm and 0.7cm of 1 ] (2);
                      \vertex[above right=0.7cm and 0.7cm of 1] (3);
                      \vertex[below left=0.7cm and 0.7cm of 1] (4);
                      \vertex[below right=0.7cm and 0.7cm of 1] (5);

                      \diagram {
                         (2) -- [gluon] (1),
                         (1) -- [gluon] (3),
                         (1) -- [scalar] (4),
                         (1) -- [scalar] (5),
                        };
                    \end{feynman}
            \end{tikzpicture}}},
            \raisebox{-0.4cm}{
            \resizebox{1.2cm}{!}{
            \begin{tikzpicture}
                    \tikzstyle{every node}=[font=\small]
                     
                     \node[main node mini] (1) {};
                     
                    \begin{feynman}
                    \vertex[above left=0.7cm and 0.7cm of 1 ] (2);
                      \vertex[above right=0.7cm and 0.7cm of 1] (3);
                      \vertex[below left=0.7cm and 0.7cm of 1] (4);
                      \vertex[below right=0.7cm and 0.7cm of 1] (5);

                      \diagram {
                         (2) -- [gluon] (1),
                         (1) -- [gluon] (3),
                         (1) -- [ghost1] (4),
                         (5) -- [ghost1] (1),
                        };
                    \end{feynman}
            \end{tikzpicture}}}, \\
            \raisebox{-0.4cm}{
            \resizebox{1.2cm}{!}{
            \begin{tikzpicture}
                    \tikzstyle{every node}=[font=\small]
                     
                     \node[main node mini] (1) {};
                     
                    \begin{feynman}
                    \vertex[above left=0.7cm and 0.7cm of 1 ] (2);
                      \vertex[above right=0.7cm and 0.7cm of 1] (3);
                      \vertex[below left=0.7cm and 0.7cm of 1] (4);
                      \vertex[below right=0.7cm and 0.7cm of 1] (5);

                      \diagram {
                         (2) -- [ghost1] (1),
                         (1) -- [ghost1] (3),
                         (1) -- [ghost1] (4),
                         (5) -- [ghost1] (1),
                        };
                    \end{feynman}
            \end{tikzpicture}}},
            \raisebox{-0.4cm}{
            \resizebox{1.6cm}{!}{
            \begin{tikzpicture}
                    \tikzstyle{every node}=[font=\small]
                     
                     \node[main node mini] (1) {};
                     
                    \begin{feynman}
                    \vertex[left=1cm of 1 ] (2);
                      \vertex[right=1cm of 1] (3);
                      \vertex[above=1cm of 1] (4);

                      \diagram {
                         (2) -- [fermion1] (1),
                         (1) -- [fermion1] (3),
                         (1) -- [scalar] (4),
                        };
                    \end{feynman}
            \end{tikzpicture}}},
            \raisebox{-0.4cm}{
            \resizebox{1.6cm}{!}{
            \begin{tikzpicture}
                    \tikzstyle{every node}=[font=\small]
                     
                     \node[main node mini] (1) {};
                     
                    \begin{feynman}
                    \vertex[left=1cm of 1 ] (2);
                      \vertex[right=1cm of 1] (3);
                      \vertex[above=1cm of 1] (4);

                      \diagram {
                         (2) -- [gluon] (1),
                         (1) -- [gluon] (3),
                         (1) -- [scalar] (4),
                        };
                    \end{feynman}
            \end{tikzpicture}}},
            \raisebox{-0.4cm}{
            \resizebox{1.6cm}{!}{
            \begin{tikzpicture}
                    \tikzstyle{every node}=[font=\small]
                     
                     \node[main node mini] (1) {};
                     
                    \begin{feynman}
                    \vertex[left=1cm of 1 ] (2);
                      \vertex[right=1cm of 1] (3);
                      \vertex[above=1cm of 1] (4);

                      \diagram {
                         (2) -- [ghost1] (1),
                         (1) -- [ghost1] (3),
                         (1) -- [scalar] (4),
                        };
                    \end{feynman}
            \end{tikzpicture}}}\Bigg\}, 
\end{aligned}
\end{equation*}
where the last two vertices could be seen as an outlier as it has an \textit{apparent} dod of $1$, while an exact investigation of its \glss{UV} behaviour reveals it is logarithmic. All other vertices (e.g. the vertex connecting one (three) gluon with three (one) eletro-weak bosons) are identically zero due to their colour structure. The linearly-divergent vertices all involve the interaction of bosons and ghosts, i.e.
\begin{equation}
V_1=\Bigg\{
            \raisebox{-0.4cm}{
            \resizebox{1.4cm}{!}{
            \begin{tikzpicture}
                    \tikzstyle{every node}=[font=\small]
                     
                     \node[main node mini] (1) {};
                     
                    \begin{feynman}
                    \vertex[left=1cm of 1 ] (2);
                      \vertex[right=1cm of 1] (3);
                      \vertex[above=1cm of 1] (4);

                      \diagram {
                         (2) -- [gluon] (1),
                         (1) -- [gluon] (3),
                         (1) -- [gluon] (4),
                        };
                    \end{feynman}
            \end{tikzpicture}}},
            \raisebox{-0.4cm}{
            \resizebox{1.6cm}{!}{
            \begin{tikzpicture}
                    \tikzstyle{every node}=[font=\small]
                     
                     \node[main node mini] (1) {};
                     
                    \begin{feynman}
                    \vertex[left=1cm of 1 ] (2);
                      \vertex[right=1cm of 1] (3);
                      \vertex[above=1cm of 1] (4);

                      \diagram {
                         (2) -- [gluon] (1),
                         (1) -- [gluon] (3),
                         (1) -- [boson] (4),
                        };
                    \end{feynman}
            \end{tikzpicture}}},
            \raisebox{-0.4cm}{
            \resizebox{1.6cm}{!}{
            \begin{tikzpicture}
                    \tikzstyle{every node}=[font=\small]
                     
                     \node[main node mini] (1) {};
                     
                    \begin{feynman}
                    \vertex[left=1cm of 1 ] (2);
                      \vertex[right=1cm of 1] (3);
                      \vertex[above=1cm of 1] (4);

                      \diagram {
                         (2) -- [ghost1] (1),
                         (1) -- [ghost1] (3),
                         (1) -- [boson] (4),
                        };
                    \end{feynman}
            \end{tikzpicture}}},
            \raisebox{-0.4cm}{
            \resizebox{1.6cm}{!}{
            \begin{tikzpicture}
                    \tikzstyle{every node}=[font=\small]
                     
                     \node[] (init) {};
                     
                     \node[main node mini] (1) [left=0.0cm of init] {};
                     
                    \begin{feynman}
                    \vertex[left=1cm of 1 ] (2);
                      \vertex[right=1cm of 1] (3);
                      \vertex[above=1cm of 1] (4);

                      \diagram {
                         (2) -- [ghost1] (1),
                         (1) -- [ghost1] (3),
                         (1) -- [gluon] (4),
                        };
                    \end{feynman}
            \end{tikzpicture}}}
\Bigg\}.
\end{equation}
We see that the subtraction of every divergent \glss{1PI} is performed using $\hat{T}$ while for fermionic self-energy corrections (both massive and massless), $T^{\text{os}}$ should be used. For a massless self-energy correction $\Sigma$, one can easily see that $T^{\text{os}}(\Sigma)=\hat{T}(\Sigma)$. Finally, the construction of \glss{OS} counterterms for massive gauge bosons is left to future work, although a similar construction will apply in that case as well.

\input{chapters/nesting_os}

\section{Localised renormalisation}
\label{sect:localised_renormalisation}

The $\mathbf{T}$ approximation operator satisfies the first two constraints laid out in sect.~\ref{sect:constraints_on_K}. Specifically, $R(\Gamma)$, for non-exceptional external momenta, is integrable. After the $R$-operation is applied to each interference diagram (as specified in sect.~\ref{sect:UV_cts_in_LU}), the Local Unitarity representation that is constructed from such subtracted interference diagrams is also integrable. In summary, $\mathbf{T}$ satisfies all \textit{local} requirements.

On the other hand, \textit{integrated-level} constraints should also be satisfied by $\mathbf{T}$, and specifically, as laid out in sect.~\ref{sect:constraints_on_K}, we want the renormalisation operator $K=\mathbf{T}+\bar{K}$ to reproduce the hybrid $\overline{\text{MS}}$+\glss{OS} scheme, and $\bar{K}$ to be obtainable with minimal analytic work, that is through the computation of single-scale vacuum diagrams.
We will achieve these goals in this section.

\subsection{Integrated counterterms}

We start by discussing the integrated version of $\mathbf{T}(\Gamma)$, which we denote with $[\mathbf{T}(\Gamma)] = \mathbf{T}([\Gamma])$, where we consider the analytic integration of $\Gamma$ in $d=4-2 \epsilon$, which can be formally written as a Laurent series in the dimensional regulator $\epsilon$:
\begin{equation}
\label{eq:d_dimensional_integration_operator}
    [\Gamma] = \left(\frac{\mu_r^2}{4\pi e^{-\gamma_\text{E}}}\right)^{L\epsilon}\int \left( \Pi_{i=1}^{L} d^{4-2\epsilon} k_i \right) \Gamma(k_1,\ldots,k_{L}) = \sum_{k=-\infty}^{+\infty} \alpha_k \epsilon^{k},
\end{equation}
where $\mu_r$ is the renormalisation scale, $L$ is the loop count of the graph $\Gamma$ and the expansion coefficients $\alpha_k$ are tensors function of internal masses and external momenta, and the normalisation is chosen so as to facilitate enforcing $\overline{\text{MS}}$ conventions on $[\Gamma]$.

Because three operators, $T$, $\tilde{T}$, $T^{\text{os}}$ participate in the definition of $\mathbf{T}(\Gamma)$, we will devote one section for each operator's integrated counterpart, $[T(\Gamma)]$, $[\tilde{T}(\Gamma)]$, $[T^{\text{os}}(\Gamma)]$. For a notational simplicity, we will often write $T(\Gamma)$ as $T \Gamma$.

\subsubsection{Integrated UV counterterms}

\label{sect:integrated_uv_cts}
We start by discussing $T(\Gamma)$. One particular advantage of the local operator $T$ is that its analytic computation $[T(\Gamma)]$ only involves single-scale massive tensor vacuum graphs and it can be performed in $d=4-2\epsilon$ dimensional Minkowski space using traditional techniques. Specifically, each local counterterm originating from $T(\Gamma)$ is of the form
\begin{equation}
    T(\Gamma)=\int \left[\prod_{i=1}^L \frac{\mathrm{d}^4 k_i}{(2\pi)^4}\right] \frac{N(\{p_j\},\{k_i\},\mathbf{m})}{\prod_{e\in\mathbf{e}}(q_e(\{k_i\})^2-m_{\text{uv}}^2)^{\alpha_e}},
\end{equation}
where $q(\{k_i\})$ is a linear combination of the loop momenta which does not depend on the external momenta $\{p_j\}$ or the masses $\mathbf{m}$, and $N$ is a polynomial, tensorial numerator in its input. 
Our current implementation offers full support for these computations up to and including three-loop amplitudes. Note that in principle, this can be extended to four loops~\cite{Pikelner:2017tgv} and beyond. Below we sketch the construction of our automated setup.

The first step is to tensor reduce the analytic integrand in order to obtain scalar integrands. Since the integral is a vacuum bubble, all tensor structures reduce to combinations of the metric. For example, for a rank two integral:
\begin{equation}
  \Gamma^{\mu \nu} = A g^{\mu \nu} = \frac{\Gamma^{\alpha}_{\ \alpha}}{d} g^{\mu \nu}
\end{equation}
and for a rank four integral (odd ranks are 0 as there is no possible tensor basis):
\begin{align}
  \begin{split}
  \Gamma^{\mu \nu \rho \sigma} =& A g^{\mu \nu} g^{\rho \sigma} + B g^{\mu \rho} g^{\nu \sigma} + C g^{\mu \sigma} g^{\nu \rho} = \frac{1}{(-1 + d) d (2 + d)}\biggl(\\
  &+\left((1+d) \Gamma^{\alpha\ \beta}_{\ \alpha\ \beta} - \Gamma^{\alpha \beta\ }_{\ \ \alpha \beta} - \Gamma^{\alpha \beta\ }_{\ \ \beta \alpha} \right) g^{\mu \nu} g^{\rho \sigma} \\
  &+\left((1+d) \Gamma^{\alpha \beta\ }_{\ \ \alpha \beta} - \Gamma^{\alpha\ \beta}_{\ \alpha\ \beta} - \Gamma^{\alpha \beta\ }_{\ \ \beta \alpha} \right) g^{\mu \rho} g^{\nu \sigma}\\
  &+\left((1+d) \Gamma^{\alpha \beta\ }_{\ \ \beta \alpha} - \Gamma^{\alpha\ \beta}_{\ \alpha\ \beta} - \Gamma^{\alpha \beta\ }_{\ \ \alpha \beta} \right) g^{\mu \sigma} g^{\nu \rho}
  \biggr) \;.
  \end{split} 
\end{align}
With increasing tensor rank, the number of possible metric structures $n$ grows exponentially (rank 6 gives $n=15$, rank 8 gives $n=105$). A naive tensor reduction algorithm would have to solve an $n \cross n$ linear system, which is slow for high rank (although this only needs to be performed once when generating the \glss{LU} integrand for a particular process). An improved algorithm that exploits the underlying symmetry group and symmetries of the momenta that the indices belong to was derived in ref.~\cite{Ruijl:2018poj} and is used in our implementation.

All up to three-loop vacuum graphs are linear combinations (with tensorial coefficients) of the following three topologies:
\begin{equation}
    \begin{tikzpicture}
        \begin{feynman}

            \tikzfeynmanset{every vertex={empty dot,minimum size=0mm}}
        \vertex (a1);

        \vertex[right=1cm of a1] (a3);

        \tikzfeynmanset{every vertex={empty dot,minimum size=0mm}}

        \vertex[right=0.5cm of a1] (a2);
        \vertex[above=0.5cm of a2] (a4);
        \vertex[below=0.5cm of a2] (a5);

        \vertex[left=0.433cm of a2] (b1);
        \vertex[left=0.433cm of a2] (b2);

        \tikzfeynmanset{every vertex={dot,minimum size=0.8mm}}
        \vertex[below right=0.71cm of a1] (b3);

        \diagram*[large]{
        (a1)--[quarter left](a4) -- [quarter left](a3),
        (a5) --[quarter right](a3),
        (a5)--[quarter left](a1),
        };
        \end{feynman}

    \end{tikzpicture}
    \quad
    \begin{tikzpicture}
        \begin{feynman}

        \tikzfeynmanset{every vertex={dot,minimum size=0.8mm}}
        \vertex (a1);
        \vertex[right=1cm of a1] (a3);

        \tikzfeynmanset{every vertex={dot,minimum size=0mm}}

        \vertex[right=0.5cm of a1] (a2);
        \vertex[above=0.5cm of a2] (a4);
        \vertex[below=0.5cm of a2] (a5);

        \vertex[left=0.433cm of a2] (b1);
        \vertex[left=0.433cm of a2] (b2);

        \diagram*[large]{
          (a1)--[quarter left](a4) -- [quarter left](a3),
          (a5) --[quarter right](a3),
          (a5)--[quarter left](a1),
          (a1) -- (a3)
          };
        \end{feynman}

    \end{tikzpicture}
    \quad
    \begin{tikzpicture}
        \begin{feynman}

        \tikzfeynmanset{every vertex={dot,minimum size=0.8mm}}
        \vertex (a1);
        \vertex[right=1cm of a1] (a3);
        \vertex[right=0.5cm of a1] (a2);
        \vertex[below=0.5cm of a2] (a5);

        \tikzfeynmanset{every vertex={dot,minimum size=0mm}}

        \vertex[above=0.5cm of a2] (a4);

        \vertex[left=0.433cm of a2] (b1);
        \vertex[left=0.433cm of a2] (b2);

        \diagram*[large]{
          (a1)--[quarter left](a4) -- [quarter left](a3),
          (a5) --[quarter right](a3),
          (a5)--[quarter left](a1),
          (a1) -- (a3),
          (a2) -- (a5),
          };
        \end{feynman}

    \end{tikzpicture}
\end{equation}
since all other topologies can be reproduced from the above three by setting certain edge powers to zero. For the three topologies, the denominator momenta form a complete basis for scalar products. As a result, all scalar products in the numerator can be rewritten as a polynomial in the propagator denominators.

The resulting integrals with numerator 1 and various powers of the denominator can be reduced to a set of master integrals using integration-by-parts (IBP) identities~\cite{Chetyrkin:1981qh}. We use LiteRed~\cite{Lee:2012cn,Lee:2013mka} to generate a parametric reduction for each of the three topologies.
The resulting 7 master integrals have been taken from the literature~\cite{Kniehl:2017ikj,Martin:2016bgz} with sufficiently deep expansions in $\epsilon$ for three-loop computations.
The IBP reduction and master integral computation has been verified by comparing various numerator and denominator configurations with numerical results from pySecDec~\cite{Borowka:2017idc}.

\subsubsection{Integrated spurious soft subtraction counterterms}
\label{sect:integrated_soft}
In massless gauge theories, all soft counterterms integrate to zero. This means that at the integrated level,
\begin{equation}
    [\hat{T}_{\text{dod}}\Gamma]=[T_{\text{dod}}\Gamma]-[T_{\text{dod}}\tilde{T}_{\text{dod}-1}\Gamma].
\end{equation}
which only requires the analytic integration of one-scale vacuum bubbles.

The reason why $[\tilde{T}_{\text{dod}-1}\Gamma]=0$ can be traced back to gauge invariance. According to sect.~\ref{sect:summary_bold_T}, we only need to study the cases in which $\Gamma$ is a gluonic, fermionic or ghost self-energy and that in which $\Gamma$ is a three vertex connecting bosons and ghosts. If 
\begin{itemize}
    \item $\Gamma$ is a self-energy, then $[\tilde{T}_{\text{dod}-1}\Gamma]=0$ for otherwise radiative correction would generate a mass for the gluon, massless quark or ghost, which is protected against by gauge invariance.
    \item $\Gamma$ is a gluonic three-vertex, then $[\tilde{T}_{\text{dod}-1}\Gamma]=0$ is a tensor with three indices and does not depend on any four-momentum. Thus it allows no Lorentz decomposition and it must be a vanishing tensor. For the same reason, if $\Gamma$ is a vertex at which a ghost, an anti-ghost and a gluon meet, then $[\tilde{T}_{\text{dod}-1}\Gamma]=0$.
    \item $\Gamma$ is a vertex coupling three gauge bosons (of any type), we can iterate the argument that we introduced for the three gluons: $[\tilde{T}_{\text{dod}-1}\Gamma]=0$ is a tensor with three indices and that does not depend on any four momentum, and thus it is zero at the integrated level.
\end{itemize}

\subsubsection{Integrated OS counterterms}
\label{sect:integrated_os_cts}
For the \glss{OS} subtraction operator we will show that while $[T^{\text{os}}\Sigma]$ cannot be fully written in terms of one-scale vacuum diagrams, $[T^{\text{os}}\Sigma]-\delta m^{\text{os}}$ can, where $\delta m^{\text{os}}$ is the \glss{OS} renormalised mass correction.
Writing the expression of $T^{\text{os}}$ in terms of $T$ and $T^{\text{os}\pm}$, we get
\begin{equation}
    [T^{\text{os}}\Sigma]=\frac{1}{2}([T^{\text{os}+}\Sigma]+[T^{\text{os}-}\Sigma])-\frac{1}{2}([T(T^{\text{os}+}\Sigma)]+[T[(T^{\text{os}-}\Sigma)])+[T\Sigma],
\end{equation}
which shows we only need to discuss $[T^{\text{os}\pm} \Sigma]$, as all other integrated counterterms are of the form discussed in sect.~\ref{sect:integrated_uv_cts}. It is also clear that $[T^{\text{os}\pm} \Sigma]$ cannot be expressed as a one-scale vacuum diagram and that it is not identically zero, unlike soft counterterms. Instead, $T^{\text{os}\pm}$ corresponds to the self-energy being evaluated at an on-shell momentum, and being projected on the subspaces identified by $P_{\pm}$. Thus, it is intimately related with the on-shell behaviour of the resummed fermionic propagator. We now consider the Lorentz decomposition of a fermionic self-energy:
\begin{equation}
    \Sigma(p,\mathbf{m})=\slashed{p} \Sigma_\psi(p^2,\mathbf{m})+m \mathds{1} \Sigma_m(p^2,\mathbf{m}).
\end{equation}
$\Sigma_\psi$ and $\Sigma_m$ are scalar functions of $p^2$ and of the mass vector $\mathbf{m}$. Evaluating this expression for $p=p^{os}=(m,0,0,0)$, we obtain
\begin{equation}
    \Sigma(p^{os},\mathbf{m})=\slashed{p}^{os} \Sigma_\psi(m^2,\mathbf{m})+m \mathds{1} \Sigma_m(m^2,\mathbf{m}).
\end{equation}
We have that $\slashed{p}^{os}=m\gamma^0$. The value of $\Sigma_\psi(m^2,\mathbf{m})$ and $\Sigma_m(m^2,\mathbf{m})$ defines the renormalisation constants of the field and mass in the \glss{OS} scheme. Inserting the decomposition of $\Sigma(p,\mathbf{m})$ into $[T^{\text{os}\pm}[\Sigma]]$, we obtain
\begin{equation}
    \frac{1}{2}([T^{\text{os}+}\Sigma]+[T^{\text{os}-}\Sigma])=m\left([\Sigma_\psi(m^2,\mathbf{m})]+[\Sigma_m(m^2,\mathbf{m})]\right)=\delta m^{\text{os}}.
\end{equation}
which agrees with the renormalisation expressions given in e.g. ref.~\cite{Denner:1991kt}.
In other words, the counterterms in $T^{\text{os}}[\Sigma]$, whose integration involves the computation of multi-scale integrals, actually reproduce the \glss{OS} mass correction to the fermionic propagator.  Unfolding the action of $T$, we can finally write a compact and more explicit formula for the full integrated $[T^{\text{os}}[\Sigma]]$:
\begin{equation}
\label{eq:integrated_os}
    [T^{\text{os}} \Sigma ]=\delta m^{\text{os}}+(p-\gamma^0 p^{os})^\mu \left[\frac{\partial}{\partial p^\mu} \Sigma(p,\mathbf{m}) \Bigg|_{\substack{\hspace{-0.1cm}p=0, \\ \mathbf{m}=0, \\ \hspace{-0.2cm}m_{\text{uv}}}}\right].
\end{equation}
The last term on the right-hand side of eq.~\eqref{eq:integrated_os} is a one-scale vacuum diagram, and the discussion of sect.~\ref{sect:integrated_uv_cts} applies.
This concludes our discussion of the integrated version of $T^{\text{os}}$. We have shown that, up to single-scale vacuum integrals, $[T^{\text{os}}]$ is the correction to the fermion mass in the \glss{OS} scheme. As discussed in the next sect.~\ref{sec:MSbar_and_OS_integrated_UVCT}, we will use this fact to automatically reproduce the hybrid $\overline{\text{MS}}$ + \glss{OS} scheme.

\subsection{Localised renormalisation in the hybrid $\overline{\text{MS}}$ + OS scheme}
\label{sec:MSbar_and_OS_integrated_UVCT}

Having determined what the integrated version of any counterterm $[\mathbf{T} (\Gamma) ]$ is, we are ready to discuss the value $\bar{K}(\Gamma)$ should take in order to reproduce the $\overline{\text{MS}}$ + \glss{OS} scheme at the integrated level. $\bar{K}(\Gamma)$ is thus determined by constraining $K(\Gamma)=\mathbf{T}(\Gamma)+\bar{K}(\Gamma)$ to reproduce the desired scheme. We will show that
\begin{itemize}
    \item $\bar{K}(\Gamma)$ can be entirely written in terms of integrated one-scale vacuum bubbles with the $\overline{\text{MS}}$ integration measure, showing that the constraint of minimal analytic work is satisfied.
    \item Given the above $\bar{K}(\Gamma)$, the mass is renormalised in the \glss{OS} scheme directly through the $\delta m^{\text{os}}$ piece participating in the definition of $T^{\text{os}}$, whereas fields and couplings are renormalised in $\overline{\text{MS}}$.
    \item Couplings can be redefined in order to reproduce \glss{OS} renormalisation of fields and $\overline{\text{MS}}$ renormalisation of vertices.
\end{itemize}
Furthermore, because $K(\Gamma)$ is required to have a local representation, much like $\mathbf{T}(\Gamma)$, $\bar{K}(\Gamma)$ also needs to have a local representation. Therefore, we will talk of localised renormalisation, for three reasons: a) renormalisation counterterms are localised \textit{within the graph} by the $R$-operation, b) $\overline{\text{MS}}$ counterterms will be localised by multiplying them with convergent, normalised tadpoles, c) \glss{OS} mass counterterms will be localised by the definition of $T^{\text{os}}$ itself, which provides a local representation of $\delta m^{os}$. 

In summary, the full, carefully-constructed $K$ will satisfy all four constraints laid out in sect.~\ref{sect:constraints_on_K}.

\subsubsection{Field and coupling renormalization in the $\overline{\text{MS}}$ scheme}
\label{sec:msbar}
\label{sec:integrated_UV_CT}

In this section we will determine $\bar{K}$ by imposing the constraint that fields and couplings are renormalised in the $\overline{\text{MS}}$ scheme. In the next section, we will argue that this also reproduces \glss{OS} renormalisation of masses. We will start by determining the $K^{\overline{\text{MS}}}$ operation, which renormalises a graph $\Gamma$ in the $\overline{\text{MS}}$. Then, we will use $K^{\overline{\text{MS}}}$ to construct $\bar{K}$.

We denote by $K^{\overline{\text{MS}}}([\Gamma])$ the $\overline{\text{MS}}$ implementation of the $K$ operator acting on the $d$-dimensional integral of the graph which is defined so as to remove all the poles in the Laurent series of $[\Gamma]$:
\begin{equation}
\label{eq:IntegratedMSbarDef}
K^{\overline{\text{MS}}}([\Gamma]) := \sum_{k=-\infty}^{-1} \alpha_k \epsilon^{k}.
\end{equation}
Our objective is to build the local equivalent $K^{\overline{\text{MS}}}(\Gamma)$ of the integrated-level definition $K^{\overline{\text{MS}}}([\Gamma])$ above so that our subtraction terms both locally remove \glss{UV} singularities and also immediately yield results where fields and couplings are renormalised in $\overline{\text{MS}}$. We let the local version of the $\overline{\text{MS}}$ operator be the sum of a subtraction operator $T$, that captures the UV divergent part of a graph, plus a finite part $\bar{K}^{\overline{\text{MS}}}$, so that
\begin{equation}
\label{eq:splitting_K_MSbar}
    K^{\overline{\text{MS}}}(\Gamma)=T(\Gamma)+\bar{K}^{\overline{\text{MS}}}(\Gamma) \,.
\end{equation}
In the following, we will fix $T$, the subtraction operator (which can be chosen arbitrarily, for the purposes of this section), and study the finite part $\bar{K}^{\overline{\text{MS}}}$ that is required for $K^{\overline{\text{MS}}}$ to be a faithful local representation of the $\overline{\text{MS}}$ renormalisation operator. In particular, the relation between $K^{\overline{\text{MS}}}(\Gamma)$ with $K^{\overline{\text{MS}}}([\Gamma])$ is simply stated: analytically integrating $K^{\overline{\text{MS}}}(\Gamma)$ should yield $K^{\overline{\text{MS}}}([\Gamma])$, i.e.
\begin{equation}
\label{eq:local_ms_equal_integrated_ms}
    K^{\overline{\text{MS}}}([\Gamma])=[K^{\overline{\text{MS}}}(\Gamma)]=[T(\Gamma)]+[\bar{K}^{\overline{\text{MS}}}(\Gamma)].
\end{equation}

Now observe that the local counterterm $Z(\gamma)$ in the $R$-operation as defined in eq.~\ref{eq:UV_master_equation} captures the divergence of all loop momenta of $\gamma$ going to infinity, minus its \glss{UV} subdivergences. The poles in $\epsilon$ from the $d=4-2\epsilon$ dimensional integral of $Z(\gamma)$, that is $[Z(\gamma)]$, have coefficients that are polynomial in the external momenta and internal masses~\cite{Bogoliubov:1957gp}. The same must hold for $K(\Gamma)$.
Thus, we obtain the following key relation:
\begin{equation}
\label{eq:KmsbarSameForT}
    K^{\overline{\text{MS}}}([\Gamma]) = K^{\overline{\text{MS}}}([ T(\Gamma)])\;.
\end{equation}
We are finally ready to clarify the effect of the constraint from eq.~\eqref{eq:local_ms_equal_integrated_ms} on $\bar{K}^{\overline{\text{MS}}}$. We can simply solve for $[\bar{K}^{\overline{\text{MS}}}(\Gamma)]$ and use eq.~\eqref{eq:KmsbarSameForT} to express the argument of $K^{\overline{\text{MS}}}$ in terms of the approximation $T(\Gamma)$
\begin{equation}
\label{eq:SchemeTransitionDefiningEq}
    [\bar{K}^{\overline{\text{MS}}}(\Gamma)]=K^{\overline{\text{MS}}}([\Gamma])-[T(\Gamma)]=(K^{\overline{\text{MS}}}-\mathds{1})[T(\Gamma)] \,.
\end{equation}
Any local operator $K^{\overline{\text{MS}}}(\Gamma)$ as in eq.~\eqref{eq:splitting_K_MSbar} for fixed arbitrary subtraction operator $T(\Gamma)$ must have its local renormalisation piece $\bar{K}^{\overline{\text{MS}}}(\Gamma)$ satisfying the integrated level constraint of eq.~\eqref{eq:SchemeTransitionDefiningEq}. The action of the operator $\left(\mathds{1}-K^{\overline{\text{MS}}}\right)$ can be understood as removing the pole parts in $\epsilon$ of the analytic $d$-dimensional integral $[ T(\Gamma)]$, as it is the complement of eq.~\eqref{eq:IntegratedMSbarDef}:
\begin{equation}
\label{eq:ComplementMSbarDef}
\left(\mathds{1}-K^{\overline{\text{MS}}}\right) \left(\,[\Gamma]\,\right) := \alpha_0 + \alpha_1 \epsilon + \ldots \;.
\end{equation}

Given the integrated-level constraint of eq.~\eqref{eq:SchemeTransitionDefiningEq}, we now detail the construction of a local operator $\bar{K}^{\overline{\text{MS}}}(\Gamma)$ that satisfies it. In practice, this means finding a function of the loop momenta $k_1,...,k_L$ of the graph $\Gamma$ whose integral yields the right-hand side of eq.~\eqref{eq:SchemeTransitionDefiningEq}, which is assumed as input. The simplest way to achieve this is to take the integrated level result and multiply it by any function of the loop momenta that is normalised to one. For example, we choose to define the \emph{local} version $\bar{K}(\Gamma)$ of the integrated quantity $[\bar{K}(\Gamma)]$ using a scalar vacuum graph as normalisation factor:
\begin{equation}
\label{eq:KbarLocalisation}
\bar{K}^{\overline{\text{MS}}}(\Gamma)(k_1, \ldots, k_{L}) = [\bar{K}^{\overline{\text{MS}}}(\Gamma)]\;\left( \Pi_{i=1}^{L}N(k_i)\right),\quad \int d^4 k N(k) = \int d^4 k \frac{2\mathrm{i}(4 \pi)^2 m_{\text{UV}}^2}{(k^2-m_{\text{UV}}^2)^3} = 1.
\end{equation}

With this final result, we have constructed the subtraction operator $K^{\overline{\text{MS}}}(\Gamma)$, which is the local equivalent to the traditional integrated-level $\overline{\text{MS}}$ subtraction operator $K^{\overline{\text{MS}}}([\Gamma])$, but with the additional merit of locally removing \glss{UV} divergences. Its final expression reads:
\begin{equation}
    \label{eq:FinalLocalMSbarKExpression}
    K^{\overline{\text{MS}}}(\Gamma) = T_\text{dod}(\Gamma) - (\mathds{1}-K^{\overline{\text{MS}}}) \left(\,  [T_\text{dod}(\Gamma)] \,\right) \;\left( \Pi_{i=1}^{n_l}N(k_i)\right).
\end{equation}

We stress that the integrated counterterm $[\bar{K}^{\overline{\text{MS}}}(\Gamma)]$ captures more than just scheme-dependent contributions; it also reproduces the rational terms~\cite{Lang:2020nnl,Pozzorini:2020hkx} (both $R_1$ and $R_2$, in the one-loop terminology of ref.~\cite{Ossola:2008xq}) of purely $(d-4)$-dimensional origin which cannot possibly be captured by the numerical four-dimensional integration of $\Gamma-T_\text{dod}(\Gamma)$, and are thus traditionally cumbersome to account for in purely numerical approaches. In other words, the choice of $K(\Gamma)=T_\text{dod}(\Gamma)$  would by itself \emph{not} correspond to a consistent renormalisation scheme because the local subtraction operator violates the gauge symmetry of the \glss{SM}~\cite{Blaschke:2013cba} and would effectively introduce new terms to the Lagrangian.
This fact is made clear when considering for example the leading-order contribution to the loop-induced process $g g \rightarrow H H$ which receives no contribution from any renormalisation counterterm. Yet, this process contains logarithmically \glss{UV} divergent graphs, which in turn induce rational and $m_\text{UV}$-independent finite contributions from $[\bar{K}(\Gamma)]$.
As argued in e.g. ref.~\cite{Weinzierl:2014iaa}, such contributions could never be captured by a computation without the introduction of any \glss{UV} regulator.

In our construction of the $R$-operator, we use the `t Hooft Veltman (HV) conventions for dimensional regularisation (see details of this convention in ref.~\cite{Signer:2008va}), meaning that indices and momenta external to $\Gamma$ are treated strictly in 4 dimensions (as they are numerically integrated), whereas internal ones are in $d=4-2\epsilon$ dimensions. This implies that the $d$-dimensional conventions are effectively different for each term generated by the $R$-operation, since the distinction between internal and external is governed by which part of the graph is in argument of the $K$-operator and the remaining graph. However, since the $\left(\mathds{1}-K^{\overline{\text{MS}}}\right)$ removes the poles, a pole can never hit the remaining graph computed in 4 dimensions. Thus, because the cross-section is finite for $d=4$, using HV dimensional regularisation yields the same result as if it were computed by keeping \emph{all} momenta and indices in $d=4-2\epsilon$ dimension (i.e. Conventional Dimensional Regularisation (CDR)) throughout the implementation of the $R$-operation.

\subsubsection{Summary: the $\bar{K}$ operator}

This completes our construction of $R^{(\overline{\text{MS}})}$, a local \glss{UV} subtraction procedure that automatically yields results renormalised in $\overline{\text{MS}}$. We can now apply it to our specific case to achieve $\overline{\text{MS}}$ renormalisation of fields and couplings; given a graph $\Gamma$, we define
\begin{equation}
    \bar{K}(\Gamma)=-\left( \Pi_{i=1}^{n_l}N(k_i)\right)(\mathds{1}-K^{\overline{\text{MS}}})[T(\Gamma)-T (\hat{T}( \Gamma))]
\end{equation}
if $\Gamma\in V_0, \ V_1$ (as defined in sect.~\ref{sect:summary_bold_T}) or if $\Gamma$ is a ghost or gluonic self-energy. For a massive fermionic self-energy $\Sigma$, the $\hat{T}$ operator inside $\bar{K}$ is substituted with $T^{\text{os}}$:
\begin{equation}
\label{eq:k_bar_on_massive_se}
    \bar{K}(\Sigma)=-\left( \Pi_{i=1}^{n_l}N(k_i)\right)(\mathds{1}-K^{\overline{\text{MS}}})[T(\Sigma)-T (T^{\text{os}}(\Sigma))].
\end{equation}
This concludes the explanation of our choice for $\bar{K}$.

\subsubsection{Mass and gauge field renormalisation in the OS scheme}
\label{sec:os_mass_renormalisation}
Having now constructed $\bar{K}$ so as to reproduce $\overline{\text{MS}}$-renormalised cross-sections, we set ourselves to reproducing the common \glss{OS} renormalisation conditions for particle masses. The key observation is that in eq.~\eqref{eq:k_bar_on_massive_se}, we omitted the $\overline{\text{MS}}$ renormalisation of $T^{\text{os}}(\Sigma)$, which does contribute to the definition of $K$. The reason behind this omission is that the integrated version of $T^{\text{os}}(\Sigma)$ automatically reproduces the renormalised mass correction in the \glss{OS} scheme, as we will now demonstrate. We observe that for a fermionic self-energy $\Sigma$, the renormalisation operator yields
\begin{equation}
    [K(\Sigma)]=\delta m^{\text{os}}+\slashed{p} \, \delta Z^{\overline{\text{MS}}},
\end{equation}
having used the definition of $\bar{K}$ from the previous section and the integrated level result of sect.~\eqref{sect:integrated_os_cts}. In short, our definition of $\bar{K}$ and $\mathbf{T}$ automatically reproduces \glss{OS} renormalisation conditions for particle masses, and $\overline{\text{MS}}$ renormalisation for all fields and vertices~\cite{Denner:1991kt}.

We now briefly explain how to achieve \glss{OS} renormalisation of not only masses but also gauge fields, with $\overline{\text{MS}}$ renormalisation of vertices. The conversion to such a scheme is especially straightforward because we have already renormalised the masses in the \glss{OS} scheme, and we would now only have to change the renormalisation of gauge fields from $\overline{\text{MS}}$ to the \glss{OS} scheme. To this end, observe that the renormalisation constants of the coupling, gauge fields and vertices are related through the Slavnov-Taylor identities.
This implies that when dressing all vertices with the renormalisation counterterm of their respective coupling, the dependence on the gauge field renormalisation constants completely disappears. This is not immediately manifest in traditional computations that consider truncated amplitudes, because in that case the gauge field renormalisation constants must be added for each external gauge boson of the amplitude considered, in accordance with the \glss{LSZ} reduction formula (see discussion around eq.~2.83 of ref.~\cite{Alwall:2014hca}).
In contrast, the \glss{LU} representation explicitly retains the self-energy corrections of external particles since their local expression is necessary for \glss{IR} cancellations. %
This implies that within \glss{LU}, changes in both coupling and gauge field renormalisation conditions can be captured at once by only adjusting the expression of the renormalised couplings. More precisely, transitioning from pure $\overline{\text{MS}}$ to the hybrid $\overline{\text{MS}}$+\glss{OS} scheme for the strong coupling $g_s$ and gluon field renormalisation $Z_g$ can be achieved using the following substitution:
\begin{equation}
\label{eq:couplingSchemeTransition}
    g_s^{\overline{\text{MS}}} \rightarrow g_s^{\overline{\text{MS}}+\mathrm{OS}}=g_s^{\overline{\text{MS}}}\left(\frac{Z_g^{\mathrm{OS}}}{Z_g^{\overline{\text{MS}}}}\right)^{-\frac{3}{2}}\,,
\end{equation}
where the quantity in parenthesis is finite, process independent and thus only needs to be computed once. In practice, such hybrid scheme for coupling renormalisation is often referred to as a \emph{decoupling scheme} because it removes the \emph{explicit} dependence from massive particles in the running of the coupling, see refs.~\cite{Collins:1978wz,Bernreuther:1981sg,Prosperi:2006hx}.
Since each supergraph necessarily factorises a specific combination of couplings that our \glss{LU} construction readily renormalises in $\overline{\text{MS}}$ by default, then the substitution of eq.~\eqref{eq:couplingSchemeTransition}, if desired, is trivial to perform in a final post-processing step, completely independently of the $R$-operation.

\section{Gauge invariance in Unitarity methods}

\label{sec:gauge_invariance}
\begin{figure}[ht!]
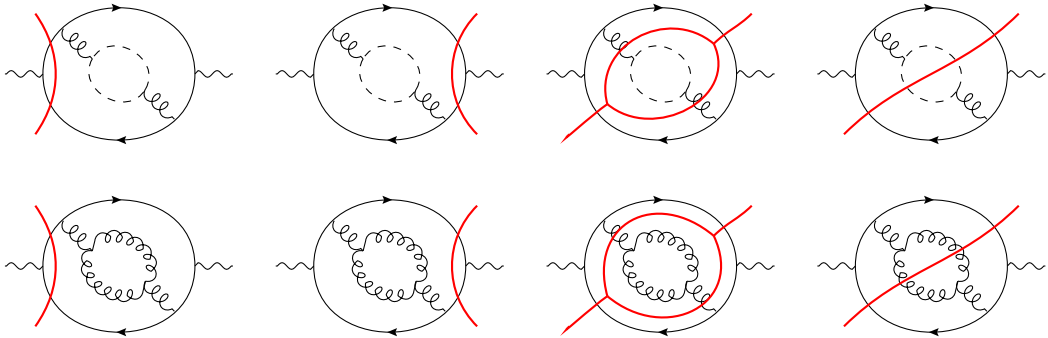

\begin{tabular}{cccc}
\input{diagrams/interferences_example1/diag1} &
\input{diagrams/interferences_example1/diag2} &
\input{diagrams/interferences_example1/diag3} &
\input{diagrams/interferences_example1/diag4} \\
\end{tabular}

\begin{tabular}{cccc}
\input{diagrams/interferences_example2/diag1} &
\input{diagrams/interferences_example2/diag2} &
\input{diagrams/interferences_example2/diag3} &
\input{diagrams/interferences_example2/diag4} \\
\end{tabular}
\caption{\label{fig:UnitaryCutsGhosts} Enumeration of all unitarity cuts, i.e. Cutkosky cuts, involved in a particular one-loop forward-scattering amplitude involving \glss{QCD} ghosts. This illustrates that when considering a common numerator for all cuts (e.g. with all propagators in the Feynman gauge), then the manifest realisation of unitarity requires to consider cuts traversing ghost lines. In other words, ghosts must be considered as external states too.}
\end{figure}

\textit{Quantum theory of gravitation} by R.P. Feynman~\cite{Feynman:1963ax} is widely considered to be one of the first papers originating the idea of ghosts, fictitious particles needed to restore gauge invariance in Yang-Mills theories. One of the most interesting aspects of the dissertation concerns the mathematical construction that he used to discern whether quantities were gauge invariant or not. Using a version of the \glss{FT} theorem, Feynman identified that the difference between polarisation sums for external physical particles and cutting rules for internal particles to be the main obstacle to the quantisation of Yang-Mills theories. In QED such difference disappears in light of the Ward identity, but not in Yang-Mills theories that feature self-interactions between the gauge vector bosons. This question has close ties to Unitarity, since the optical theorem requires to consider all possible Cutkosky cuts, i.e. thresholds, crossing internal edges, including ghosts (see fig.~\ref{fig:UnitaryCutsGhosts}). The role of the fictitious ghost degrees of freedom then becomes clear when investigating how such cuts of internal virtual propagators can reproduce physical polarisation sums.

In our formalism, the lack of Unitarity in the naive quantisation of non-abelian gauge theories manifests itself in the break-down of the \glss{KLN} \gls{IR} cancellation pattern, and thus the failure of setting up the \glss{LU} representation of eq.~\eqref{eq:LU_formula} for the cross-section of the scattering processes sensitive to the non-abelian nature of the theory. Indeed, the equivalence between the \glss{LU} representation and the traditional S-matrix formalism requires that the numerator of each interference diagram must correspond to the numerator obtained by operating Cutkosky cut on the corresponding supergraph. In other terms, it must correspond to the numerator of a forward-scattering amplitude in the limit of the cut particles going on-shell. This requirement is no longer fulfilled as soon as particles appearing as an external state in the interference diagram has a physical polarisation sum which differs from the numerator of its propagator in the on-shell limit. 

More specifically, when constructing the \glss{LU} representation, the numerator of the propagator of a gauge vector boson in the Feynman gauge is always $g_{\mu\nu}$, irrespective of whether it is crossed by a Cutkosky cut or not. When understood as coming from a polarisation sum, this same numerator can only be recovered when retaining in the sum both physical and unphysical degrees of freedom of the gauge vector.
This is in tension with the fact that only physical, i.e. transverse, polarisations must be considered for the asymptotic states. 

This tension can easily be resolved with the proper handling of the fictitious ghost degrees of freedom.
We now discuss this in more details by considering the case of the gluon in the Feynman gauge.
When traversed by a Cutkosky cut, the numerator of the gluon propagator remains unchanged and reads:
\begin{equation}
\label{eq:spin_sum_internal}
 \raisebox{-0.5cm}{\resizebox{4cm}{!}{
\begin{tikzpicture}
    \tikzstyle{every node}=[font=\small]
    \begin{feynman}
      \vertex (a1);
      \vertex[right=3cm of a1] (a3);
      
      \node[main node5] (L1) [right=0cm of a1];
      \node[main node5] (L2) [right=0cm of a3];
      
      \node[] (Label1) [above=0.2cm of L1] {$\mu$};
      \node[] (Label2) [above=0.2cm of L2] {$\nu$};
      
      \vertex [above right=0.5cm and 1.cm of a1] (cut1);
      \vertex [below right=0.5cm and 1.2cm of a1] (cut2);

      \diagram* {
        (a1) --[gluon, momentum={\(p\)}] (a3),
      };
      \draw[thick, red] (cut1) to[out=-70, in=100] (cut2);
    \end{feynman}
  \end{tikzpicture}}} 
= \ g^{\mu\nu} \delta^{(+)}(p^2)=\delta^{(+)}(p^2)\sum_{\lambda=0}^3 \raisebox{-0.2cm}{ \resizebox{4cm}{!}{\begin{tikzpicture}
    \tikzstyle{every node}=[font=\small]
    \begin{feynman}
      \vertex (a1);
      \node[main node5] (L1) [right=0cm of a1];
      \vertex[right=1.3cm of a1] (a3);
      \vertex[right=1.7cm of a1] (a4);
      \vertex[right=3.0cm of a1] (a5);
      \node[main node5] (L2) [right=0cm of a5];
      
      \node[] (Label1) [above left=0.2cm and 0.2cm of L1] {$\mu,\lambda$};
      \node[] (Label2) [above right=0.2cm and 0.2cm of L2] {$\nu,\lambda$};
      
      \vertex [above right=0.5cm and 0.9cm of a1] (cut1);
      \vertex [below right=0.5cm and 1.1cm of a1] (cut2);

      \diagram* {
        (a1) --[gluon, momentum={\(p\)}] (a3),
        (a4) --[gluon, momentum={\(p\)}] (a5),
      };
    \end{feynman}
  \end{tikzpicture}}},
\end{equation}
which can be understood as coming from the sum over the two physical, i.e. transverse ($\lambda=1,2$), and unphysical, i.e. longitudinal ($\lambda=0,3$), polarisation eigenstates of the gluon.
In contrast, the sum over its physical polarisation states only with defining vector $n^\mu$ and $p \cdot n \ne 0$, reads:
\begin{equation}
\label{eq:spin_sum_external}
 \sum_{\lambda=1}^2 
\raisebox{-0.2cm}{ \resizebox{4cm}{!}{\begin{tikzpicture}
    \tikzstyle{every node}=[font=\small]
    \begin{feynman}
      \vertex (a1);
      \node[main node5] (L1) [right=0cm of a1];
      \vertex[right=1.3cm of a1] (a3);
      \vertex[right=1.7cm of a1] (a4);
      \vertex[right=3.0cm of a1] (a5);
      \node[main node5] (L2) [right=0cm of a5];
      
      \node[] (Label1) [above left=0.2cm and 0.2cm of L1] {$\mu,\lambda$};
      \node[] (Label2) [above right=0.2cm and 0.2cm of L2] {$\nu,\lambda$};
      
      \vertex [above right=0.5cm and 0.9cm of a1] (cut1);
      \vertex [below right=0.5cm and 1.1cm of a1] (cut2);

      \diagram* {
        (a1) --[gluon, momentum={\(p\)}] (a3),
        (a4) --[gluon, momentum={\(p\)}] (a5),
      };
    \end{feynman}
  \end{tikzpicture}}} = %
\delta^{(+)}(p^2) \left(g^{\mu\nu}-\frac{p^\mu n^\nu+p^\nu n^\mu }{p\cdot n} + \frac{n^2 p^\mu p^\nu}{(p\cdot n)^2} \right).
\end{equation}
The inclusion of ghosts within loops only partially solves the issue. The perturbative expansion generated in this fashion is gauge invariant and unitary. However, unitarity is still not realised at the local level. This is made clear when explicitly computing each side of the identity given by the optical theorem and observing that the asymmetry caused by the differences between eq.~\eqref{eq:spin_sum_internal} and eq.~\eqref{eq:spin_sum_external} renders any local realisation of the identity impossible.

It should therefore not come as a surprise that our solution for the \glss{LU} construction is to allow for final-state ghosts, and at the same time use the polarisation sum over physical and unphysical external gluon polarisations (eq.~\eqref{eq:spin_sum_internal}). In other word, all Cutkosky cut of the forward-scattering diagram are considered irrespectively of their particle content, thus implying that ghosts can appear as external states. 
This essentially corresponds to applying the following sketched identity
\begin{equation}
\label{eq:naive_ghost_identity}
    \sum_{\lambda=1}^2 \raisebox{-0.2cm}{ \resizebox{4cm}{!}{\begin{tikzpicture}
    \tikzstyle{every node}=[font=\small]
    \begin{feynman}
      \vertex (a1);
      \node[main node5] (L1) [right=0cm of a1];
      \vertex[right=1.3cm of a1] (a3);
      \vertex[right=1.7cm of a1] (a4);
      \vertex[right=3.0cm of a1] (a5);
      \node[main node5] (L2) [right=0cm of a5];
      
      \node[] (Label1) [above left=0.2cm and 0.2cm of L1] {$\mu,\lambda$};
      \node[] (Label2) [above right=0.2cm and 0.2cm of L2] {$\nu,\lambda$};
      
      \vertex [above right=0.5cm and 0.9cm of a1] (cut1);
      \vertex [below right=0.5cm and 1.1cm of a1] (cut2);

      \diagram* {
        (a1) --[gluon, momentum={\(p\)}] (a3),
        (a4) --[gluon, momentum={\(p\)}] (a5),
      };
    \end{feynman}
  \end{tikzpicture}}} \sim \left( \sum_{\lambda=0}^3 
\raisebox{-0.2cm}{ \resizebox{4cm}{!}{\begin{tikzpicture}
    \tikzstyle{every node}=[font=\small]
    \begin{feynman}
      \vertex (a1);
      \node[main node5] (L1) [right=0cm of a1];
      \vertex[right=1.3cm of a1] (a3);
      \vertex[right=1.7cm of a1] (a4);
      \vertex[right=3.0cm of a1] (a5);
      \node[main node5] (L2) [right=0cm of a5];
      
      \node[] (Label1) [above left=0.2cm and 0.2cm of L1] {$\mu,\lambda$};
      \node[] (Label2) [above right=0.2cm and 0.2cm of L2] {$\nu,\lambda$};
      
      \vertex [above right=0.5cm and 0.9cm of a1] (cut1);
      \vertex [below right=0.5cm and 1.1cm of a1] (cut2);

      \diagram* {
        (a1) --[gluon, momentum={\(p\)}] (a3),
        (a4) --[gluon, momentum={\(p\)}] (a5),
      };
    \end{feynman}
  \end{tikzpicture}}} \right) + \raisebox{-0.2cm}{ \resizebox{2.5cm}{!}{\begin{tikzpicture}
    \tikzstyle{every node}=[font=\small]
    \begin{feynman}
      \vertex (a1);
      \node[main node5] (L1) [right=0cm of a1];
      \vertex[right=1.3cm of a1] (a3);
      \vertex[right=1.7cm of a1] (a4);
      \vertex[right=3.0cm of a1] (a5);
      \node[main node5] (L2) [right=0cm of a5];

      \vertex [above right=0.5cm and 0.9cm of a1] (cut1);
      \vertex [below right=0.5cm and 1.1cm of a1] (cut2);

      \diagram* {
        (a1) --[ghost1, momentum={\(p\)}] (a3),
        (a4) --[ghost1, momentum={\(p\)}] (a5),
      };
    \end{feynman}
  \end{tikzpicture}}}
\end{equation}
to \emph{all} gluon propagators, whether crossed by a Cutkosky cut or not.
This removes the asymmetry previously mentioned and thus restores the local realisation with \glss{LU} of the \glss{KLN} \glss{IR} cancellation pattern.
As a side note, we explicitly tested that the equivalence between considering the spin sum of eq.~\eqref{eq:spin_sum_external} for external particles and that of eq.~\eqref{eq:spin_sum_internal} but complemented with external ghosts can even be made \emph{local} for the particular cases of all \glss{LO} Cutkosky cuts of the process $\gamma^\star \rightarrow t \bar{t} g g g$. Local agreement per cut between the two expressions required the symmetrisation of both the ghost Feynman rules w.r.t the orientation of the ghost line and also of the momenta assignments of external gluons/ghosts.
This local equivalence is however not necessary for the \glss{LU} construction, because supergraphs involving ghosts and gluons can safely be considered as completely independent and their respective \glss{LU} representations will be locally \glss{IR}-finite by themselves.
This is fortunate, as the local realisation of gauge symmetry is in general very challenging and it involves further considerations than the role of ghosts only. We however note that such localisation, even partial, could help mitigate the potential presence of large numerical gauge cancellations between supergraphs. While such large cancellations are poised to happen for particular scatterings and observables, we have yet to encounter a case where this issue is a limiting factor.

The precise justification for eq.~\eqref{eq:naive_ghost_identity} involves considering the \glss{BRST}~\cite{tyutin2008gauge,brst} symmetry of the Lagrangian, but for our purpose it suffices to say that its validity for internal closed gluon loops of amplitudes is well-established~\cite{Kugo:1979gm} and the extension to closed loops of forward scattering diagrams, and applied irrespective of the Cutkosky cut follows analogously.

\FloatBarrier
\newpage

\section{Results}
\label{sec:results}

The generalisation of the \glss{LU} representation in the presence of raised propagators and its renormalisation procedure presented in this work is necessary for tackling the computation of any complete physical cross-section beyond the \glss{NLO} accuracy. These results stand as the first \glss{NNLO} cross-sections computed fully numerically in momentum space.
The aim of this section is therefore to demonstrate the viability and correctness of this generalisation. We also illustrate \glss{LU} applications at \glss{NLO} for $1 \rightarrow 3$ processes with final-state observable density functions that act both on the kinematics and on the selection of Cutkosky cuts considered. Such non-trivial examples were lacking in our original publication of ref.~\cite{2021}, which introduced \glss{LU}.
Finally we conclude this section with a study of \glss{UV} and \glss{IR} limits of individual supergraphs contributing up to \glss{N3LO}. This highlights our automated testing procedures and provides and overview of the current performance of our implementation.

All results within this section are obtained in the \glss{SM} with the following choice of parameters:

\begin{table}[h!]
\begin{center}
{\setlength\doublerulesep{1.5pt}   %
 \aboverulesep=0ex 
 \belowrulesep=-0.2ex
\begin{tabular}{ll|ll|ll}

\toprule[1pt]\midrule[0.3pt]
	Parameter &  \multicolumn{1}{l}{value} & Parameter & \multicolumn{1}{l}{value} & Parameter & value 
\\	\hline	\midrule[0.5pt]
	$m_t^{(\overline{\text{MS}})}(\mu_r^2)=m_t^{(\text{OS})}$ & $\mathtt{173.0}$ & $\Gamma_{t}$ & $\mathtt{0.0}$ & $y_t^{(\overline{\text{MS}})}$ & $m_t^{(\overline{\text{MS}})} \frac{\sqrt{2}}{\text{vev}}$
\\[0.2pt]
	$m_H$ & $\mathtt{125.0}$ & $m_Z$ & $\mathtt{91.188}$ & $G_F$ & $\mathtt{1.16639\;\cdot\;10^{-5}}$
\\[0.2pt]
	$\alpha_s^{(\overline{\text{MS}})}(\mu_r^2)=\alpha_s(m_Z^2)$ & $\mathtt{0.118}$ & $\alpha_{\text{QED}}^{-1}$ & $\mathtt{132.507}$ & $n_f$ & $1$ $(Q_q= \mathtt{1/3})$
\\[0.1cm]
\midrule[0.3pt]\bottomrule[1pt]
\end{tabular}
}
\end{center}
\caption{\label{tab:SMparameters} Common \gls{SM} parameters used for obtaining all numerical results presented in sect.~\ref{sec:results}. Dimensionful parameters are given in GeV.
Notice that for simplicity, $\alpha_s^{(\overline{\text{MS}})}$, $m_t^{(\overline{\text{MS}})}$ and $y_t^{(\overline{\text{MS}})}$ running are disabled for the demonstrative results of this section.
}
\end{table}

Throughout this section, we will detail results for individual supergraphs which are not individually gauge-invariant, and are computed in the Feynman gauge.
Whenever $n_f=1$ contributions are reported, it refers only to the contribution from one massless fermion to the gluon self-energy, i.e. there is always a single quark connected to the photon, and with an electric charge of $Q_q=1/3$.
Finally, note that depending on what is most convenient to compare against we report result from either $e^+ e^- \rightarrow \gamma \rightarrow X$ or $\gamma^\star \rightarrow X$, with $p_{e^+}=(E/2,0,0,E/2),p_{e^-}=(E/2,0,0,-E/2)$ and $p_{\gamma^\star}=(E,0,0,0)$.
Given that we are only considering \glss{QCD} corrections, the \emph{inclusive} cross-sections of both processes can be related at any perturbative order through the following simple relation reproducing the effect of the closed lepton trace:
\begin{equation}
    \sigma_{e^+ e^- \rightarrow \gamma \rightarrow X} = \frac{2}{3}\frac{g_e^2 Q_e^2}{E^3} \sigma_{\gamma^\star \rightarrow X},
\end{equation}
with $g_e^2 Q_e^2 = 4\pi \alpha_{\text QED}$ for the coupling strength of the electron to the photon. The conventional conversion factor of $1 [\text{pb}] = 0.389379304\cdot 10^9\;[\text{GeV}^{-2}]$ is also applied in the report of our results for $\sigma_{e^+ e^- \rightarrow \gamma \rightarrow X}$.

In all cases, we checked that the result for each individual supergraph is independent of the arbitrary choice of $m_\text{UV}$ introduced in the local \glss{UV} subtraction counterterms, and which we set equal to $\mu_r$ by default.

\subsection{Inclusive NNLO cross-section of $\gamma^\star \rightarrow jj$ and $\gamma^\star \rightarrow t\bar{t}$}

We start by presenting \glss{NNLO} inclusive cross-sections for the production of a pair of jets and a pair of massive quarks from an off-shell vector current, i.e photon. The observable function is in this case simply the identity function and the contribution from all Cutkosky cuts of the supergraphs will be retained. In such a case, the introduction of a contour deformation is not necessary since all non-pinched threshold singularities are guaranteed to also cancel in virtue of the general \glss{LU} pair-wise threshold algebraic cancellation mechanism.

In order to facilitate comparison with the analytic computation of the R-ratio given in ref.~\cite{Herzog:2017dtz}, we choose here to produce results for the inclusive cross-section of $\gamma^\star(p) \rightarrow jj$ up to order $\mathcal{O}(\alpha_s^2)$ and renormalised completely in $\overline{\text{MS}}$ at $p^2=\mu_r^2=(400\;\text{GeV})^2$. We consider a single massless quark flavour and the running of $\alpha_s$ is disabled and kept fixed at the value given in tab.~\ref{tab:SMparameters}. For convenience, we report here the analytical result given in eqs.~(5.1) and~(5.2) of ref.~\cite{Herzog:2017dtz}:
\begin{eqnarray}
    &&\sigma^{\overline{\text{MS}}}_{\gamma^\star \rightarrow j j} (p^2,\mu_r^2=p^2) = \sigma_{\gamma^\star \rightarrow d \bar{d}}^{(\text{LO})}(p^2) \\
    &&  \times  \left[ 1 + \left(3 C_F\right) \;\frac{\alpha_s}{4\pi} + \left( -\frac{3}{2}C_F^2 + C_A C_F \left[ \frac{123}{2}-44 \zeta_3 \right] -C_F n_f \left[ 11 - 8 \zeta_3 \right] \right)\;\left(\frac{\alpha_s}{4\pi}\right)^2 +\mathcal{O}(\alpha_s^3)\right] \nonumber
\end{eqnarray}

The \glss{NNLO} cross-section for $\gamma^\star(p) \rightarrow t\bar{t}$ has been computed completely analytically using the optical theorem in refs.~\cite{Chetyrkin:1996cf,Chetyrkin:1996yp} (and at $\mathcal{O}(\alpha_s^3)$ in ref.~\cite{Kiyo:2009gb}). The computation of the $n_f$ part is exact, whereas the rest of the contributions are computed using a Pad\'e approximant interpolating between the threshold and high-energy regimes. The benchmark results reported in tab.~\ref{tab:NNLOresOS} are computed from the analytical formulae presented in refs.~\cite{Chetyrkin:1996cf,Chetyrkin:1996yp} that are too long to be reported here. Differential and numerical \glss{NNLO} results obtained using traditional phase-space subtraction methods have been presented more recently in e.g. refs.~\cite{Chen:2016zbz,Bernreuther:2016ccf,Wang:2020ell,Chen:2021emn}.

\begin{figure}
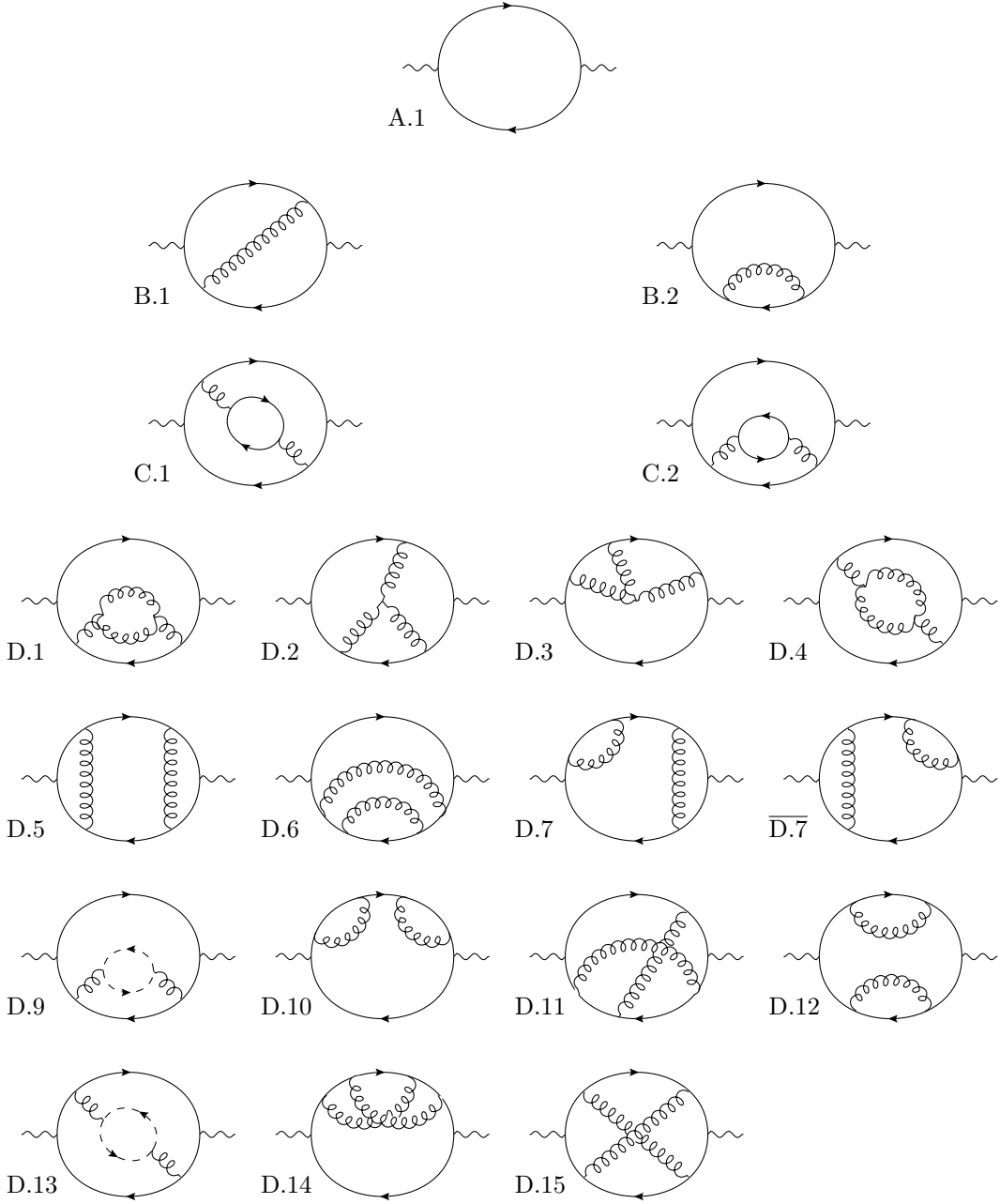

  \centering
  \begin{tabular}{cccc}
    \multicolumn{4}{c}{%
  \setbox1=\hbox{\input{diagrams/supergraphs_jj_LO/diag1}}%
  \leavevmode\rlap{\usebox1}%
  \rlap{\hspace*{-0.2cm}\raisebox{\dimexpr\ht1-3\baselineskip-2.0cm}{\small A.1}}%
  \phantom{\usebox1}%
}
    \\
    \multicolumn{2}{c}{%
  \setbox1=\hbox{\input{diagrams/supergraphs_jj_NLO/diag1}}%
  \leavevmode\rlap{\usebox1}%
  \rlap{\hspace*{-0.2cm}\raisebox{\dimexpr\ht1-3\baselineskip-2.0cm}{\small B.1}}%
  \phantom{\usebox1}%
} &
    \multicolumn{2}{c}{%
  \setbox1=\hbox{\input{diagrams/supergraphs_jj_NLO/diag2}}%
  \leavevmode\rlap{\usebox1}%
  \rlap{\hspace*{-0.2cm}\raisebox{\dimexpr\ht1-3\baselineskip-2.0cm}{\small B.2}}%
  \phantom{\usebox1}%
}
    \\
    \multicolumn{2}{c}{%
  \setbox1=\hbox{\input{diagrams/supergraphs_jj_N2LO/diag8}}%
  \leavevmode\rlap{\usebox1}%
  \rlap{\hspace*{-0.2cm}\raisebox{\dimexpr\ht1-3\baselineskip-2.0cm}{\small C.1}}%
  \phantom{\usebox1}%
} & %
    \multicolumn{2}{c}{%
  \setbox1=\hbox{\input{diagrams/supergraphs_jj_N2LO/diag12}}%
  \leavevmode\rlap{\usebox1}%
  \rlap{\hspace*{-0.2cm}\raisebox{\dimexpr\ht1-3\baselineskip-2.0cm}{\small C.2}}%
  \phantom{\usebox1}%
} %
    \\
  \setbox1=\hbox{\input{diagrams/supergraphs_jj_N2LO/diag13}}%
  \leavevmode\rlap{\usebox1}%
  \rlap{\hspace*{-0.2cm}\raisebox{\dimexpr\ht1-3\baselineskip-2.0cm}{\small D.1}}%
  \phantom{\usebox1}%
 & %
  \setbox1=\hbox{\input{diagrams/supergraphs_jj_N2LO/diag5}}%
  \leavevmode\rlap{\usebox1}%
  \rlap{\hspace*{-0.2cm}\raisebox{\dimexpr\ht1-3\baselineskip-2.0cm}{\small D.2}}%
  \phantom{\usebox1}%
 & %
  \setbox1=\hbox{\input{diagrams/supergraphs_jj_N2LO/diag16}}%
  \leavevmode\rlap{\usebox1}%
  \rlap{\hspace*{-0.2cm}\raisebox{\dimexpr\ht1-3\baselineskip-2.0cm}{\small D.3}}%
  \phantom{\usebox1}%
 & %
  \setbox1=\hbox{\input{diagrams/supergraphs_jj_N2LO/diag9}}%
  \leavevmode\rlap{\usebox1}%
  \rlap{\hspace*{-0.2cm}\raisebox{\dimexpr\ht1-3\baselineskip-2.0cm}{\small D.4}}%
  \phantom{\usebox1}%
    \\
  \setbox1=\hbox{\input{diagrams/supergraphs_jj_N2LO/diag4}}%
  \leavevmode\rlap{\usebox1}%
  \rlap{\hspace*{-0.2cm}\raisebox{\dimexpr\ht1-3\baselineskip-2.0cm}{\small D.5}}%
  \phantom{\usebox1}%
 & %
  \setbox1=\hbox{\input{diagrams/supergraphs_jj_N2LO/diag6}}%
  \leavevmode\rlap{\usebox1}%
  \rlap{\hspace*{-0.2cm}\raisebox{\dimexpr\ht1-3\baselineskip-2.0cm}{\small D.6}}%
  \phantom{\usebox1}%
 & %
  \setbox1=\hbox{\input{diagrams/supergraphs_jj_N2LO/diag14bis}}%
  \leavevmode\rlap{\usebox1}%
  \rlap{\hspace*{-0.2cm}\raisebox{\dimexpr\ht1-3\baselineskip-2.0cm}{\small D.7}}%
  \phantom{\usebox1}%
 & %
  \setbox1=\hbox{\input{diagrams/supergraphs_jj_N2LO/diag14}}%
  \leavevmode\rlap{\usebox1}%
  \rlap{\hspace*{-0.2cm}\raisebox{\dimexpr\ht1-3\baselineskip-2.0cm}{\small $\overline{\text{D.7}}$}}%
  \phantom{\usebox1}%
    \\
  \setbox1=\hbox{\input{diagrams/supergraphs_jj_N2LO/diag11}}%
  \leavevmode\rlap{\usebox1}%
  \rlap{\hspace*{-0.2cm}\raisebox{\dimexpr\ht1-3\baselineskip-2.0cm}{\small D.9}}%
  \phantom{\usebox1}%
 & %
  \setbox1=\hbox{\input{diagrams/supergraphs_jj_N2LO/diag1}}%
  \leavevmode\rlap{\usebox1}%
  \rlap{\hspace*{-0.2cm}\raisebox{\dimexpr\ht1-3\baselineskip-2.0cm}{\small D.10}}%
  \phantom{\usebox1}%
 & %
  \setbox1=\hbox{\input{diagrams/supergraphs_jj_N2LO/diag15}}%
  \leavevmode\rlap{\usebox1}%
  \rlap{\hspace*{-0.2cm}\raisebox{\dimexpr\ht1-3\baselineskip-2.0cm}{\small D.11}}%
  \phantom{\usebox1}%
 & %
  \setbox1=\hbox{\input{diagrams/supergraphs_jj_N2LO/diag2}}%
  \leavevmode\rlap{\usebox1}%
  \rlap{\hspace*{-0.2cm}\raisebox{\dimexpr\ht1-3\baselineskip-2.0cm}{\small D.12}}%
  \phantom{\usebox1}%
    \\
  \setbox1=\hbox{\input{diagrams/supergraphs_jj_N2LO/diag10}}%
  \leavevmode\rlap{\usebox1}%
  \rlap{\hspace*{-0.2cm}\raisebox{\dimexpr\ht1-3\baselineskip-2.0cm}{\small D.13}}%
  \phantom{\usebox1}%
 & %
  \setbox1=\hbox{\input{diagrams/supergraphs_jj_N2LO/diag7}}%
  \leavevmode\rlap{\usebox1}%
  \rlap{\hspace*{-0.2cm}\raisebox{\dimexpr\ht1-3\baselineskip-2.0cm}{\small D.14}}%
  \phantom{\usebox1}%
 & %
  \setbox1=\hbox{\input{diagrams/supergraphs_jj_N2LO/diag3}}%
  \leavevmode\rlap{\usebox1}%
  \rlap{\hspace*{-0.2cm}\raisebox{\dimexpr\ht1-3\baselineskip-2.0cm}{\small D.15}}%
  \phantom{\usebox1}%
 & %
    \\
  \end{tabular}
  \caption{\label{tab:jjjSGs}All contributing supergraphs to the class of processes $\gamma^\star \rightarrow j j + X$ and $\gamma^\star \rightarrow t\bar{t}+ X$ at \glss{LO} (A.1), \glss{NLO} (B.i) and \glss{NNLO} (C.i and D.i). The outer fermion loop is set massless for $\gamma^\star \rightarrow j j+X$ and massive with mass $m_t$ for $\gamma^\star \rightarrow t\bar{t}+X$. The \glss{NNLO} contribution proportional to $n_f$ only comes from the two supergraphs C.1 and C.2, when the inner fermion loop is always considered massless. We note that supergraphs D.7 and $\overline{\text{D.7}}$ are isomorphic, but we chose to report them separately here because the symmetry from exchanging the two identical external off-shell photons was not considered at generation time.
  }
\end{figure}

We show in tab.~\ref{tab:NNLOres} the contribution from each of the seventeen supergraphs listed in fig.~\ref{tab:jjjSGs}, which make up the cross-section up to order $\mathcal{O}(\alpha_s^2)$ for this class of processes and renormalised completely in $\overline{\text{MS}}$.
In the second column of tab.~\ref{tab:NNLOres}, we also present similar results but for the process $\gamma^\star \rightarrow t\bar{t}$ at $p^2=\mu_r^2=(600\;\text{GeV})^2$.
Cross-sections for external particles whose mass is renormalised in $\overline{\text{MS}}$ are typically not readily available in the literature since the usual reduction formula of the cross-section in terms of truncated Green's functions is not directly applicable and amplitude derivatives arise\footnote{
Though of little phenomenological relevance, the difficulty in computing the \glss{NNLO} cross-section $\sigma_{\gamma^\star \rightarrow t \bar{t}}^{(\overline{\text{MS}})}$ renormalised entirely in $\overline{\text{MS}}$ makes it of theoretical interest and this is why we decided to report it as well in this section.}. This is the reason why no comparison benchmark is given in this scheme.

However, as discussed in sect.~\ref{sec:MSbar_and_OS_integrated_UVCT}, our final implementation of the $R$-operation automatically produces cross-sections in the hybdrid $\overline{\text{MS}}$+\glss{OS} scheme, which we denote $\sigma_{\gamma^\star \rightarrow t \bar{t}}^{\left[\alpha_s^{(\overline{\text{MS}})},m_t^{(\text{OS})}\right]}$ and where fermion masses (but not fields) are renormalised in the \glss{OS} scheme. We thus also report in tab.~\ref{tab:NNLOresOS} results in this more common hybdrid scheme for both $p^2=(400\;\text{GeV})^2$ and $p^2=(3000\;\text{GeV})^2$ (with $\mu_r^2=m_t^2$), and compare them to the analytical benchmark from refs.~\cite{Chetyrkin:1996cf,Chetyrkin:1996yp}.

At \glss{NLO}, we also verified that we could obtain the same hybrid scheme result from adding to our pure $\overline{\text{MS}}$ result the well-known contribution from the one-loop mass renormalisation counterterm $\delta m_t^{\alpha_s}$:
\begin{equation}
\label{eq:deltamtOneLoop}
    \delta m_t^{\alpha_s,(OS)}-\delta m_t^{\alpha_s,(\overline{\text{MS}})} = C_F \frac{\alpha_s}{(4\pi)^2} \left(4 + 3 \log{\left(\mu_r^2/m_t^2\right)} \right).
\end{equation}
This serves as further confirmation that the integrated counterpart of our \glss{OS} subtraction operator presented in sect.~\ref{sect:integrated_os_cts} does indeed reproduce the traditional \glss{NLO} \glss{QCD} top quark mass renormalisation counterterm in the \glss{OS} scheme.

In all cases, we find agreement at, or better than, the per-mil level, similar to the Monte-Carlo accuracy of the integration of most individual supergraphs, with the exception of those with a central value accidentally close to zero.
We note however that there is a small tension in \glss{NNLO} cross-section of $\sigma_{\gamma^\star \rightarrow t \bar{t}}^{\left[\alpha_s^{(\overline{\text{MS}})},m_t^{(\text{OS})}\right]}$ for $p^2=\mu_r^2=(400\;\text{GeV})^2$. 
We attribute this to the fact that the result from ref.~\cite{Chetyrkin:1996cf} is not exact, but instead an interpolation between two asymptotic regimes of the top quark velocity $v$ (threshold production at $v=0$ and high-energy limit at $v=1$).
Indeed, $p^2=(400\;\text{GeV})^2$ corresponds to a top quark velocity of $v=0.5$, which is therefore not expected to be perfectly captured by expansions around any of the two regimes.
This hypothesis is supported by the observation that we find better agreement for the pure $n_f$ contribution (for which the result given in refs.~\cite{Chetyrkin:1996cf,Chetyrkin:1996yp} is exact) and also in the high-energy limit for $p^2=(3000\;\text{GeV})^2$.

\FloatBarrier

\renewcommand{\arraystretch}{1.15}
\begin{table}[ht!]
\vspace{-1.5cm}
\begin{adjustwidth}{-1.5cm}{-1.5cm}

\begin{center}
{\setlength\doublerulesep{1.5pt}   %
 \aboverulesep=0.06ex 
 \belowrulesep=-0.2ex
\begin{tabular}{c|c|cc|cc}
\toprule[1pt]\midrule[0.3pt]
	\multirow{3}{*}{SG id}
	& \multirow{3}{*}{$\Xi$} 
	& \multirow{2}{*}{$\sigma_{\gamma^\star \rightarrow j j}^{(\overline{\text{MS}})}$ [$\text{GeV}^{-2}$]}
	& \multirow{3}{*}{$\Delta$ [\%]}
	& \multirow{2}{*}{$\sigma_{\gamma^\star \rightarrow t \bar{t}}^{(\overline{\text{MS}})}$ [$\text{GeV}^{-2}$]}
	& \multirow{3}{*}{$\Delta$ [\%]}
\\ 
& & & & &
\\ 
& & $p_{\gamma^\star}^2=\mu_r^2=(400\;\text{GeV})^2$ & & $p_{\gamma^\star}^2=\mu_r^2=(600\;\text{GeV})^2$ &
\\\hline	
\multicolumn{1}{c}{\bf LO} & \multicolumn{5}{c}{$\mathcal{O}(\alpha_s^0)$} \\
\midrule
	A.1 & \getRes{ajjLOSGQG000mplcty} & \getRes{ajjLOSGQG000} & \getRes{ajjLOSGQG000delta} & \getRes{attxLOSGQG000} & \getRes{attxLOSGQG000delta}
\\
\cdashlinelr{1-6}
\multicolumn{2}{c|}{Total} & \getRes{ajjLOTotal} & \getRes{ajjLOTotaldelta} & \getRes{attxLOTotal} & \getRes{attxLOTotaldelta}
\\
\midrule
\multicolumn{1}{c}{\bf NLO} & \multicolumn{5}{c}{$\mathcal{O}(\alpha_s)$} \\
\midrule
	B.1 & \getRes{ajjNLOSGQG000mplcty} & \getRes{ajjNLOSGQG000} & \getRes{ajjNLOSGQG000delta} & \getRes{attxNLOSGQG000} & \getRes{attxNLOSGQG000delta}
\\
	B.2 & \getRes{ajjNLOSGQG001mplcty} & \getRes{ajjNLOSGQG001} & \getRes{ajjNLOSGQG001delta} & \getRes{attxNLOSGQG001} & \getRes{attxNLOSGQG001delta}
\\
\cdashlinelr{1-6}
\multicolumn{2}{c|}{Total} & \getRes{ajjNLOTotal} & \getRes{ajjNLOTotaldelta} & \getRes{attxNLOTotal} & \getRes{attxNLOTotaldelta}
\\
\multicolumn{2}{c|}{Benchmark} & \getRes{ajjNLOTarget} & \getRes{ajjNLOTargetdelta} & \multicolumn{2}{c}{N/A}
\\
\midrule
\multicolumn{1}{c}{\bf NNLO} & \multicolumn{5}{c}{$\mathcal{O}(\alpha_s^2)$ ($n_f=1$ contribution)} \\
\midrule
	C.1 & \getRes{ajjNNLOnf1SGQG015mplcty} & \getRes{ajjNNLOnf1SGQG015} & \getRes{ajjNNLOnf1SGQG015delta} & \getRes{attxNNLOnf1SGQG015} & \getRes{attxNNLOnf1SGQG015delta}
\\
	C.2 & \getRes{ajjNNLOnf1SGQG025mplcty} & \getRes{ajjNNLOnf1SGQG025} & \getRes{ajjNNLOnf1SGQG025delta} & \getRes{attxNNLOnf1SGQG026} & \getRes{attxNNLOnf1SGQG026delta}
\\
\cdashlinelr{1-6}
\multicolumn{2}{c|}{Total} & \getRes{ajjNNLOnf1Total} & \getRes{ajjNNLOnf1Totaldelta} & \getRes{attxNNLOnf1Total} & \getRes{attxNNLOnf1Totaldelta}
\\
\multicolumn{2}{c|}{Benchmark} & \getRes{ajjNNLOnf1Target} & \getRes{ajjNNLOnf1Targetdelta} & \multicolumn{2}{c}{N/A}
\\
\midrule
\multicolumn{1}{c}{\bf NNLO} & \multicolumn{5}{c}{$\mathcal{O}(\alpha_s^2)$ (all other contributions)} \\
\midrule
	D.1 & \getRes{ajjNNLOSGQG023mplcty} & \getRes{ajjNNLOSGQG023} & \getRes{ajjNNLOSGQG023delta} & \getRes{attxNNLOSGQG024} & \getRes{attxNNLOSGQG024delta}
\\
	D.2 & \getRes{ajjNNLOSGQG006mplcty} & \getRes{ajjNNLOSGQG006} & \getRes{ajjNNLOSGQG006delta} & \getRes{attxNNLOSGQG006} & \getRes{attxNNLOSGQG006delta}
\\
	D.3 & \getRes{ajjNNLOSGQG016mplcty} & \getRes{ajjNNLOSGQG016} & \getRes{ajjNNLOSGQG016delta} & \getRes{attxNNLOSGQG017} & \getRes{attxNNLOSGQG017delta}
\\
	D.4 & \getRes{ajjNNLOSGQG013mplcty} & \getRes{ajjNNLOSGQG013} & \getRes{ajjNNLOSGQG013delta} & \getRes{attxNNLOSGQG013} & \getRes{attxNNLOSGQG013delta}
\\
	D.5 & \getRes{ajjNNLOSGQG003mplcty} & \getRes{ajjNNLOSGQG003} & \getRes{ajjNNLOSGQG003delta} & \getRes{attxNNLOSGQG003} & \getRes{attxNNLOSGQG003delta}
\\
	D.6 & \getRes{ajjNNLOSGQG022mplcty} & \getRes{ajjNNLOSGQG022} & \getRes{ajjNNLOSGQG022delta} & \getRes{attxNNLOSGQG023} & \getRes{attxNNLOSGQG023delta}
\\
	D.7 & \getRes{ajjNNLOSGQG011mplcty} & \getRes{ajjNNLOSGQG011} & \getRes{ajjNNLOSGQG011delta} & \getRes{attxNNLOSGQG011} & \getRes{attxNNLOSGQG011delta}
\\
	$\overline{\text{D.7}}$ & \getRes{ajjNNLOSGQG009mplcty} & \getRes{ajjNNLOSGQG009} & \getRes{ajjNNLOSGQG009delta} & \getRes{attxNNLOSGQG009} & \getRes{attxNNLOSGQG009delta}
\\
	D.9 & \getRes{ajjNNLOSGQG024mplcty} & \getRes{ajjNNLOSGQG024} & \getRes{ajjNNLOSGQG024delta} & \getRes{attxNNLOSGQG025} & \getRes{attxNNLOSGQG025delta}
\\
	D.10 &\getRes{ajjNNLOSGQG020mplcty} & \getRes{ajjNNLOSGQG020} & \getRes{ajjNNLOSGQG020delta}  & \getRes{attxNNLOSGQG021} & \getRes{attxNNLOSGQG021delta}
\\
	D.11 &\getRes{ajjNNLOSGQG005mplcty} & \getRes{ajjNNLOSGQG005} & \getRes{ajjNNLOSGQG005delta}  & \getRes{attxNNLOSGQG005} & \getRes{attxNNLOSGQG005delta}
\\
	D.12 &\getRes{ajjNNLOSGQG001mplcty} & \getRes{ajjNNLOSGQG001} & \getRes{ajjNNLOSGQG001delta}  & \getRes{attxNNLOSGQG001} & \getRes{attxNNLOSGQG001delta}
\\
	D.13 &\getRes{ajjNNLOSGQG014mplcty} & \getRes{ajjNNLOSGQG014} & \getRes{ajjNNLOSGQG014delta}  & \getRes{attxNNLOSGQG014} & \getRes{attxNNLOSGQG014delta}
\\
	D.14 &\getRes{ajjNNLOSGQG017mplcty} & \getRes{ajjNNLOSGQG017} & \getRes{ajjNNLOSGQG017delta}  & \getRes{attxNNLOSGQG018} & \getRes{attxNNLOSGQG018delta}
\\
	D.15 &\getRes{ajjNNLOSGQG000mplcty} & \getRes{ajjNNLOSGQG000} & \getRes{ajjNNLOSGQG000delta}  & \getRes{attxNNLOSGQG000} & \getRes{attxNNLOSGQG000delta}
\\
\cdashlinelr{1-6}
\multicolumn{2}{c|}{Total} & \getRes{ajjNNLOTotal} & \getRes{ajjNNLOTotaldelta} & \getRes{attxNNLOTotal} & \getRes{attxNNLOTotaldelta}
\\
\multicolumn{2}{c|}{Benchmark} & \getRes{ajjNNLOTarget} & \getRes{ajjNNLOTargetdelta} & \multicolumn{2}{c}{N/A}
\\
\midrule[0.3pt]\bottomrule[1pt]
\end{tabular}
}
\end{center}
\end{adjustwidth}
\caption{\label{tab:NNLOres} Contributions from individual supergraphs listed in tab.~\ref{tab:jjjSGs} to the cross-section up to order $\mathcal{O}(\alpha_s^2)$ (\glss{LO}+\glss{NLO}+\glss{NNLO}) for the processes $\gamma^\star\rightarrow j j$ and $\gamma^\star\rightarrow t \bar{t}$ with $n_f=1$ and $Q_q=1/3$ and \gls{SM} parameters as given in table.~\ref{tab:SMparameters}. $\Xi$ denotes the symmetry factor of the supergraph (included in the result reported). $\Delta$ reports the relative Monte-Carlo error, except for the ``Benchmark'' entry~\cite{Herzog:2017dtz} where it refers to the relative difference w.r.t the \glss{LU} result instead.}
\end{table}
\renewcommand{\arraystretch}{1.}

\renewcommand{\arraystretch}{1.15}
\begin{table}[ht!]
\vspace{-1.5cm}
\begin{adjustwidth}{-1.5cm}{-1.5cm}

\begin{center}
{\setlength\doublerulesep{1.5pt}   %
 \aboverulesep=0.06ex 
 \belowrulesep=-0.2ex
\begin{tabular}{c|c|cc|cc}
\toprule[1pt]\midrule[0.3pt]
	\multirow{3}{*}{SG id}
	& \multirow{3}{*}{$\Xi$} 
	& \multirow{2}{*}{$\sigma_{\gamma^\star \rightarrow t \bar{t}}^{\left[\alpha_s^{(\overline{\text{MS}})},m_t^{(\text{OS})}\right]}$ [$\text{GeV}^{-2}$]}
	& \multirow{3}{*}{$\Delta$ [\%]}
	& \multirow{2}{*}{$\sigma_{\gamma^\star \rightarrow t \bar{t}}^{\left[\alpha_s^{(\overline{\text{MS}})},m_t^{(\text{OS})}\right]}$ [$\text{GeV}^{-2}$]}
	& \multirow{3}{*}{$\Delta$ [\%]}
\\ 
& & & & &
\\ 
& & $\mu_r^2=m_t^2,\;p_{\gamma^\star}^2=(400\;\text{GeV})^2$ & & $\mu_r^2=m_t^2,\;p_{\gamma^\star}^2=(3000\;\text{GeV})^2$ &
\\\hline	
\multicolumn{1}{c}{\bf LO} & \multicolumn{5}{c}{$\mathcal{O}(\alpha_s^0)$} \\
\midrule
	A.1 & \getRes{attxOSbench1LOSGQG000mplcty} & \getRes{attxOSbench1LOSGQG000} & \getRes{attxOSbench1LOSGQG000delta} & \getRes{attxOSbenchHELOSGQG000} & \getRes{attxOSbenchHELOSGQG000delta}
\\
\cdashlinelr{1-6}
\multicolumn{2}{c|}{Total} & \getRes{attxOSbench1LOTotal} & \getRes{attxOSbench1LOTotaldelta} & \getRes{attxOSbenchHELOTotal} & \getRes{attxOSbenchHELOTotaldelta}
\\
\midrule
\multicolumn{1}{c}{\bf NLO} & \multicolumn{5}{c}{$\mathcal{O}(\alpha_s)$} \\
\midrule
	B.1 & \getRes{attxOSbench1NLOSGQG000mplcty} & \getRes{attxOSbench1NLOSGQG000} & \getRes{attxOSbench1NLOSGQG000delta} & \getRes{attxOSbenchHENLOSGQG000} & \getRes{attxOSbenchHENLOSGQG000delta}
\\
	B.2 & \getRes{attxOSbench1NLOSGQG001mplcty} & \getRes{attxOSbench1NLOSGQG001} & \getRes{attxOSbench1NLOSGQG001delta} & \getRes{attxOSbenchHENLOSGQG001} & \getRes{attxOSbenchHENLOSGQG001delta}
\\
\cdashlinelr{1-6}
\multicolumn{2}{c|}{Total} & \getRes{attxOSbench1NLOTotal} & \getRes{attxOSbench1NLOTotaldelta} & \getRes{attxOSbenchHENLOTotal} & \getRes{attxOSbenchHENLOTotaldelta}
\\
\multicolumn{2}{c|}{Benchmark} & \getRes{attxOSbench1NLOTarget} & \getRes{attxOSbench1NLOTargetdelta} & \getRes{attxOSbenchHENLOTarget} & \getRes{attxOSbenchHENLOTargetdelta}
\\
\midrule
\multicolumn{1}{c}{\bf NNLO} & \multicolumn{5}{c}{$\mathcal{O}(\alpha_s^2)$ ($n_f=1$ contribution)} \\
\midrule
	C.1 & \getRes{attxOSbench1NNLOnf1SGQG015mplcty} & \getRes{attxOSbench1NNLOnf1SGQG015} & \getRes{attxOSbench1NNLOnf1SGQG015delta} & \getRes{attxOSbenchHENNLOnf1SGQG015} & \getRes{attxOSbenchHENNLOnf1SGQG015delta}
\\
	C.2 & \getRes{attxOSbench1NNLOnf1SGQG026mplcty} & \getRes{attxOSbench1NNLOnf1SGQG026} & \getRes{attxOSbench1NNLOnf1SGQG026delta} & \getRes{attxOSbenchHENNLOnf1SGQG026} & \getRes{attxOSbenchHENNLOnf1SGQG026delta}
\\
\cdashlinelr{1-6}
\multicolumn{2}{c|}{Total} & \getRes{attxOSbench1NNLOnf1Total} & \getRes{attxOSbench1NNLOnf1Totaldelta} & \getRes{attxOSbenchHENNLOnf1Total} & \getRes{attxOSbenchHENNLOnf1Totaldelta}
\\
\multicolumn{2}{c|}{Benchmark} & \getRes{attxOSbench1NNLOnf1Target} & \getRes{attxOSbench1NNLOnf1Targetdelta} & \getRes{attxOSbenchHENNLOnf1Target} & \getRes{attxOSbenchHENNLOnf1Targetdelta}
\\
\midrule
\multicolumn{1}{c}{\bf NNLO} & \multicolumn{5}{c}{$\mathcal{O}(\alpha_s^2)$ (all other contributions)} \\
\midrule
	D.1 & \getRes{attxOSbench1NNLOSGQG024mplcty} & \getRes{attxOSbench1NNLOSGQG024} & \getRes{attxOSbench1NNLOSGQG024delta} & \getRes{attxOSbenchHENNLOSGQG024} & \getRes{attxOSbenchHENNLOSGQG024delta}
\\
	D.2 & \getRes{attxOSbench1NNLOSGQG006mplcty} & \getRes{attxOSbench1NNLOSGQG006} & \getRes{attxOSbench1NNLOSGQG006delta} & \getRes{attxOSbenchHENNLOSGQG006} & \getRes{attxOSbenchHENNLOSGQG006delta}
\\
	D.3 & \getRes{attxOSbench1NNLOSGQG017mplcty} & \getRes{attxOSbench1NNLOSGQG017} & \getRes{attxOSbench1NNLOSGQG017delta} & \getRes{attxOSbenchHENNLOSGQG017} & \getRes{attxOSbenchHENNLOSGQG017delta}
\\
	D.4 & \getRes{attxOSbench1NNLOSGQG013mplcty} & \getRes{attxOSbench1NNLOSGQG013} & \getRes{attxOSbench1NNLOSGQG013delta} & \getRes{attxOSbenchHENNLOSGQG013} & \getRes{attxOSbenchHENNLOSGQG013delta}
\\
	D.5 & \getRes{attxOSbench1NNLOSGQG003mplcty} & \getRes{attxOSbench1NNLOSGQG003} & \getRes{attxOSbench1NNLOSGQG003delta} & \getRes{attxOSbenchHENNLOSGQG003} & \getRes{attxOSbenchHENNLOSGQG003delta}
\\
	D.6 & \getRes{attxOSbench1NNLOSGQG023mplcty} & \getRes{attxOSbench1NNLOSGQG023} & \getRes{attxOSbench1NNLOSGQG023delta} & \getRes{attxOSbenchHENNLOSGQG023} & \getRes{attxOSbenchHENNLOSGQG023delta}
\\
	D.7 & \getRes{attxOSbench1NNLOSGQG011mplcty} & \getRes{attxOSbench1NNLOSGQG011} & \getRes{attxOSbench1NNLOSGQG011delta} & \getRes{attxOSbenchHENNLOSGQG011} & \getRes{attxOSbenchHENNLOSGQG011delta}
\\
	$\overline{\text{D.7}}$ & \getRes{attxOSbench1NNLOSGQG009mplcty} & \getRes{attxOSbench1NNLOSGQG009} & \getRes{attxOSbench1NNLOSGQG009delta} & \getRes{attxOSbenchHENNLOSGQG009} & \getRes{attxOSbenchHENNLOSGQG009delta}
\\
	D.9 & \getRes{attxOSbench1NNLOSGQG025mplcty} & \getRes{attxOSbench1NNLOSGQG025} & \getRes{attxOSbench1NNLOSGQG025delta} & \getRes{attxOSbenchHENNLOSGQG025} & \getRes{attxOSbenchHENNLOSGQG025delta}
\\
	D.10 &\getRes{attxOSbench1NNLOSGQG021mplcty} & \getRes{attxOSbench1NNLOSGQG021} & \getRes{attxOSbench1NNLOSGQG021delta}  & \getRes{attxOSbenchHENNLOSGQG021} & \getRes{attxOSbenchHENNLOSGQG021delta}
\\
	D.11 &\getRes{attxOSbench1NNLOSGQG005mplcty} & \getRes{attxOSbench1NNLOSGQG005} & \getRes{attxOSbench1NNLOSGQG005delta}  & \getRes{attxOSbenchHENNLOSGQG005} & \getRes{attxOSbenchHENNLOSGQG005delta}
\\
	D.12 &\getRes{attxOSbench1NNLOSGQG001mplcty} & \getRes{attxOSbench1NNLOSGQG001} & \getRes{attxOSbench1NNLOSGQG001delta}  & \getRes{attxOSbenchHENNLOSGQG001} & \getRes{attxOSbenchHENNLOSGQG001delta}
\\
	D.13 &\getRes{attxOSbench1NNLOSGQG014mplcty} & \getRes{attxOSbench1NNLOSGQG014} & \getRes{attxOSbench1NNLOSGQG014delta}  & \getRes{attxOSbenchHENNLOSGQG014} & \getRes{attxOSbenchHENNLOSGQG014delta}
\\
	D.14 &\getRes{attxOSbench1NNLOSGQG017mplcty} & \getRes{attxOSbench1NNLOSGQG017} & \getRes{attxOSbench1NNLOSGQG017delta}  & \getRes{attxOSbenchHENNLOSGQG018} & \getRes{attxOSbenchHENNLOSGQG018delta}
\\
	D.15 &\getRes{attxOSbench1NNLOSGQG000mplcty} & \getRes{attxOSbench1NNLOSGQG000} & \getRes{attxOSbench1NNLOSGQG000delta}  & \getRes{attxOSbenchHENNLOSGQG000} & \getRes{attxOSbenchHENNLOSGQG000delta}
\\
\cdashlinelr{1-6}
\multicolumn{2}{c|}{Total} & \getRes{attxOSbench1NNLOTotal} & \getRes{attxOSbench1NNLOTotaldelta} & \getRes{attxOSbenchHENNLOTotal} & \getRes{attxOSbenchHENNLOTotaldelta}
\\
\multicolumn{2}{c|}{Benchmark} & \getRes{attxOSbench1NNLOTarget} & \getRes{attxOSbench1NNLOTargetdelta} & \getRes{attxOSbenchHENNLOTarget} & \getRes{attxOSbenchHENNLOTargetdelta}
\\
\midrule[0.3pt]\bottomrule[1pt]
\end{tabular}
}
\end{center}
\end{adjustwidth}
\caption{\label{tab:NNLOresOS} Contributions from individual supergraphs listed in tab.~\ref{tab:jjjSGs} to the cross-section up to order $\mathcal{O}(\alpha_s^2)$ for the process $\gamma^\star\rightarrow t \bar{t}$ and two different energies $p_{\gamma^\star}^2=(400\;\text{GeV})^2$ and $p_{\gamma^\star}^2=(360\;\text{GeV})^2$. The strong coupling is renormalised in $\overline{\text{MS}}$ with $\mu_r^2=m_t^2$ and the top mass in the \glss{OS} scheme. $\Xi$ denotes the symmetry factor of the supergraph (included in the result reported). $\Delta$ reports the relative Monte-Carlo error, except for the ``Benchmark'' entry~\cite{Chetyrkin:1996cf,Chetyrkin:1996yp} where it refers to the relative difference w.r.t the \glss{LU} result instead.}
\end{table}
\renewcommand{\arraystretch}{1.}

\FloatBarrier

We find no sign of large gauge cancellation given that the maximal supergraph contribution (in absolute value) remains of the same order of magnitude as the total cross-section.
However, we find a mild hierarchy between individual supergraph contributions, spanning two orders of magnitude at \glss{NNLO}. These hierarchies typically become stronger when more supergraphs contribute (e.g. fig.~12 of ref.~\cite{2021} which shows the relative \gls{LO} contribution from all 104 supergraphs from the process $e^+ e^- \rightarrow t\bar{t} g/g_h g/\bar{g_h} g$.). We take advantage of such hierarchies by integrating all supergraphs together with the integrator {\small \sc Havana} (see sect.~5.4.2 of ref.~\cite{2021}) that performs a discrete importance sampling over them.
This implies that the sampling statistics for obtaining each of the supergraph cross-sections shown in tab.~\ref{tab:NNLOres} covers a wide range of values. For instance, at \glss{NNLO} supergraph D.1 required the largest statistics ($2.1\cdot 10^9$ points), whereas supergraph D.15 received the smallest amount of sample points ($0.2\cdot 10^9$). Thus, discrete importance sampling over supergraphs optimises the allocation of computation time amongst supergraphs, accounting for \emph{both} their contribution relative to the total cross-section but also the differences in the variance of the corresponding \glss{LU} integrands. We anticipate that the use of this integration strategy will become increasingly more important as the number of contributing supergraphs grows larger for more complicated applications.  

We also report in tab.~\ref{tab:ttxNNLOheavyfermion} on the contribution analogous to that of the $n_f$ contribution at \glss{NNLO}, but instead for $\gamma^\star\rightarrow t \bar{t}$ and stemming from a nested closed \emph{massive} fermion loop with mass $m_t$. Again, we do not provide a benchmark comparison for this contribution and we include it solely to demonstrate that, also beyond \glss{NLO}, the \glss{LU} representation and its Monte-Carlo integration can easily accommodate additional masses.

We checked that the complete \glss{NLO} cross-section of $\gamma^\star \rightarrow j j$ and $\gamma^\star \rightarrow t \bar{t}$ is independent of $\mu_r$ (as expected since this process receives no contribution from renormalisation counterterms). This test only holds for the sum of all supergraph contributions and it therefore verifies that the expected gauge cancellations across them take place. For instance, it confirms that the $\delta Z_t^{\overline{\text{MS}}}$ contributions generated through our automated renormalisation procedure described in sect.~\ref{sect:localised_renormalisation} properly cancel out between the double-triangle (B.1) and self-energy (B.2) supergraphs.
We also verified that we reproduce the correct $\mu_r$ dependence at \glss{NNLO}, as predicted by the analytical expression given in ref.~\cite{Chetyrkin:1996cf}.
In general, a similar test can always be performed by comparing the $\mu_r$-dependence of the \glss{LU} cross-section with the logarithmic dependence reconstructed from the renormalisation group flow.

\begin{figure}[h!]
\begin{center}
\begin{minipage}{0.97\linewidth}
\centering
\floatbox[{\capbeside\thisfloatsetup{capbesideposition={right,center},capbesidewidth=4cm}}]{table}[\FBwidth]
{
\caption{\label{tab:ttxNNLOheavyfermion} \glss{NNLO} contribution to $\gamma^\star\rightarrow t \bar{t}$ from a massive nested fermion loop with mass $m_t$. Same conventions as for tab.~\ref{tab:NNLOresOS}.
The two supergraphs shown are denoted with a star so as to stress that the internal closed quark loop line is massive.}
}
{
\renewcommand{\arraystretch}{1.15}
{\setlength\doublerulesep{1.5pt}   %
 \aboverulesep=0.06ex 
 \belowrulesep=-0.2ex
\begin{tabular}{c|c|cc}
\toprule[1pt]\midrule[0.3pt]
	\multirow{3}{*}{SG id}
	& \multirow{3}{*}{$\Xi$}
	& \multirow{2}{*}{$\sigma_{\gamma^\star \rightarrow t \bar{t}}^{\left[\alpha_s^{(\overline{\text{MS}})},m_t^{(\text{OS})}\right]}$ [$\text{GeV}^{-2}$]}
	& \multirow{3}{*}{$\Delta$ [\%]}
\\ 
& & &
\\ 
& & $\mu_r^2=m_t^2,\;p_{\gamma^\star}^2=(3000\;\text{GeV})^2$ &
\\\hline
\multicolumn{1}{c}{\bf NNLO} & \multicolumn{3}{c}{$\mathcal{O}(\alpha_s^2)$ (heavy nested fermion loops)} \\
\midrule
	C.1$^\star$ & \getRes{attxOSbenchHENNLOheavyfermionloopSGQG016mplcty} & \getRes{attxOSbenchHENNLOheavyfermionloopSGQG016} & \getRes{attxOSbenchHENNLOheavyfermionloopSGQG016delta}
\\
	C.2$^\star$ & \getRes{attxOSbenchHENNLOheavyfermionloopSGQG027mplcty} & \getRes{attxOSbenchHENNLOheavyfermionloopSGQG027} & \getRes{attxOSbenchHENNLOheavyfermionloopSGQG027delta}
\\
\cdashlinelr{1-4}
\multicolumn{2}{c|}{Total} & \getRes{attxOSbenchHENNLOheavyfermionloopTotal} & \getRes{attxOSbenchHENNLOheavyfermionloopTotaldelta}
\\
\multicolumn{2}{c|}{Benchmark} & \multicolumn{2}{c}{N/A}
\\
\midrule[0.3pt]\bottomrule[1pt]
\end{tabular}
}
\renewcommand{\arraystretch}{1.}
}
\end{minipage}\hfill
\end{center}
\end{figure}

\renewcommand{\arraystretch}{1.}

\FloatBarrier
\newpage
\subsection{Semi-inclusive NLO cross-section of $e^+ e^- \rightarrow \gamma \rightarrow jjj$}
\label{sec:epem_jjj_nlo}
We turn to the differential \glss{NLO} cross-section of $e^+ e^- \rightarrow \gamma \rightarrow jjj$, which was already computed with a similar numerical method in refs.~\cite{LTD_Soper_1,LTD_Soper_2} and that served as an inspiration for our original work on \glss{LU}.
The supergraphs contributing to this process up to \glss{NLO} are identical to those contributing to $e^+ e^- \rightarrow \gamma \rightarrow jj$ up to \glss{NNLO}.
The difference being that in this section Cutkosky cuts traversing only two edges and leaving a two-loop amplitude on either side are removed.
We view this removal of Cutkosky cuts as stemming from the final-state observable density function, which encodes the process definition in this way. It also implies that each supergraph is only \glss{IR}-finite when the observable density selects \glss{IR}-safe kinematics. Another important consequence of removing the Cutkosky cuts not satisfying our process definition is that it also breaks the pair-wise cancellation of \emph{non-pinched} thresholds. Therefore, contrary to the case of inclusive $1\rightarrow X$ cross-sections, we must consider the deformation discussed in sect.~5.3.4 of ref.~\cite{2021} and whose generic construction is presented in ref.~\cite{2020}.

Contrary to the simpler moments of an event shape computed in refs.~\cite{LTD_Soper_1,LTD_Soper_2}, we choose a more modern and complex selector that requires at least three clustered jets to be resolved, using the anti-kT~\cite{Ellis:1993tq,Cacciari:2006sm} clustering algorithm with the following parameters:
\begin{equation}
\label{jjjObservable}
\Delta R = 0.7,\quad p_t(j) > 50\;\text{GeV},\quad n_\text{jets} = 3\,.
\end{equation}
In tab.~\ref{tab:NLOjjjres}, we report results in the pure $\overline{\text{MS}}$ scheme for $s=(p_{e^+}+p_{e^-})^2=(1000\;\text{GeV})^2$ and at $\mu_r^2 = m_Z^2$. 
Supergraph D.2, resp. D.12, required the largest, resp. smallest, number of sample points ($9.1\cdot 10^9$, resp $0.3\cdot 10^9$). After multiple adaptive iterations, the overall sample generation efficiency reached 81\% (fraction of sample points for each Cutkosky cuts passing the selection criterion of eq.~\eqref{jjjObservable}).

\renewcommand{\arraystretch}{1.15}
\begin{table}[ht!]

\begin{adjustwidth}{-1.5cm}{-1.5cm}

\begin{center}
{\setlength\doublerulesep{1.5pt}   %
 \aboverulesep=0.06ex 
 \belowrulesep=-0.2ex
\begin{tabular}{c|c|cc|c}
\toprule[1pt]\midrule[0.3pt]
	\multirow{3}{*}{SG id}
	& \multirow{3}{*}{$\Xi$} 
	& \multirow{2}{*}{$\Re\left[\sigma_{e^+ e^- \rightarrow \gamma \rightarrow j j j}^{(\overline{\text{MS}})}\right]$ [pb]}
	& \multirow{2}{*}{$\Delta$ [\%]}
	& \multirow{2}{*}{$\Im\left[\sigma_{e^+ e^- \rightarrow \gamma \rightarrow j j j}^{(\overline{\text{MS}})}\right]$ [pb]}
\\ 
& & & &
\\ 
& & \multicolumn{3}{c}{$s=(p_{e^+}+p_{e^-})^2=(1000\;\text{GeV})^2$, $\mu_r^2=m_Z^2$} 
\\\hline	
\multicolumn{1}{c}{\bf LO} & \multicolumn{4}{c}{$\mathcal{O}(\alpha_s)$} \\
\midrule
	B.1 & \getRes{epemjjjLOSGQG000mplcty} & \getRes{epemjjjLOSGQG000} & \getRes{epemjjjLOSGQG000delta} & $0$
\\
	B.2 & \getRes{epemjjjLOSGQG002mplcty} & \getRes{epemjjjLOSGQG002} & \getRes{epemjjjLOSGQG002delta} & $0$
\\
\cdashlinelr{1-5}
\multicolumn{2}{c|}{Total} & \getRes{epemjjjLOTotal} & \getRes{epemjjjLOTotaldelta} & $0$
\\
\multicolumn{2}{c|}{Benchmark} & \getRes{epemjjjLOTarget} & \getRes{epemjjjLOTargetdelta} & $0$
\\
\midrule
\multicolumn{1}{c}{\bf NLO} & \multicolumn{4}{c}{$\mathcal{O}(\alpha_s^2)$ ($n_f=1$ contribution)} \\
\midrule
	C.1 & \getRes{epemjjjNLOnf1ReSGQG015mplcty} & \getRes{epemjjjNLOnf1ReSGQG015} & \getRes{epemjjjNLOnf1ReSGQG015delta} & \getRes{epemjjjNLOnf1ImSGQG015}
\\
	C.2 & \getRes{epemjjjNLOnf1ReSGQG036mplcty} & \getRes{epemjjjNLOnf1ReSGQG036} & \getRes{epemjjjNLOnf1ReSGQG036delta} & \getRes{epemjjjNLOnf1ImSGQG036}
\\
\cdashlinelr{1-5}
\multicolumn{2}{c|}{Total} & \getRes{epemjjjNLOnf1ReTotal} & \getRes{epemjjjNLOnf1ReTotaldelta} & \getRes{epemjjjNLOnf1ImTotal}
\\
\multicolumn{2}{c|}{Benchmark} & \getRes{epemjjjNLOnf1ReTarget} & \getRes{epemjjjNLOnf1ReTargetdelta} & $0$
\\
\midrule
\multicolumn{1}{c}{\bf NLO} & \multicolumn{4}{c}{$\mathcal{O}(\alpha_s^2)$ (all other contributions)} \\
\midrule
	D.1 & \getRes{epemjjjNLOReSGQG034mplcty} & \getRes{epemjjjNLOReSGQG034} & \getRes{epemjjjNLOReSGQG034delta} & \getRes{epemjjjNLOImSGQG034}
\\
	D.2 & \getRes{epemjjjNLOReSGQG006mplcty} & \getRes{epemjjjNLOReSGQG006} & \getRes{epemjjjNLOReSGQG006delta} & \getRes{epemjjjNLOImSGQG006}
\\
	D.3 & \getRes{epemjjjNLOReSGQG027mplcty} & \getRes{epemjjjNLOReSGQG027} & \getRes{epemjjjNLOReSGQG027delta} & \getRes{epemjjjNLOImSGQG027}
\\
	D.4 & \getRes{epemjjjNLOReSGQG013mplcty} & \getRes{epemjjjNLOReSGQG013} & \getRes{epemjjjNLOReSGQG013delta} & \getRes{epemjjjNLOImSGQG013}
\\
	D.5 & \getRes{epemjjjNLOReSGQG003mplcty} & \getRes{epemjjjNLOReSGQG003} & \getRes{epemjjjNLOReSGQG003delta} & \getRes{epemjjjNLOImSGQG003}
\\
	D.6 & \getRes{epemjjjNLOReSGQG033mplcty} & \getRes{epemjjjNLOReSGQG033} & \getRes{epemjjjNLOReSGQG033delta} & \getRes{epemjjjNLOImSGQG033}
\\
	D.7 & \getRes{epemjjjNLOReSGQG011mplcty} & \getRes{epemjjjNLOReSGQG011} & \getRes{epemjjjNLOReSGQG011delta} & \getRes{epemjjjNLOImSGQG011}$\;\pm \mathtt{0.18}\%$
\\
	$\overline{\text{D.7}}$ & \getRes{epemjjjNLOReSGQG009mplcty} & \getRes{epemjjjNLOReSGQG009} & \getRes{epemjjjNLOReSGQG009delta} & \getRes{epemjjjNLOImSGQG009}$\;\pm \mathtt{0.18}\%$
\\
	D.9 & \getRes{epemjjjNLOReSGQG035mplcty} & \getRes{epemjjjNLOReSGQG035} & \getRes{epemjjjNLOReSGQG035delta} & \getRes{epemjjjNLOImSGQG035}
\\
	D.10 &\getRes{epemjjjNLOReSGQG031mplcty} & \getRes{epemjjjNLOReSGQG031} & \getRes{epemjjjNLOReSGQG031delta} & \getRes{epemjjjNLOImSGQG031}
\\
	D.11 &\getRes{epemjjjNLOReSGQG005mplcty} & \getRes{epemjjjNLOReSGQG005} & \getRes{epemjjjNLOReSGQG005delta} & \getRes{epemjjjNLOImSGQG005}
\\
	D.12 &\getRes{epemjjjNLOReSGQG001mplcty} & \getRes{epemjjjNLOReSGQG001} & \getRes{epemjjjNLOReSGQG001delta} & \getRes{epemjjjNLOImSGQG001}
\\
	D.13 &\getRes{epemjjjNLOReSGQG014mplcty} & \getRes{epemjjjNLOReSGQG014} & \getRes{epemjjjNLOReSGQG014delta} & \getRes{epemjjjNLOImSGQG014}
\\
	D.14 &\getRes{epemjjjNLOReSGQG028mplcty} & \getRes{epemjjjNLOReSGQG028} & \getRes{epemjjjNLOReSGQG028delta} & \getRes{epemjjjNLOImSGQG028}
\\
	D.15 &\getRes{epemjjjNLOReSGQG000mplcty} & \getRes{epemjjjNLOReSGQG000} & \getRes{epemjjjNLOReSGQG000delta} & \getRes{epemjjjNLOImSGQG000}
\\
\cdashlinelr{1-5}
\multicolumn{2}{c|}{Total} & \getRes{epemjjjNLOReTotal} & \getRes{epemjjjNLOReTotaldelta} & \getRes{epemjjjNLOImTotal}
\\
\multicolumn{2}{c|}{Benchmark} & \getRes{epemjjjNLOReTarget} & \getRes{epemjjjNLOReTargetdelta} & $0$
\\
\midrule[0.3pt]\bottomrule[1pt]
\end{tabular}
}
\end{center}
\end{adjustwidth}
\caption{\label{tab:NLOjjjres} Contributions from individual supergraphs listed in tab.~\ref{tab:jjjSGs} (external current $e^+ e^- \rightarrow \gamma$ is not shown there) to the semi-inclusive cross-section up to order $\mathcal{O}(\alpha_s^2)$ (\glss{LO}+\glss{NLO}) for the processes $e^+ e^- \rightarrow \gamma \rightarrow j j j$ with $n_f=1$ and $Q_q=1/3$ and \gls{SM} parameters as given in table.~\ref{tab:SMparameters} and the fiducial cuts discussed in eq.~\eqref{jjjObservable}. $\Xi$ denotes the symmetry factor of the supergraph (included in the result reported). $\Delta$ reports the relative Monte-Carlo error, except for the ``Benchmark'' entry where it refers to the relative difference w.r.t the \glss{LU} result instead. The benchmark results are obtained in this case with \gls{MG5aMC}~\cite{Alwall:2014hca} and are themselves subject to a 1\% MC uncertainty.}
\end{table}
\renewcommand{\arraystretch}{1.}

\FloatBarrier

In general, we find a significantly poorer integration convergence than for the inclusive \glss{NNLO} counterpart, and we could not reach percent-level accuracy. We attribute this mainly to two factors. First, \glss{IR} singularities are now regularised through a combined effects of the \glss{LU} pair-wise cancellation and the clustering observable, and the latter leaves regions of phase-space with collinear enhancements inducing $\log\left(\Delta R\right)$ cross-section contributions that can be difficult to accurately reproduce numerically.
This is expected to be improved by a smarter parameterisation of the spatial part of the supergraph loop momenta and better sampling strategy (see sect.~5.4 of ref.~\cite{2021}).
Second, and likely more important, the introduction of a contour deformation can increase the variance significantly, especially for massless processes (e.g. see visualisations of figs.~13-16 in ref.~\cite{2021}). A more precise understanding of the behaviour of the contour deformation close to \glss{IR} limits can help mitigate this problem and this will be the focus of future work. We note that a promising alternative is to remove the need for a deformation altogether, by instead subtracting non-pinched thresholds locally~\cite{Kermanschah:2021wbk}.

The introduction of a contour deformation generates an imaginary part for the integrand, but unitarity guarantees that, at the integrated level, any observable remains real. More pragmatically, the imaginary contribution from each Cutkosky cut with the amplitude graph $\Gamma_L$ to its right and complex-conjugated amplitude graph $\Gamma_R^\dagger$ to its left is cancelled by its complex-conjugated partner identifying the two graphs $\Gamma_R$ and $\Gamma_L^\dagger$ instead.
For supergraphs that are left-right symmetric, these pairs of Cutkosky cuts related by unitary, i.e. by overall complex-conjugation, are all contained within the same graph. Fig.~\ref{tab:jjjSGs} reveals that the only supergraph contributing to $e^+ e^- \rightarrow \gamma \rightarrow jjj$ that is \emph{not} left-right symmetric is D.7, whose symmetric partner we call $\overline{\text{D}.7}$. In the second column of fig.~\ref{tab:NLOjjjres}, we report the imaginary result of the \glss{LU} cross-section and, as expected, we find that it is zero (within the Monte-Carlo accuracy) for all supergraphs except D.7 and $\overline{\text{D}.7}$ that are exactly opposite of each other.
Supergraphs for which the central value of the imaginary part of the integral is extremely small corresponds to cases where no deformation was necessary or where the imaginary part of the integrand cancels locally for our choice of deformation.

\newpage
\subsection{(Semi-)inclusive NLO cross-section of $e^+ e^- \rightarrow \gamma \rightarrow t\bar{t} H$}

\begin{figure}[ht!]
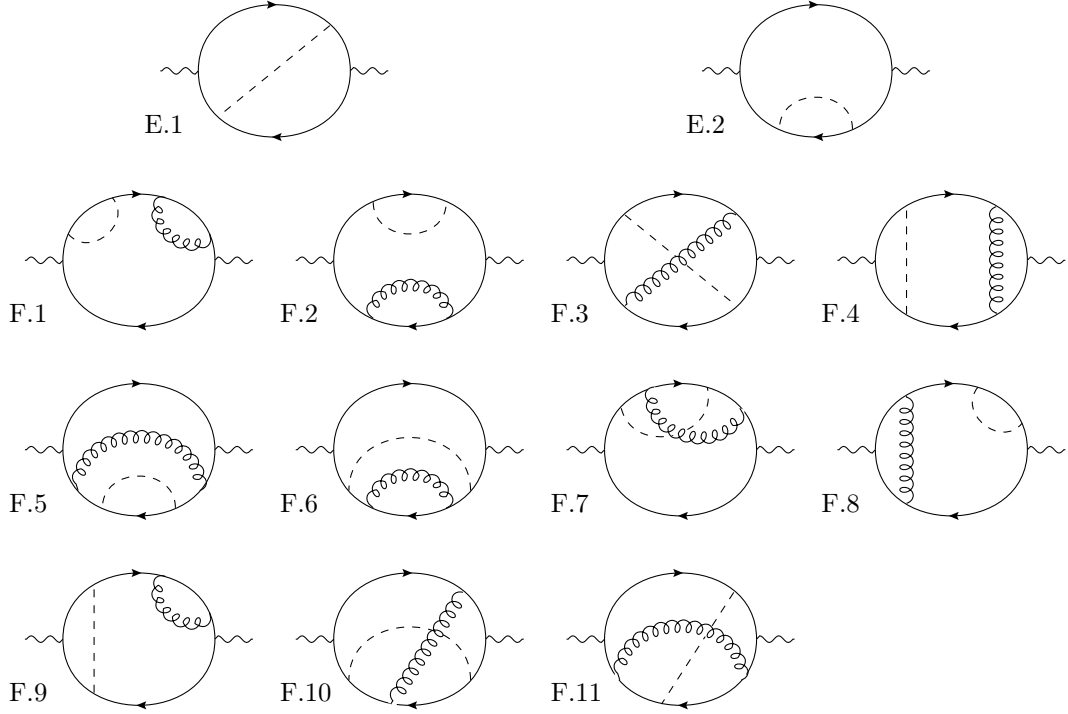

  \centering
  \begin{tabular}{cccc}
    \multicolumn{2}{c}{%
  \setbox1=\hbox{\input{diagrams/supergraphs_ttH_LO/diag1}}%
  \leavevmode\rlap{\usebox1}%
  \rlap{\hspace*{-0.2cm}\raisebox{\dimexpr\ht1-3\baselineskip-2.0cm}{\small E.1}}%
  \phantom{\usebox1}%
} &
    \multicolumn{2}{c}{%
  \setbox1=\hbox{\input{diagrams/supergraphs_ttH_LO/diag2}}%
  \leavevmode\rlap{\usebox1}%
  \rlap{\hspace*{-0.2cm}\raisebox{\dimexpr\ht1-3\baselineskip-2.0cm}{\small E.2}}%
  \phantom{\usebox1}%
}
    \\
  \setbox1=\hbox{\input{diagrams/supergraphs_ttH_NLO/diag1}}%
  \leavevmode\rlap{\usebox1}%
  \rlap{\hspace*{-0.2cm}\raisebox{\dimexpr\ht1-3\baselineskip-2.0cm}{\small F.1}}%
  \phantom{\usebox1}%
 & %
  \setbox1=\hbox{\input{diagrams/supergraphs_ttH_NLO/diag2}}%
  \leavevmode\rlap{\usebox1}%
  \rlap{\hspace*{-0.2cm}\raisebox{\dimexpr\ht1-3\baselineskip-2.0cm}{\small F.2}}%
  \phantom{\usebox1}%
 & %
  \setbox1=\hbox{\input{diagrams/supergraphs_ttH_NLO/diag3}}%
  \leavevmode\rlap{\usebox1}%
  \rlap{\hspace*{-0.2cm}\raisebox{\dimexpr\ht1-3\baselineskip-2.0cm}{\small F.3}}%
  \phantom{\usebox1}%
 & %
  \setbox1=\hbox{\input{diagrams/supergraphs_ttH_NLO/diag4}}%
  \leavevmode\rlap{\usebox1}%
  \rlap{\hspace*{-0.2cm}\raisebox{\dimexpr\ht1-3\baselineskip-2.0cm}{\small F.4}}%
  \phantom{\usebox1}%
    \\
  \setbox1=\hbox{\input{diagrams/supergraphs_ttH_NLO/diag6}}%
  \leavevmode\rlap{\usebox1}%
  \rlap{\hspace*{-0.2cm}\raisebox{\dimexpr\ht1-3\baselineskip-2.0cm}{\small F.5}}%
  \phantom{\usebox1}%
 & %
  \setbox1=\hbox{\input{diagrams/supergraphs_ttH_NLO/diag6_1}}%
  \leavevmode\rlap{\usebox1}%
  \rlap{\hspace*{-0.2cm}\raisebox{\dimexpr\ht1-3\baselineskip-2.0cm}{\small F.6}}%
  \phantom{\usebox1}%
 & %
  \setbox1=\hbox{\input{diagrams/supergraphs_ttH_NLO/diag7}}%
  \leavevmode\rlap{\usebox1}%
  \rlap{\hspace*{-0.2cm}\raisebox{\dimexpr\ht1-3\baselineskip-2.0cm}{\small F.7}}%
  \phantom{\usebox1}%
 & %
  \setbox1=\hbox{\input{diagrams/supergraphs_ttH_NLO/diag14}}%
  \leavevmode\rlap{\usebox1}%
  \rlap{\hspace*{-0.2cm}\raisebox{\dimexpr\ht1-3\baselineskip-2.0cm}{\small F.8}}%
  \phantom{\usebox1}%
    \\
  \setbox1=\hbox{\input{diagrams/supergraphs_ttH_NLO/diag14_1}}%
  \leavevmode\rlap{\usebox1}%
  \rlap{\hspace*{-0.2cm}\raisebox{\dimexpr\ht1-3\baselineskip-2.0cm}{\small F.9}}%
  \phantom{\usebox1}%
 & %
  \setbox1=\hbox{\input{diagrams/supergraphs_ttH_NLO/diag15}}%
  \leavevmode\rlap{\usebox1}%
  \rlap{\hspace*{-0.2cm}\raisebox{\dimexpr\ht1-3\baselineskip-2.0cm}{\small F.10}}%
  \phantom{\usebox1}%
 & %
  \setbox1=\hbox{\input{diagrams/supergraphs_ttH_NLO/diag15_1}}%
  \leavevmode\rlap{\usebox1}%
  \rlap{\hspace*{-0.2cm}\raisebox{\dimexpr\ht1-3\baselineskip-2.0cm}{\small F.11}}%
  \phantom{\usebox1}%
 & %
    \\
  \end{tabular}
  \caption{\label{tab:ttHSGs}All contributing supergraphs to the class of processes $e^+ e^- \rightarrow \gamma \rightarrow t \bar{t} H + X$ at LO (E.1) and NLO (F.i). The external current $e^+ e^- \rightarrow \gamma$ is not shown. We note that we do not show the isomorphic partners $\overline{\text{F.1}}$, $\overline{\text{F.4}}$, $\overline{\text{F.7}}$, $\overline{\text{F.8}}$ and $\overline{\text{F.9}}$ obtained by a mirror symmetry around the vertical central axis of the corresponding graph F.i (see explicit example with graphs D.7 and $\overline{\text{D.7}}$ in tab.~\ref{tab:jjjSGs}).
  }
\end{figure}

In light of the relatively poor convergence of the semi-inclusive cross-section of $e^+ e^- \rightarrow \gamma \rightarrow j j j$, it is interesting to consider a different type of \glss{NLO}-accurate cross-section for a process whose definition still requires the removal of Cutkosky cuts (so that a contour deformation is still required) but which is free of phase-space \glss{IR} singularities at \glss{LO}, so that it can be computed inclusively as well.
To this end, we choose the process $e^+ e^- \rightarrow \gamma \rightarrow t\bar{t} H$ for which we list the contributing supergraphs up to \glss{NLO} in fig.~\ref{tab:ttHSGs}. This process has the additional benefit of involving supergraphs with a self-energy correcting an \emph{internal} top quark propagator (supergraphs F.1 and F.9), that typically require a contribution from the top quark mass \glss{OS} renormalisation counterterm $\delta m_t$ and that our careful definition of local \glss{UV} counterterms must be able to automatically reproduce (see sect.~\ref{sect:integrated_os_cts}).
For this reason, we will show results with the top quark mass renormalised in the \glss{OS} scheme. All other quantities (top quark field, top quark Yukawa coupling and $\alpha_s$) are renormalised in $\overline{\text{MS}}$ and their related running is ignored. We choose $\sqrt{s}=p_{e^+}^0 + p_{e^-}^0=1000\;\text{GeV}$ and $\mu_r = m_Z$.
In order to explore potential degradation of the Monte-Carlo convergence in the presence of a complicated observable selector function, we also show results for the semi-inclusive cross-section defined over a fiducial volume characterised by the following acceptance cuts:
\begin{eqnarray}
    \label{tthFiducialVolume}
    200 \text{GeV} < E_t < 600 \text{GeV}&,\;\;& -0.8 < \cos\left({\theta_t}\right) < 0.8 \nonumber\\
    250 \text{GeV} < E_H < 500 \text{GeV}&,\;\;& -0.6 < \cos\left({\theta_H}\right) < 0.6\;,
\end{eqnarray}
where $E_X$ denotes the energy component of the momentum carried by \emph{any} particle $X$, and the polar angle $\theta_X$ is defined as $\cos\left({\theta_X}\right) = \vec{p}_X\cdot \hat{e}_z / |\vec{p}_X|$, $\hat{e}_z=(0,0,1)$.

We find a numerical convergence significantly better for this process than for $e^+ e^- \rightarrow \gamma \rightarrow jjj$. The largest number of sample points for the inclusive, resp. semi-inclusive, cross-section is $1.5\cdot 10^9$, resp. $4 \cdot 10^9$, for supergraph F.3 and the lowest number is $0.16\cdot 10^9$, resp. $0.95\cdot 10^9$, for supergraph F.5. The overall sample generation efficiency in the semi-inclusive case is 38\%.
We attribute the improved convergence to a better behaviour of our choice of deformation for the simpler structure of \glss{IR} singularities featured in this process. This hypothesis is reinforced by the observation that imposing complicated fiducial cuts has only a minor impact on the convergence. 
We find agreement well below the percent level, thus offering sufficient resolution for establishing the validity of our procedure for reproducing the \glss{OS} mass renormalisation counterterm. 
Moreover, we verified again that we could obtain the same result starting from the pure $\overline{\text{MS}}$ implementation of $e^+ e^- \rightarrow \gamma \rightarrow t\bar{t} H$ but complemented with the insertion of explicit integrated-level expression of $\delta m_t^{\alpha_s}$ given in eq.~\ref{eq:deltamtOneLoop}.

In tab.~\ref{tab:NLOttHres}, we only show the result for the real part of the cross-section. This is because we grouped together the result from all supergraphs that are not themselves left-right symmetric, so that the imaginary part is always zero, and we verified that this is the case, within the Monte-Carlo uncertainty (similarly as what is explicitly shown in the second column of tab.~\ref{tab:NLOjjjres}).

\renewcommand{\arraystretch}{1.15}
\begin{table}[ht!]

\begin{adjustwidth}{-1.5cm}{-1.5cm}

\begin{center}
{
\setlength\doublerulesep{1.5pt}   %
\aboverulesep=0.06ex 
\belowrulesep=-0.2ex
 
\begin{tabular}{c|c|cc|cc}
\toprule[1pt]\midrule[0.3pt]
	\multirow{5}{*}{SG id}
	& \multirow{5}{*}{$\Xi$}
	& \multicolumn{4}{c}{\multirow{3}{*}{$\Re\left[\sigma_{e^+ e^- \rightarrow \gamma \rightarrow t \bar{t} H}^{\left[ y_t^{(\overline{\text{MS}})},\;m_t^{(\text{OS})}\right] }\right]$}}
\\ 
& & \multicolumn{4}{c}{}
\\ 
& & \multicolumn{4}{c}{}
\\ 
& & \multicolumn{4}{c}{$s=(p_{e^+}+p_{e^-})^2=(1000\;\text{GeV})^2$, $\mu_r^2=m_Z^2$} 
\\
    & & Inclusive xs [pb] & $\Delta$ [\%] & Semi-inclusive xs [pb] & $\Delta$ [\%]
\\\hline
\multicolumn{1}{c}{\bf LO} & \multicolumn{5}{c}{$\mathcal{O}(\alpha_s)$} \\
\midrule
	E.1 & \getRes{epemttxhLOinclSGQG000mplcty} & \getRes{epemttxhLOinclSGQG000} & \getRes{epemttxhLOinclSGQG000delta} & \getRes{epemttxhLOsinclSGQG000} & \getRes{epemttxhLOsinclSGQG000delta}
\\
	E.2 & \getRes{epemttxhLOinclSGQG002mplcty} & \getRes{epemttxhLOinclSGQG002} & \getRes{epemttxhLOinclSGQG002delta} & \getRes{epemttxhLOsinclSGQG002} &\getRes{epemttxhLOsinclSGQG002delta}
\\
\cdashlinelr{1-6}
\multicolumn{2}{c|}{Total} & \getRes{epemttxhLOinclTotal} & \getRes{epemttxhLOinclTotaldelta} & \getRes{epemttxhLOsinclTotal} & \getRes{epemttxhLOsinclTotaldelta}
\\
\multicolumn{2}{c|}{Benchmark} & \getRes{epemttxhLOinclTarget} & \getRes{epemttxhLOinclTargetdelta} & \getRes{epemttxhLOsinclTarget} & \getRes{epemttxhLOsinclTargetdelta}
\\
\midrule
\multicolumn{1}{c}{\bf NLO} & \multicolumn{5}{c}{$\mathcal{O}(\alpha_s^2)$} \\
\midrule
	F.1+$\overline{\text{F.1}}$ & \getRes{epemttxhNLOinclF000001mplcty} & \getRes{epemttxhNLOinclF000001} & \getRes{epemttxhNLOinclF000001delta} & \getRes{epemttxhNLOsinclF000001} & \getRes{epemttxhNLOsinclF000001delta}
\\
	F.2 & \getRes{epemttxhNLOinclF000002mplcty} & \getRes{epemttxhNLOinclF000002} & \getRes{epemttxhNLOinclF000002delta} & \getRes{epemttxhNLOsinclF000002} & \getRes{epemttxhNLOsinclF000002delta}
\\
	F.3 & \getRes{epemttxhNLOinclF000003mplcty} & \getRes{epemttxhNLOinclF000003} & \getRes{epemttxhNLOinclF000003delta} & \getRes{epemttxhNLOsinclF000003} & \getRes{epemttxhNLOsinclF000003delta}
\\
	F.4+$\overline{\text{F.4}}$ & \getRes{epemttxhNLOinclF000004mplcty} & \getRes{epemttxhNLOinclF000004} & \getRes{epemttxhNLOinclF000004delta} & \getRes{epemttxhNLOsinclF000004} & \getRes{epemttxhNLOsinclF000004delta}
\\
	F.5 & \getRes{epemttxhNLOinclF000005mplcty} & \getRes{epemttxhNLOinclF000005} & \getRes{epemttxhNLOinclF000005delta} & \getRes{epemttxhNLOsinclF000005} & \getRes{epemttxhNLOsinclF000005delta}
\\
	F.6 & \getRes{epemttxhNLOinclF000006mplcty} & \getRes{epemttxhNLOinclF000006} & \getRes{epemttxhNLOinclF000006delta} & \getRes{epemttxhNLOsinclF000006} & \getRes{epemttxhNLOsinclF000006delta}
\\
	F.7+$\overline{\text{F.7}}$ & \getRes{epemttxhNLOinclF000007mplcty} & \getRes{epemttxhNLOinclF000007} & \getRes{epemttxhNLOinclF000007delta} & \getRes{epemttxhNLOsinclF000007} & \getRes{epemttxhNLOsinclF000007delta}
\\
	F.8+$\overline{\text{F.8}}$ & \getRes{epemttxhNLOinclF000008mplcty} & \getRes{epemttxhNLOinclF000008} & \getRes{epemttxhNLOinclF000008delta} & \getRes{epemttxhNLOsinclF000008} & \getRes{epemttxhNLOsinclF000008delta}
\\
	F.9+$\overline{\text{F.9}}$ & \getRes{epemttxhNLOinclF000009mplcty} & \getRes{epemttxhNLOinclF000009} & \getRes{epemttxhNLOinclF000009delta} & \getRes{epemttxhNLOsinclF000009} & \getRes{epemttxhNLOsinclF000009delta}
\\
	F.10 & \getRes{epemttxhNLOinclF000010mplcty} & \getRes{epemttxhNLOinclF000010} & \getRes{epemttxhNLOinclF000010delta} & \getRes{epemttxhNLOsinclF000010} & \getRes{epemttxhNLOsinclF000010delta}
\\
	F.11 & \getRes{epemttxhNLOinclF000011mplcty} & \getRes{epemttxhNLOinclF000011} & \getRes{epemttxhNLOinclF000011delta} & \getRes{epemttxhNLOsinclF000011} & \getRes{epemttxhNLOsinclF000011delta}
\\
\cdashlinelr{1-6}
\multicolumn{2}{c|}{Total} & \getRes{epemttxhNLOinclTotal} & \getRes{epemttxhNLOinclTotaldelta} & \getRes{epemttxhNLOsinclTotal} & \getRes{epemttxhNLOsinclTotaldelta}
\\
\multicolumn{2}{c|}{Benchmark} & \getRes{epemttxhNLOinclTarget} & \getRes{epemttxhNLOinclTargetdelta} & \getRes{epemttxhNLOsinclTarget} & \getRes{epemttxhNLOsinclTargetdelta}
\\
\midrule[0.3pt]\bottomrule[1pt]
\end{tabular}
}
\end{center}
\end{adjustwidth}
\caption{\label{tab:NLOttHres} Contributions from individual supergraphs listed in tab.~\ref{tab:ttHSGs} to the cross-section up to order $\mathcal{O}(\alpha_s^2)$ (LO+NLO) for the processes $e^+ e^- \rightarrow \gamma \rightarrow t \bar{t} H$ with the \gls{SM} parameters as given in table.~\ref{tab:SMparameters} except that the top-quark mass is renormalised in the OS scheme here, and not in $\overline{\text{MS}}$. We show results for both the fully inclusive cross-section and for a semi-inclusive one with fiducial cuts given in Eqs.~\ref{tthFiducialVolume}. The entries labelled "F.i+$\overline{\text{F.i}}$" report the result for the \emph{sum} of the contribution from both isomorphic supergraphs F.i and $\overline{\text{F.i}}$.
$\Xi$ denotes the symmetry factor of the supergraph (included in the result reported). $\Delta$ reports the relative Monte-Carlo error, except for the ``Benchmark'' entry where it refers to the relative difference w.r.t the LU result instead. The benchmark results are obtained in this case from \gls{MG5aMC}~\cite{Alwall:2014hca} and are themselves subject to a 0.1\% MC uncertainty.}
\end{table}
\renewcommand{\arraystretch}{1.}

\FloatBarrier
\newpage
\subsection{Code performance and example of specific supergraphs up to N3LO}
\label{sec:code_performance_and_n3lo}

In this section, we will provide more details on the current performance of our implementation of the \glss{LU} representation of differential cross-sections in a private computer code named $\alpha\text{Loop}$. Ultimately, the objective is to minimise the total computational time for computing a given observable of a given scattering process up to a target relative accuracy and at a set perturbative order. However, such an inclusive metric aggregates the performance of many different aspects of an implementation and it is therefore not particularly insightful. We therefore find it useful to separate optimisations impacting this overall performance into two classes:
\begin{itemize}
    \item {\bf Integrator optimisations} aim at reducing the number of sample points that are necessary to reach a certain accuracy.
    \item {\bf Integrand optimisations} aim at reducing the time necessary for evaluating a given sample point. We also include peripheral concepts such as generation timing, numerical stability and memory footprint (RAM and disk) under this umbrella term of integrand performance. 
\end{itemize}

Integrator performance is mostly driven by the choice of adaptive sampling algorithms, as well as the various parameterisations considered when building the overall integrand (e.g. typically within a multi-channeling approach). The improvements foreseen in sect.~5.4 of ref.~\cite{2021} have not yet been fully explored and as such it is too early to present details of our current integrator performance. We will limit ourselves here to stating that the results presented in sect.~\ref{sec:results} were obtained in less than hundred thousand CPU hours per process.

Integrand performance is mostly driven by the many design choices entering the computer implementation of the \glss{LU} representation. Only our eventual publication of the code can give a detailed account of all these choices, but they mostly relate to efficient graph manipulations and isomorphisms, {\sc\small form}-optimised implementation of the \glss{cLTD}~\cite{Capatti:2020ytd} representation, leveraging partial factorisation in the structure of the subtraction terms generated by the $R$-operation and finally the use of dual numbers for the exact numerical computation of derivatives.
We note that the run-time efficiency of the contour deformation also crucially depends on optimisations in its implementation. For the most part, those have already been presented in details, both qualitatively and quantitatively, in ref.~\cite{2020}, so that we will only consider here the profiling of \glss{LU} integrands evaluated with real kinematics.

Contrary to integrator performance, our current implementation in $\alpha\text{Loop}$ is good enough to warrant the publication of integrand performance benchmarks that are useful as an anchor point to assess future improvements (by ourselves but hopefully also independent groups seeking to replicate and improve upon our work). Together with these detailed integrand performance statistics, we also present applications of our automated testing suite to investigate the quality of cancellations in various \glss{IR} and \glss{UV} limits of the \glss{LU} integrand.

\begin{figure}[ht!]
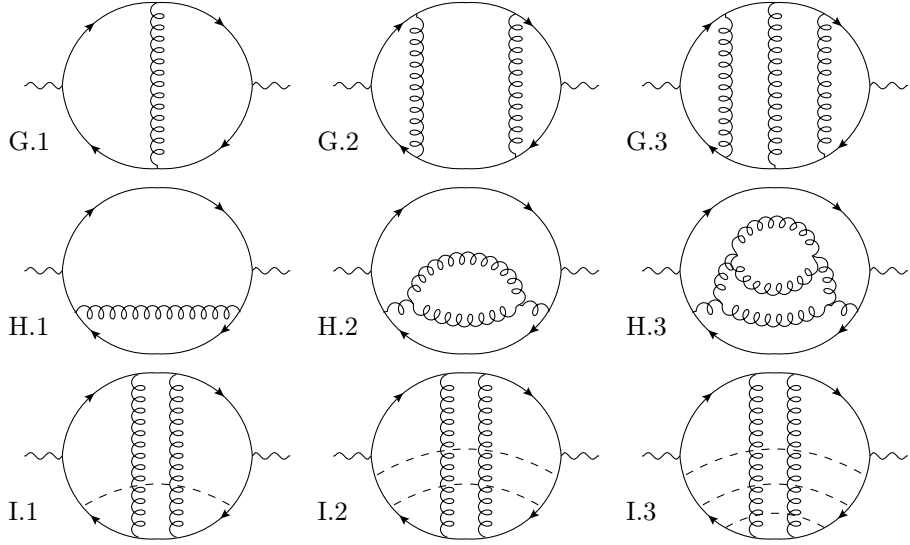

  \centering
  \begin{tabular}{ccc}
  \setbox1=\hbox{\input{diagrams/supergraphs_DT_class/1L}}%
  \leavevmode\rlap{\usebox1}%
  \rlap{\hspace*{-0.2cm}\raisebox{\dimexpr\ht1-3\baselineskip-2.0cm}{\small G.1}}%
  \phantom{\usebox1}%
 &
  \setbox1=\hbox{\input{diagrams/supergraphs_DT_class/2L}}%
  \leavevmode\rlap{\usebox1}%
  \rlap{\hspace*{-0.2cm}\raisebox{\dimexpr\ht1-3\baselineskip-2.0cm}{\small G.2}}%
  \phantom{\usebox1}%
 &
  \setbox1=\hbox{\input{diagrams/supergraphs_DT_class/3L}}%
  \leavevmode\rlap{\usebox1}%
  \rlap{\hspace*{-0.2cm}\raisebox{\dimexpr\ht1-3\baselineskip-2.0cm}{\small G.3}}%
  \phantom{\usebox1}%

    \\
  \setbox1=\hbox{\input{diagrams/supergraphs_SE_class/1L}}%
  \leavevmode\rlap{\usebox1}%
  \rlap{\hspace*{-0.2cm}\raisebox{\dimexpr\ht1-3\baselineskip-2.0cm}{\small H.1}}%
  \phantom{\usebox1}%
 &
  \setbox1=\hbox{\input{diagrams/supergraphs_SE_class/2L}}%
  \leavevmode\rlap{\usebox1}%
  \rlap{\hspace*{-0.2cm}\raisebox{\dimexpr\ht1-3\baselineskip-2.0cm}{\small H.2}}%
  \phantom{\usebox1}%
 &
  \setbox1=\hbox{\input{diagrams/supergraphs_SE_class/3L}}%
  \leavevmode\rlap{\usebox1}%
  \rlap{\hspace*{-0.2cm}\raisebox{\dimexpr\ht1-3\baselineskip-2.0cm}{\small H.3}}%
  \phantom{\usebox1}%

    \\
  \setbox1=\hbox{\input{diagrams/supergraphs_multi_H_class/1H}}%
  \leavevmode\rlap{\usebox1}%
  \rlap{\hspace*{-0.2cm}\raisebox{\dimexpr\ht1-3\baselineskip-2.0cm}{\small I.1}}%
  \phantom{\usebox1}%
 &
  \setbox1=\hbox{\input{diagrams/supergraphs_multi_H_class/2H}}%
  \leavevmode\rlap{\usebox1}%
  \rlap{\hspace*{-0.2cm}\raisebox{\dimexpr\ht1-3\baselineskip-2.0cm}{\small I.2}}%
  \phantom{\usebox1}%
 &
  \setbox1=\hbox{\input{diagrams/supergraphs_multi_H_class/3H}}%
  \leavevmode\rlap{\usebox1}%
  \rlap{\hspace*{-0.2cm}\raisebox{\dimexpr\ht1-3\baselineskip-2.0cm}{\small I.3}}%
  \phantom{\usebox1}%

    \\
  \end{tabular}
  \caption{\label{fig:complexity_scaling_supergraphs} Example supergraphs for increasing $\text{N}^k\text{LO}$ perturbative order, with $k=1,2,3$ for the "double-triangle" class G.x and the "nested self-energy" class H.x.
  The class of supergraphs I.x corresponds to \glss{NNLO} contributions to the process $\gamma^\star \rightarrow t \bar{t} + N \times H$ where $N=1,2,3$.
  }
\end{figure}

The complexity of the \glss{LU} representation of the $\text{N}^k\text{LO}$ correction to a $1\rightarrow N$ process scales both in $k$ and $N$. One may naively think that since \glss{LU} combines phase-space and loop integrals, there should be no distinction between the complexity growth in these two scaling parameters. This is true for quantities such as the total number of contributing supergraphs and the dimensionality $3\times(N+k-1)$ of the \glss{LU} Monte-Carlo integral. Yet, the scaling of the overall complexity of the \glss{LU} representation remains far worse in $k$ than $N$ for the following four reasons:
\begin{enumerate}
    \item The number of Cutkosky cuts that must be considered in any supergraph is bounded by the number of ways one can distribute $n_\text{left}$ and $n_\text{right}$ loops on either side of a Cutkosky cut that will have an additional multiplicity of $k-n_\text{left}-n_\text{right}$, and counting each occurrence twice to account for complex conjugation. This means that the maximum number of Cutkosky cuts in any $\text{N}^k\text{LO}$ supergraph contributing to a $1\rightarrow N$ process is $(k+2)(k+1)$, irrespective of $N$.
    \item The numerical severity of cancellations between various terms of the \glss{LU} integrand in the \glss{IR} and \glss{UV} limits only depends on $k$. In other words, the more divergent limits can be reached simultaneously, the more challenging it is to keep the \glss{LU} integrand numerically stable everywhere in the integration space.
    \item Most importantly, the complexity of the \glss{UV} subtraction grows exponentially with $k$ and not at all with $N$. This is because the number of subtraction terms generated by the $R$-operator is dictated by its maximum recursion depth, which is the maximum loop count $k$ that can appear in any amplitude on either side of a Cutkosky cut.
    \item Even without \glss{UV} subtraction, the \glss{LU} integrand still involves the \glss{cLTD} expression of multi-loop amplitudes, whose complexity scales with both $N$ and $k$, but was shown to be milder in $N$ (see ref.~\cite{Capatti:2020ytd}). 
\end{enumerate}
In order to investigate separately the scaling of our implementation in both $k$ and $N$, we choose to study the three different series of characteristic supergraphs shown in fig.~\ref{fig:complexity_scaling_supergraphs}. The first two series are contributions to $\gamma^\star \rightarrow j j$ (so $N=2$) at $\text{N}^k\text{LO}$, with $k=1,2,3$. The third series are contributions to $\gamma^\star \rightarrow t \bar{t} + N \times H$ at \glss{NNLO} (so $k=2$) with $N=1,2,3$.

We start by reporting in tab.~\ref{tab:code_performance} the following key metrics for each of the nine supergraphs of fig.~\ref{fig:complexity_scaling_supergraphs}:

\begin{itemize}
    \item $t_\text{gen}$ [s] : Time spent in {\sc\small form} for generating the source code of this supergraph (compilation time is typically less than $t_\text{gen}$ with sufficiently many cores).
    \item $M_\text{disk}$ [MB] : Size of the compiled shared library on disk.
    \item $N_\text{sg}$ [-] : Total number of non-isomorphic supergraphs with a unique contribution to the process this supergraph belongs to. Distinct supergraphs related by a symmetry interchanging the two external photons are counted twice.
    \item $N_\text{cuts}$ [-] : Total number of Cutkosky cuts for this supergraph.
    \item $t_\text{eval}$ [ms] : Evaluation time for evaluating the \glss{LU} integrand for all Cutkosky cuts.
    \item $t_\text{eval}^{(\text{f128})}$ [ms] : Evaluation time for evaluating the \glss{LU} integrand in quadruple precision.
\end{itemize}

Note that the timings $t_\text{eval}$ and $t_\text{eval}^{(\text{f128})}$ include numerical stability tests, which at least double the evaluation time. Also note that statistics provided here have qualitative merits only, so we do not specify the hardware that ran the tests.
\begin{table}[h!]
\begin{center}
{\setlength\doublerulesep{1.5pt}   %
 \aboverulesep=0ex 
 \belowrulesep=0ex
\begin{tabular}{cll|cccc|cccc}
\toprule[1pt]
	SG & proc. & order & $t_\text{gen}$ [s] & $M_\text{disk}$ [MB] & $N_\text{sg}$ [-] & $N_\text{cuts}$ [-] &  $t_\text{eval}$ [ms] & $t_\text{eval}^{(\text{f128})}$ [ms]
\\ \midrule[0.5pt]
    G.1 & $1\rightarrow 2$ & \glss{NLO} & $\mathtt{0.1}$ & $\mathtt{0.13}$ & $\mathtt{2}$ & $\mathtt{4}$ & $\mathtt{0.004}$ & $\mathtt{0.13}$ \\
    G.2 & $1\rightarrow 2$ &  \glss{NNLO} & $\mathtt{4.7}$ & $\mathtt{3.0}$ & $\mathtt{17}$ & $\mathtt{9}$ & $\mathtt{0.04}$ & $\mathtt{2.1}$ \\
    G.3 & $1\rightarrow 2$ &  \glss{N3LO} & $\mathtt{36K}$ & $\mathtt{509}$ & $\mathtt{220}$ & $\mathtt{16}$ & $\mathtt{17.6}$ & $\mathtt{281}$
\\ \midrule[0.5pt]
    H.1 & $1\rightarrow 2$ &  \glss{NLO} & $\mathtt{0.07}$ & $\mathtt{0.12}$ & $\mathtt{2}$ & $\mathtt{2}$ & $\mathtt{0.006}$ & $\mathtt{0.14}$ \\
    H.2 & $1\rightarrow 2$ &  \glss{NNLO} & $\mathtt{1.5}$ & $\mathtt{1.3}$ & $\mathtt{17}$ & $\mathtt{3}$ & $\mathtt{0.056}$ & $\mathtt{1.9}$ \\
    H.3 & $1\rightarrow 2$ &  \glss{N3LO} & $\mathtt{255}$ & $\mathtt{43}$ & $\mathtt{220}$ & $\mathtt{4}$ & $\mathtt{2.35}$ & $\mathtt{56}$
\\ \midrule[0.5pt]
    I.1 & $1\rightarrow 3$ &  \glss{NNLO} & $\mathtt{126}$ & $\mathtt{22}$ & $\mathtt{266}$ & $\mathtt{9}$ & $\mathtt{0.32}$ & $\mathtt{12.4}$ \\
    I.2 & $1\rightarrow 4$ &  \glss{NNLO} & $\mathtt{1.9K}$ & $\mathtt{120}$ & $\mathtt{4492}$ & $\mathtt{9}$ & $\mathtt{4.4}$ & $\mathtt{67}$ \\
    I.3 & $1\rightarrow 5$ &  \glss{NNLO} & $\mathtt{36K}$ & $\mathtt{20K}$ & $\mathcal{O}\left(\mathtt{100K}\right)$ & $\mathtt{9}$ & $\mathtt{3.6K}$ & $\mathtt{17.3K}$ \\
\midrule[0.3pt]\bottomrule[1pt]
\end{tabular}
}
\end{center}
\caption{\label{tab:code_performance} Performance for characteristic supergraphs of fig.~\ref{fig:complexity_scaling_supergraphs} corresponding to $\text{N}^k\text{LO}$ corrections for $1\rightarrow N \times X$ processes, with $k=1,2,3$ and $N=1,2,3$. See text for details.}
\end{table}

Despite the minimal sample size in the progression in the perturbative order $k$ and process multiplicity $N$, we see that as expected the complexity growth in these parameters is rather steep. Similarly, the growth in the number of unique supergraphs is factorial despite the grouping into isomorphic sets. For these reasons, a rough rule of thumbs for what our current implementation of \glss{LU} can accommodate is any contribution with $k+N \le 6$, that is 5-loop supergraphs. 

For instance, I.3 is a single 6-loop supergraph whose generation proved challenging to complete, even though it requires no \glss{UV} counterterm. This is because it contains a two-loop six-point integral with a rank-7 numerator for which the \glss{cLTD} representation involves many terms.
Note that for such a supergraph, and in general for higher multiplicity processes, the original \glss{LTD} representation can be superior, also because in that case stability in the \glss{UV} regime is not as important. In the future we plan on using a combination of both representation to improve on run time.
We stress that this current \emph{practical} limitation $k+N \le 6$ is specific to our implementation in $\alpha\text{Loop}$ and should not be considered as a limitation inherent to \glss{LU}.
Our aim with this work is to establish the complete generality of \glss{LU} and demonstrate it with a first implementation already capable of computing cross-sections at or beyond the state-of-the-art.
Conceptually, \glss{LU} is now a mature approach for processes with final-state \glss{IR} singularities. Its practical application to the fully numerical computation of (differential) cross-sections is now ready to be incrementally improved by future work, starting from the baseline performance presented here.

We developed an automated testing framework in $\alpha\text{Loop}$ of the \glss{LU} representation of individual supergraphs. These tests involve the exhaustive enumeration of all \glss{UV} and \glss{IR} limits, followed by successive numerical evaluations of the \glss{LU} integrand for sample points progressively approaching each limit so as to test the expected local cancellation pattern. In particular, investigating the scaling of the various terms with the approach parameter $\lambda$ allows to numerically reconstruct the power of their asymptotic $\lambda^\xi$ behaviour so as to verify integrability and theoretical expectations.
We note that the \glss{LU} integrand is ultimately expressed with inputs in the unit hypercube, so that a conformal map $x\in[0,1]\rightarrow r\in[0,\infty]$ must be used.
The logarithmic map $r=-Q x\log(1-x)$ is convenient because its Jacobian scales like the measure $dr$ for \emph{both} $r\rightarrow 0$ (soft) and $r\rightarrow \infty$ (\glss{UV}). For production runs however, we instead currently use the map $Q \frac{x}{1-x}$, which can yield unbounded integrable singularities, but that we found to be converging better\footnote{This is likely because the logarithmic map makes it difficult for the adaptive algorithm to adjust to the correct typical contributing region of interest for the radii of the spatial parts integrated over. Indeed the overall normalising scale $Q$ is just a rough estimate, and using the rational polynomial map $Q \frac{x}{1-x}$ makes it easier for the adaptive algorithm to adjust to the region of interest compared to the logarithm map. In the future, this issue can be solved by considering more complicated conformal maps.} when using a naive independent spherical parameterisation for each of the spatial momenta integrated over.
We now give more details about the implementation of these tests and show explicit results from a curated list of limits for the supergraph H.3, which we recall here with the relevant momenta labels:
\begin{equation}
\raisebox{-0.4cm}{
\resizebox{6cm}{!}{
    \begin{tikzpicture}
    \tikzstyle{every node}=[font=\tiny]
    \begin{feynman}
      \vertex (a1);
      \vertex[right=2.5cm of a1] (a3);
      \vertex[below left=0.2cm and -0.15cm of a1] (lab1) {$p_8$};
      \vertex[below right=0.2cm and -0.15cm of a3] (lab2) {$p_8$};
      \vertex[above right=1.1cm and 1.25cm of a1] (a2);
      \vertex[below right=1.1cm and 1.25cm of a1] (a4);

      \vertex[below right=0.55cm and 0.16cm of a1] (cm1);
      \vertex[below left=0.55cm and 0.16cm of a3] (cm2);
      
      \vertex[above right=0.2cm and 0.4cm of cm1] (m1);
      \vertex[above left=0.2cm and 0.4cm of cm2] (m2);
      
      \vertex[above right=0.65 and 0.2cm of m1] (tm1);
      \vertex[above left=0.65 and 0.2cm of m2] (tm2);
      
      \vertex[left=0.5cm of a1] (att1);
      \vertex[right=0.5cm of a3] (att2);

      \diagram* {
        (a1) --[fermion1, quarter left, edge label=\(p_7\),inner sep=1pt] (a2) --[fermion1, quarter left] (a3) -- [fermion1, quarter left,inner sep=2pt] (a4) -- [fermion1, quarter left, edge label=\(p_1\),inner sep=1pt] (a1),
        
        (cm2) -- [gluon, edge label=\(p_2\),inner sep=-3.5pt, outer sep=4pt] (m2), 
        (m1) -- [gluon, edge label=\(p_2\),inner sep=-2pt, outer sep=4pt] (cm1),
        (m1) -- [gluon, bend right, edge label'=\(p_6\),inner sep=5pt, ] (m2),
        
        (tm1) -- [gluon, bend right, edge label'=\(p_3\),inner sep=2pt] (m1),
        (m2) -- [gluon, bend right, edge label'=\(p_3\),inner sep=2pt] (tm2),
        
        (tm1) -- [gluon, half right, edge label=\(p_5\),inner sep=3pt] (tm2),
        (tm2) -- [gluon, half right, edge label=\(p_4\),inner sep=4pt] (tm1),
        
        (att1) -- [boson] (a1),
        (att2) -- [boson] (a3),
      };
    \end{feynman}
  \end{tikzpicture}

}
}
\label{fig:supergraph_for_tests}
\end{equation}

\subsubsection{Numerical tests of the local UV subtraction}
\label{sec:UV_tests}

\begin{figure}[ht!]
\begin{center}
\begin{minipage}{0.49\linewidth}
\centering
\includegraphics[width=\linewidth]{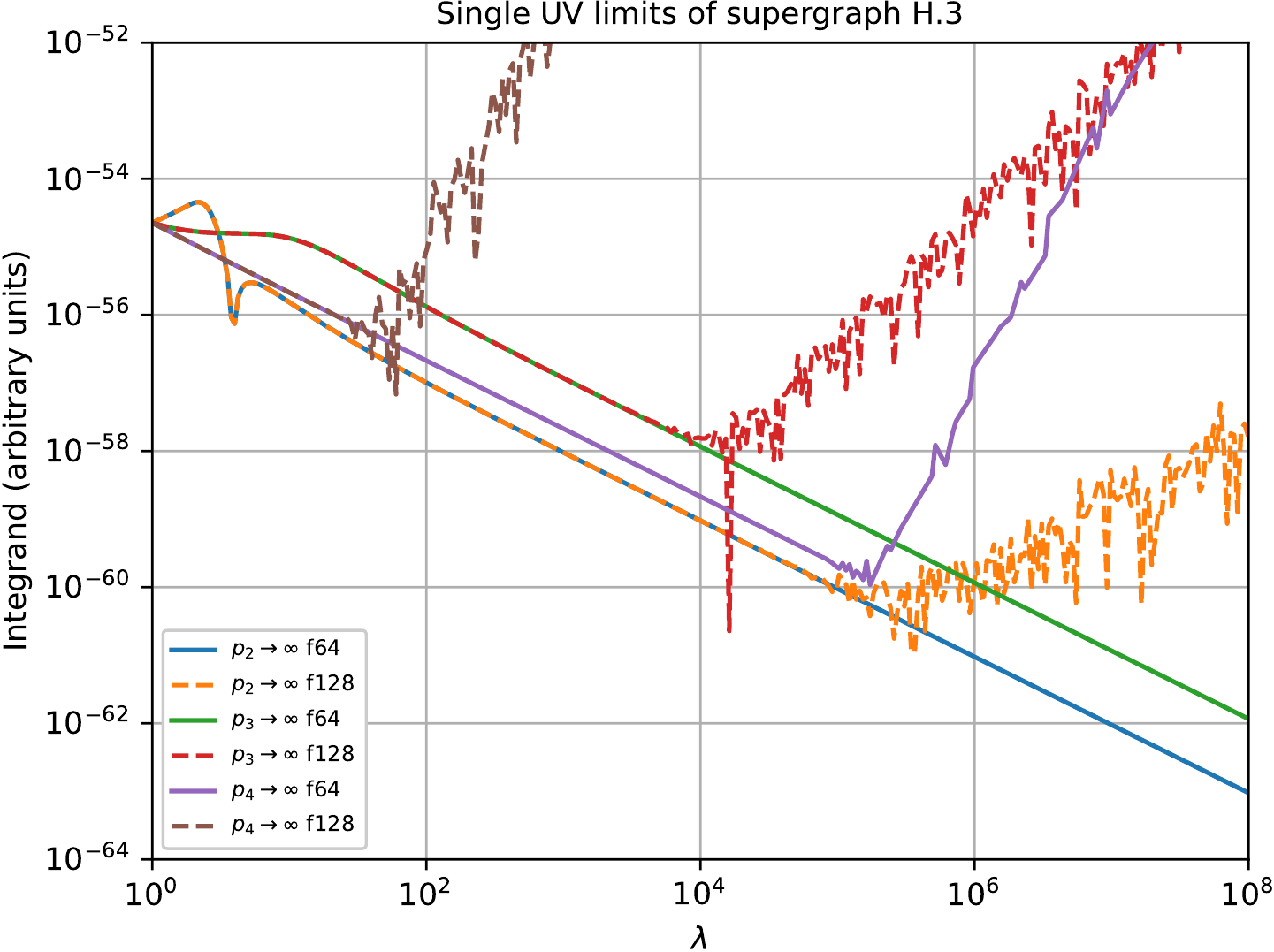}
\end{minipage}
\begin{minipage}{0.49\linewidth}
\centering
\includegraphics[width=\linewidth]{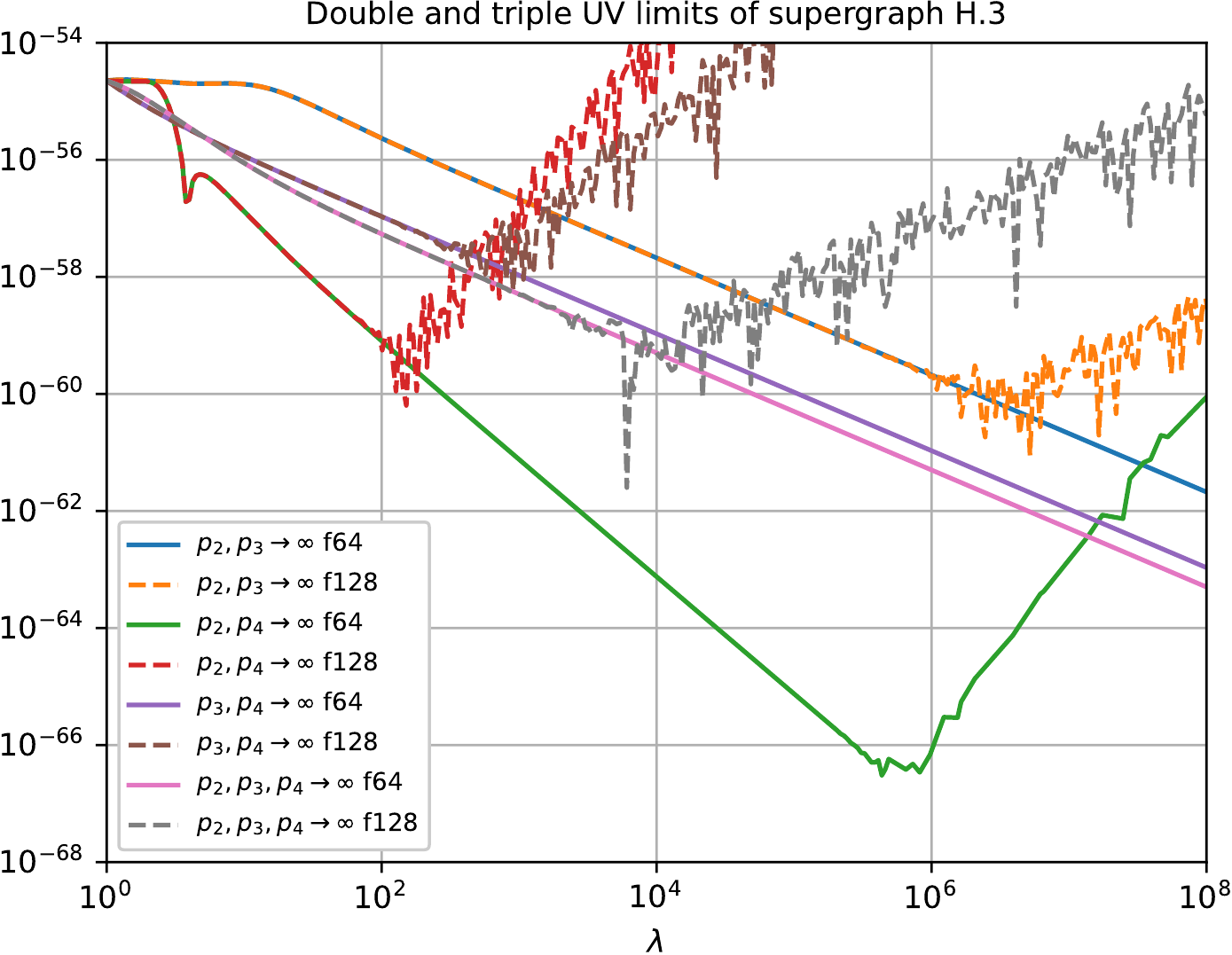}
\end{minipage}
\caption{\label{fig:UV_limits_test} Behaviour of the \glss{LU} integrand of supergraph H.3 of eq.~\ref{fig:supergraph_for_tests} for each possible subset of momenta from the loop momentum basis $\{p_1,p_2,p_3,p_4\}$ being linearly rescaled by $\lambda$ in order to approach infinity. The \glss{UV} scaling of the Minkowsi integration measure (i.e. $d p_i \rightarrow \lambda^4 d p_i$) is included in the results shown. Solid lines correspond to evaluations in quadruple precision (f128) whereas dashed ones are evaluations in double precision (f64).}
\end{center}
\end{figure}

The exhaustive enumeration of \glss{UV} limits can be achieved by considering all possible loop momentum bases of a given supergraph. For each basis, we test the behaviour of the \glss{LU} integrand when sending to infinity all possible subset of momenta in that basis, while keeping the other ones fixed.
Each three-momentum $\vec{p}_i$ in the loop momentum basis is assigned a random direction $\vec{d}_i$ normalised to the scattering energy $Q$ (i.e. $|\vec{d}_i| = Q$), and those sent to infinity are rescaled by $\lambda$, i.e, $\vec{d}_i\rightarrow \lambda \vec{d}_i$. 
The Cutkosky cuts traversing edges rescaled to infinity are exponentially dampened by the normalising $h(t)$ function, whereas all others must still converge in virtue of the local \glss{UV} subtraction procedure.
The automation of this exhaustive testing procedure proved to be an invaluable tool for verifying the correct implementation of the $R$-operation on all amplitude graphs, and it can be viewed as a local analogue to the \glss{UV} pole cancellation cross-check typically performed at the integrated level in traditional analytical amplitude computations in dimensional regularisation. 
We show in fig.~\ref{fig:UV_limits_test} an example of the application of this exhaustive testing strategy when considering the supergraph H.3 of eq.~\ref{fig:supergraph_for_tests} and the loop momentum basis $\{p_1,p_2,p_3,p_4\}$.  

\FloatBarrier

Our automated testing suite can generate plots like shown in fig.~\ref{fig:UV_limits_test}, however they are typically not rendered and instead only analysed in order to automatically extract the power $\xi$ of the leading behaviour in $\lambda^\xi$. When including the scaling of the measure, $\xi$ can be associated to the \glss{dod} of the subtracted amplitudes, and a successful \glss{UV} subtraction will yield $\xi \leq -1$, indicating that the subtracted \glss{LU} expression is bounded and integrable.
We see that \glss{UV} limits involving the innermost momentum $p_4$ are numerically less stable because they involve the cancellation of more (nested) terms of the unfolded $R$-operation.
In general, we find that the point of breakdown of numerical stability in double, resp. quadruple, precision which we denote $\lambda^\star_{\text{f64}}$, resp. $\lambda^\star_{\text{f128}}$, heuristically obeys the expected hierarchy $\lambda^\star_{\text{f128}}  \gtrsim {\lambda^\star_{\text{f64}}}^2$.
It is also apparent that being able to leverage quadruple precision as a stability rescue mechanism is often crucial, given that we find cases of $\lambda^\star_{\text{f64}}$ as low as $100$.
We stress that these results are obtained using the \glss{cLTD} representation of loop integrals, whose numerical stability in the \glss{UV} is far superior to that of its \glss{LTD} counterpart (see ref.~\cite{Capatti:2020ytd}).
The $R$-operation is designed to subtract all \glss{UV} limits down to $\xi=-1$, however accidental cancellation can yield more converging behaviours, like for the double \glss{UV} limit $p_2,p_4\rightarrow \infty$ for instance.
We stress that we show in fig.~\ref{fig:UV_limits_test} the \glss{UV} behaviour of the complete \glss{LU} integrand only, however \glss{UV} subtraction can also be investigated for the contribution of each Cutkosky cut individually.

\subsubsection{Numerical tests of local LU cancellations on IR limits}
\label{sec:IR_tests}

\begin{figure}[ht!]
\begin{center}
\begin{minipage}{0.49\linewidth}
\centering
\includegraphics[width=\linewidth]{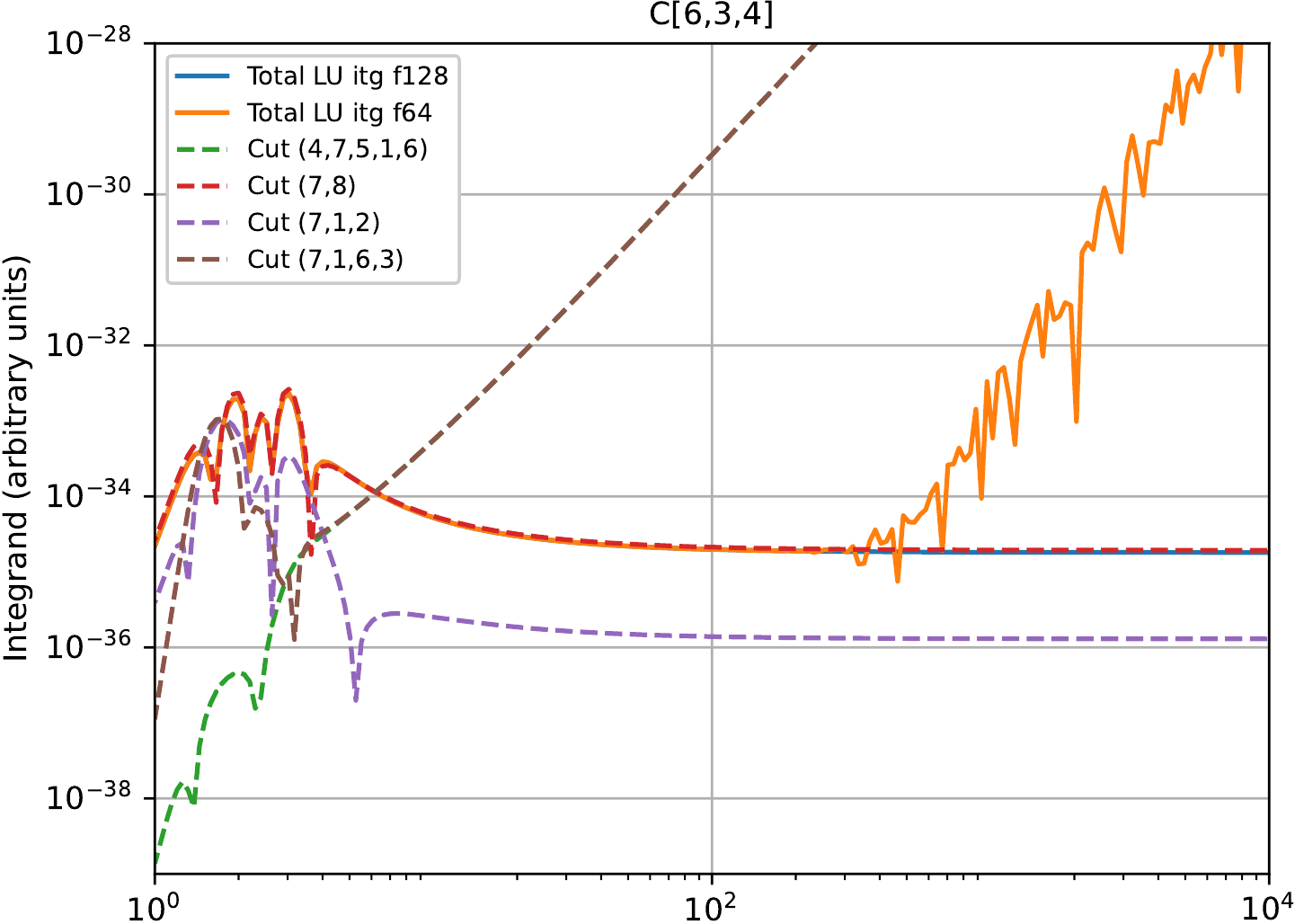}
\end{minipage}
\begin{minipage}{0.49\linewidth}
\centering
\includegraphics[width=\linewidth]{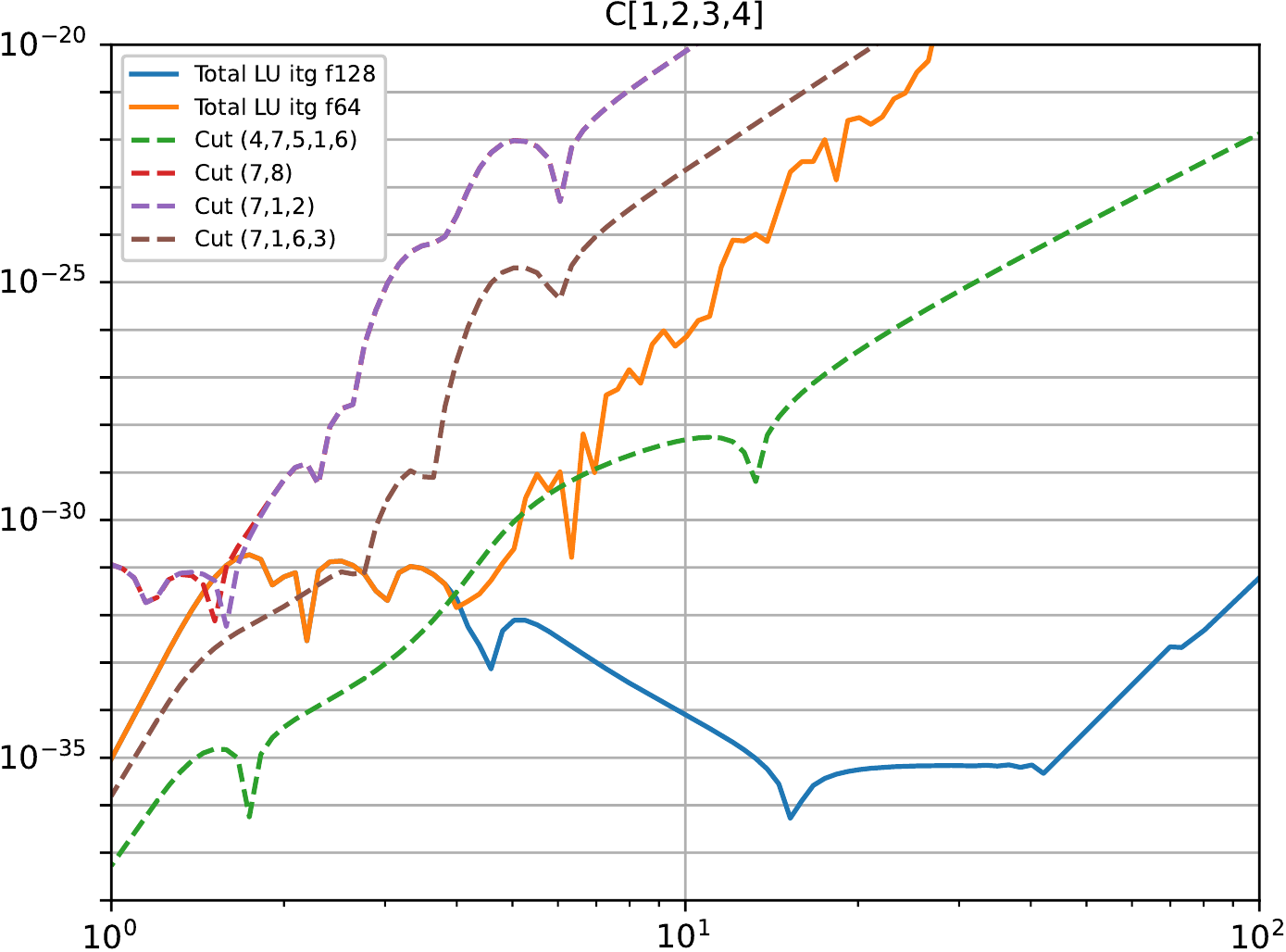}
\end{minipage}\\[0.25cm]
\begin{minipage}{0.49\linewidth}
\centering
\includegraphics[width=\linewidth]{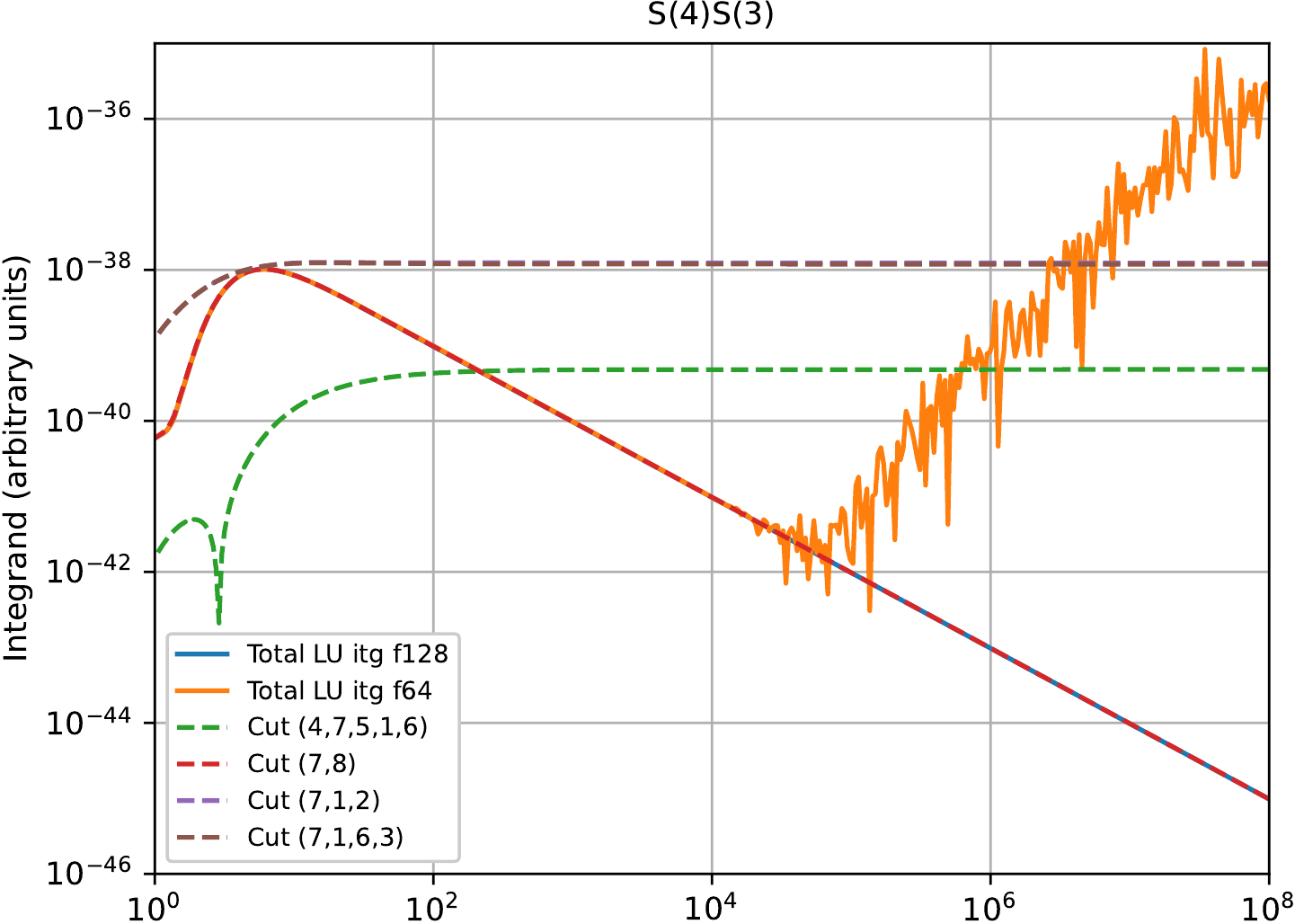}
\end{minipage}
\begin{minipage}{0.49\linewidth}
\centering
\includegraphics[width=\linewidth]{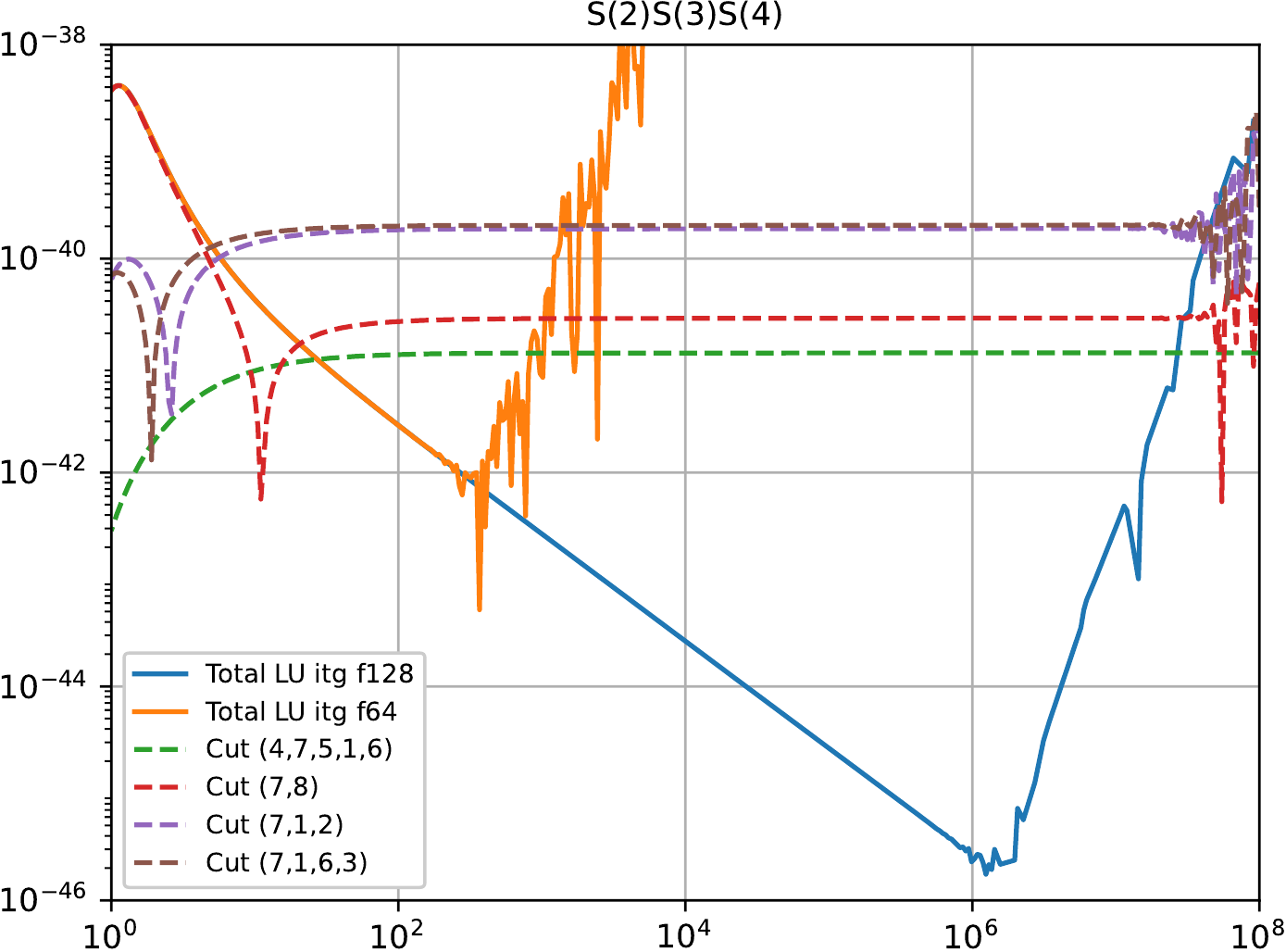}
\end{minipage}\\[0.25cm]
\begin{minipage}{0.49\linewidth}
\centering
\includegraphics[width=\linewidth]{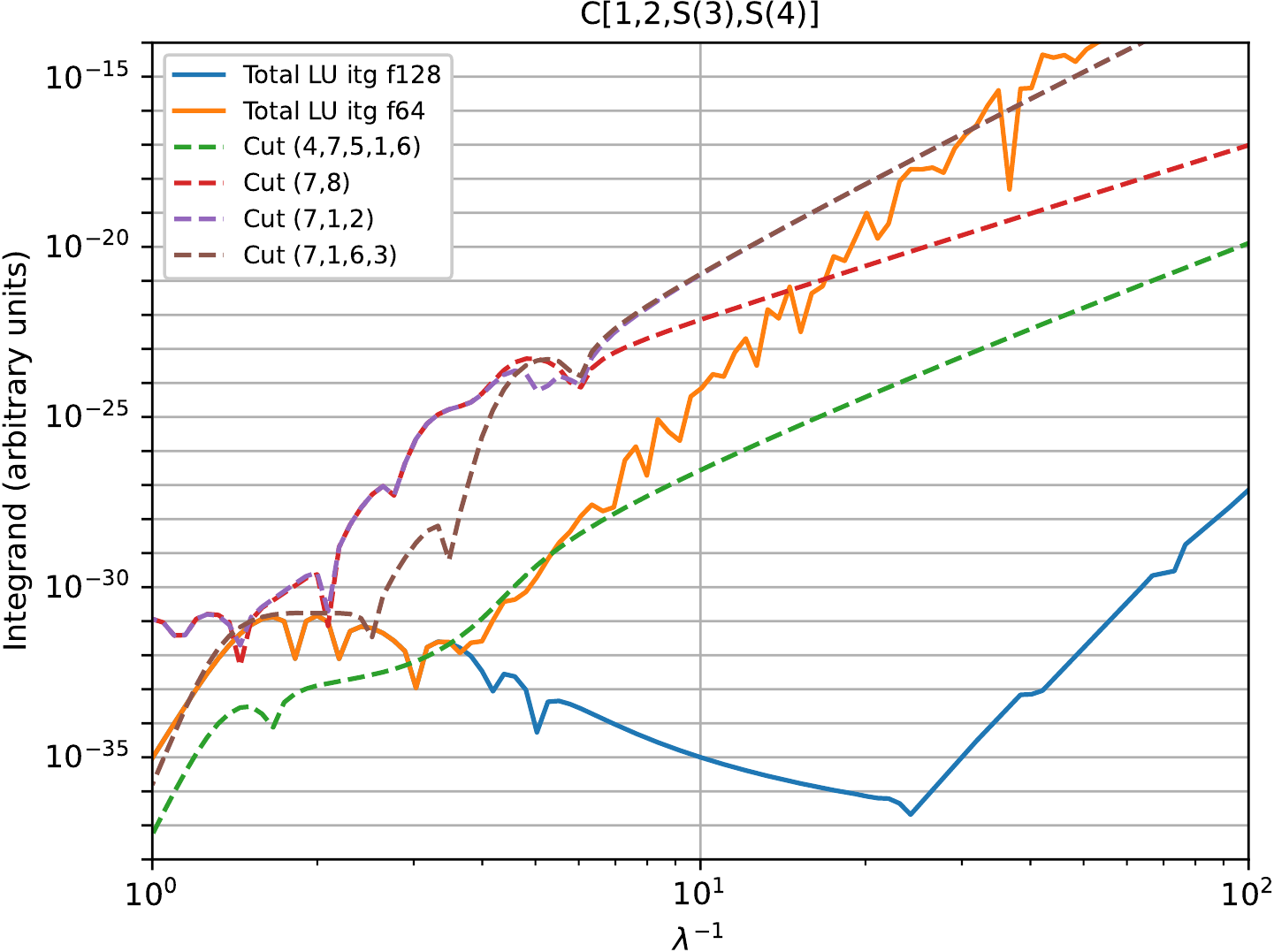}
\end{minipage}
\begin{minipage}{0.49\linewidth}
\centering
\includegraphics[width=\linewidth]{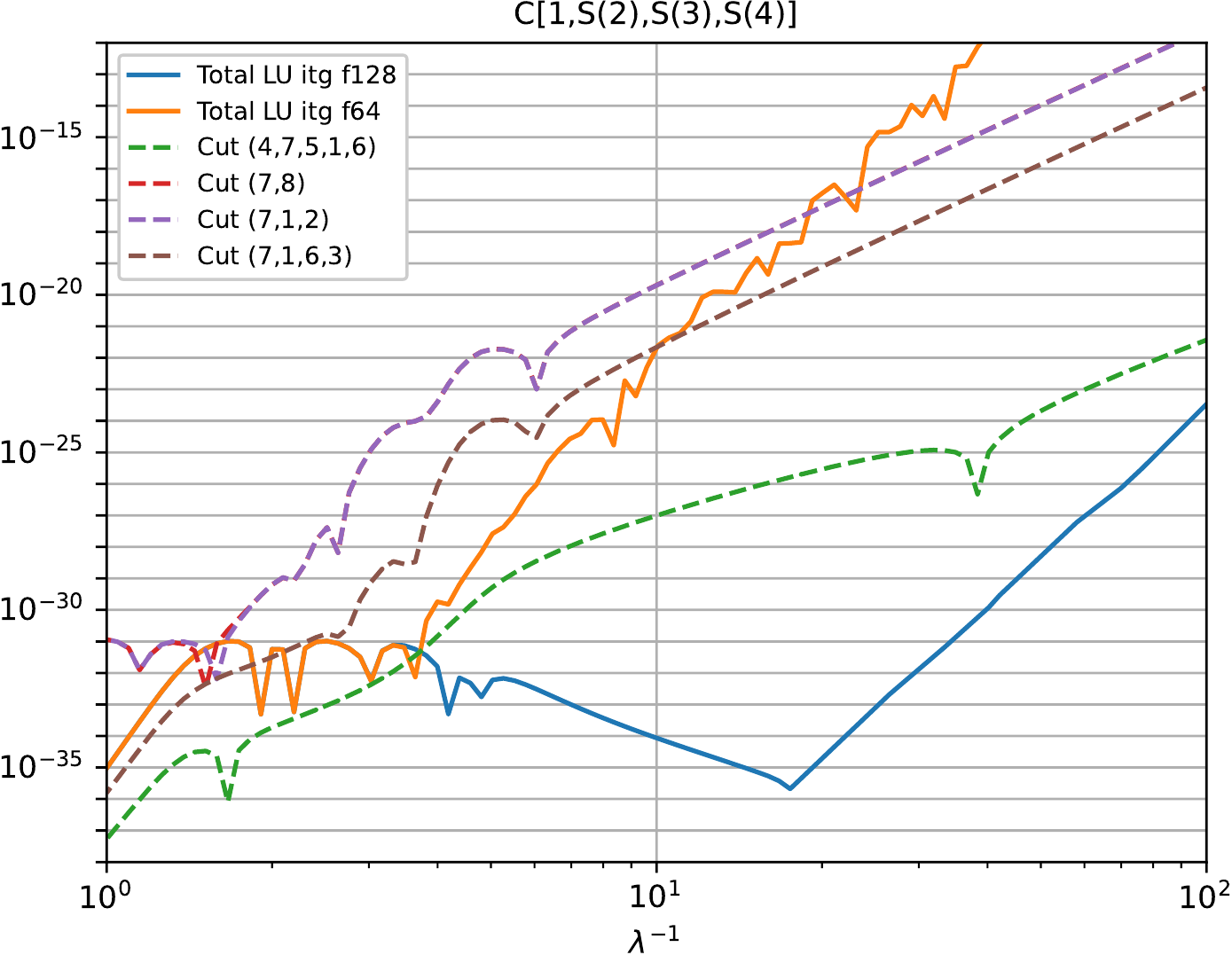}
\end{minipage}
\caption{\label{fig:IR_limits_test} Behaviour of the \glss{LU} integrand of supergraph H.3 of eq.~\ref{fig:supergraph_for_tests} for various collinear, soft and soft-collinear limits approached using a rescaling parameter $\lambda$ as shown in eq.~\eqref{eq:IR_rescaling_rules}. The soft scaling of the measure, i.e. $d\vec{p}_i \rightarrow \lambda^3 d\vec{p}_i$, is included in the lines shown for each momentum approaching a soft limit, but our paramaterisation does not reflect the collinear suppression in the measure proportional to $\cos(\theta_{ij})$. 
Dashed lines correspond to the \glss{LU} integrand for each of the four Cutkosky cuts of supergraph H.3, whereas the solid blue, resp. orange, line corresponds to the sum of all cut contributions evaluated using quadruple precision (f128) arithmetics, resp. double precision (f64) arithmetics.}
\end{center}
\end{figure}

The enumeration of all \glss{IR} limits of a supergraph is a bit more involved than for \glss{UV} limits.
One possible approach is to investigate each contributing Cutkosky cut (some may be excluded by the observable definition) and consider all \glss{IR} limits involving the massless particles in that cut. Each limit constructed in this way is identified by a unordered set of ordered sets of collinear edges, together with a list of edges going soft. For instance an \glss{IR} limit denoted {\texttt{C[1,2,3]C[4,S(5)]S(6)S(7)}} corresponds to two sets of momenta collinear to two different normalised collinear directions $\vec{d}_c^{\;(1)}$ and $\vec{d}_c^{\;(2)}$, as well as three momenta approaching a soft configuration. More specifically, this particular limit would be approached using the following parametric scaling involving seven normalised random directions $\vec{d}_i$:
\begin{eqnarray}
\label{eq:IR_rescaling_rules}
    \vec{p}_1 &=& Q \left( x^{(1)}_1 \vec{d}_{c}^{\;(1)} + \lambda\; \vec{d}^{\;(\perp 1)}_1 \right),\quad
    \vec{p}_2 = Q \left( x^{(1)}_2 \vec{d}_{c}^{\;(1)} + \lambda\; \vec{d}^{\;(\perp 1)}_2 \right), \nonumber \\
    \vec{p}_3 &=& Q \left( x^{(1)}_3 \vec{d}_{c}^{\;(1)} + \lambda\; \vec{d}^{\;(\perp 1)}_3 \right),
    \vec{p}_4 = Q \left( x^{(2)}_1 \vec{d}_{c}^{\;(2)} + \lambda\; \vec{d}^{\;(\perp 2)}_4 \right), \nonumber \\
    \vec{p}_5 &=& Q \lambda \left( x^{(2)}_2 \vec{d}_{c}^{\;(2)} + \lambda\; \vec{d}^{\;(\perp 2)}_5 \right),
    \vec{p}_6 = Q \;\lambda\; \vec{d}_6,\quad
    \vec{p}_7 = Q \;\lambda\; \vec{d}_7,
\end{eqnarray}
where we used the short-hand:
\begin{equation}
    \vec{d}^{\;(\perp i)}_j = \frac{\vec{d}_j-\left(\vec{d}_j\,\cdot\, \vec{d}_{c}^{\;(i)}\right)\;\vec{d}_j}{\left|\vec{d}_j-\left(\vec{d}_j\,\cdot\, \vec{d}_{c}^{\;(i)}\right)\;\vec{d}_j\right|},
\end{equation}
and $Q$ is the scattering energy and within each set the collinear fractions $x^{\;(i)}_j\in[0,1]$ are forced to be in descending order ($x^{\;(i)}_j>x^{\;(i)}_{j+1}$).
For soft-collinear configurations, the scaling choice of eq.~\eqref{eq:IR_rescaling_rules} implies that the transverse component of soft-collinear momenta scales like $\lambda^2$; this is however not a problem for interpreting the asymptotic scaling $\lambda^\xi$ of the \glss{LU} integrand because, as we will discuss later, it always goes to a constant ($\xi=0$) on any collinear limit. 
Notice that when applied to a massive propagator, the soft approach $\vec{p}_i = Q \lambda \vec{d}_i$ also allows us to probe the implementation of our local analogue of the mass renormalisation counterterm in the \glss{OS} scheme. In that case, only the spatial part of the momentum will approach zero while the energy component will approach the mass of the propagator, and \glss{IR} subtraction of repeated propagators is not mandatory for convergence.
Approaching \glss{IR} limits in the context of the \glss{LU} expression is considerably simpler than in the traditional context of real-emission phase-space subtraction where complicated mappings are necessary (e.g. see~\cite{DelDuca:2019ctm}). In the context of \glss{LU}, momentum conservation and on-shellness of external particles is automatically enforced by the causal flow.
We now present in fig.~\ref{fig:IR_limits_test} a curated list of interesting soft, collinear an soft-collinear limits of the supergraph H.3 of eq.~\ref{fig:supergraph_for_tests}. Note that contrary to what our automated test would do, we choose here to study limits not directly involving sets of momenta appearing in Cutkosky cuts, but instead we choose a basis involving the repeated propagators so as make it easier to study their soft limit.

\FloatBarrier

\glss{IR} finiteness in \glss{LU} involves (multiple) pair-wise cancellations among Cutkosky cuts, so that we find it useful to show in fig.~\ref{fig:IR_limits_test} results for both the four individual cuts (dashed lines) of supergraph H.3 as well as for their sum, evaluated using double precision (orange lines) and quadruple precision arithmetics (blue lines).
Our choice of parameterisation explicitly shows the soft-scaling of the measure, i.e. $d\vec{p}_i \rightarrow \lambda^3 d\vec{p}_i$, which is included in the results shown.
However, our chosen parameterisation does not manifestly exhibits the collinear suppression $\propto \cos(\theta_{ij})$ it contains, meaning that we did not align the polar axis with the collinear directions.
While it may be beneficial to do so for convergences purposes, we see that it is not strictly necessary to do so in order to demonstrate that the \glss{LU} integrand is bounded. 
Indeed, thanks to the \glss{LU} cancellation pattern, we expect that the \glss{LU} integrand goes to a \textbf{constant} for any collinear limit of any process and at any perturbative order!
This highlights a crucial difference w.r.t traditional phase-space subtraction strategies based on the factorisation of \glss{IR} singularities of amplitude, namely the fact that within \glss{LU} there remains no integrable singularities even without aligning the integration measure with collinear directions.
This overzealous cancellation of \glss{IR} singularities within \glss{LU} contributes to facilitating an efficient numerical integration since it leaves ample freedom in the choice of parameterisations that leave no integrable singularities.

As already discussed at length in this work, cancellation of soft configurations is intricate. In the absence of raised propagators, like for supergraphs of the G.x series in fig.~\ref{fig:complexity_scaling_supergraphs}, soft finiteness is directly inherited from the pair-wise collinear cancellation pattern of \glss{LU} (since soft confirgurations are end-point of collinear ones, see discussion in sect. 3.2.6 of ref.~\cite{2021}).
However, for supergraphs featuring raised propagators, like in the case of H.3 investigated here where both $p_2$ and $p_3$ are raised, the soft finiteness also involves amplitude-level cancellations with the soft counterterms part of our implementation of the $R$-operator. As already discussed these cancellations have nothing to do with the \glss{KLN} theorem, and it is for example what guarantees that the cuts $(7,1,2)$ and $(7,1,6,3)$ go to a constant by themselves for the double-soft limit {\texttt{S(3)S(4)}} and the triple-soft limit {\texttt{S(2)S(3)S(4)}}. Then, the usual pair-wise \glss{LU} cancellation mechanism offers a further suppression so that the asymptotic scaling $\lambda^\xi$ of the complete \glss{LU} integrand features $\xi \leq 1$.
We see in the triple-soft limit {\texttt{S(2)S(3)S(4)}} that the evaluation of individual cuts can themselves become numerically unstable, demonstrating that alike for \glss{UV} cancellations, the amplitude-level soft cancellations can themselves be subject to numerical instabilities. These numerical instabilities however happens considerably deeper than those stemming from cancellations across Cutkosky cuts, so that they are not of concern. 
We note again, how strikingly different the cancellation of soft singularities within \glss{LU} is compared to the traditional phase-space subtraction analogue based on the Eikonal approximation. More specifically, within \glss{LU} soft cancellations happens essentially solely due to kinematics and in virtue of these configurations being endpoints of collinear limits. And in particular, \glss{LU} soft cancellations do not rely on any details of the $\text{SU}(3)$ colour structure of amplitudes.

We note that for the deeper limits like the quadruple collinear {\texttt{C(1,2,3,4)}}, the cancellations between cuts are especially severe, spanning seventeen orders of magnitude at the breakdown point $\lambda^\star_{\text{f128}}$ of quadruple-precision evaluations.
This leaves barely enough room for observing the constant asymptotic regime in quadruple precision and clearly shows that double-precision is not sufficient for capturing the complete non-trivial dynamics appearing in the bulk of the phase-space.
Still, up to \glss{N3LO} at least, quadruple precision arithmetics appears sufficient in the sense that the results shown suggest that cutting off the integrand at $\lambda^\star_{\text{f128}}$ is unlikely to have any noticeable impact in the cross-section computed.

In $\alpha\text{Loop}$, \glss{IR} cancellation tests such as presented here are performed automatically on the exhaustive set of \glss{IR} limits relevant to each supergraph. The plots are not rendered, but instead analyzed so as to extract the asymptotic scaling power $\xi$ and confront it with the theoretical expectation discussed in this section. These tests are essential for assessing the validity of the \glss{LU} construction, both conceptually and of its implementation in $\alpha\text{Loop}$. They are analogous to the well-known consistency checks of phase-space subtraction approaches, except that in the case of \glss{LU} they also probe the \glss{IR} behaviour of the loop integrals involved.

\FloatBarrier

\section{Conclusion}
\label{sec:conclusion}

The \glss{LU} representation of differential cross-sections locally realises the cancellations of infrared singularities predicted by the \glss{KLN} theorem.
In this work we solved the two remaining conceptual challenges for computing physical cross sections beyond \glss{NLO} with \glss{LU}.

The first challenge is that self-energy corrections naturally introduce propagators with denominators raised to a higher power.
We re-framed the discussion of building the \glss{LU} representation of such configurations in the broader context of the analysis of higher-order residue contributions stemming from thresholds of forward-scattering diagrams.
We generalised the \glss{LU} construction accordingly, and from the resulting expression we identified a modification of the Cutkosky cutting rule that renders it also applicable for higher-order residues.
This generalisation involves taking derivatives of amplitudes, and we showed how this can be implemented efficiently in a numerical code using dual number representations. We demonstrated that in the \glss{OS} renormalisation scheme, these amplitude derivatives cancel at the integrated level, which explains why they do not appear in traditional computations.

The second challenge is that of subtracting and renormalising \glss{UV} divergences from the loop amplitudes of the interference graphs. We use the $R$-operation to locally subtract \glss{UV} singularity. We further refined the subtraction operator so as to also remove spurious soft singularities stemming from self-energy insertions, whose contribution at the integrated level leads to no pole. Finally, we redefine the local subtraction operator for massive propagators, such that its integrated version yields the On-Shell mass renormalisation counterterm.
We thus construct the appropriate integrated-level counterterms such that the cross-sections obtained are automatically renormalised in the commonly used hybrid $\overline{\text{MS}}$ and \glss{OS} scheme. The analytic work necessary for achieving this only involves the computation of single-scale massive vacuum bubbles.

Lastly, we address how \glss{IR}-safety in \glss{LU} requires the spin-sum rule of gluons to match the numerator of its propagator. For this to yield physical results, it is necessary to include \glss{QCD} ghosts in the final states.

The refinements of the \glss{LU} formulation introduced in this work address all remaining conceptual bottlenecks for applying it at any perturbative order and to any scattering process featuring only \glss{FSR} singularities. 
This sets the stage for its first non-trivial practical applications, of which we show a curated selection. 
At \glss{NLO}, we provide (semi-)inclusive results for $e^+ e^- \rightarrow \gamma \rightarrow jjj$ and $e^+ e^- \rightarrow \gamma \rightarrow t\bar{t} H$.
We showed the first purely numerical computation in momentum-space of an inclusive \glss{NNLO} cross-section, for the processes $e^+ e^- \rightarrow j j$ and $e^+ e^- \rightarrow t \bar{t}$. The agreement with analytic results validates our implementation. Additionally, investigation of specific supergraphs contributing up to \glss{N3LO} offer a first glance at the performance of our approach, especially in terms of its scaling in both the perturbative order and process multiplicity. This first general implementation serves as a benchmark, from which incremental progress can be achieved.

Local Unitarity will soon be mature enough to tackle the computation of challenging cross-sections not yet available through other methods.

\section{Acknowledgements}
This project has received funding from the European Research Council (ERC) under grant agreement No 694712 (PertQCD) and SNF grant No 179016. Numerical results presented in this work used computational resources from the {\sc\small lxplus} computing cluster of CERN. We thank D. Kermanschah for enlightening exchanges on the contour deformation within \glss{LU}, A. Schweizer for his insight into details of the integration of integrated \glss{UV} counterterms, A. Pelloni for his help with {\sc\small QGRAF} generation and implementation of the \glss{cLTD} expression and finally C. Anastasiou for his continuous support.
\newpage
\appendix
\section{Example use of multi-variate duals to compute partial derivatives}
\label{sect:multivariate_dual}
Let $f(x,y)=e^{x+y}\sin(xy)$; we rewrite $f$ as a composition of elementary functions by defining, on top of $s(x)=\sin(x)$ and $g_\times(x,y)=xy$, the functions $g_+(x,y)=x+y$ and $e(x)=e^{x}$. Let us now determine the value of $g_+$, $g_\times$, $s$ and $e$ when evaluated at multi-variate dual numbers $\mathbf{\bar{x}}=c_{00}+c_{10}\varepsilon_1+c_{01}\varepsilon_2+c_{11}\varepsilon_1\varepsilon_2$ and $\mathbf{\bar{x}}'=c_{00}'+c_{10}'\varepsilon_1+c_{01}'\varepsilon_2+c_{11}'\varepsilon_1\varepsilon_2$ (so $\mathbf{\bar{x}},\;\mathbf{\bar{x}}' \in \mathcal{D}\left(2;(1,1)\right)$):
\begin{align}
\begin{split}
    s\left(\bar{x}\right)=& \sin(c_{00})+\cos(c_{00})(c_{10}\varepsilon_1+c_{01}\varepsilon_2)-\varepsilon_1\varepsilon_2(c_{10}c_{01}\sin(c_{00})-c_{11}\cos(c_{00})) \;,\\
    g_+(\bar{x},\bar{x}')=& c_{00}+c_{00}'+( c_{10}'+c_{10})\varepsilon_1+( c_{01}'+ c_{01})\varepsilon_2+\left(c_{11}+ c_{11}'\right) \varepsilon_1\varepsilon_2 \;, \\
    g_\times(\bar{x},\bar{x}')=& c_{00} c_{00}'+(c_{00} c_{10}'+c_{00}' c_{10})\varepsilon_1+(c_{00} c_{01}'+c_{00}' c_{01})\varepsilon_2\\ 
    &+\left(c_{10} c_{01}'+c_{10}' c_{01} 
    +c_{00} c_{11}'+ c_{00}'c_{11}\right) \varepsilon_1\varepsilon_2 \;,\\
    e(\bar{x})=& e^{c_{00}}+e^{c_{00}}(c_{10}\varepsilon_1+c_{01}\varepsilon_2)+\varepsilon_1\varepsilon_2e^{c_{00}}(c_{10}c_{01}-c_{11}) \;.
\end{split}
\end{align}
This allows us to compute the full derivative of $f$ knowing the expansion of the elementary functions. In particular, this yields
\begin{align}
\begin{split}
    f(x+\varepsilon_1,y+\varepsilon_2)&=g_\times(e(g_+(x+\varepsilon_1,y+\varepsilon_2)), s(g_\times(x+\varepsilon_1,y+\varepsilon_2))) \\
    &=g_\times(e(x+y+\varepsilon_1+\varepsilon_2), s(xy+x\varepsilon_2+y\varepsilon_1+\varepsilon_1\varepsilon_2)) \,.
\end{split}
\end{align}
We have
\begin{align}
\begin{split}
    &e(x+y+\varepsilon_1+\varepsilon_2)=e^{x+y}(1+\varepsilon_1+\varepsilon_2+\varepsilon_1\varepsilon_2), \\
    &s(xy+x\varepsilon_2+y\varepsilon_1+\varepsilon_1\varepsilon_2)=\sin(xy)+\cos(xy)(x\varepsilon_2+y\varepsilon_1) -\varepsilon_1\varepsilon_2(xy\sin(xy)-\cos(xy)) \,,
    \end{split}
\end{align}
from which we finally obtain
\begin{align}
\begin{split}
    f(x+\varepsilon_1,y+\varepsilon_2)&=e^{x+y}\sin(xy) \\
    &+e^{x+y}(\sin(xy)+y\cos(xy))\varepsilon_1 \\
    &+e^{x+y}(\sin(xy)+x\cos(xy))\varepsilon_2 \\
    &+e^{x+y}(\sin(xy)+(x+y)\cos(xy)-xy\sin(xy)+\cos(xy))\varepsilon_1 \varepsilon_2 \;.
\end{split}
\end{align}
We see that the coefficients of the power series reproduce the partial derivatives of $f$.

\section{Example application of the $R$-operation}
\label{sect:R_example}

We consider the following diagram:
\begin{equation}
    \Pi=\raisebox{-1.3cm}{\resizebox{4cm}{!}{
\begin{tikzpicture}
        \tikzstyle{every node}=[font=\small]
         
        \begin{feynman}
        \vertex (1) {};
            
          \vertex[right=0.1cm of 1] (a1);
          
          \vertex[right=0.7cm of a1] (a2);
          \vertex[right=2cm of a2] (a3);
          
          \vertex[right=1cm of a2] (m);
          
          \vertex[above right= 0.64cm and 0.52cm of a2] (c1);
          
          \vertex[above left= 0.64cm and 0.52cm of a3] (c2);
          
          \vertex[right=0.7cm of a3] (a4);
          
          \diagram {
             (a1) -- [gluon,momentum'=$q$] (a2),
             (a3) -- [gluon] (a4),
             
             (a3) --[fermion1, half left, edge label=\(e_1\)] (a2),
             
             (a2) -- [fermion1, quarter left, edge label'=\(e_2\),momentum=$p$] (c1),
             (c2) -- [fermion1, quarter left, edge label=\(e_3\)] (a3),
             
             (c1) -- [fermion1, half left, edge label'=\(e_4\),momentum=$k$] (c2),
             (c1) -- [gluon, half right, edge label=\(e_5\)] (c2),
             
            };
        \end{feynman}
\end{tikzpicture}}}\;,
\end{equation}
where the fermion line is considered massive with mass $m$.
Since the $R$-operation requires to isolate subgraphs, we represent the graph $\Pi$ in terms of propagator and vertex functions in our code as follows:
\begin{verbatim}
 vx(-1,21,1,-p,q,p-q,7,2,4)*vx(-1,21,1,-p+q,-q,p,5,1,8)*
 vx(-1,21,1,-k,k-p,p,13,10,6)*vx(-1,21,1,-p,-k+p,k,9,11,12)*
 prop(21,in,q,2)*prop(22,out,q,1)*prop(1,virtual,p-q,5,4)*
 prop(1,virtual,p,7,6)*prop(1,virtual,p,9,8)*
 prop(21,virtual,k-p,11,10)*prop(1,virtual,k,13,12)
\end{verbatim}
where the Particle Group Data index of each particle is specified and a set of indices is provided. Then, to isolate the subgraph with loop momentum $k$, one collects all structures that contain $k$:
\begin{verbatim}
 vx(-1,21,1,-k,k-p,p,13,10,6)*vx(-1,21,1,-p,-k+p,k,9,11,12)*
 prop(21,virtual,k-p,11,10)*prop(1,virtual,k,13,12)
\end{verbatim}

In the following we will represent the isolation of the subgraphs mathematically.
After the substitution of the Feynman rules (and working in the Feynman gauge), $\Pi$ reads
\begin{equation}
    \Gamma^{\mu\nu}(q)=\frac{[(\slashed{p}+m) \gamma^\mu (\slashed{p}-\slashed{q}+m) \gamma^\nu (\slashed{p}+m)]_{ij}}{(p^2-m^2)^2 ((p-q)^2-m^2)}\gamma_{ji}(p), \quad \gamma_{ji}(p)=\frac{[\gamma^\nu (\slashed{k}+m) \gamma_\nu]_{ji}}{(k^2-m^2)(k-p)^2}.
\end{equation}
The UV divergent subgraphs of $\Pi$ are $\gamma_1=\{e_4,e_5\}, \gamma_2=\{ e_1, e_2, e_4, e_5 \}, \Gamma=\{e_1,e_2,e_3,e_4,e_5\}$. The wood in this case is very simple
\begin{equation}
    W[\Gamma]=\{\{\}, \{\gamma_1\}, \{\gamma_2\}, \{\Gamma\} \}.
\end{equation}
Thus, unfolding the $R$ formula and using the linearity of $K$, we find
\begin{equation}
    R[\Gamma]=\Gamma-K(\gamma_1)* (\Gamma \setminus \gamma_1) -K(\gamma_2)* (\Gamma \setminus \gamma_2) -K(\Gamma) + K(K(\gamma_1)* (\Gamma \setminus \gamma_1))  + K(K(\gamma_2)* (\Gamma \setminus \gamma_2)).
\end{equation}

Before continuing, we realise a special simplification: $\Gamma \setminus \gamma_2$ is a tadpole that has the shape $\mathcal{N}(k)_{ij}/ (k^2-m^2)$. The $K$ operator acting on this object leaves it identical, because in this case it consists in Taylor expanding around the external shift which is absent. As a result:
\begin{equation}
-K(\gamma_2)* (\Gamma \setminus \gamma_2) + K(K(\gamma_2)* (\Gamma \setminus \gamma_2)) = 0 \;.
\end{equation}
and thus $\gamma_2$ drops out of the wood.

Since $K=\mathbf{T}+\bar{K}$, we can always subdivide the problem of computing $K(\gamma)$ into that of computing $\mathbf{T}(\gamma)$ and $\bar{K}(\gamma)$.

\paragraph{Term $K(\Gamma)$:\\[0.6cm]}

We saw that for a generic gluonic self-energy $\Pi_g(q,\mathbf{m})$
\begin{equation}
    \mathbf{T}(\Pi_g)= \Pi_g(0,\mathbf{m})+q^\mu \frac{\partial}{\partial q^\mu } \Pi_g(q,\mathbf{m})\big|_{q=0}+q^\mu q^\nu \frac{\partial}{\partial q^\mu } \frac{\partial}{\partial q^\mu }\Pi_g(q,\mathbf{0})\big|_{\substack{\hspace{-0.05cm}q=0 \\  m_{\text{uv}}}}.
\end{equation}
In this specific case we have $\mathbf{m}=\{m\}$. Applying this formula, we find
\begin{align}
    \begin{split}
    \mathbf{T}_2(\Gamma)^{\mu\nu}&=\frac{\text{Tr}[(\slashed{k}+m)\gamma^\rho(\slashed{p}+m) \gamma^\mu (\slashed{p}-\slashed{q}+m) \gamma^\nu (\slashed{p}+m)\gamma_\rho]}{(p^2-m^2)^3 (k^2-m^2)(k-p)^2} \\
    &+\frac{ 2 (p\cdot q) \text{Tr}[(\slashed{k}+m)\gamma^\rho(\slashed{p}+m) \gamma^\mu (\slashed{p}+m) \gamma^\nu (\slashed{p}+m)\gamma_\rho]}{(p^2-m^2)^4(k^2-m^2)(k-p)^2} \\
    &+\frac{ -2 (p\cdot q) \text{Tr}[\slashed{k}\gamma^\rho\slashed{p} \gamma^\mu \slashed{q} \gamma^\nu \slashed{p}\gamma_\rho] - q^2 \text{Tr}[\slashed{k}\gamma^\rho\slashed{p} \gamma^\mu \slashed{p} \gamma^\nu \slashed{p}\gamma_\rho] }{(p^2-m_{\text{uv}}^2)^4 (k^2-m_{\text{uv}}^2)((k-p)^2-m_{\text{uv}}^2)}\\
    &+\frac{ 4(p\cdot q)^2 \text{Tr}[\slashed{k}\gamma^\rho\slashed{p} \gamma^\mu \slashed{p} \gamma^\nu \slashed{p}\gamma_\rho]}{(p^2-m_{\text{uv}}^2)^5 (k^2-m_{\text{uv}}^2)((k-p)^2-m_{\text{uv}}^2)}
    \end{split}
\end{align}

We now consider $\bar{K}(\Gamma)$. Recall that
\begin{equation}
    \bar{K}(\Gamma)=-\prod_{i=1}^n N(k_i) \left(\mathds{1}-K^{\overline{\text{MS}}}\right)[T(\Gamma)-T (\hat{T}( \Gamma))] \,,
\end{equation}
which, for this case, evaluates to
\begin{align}
    \begin{split}
    \bar{K}(\Gamma)^{\mu\nu}=&-\frac{(2\mathrm{i}(4 \pi)^2 m_{\text{UV}}^2)^2}{(p^2-m_{\text{uv}}^2)^3 (k^2-m_{\text{uv}}^2)^3} \left(\mathds{1}-K^{\overline{\text{MS}}}\right) \left( \frac{\mu^2}{4\pi e^{-\gamma_E}}\right)^{2\epsilon} \int \mathrm{d}^{4-2\epsilon}p' \, \mathrm{d}^{4-2\epsilon}k' \\ &\Bigg[ +\frac{ -2(p'\cdot q) \text{Tr}[\slashed{k}'\gamma^\rho\slashed{p}' \gamma^\mu \slashed{q} \gamma^\nu \slashed{p}'\gamma_\rho] - q^2 \text{Tr}[\slashed{k}'\gamma^\rho\slashed{p}' \gamma^\mu \slashed{p}' \gamma^\nu \slashed{p}'\gamma_\rho]}{({p'}^2-m_{\text{uv}}^2)^4 (k^2-m_{\text{uv}}^2)((k-p')^2-m_{\text{uv}}^2)}\\
    &+\frac{ 4(p'\cdot q)^2 \text{Tr}[\slashed{k}'\gamma^\rho\slashed{p}' \gamma^\mu \slashed{p}' \gamma^\nu \slashed{p}'\gamma_\rho]}{({p'}^2-m_{\text{uv}}^2)^5 ({k'}^2-m_{\text{uv}}^2)((k'-p')^2-m_{\text{uv}}^2)}\Bigg]\\
    =& -\frac{(2\mathrm{i}(4 \pi)^2 m_{\text{UV}}^2)^2}{(p^2-m_{\text{uv}}^2)^3} (\mathrm{i} 16 \pi^2)^2 \Biggl(\\
    &+q^{\mu} q^{\nu} \left( \frac{31}{162} + \frac{1}{9}\pi^2 - \frac{136}{81 \sqrt{3}} \text{Cl}_2\left(\frac{\pi}{3}\right)  + \frac{10}{9} \ln\left(\frac{\mu^2}{m_{\text{uv}}^2}\right) + \frac{4}{3} \ln\left(\frac{\mu^2}{m_{\text{uv}}^2}\right)^2\right)\\
    &- g^{\mu \nu} q^2 \left( \frac{38}{81} + \frac{1}{9}\pi^2 - \frac{136}{81 \sqrt{3}} \text{Cl}_2\left(\frac{\pi}{3}\right) + \frac{16}{9} \ln\left(\frac{\mu^2}{m_{\text{uv}}^2}\right) + \frac{4}{3} \ln\left(\frac{\mu^2}{m_{\text{uv}}^2}\right)^2 \right) \Biggr) \;,
    \end{split}
\end{align}
where $\text{Cl}_2$ is the Clausen function of order 2.

\paragraph{Term $K(\gamma_1)$:\\[0.6cm]}

 Recall that, for a fermionic self-energy $\Sigma(p,\mathbf{m})$, one has
\begin{equation}
    \mathbf{T}\Sigma(p,\mathbf{m})=\frac{(1+\gamma^0)}{2}\Sigma(p^{os},\mathbf{m})+\frac{(1-\gamma^0)}{2}\Sigma(-p^{os},\mathbf{m})+(p-\gamma^0 p^{os})^\mu \frac{\partial}{\partial p^\mu }\Sigma(p,\mathbf{0})\big|_{\substack{\hspace{-0.05cm}p=0 \\  m_{\text{uv}}}}.
\end{equation}
In this case, the only internal mass is $m$, so $\mathbf{m}=\{m\}$. The subscript $ m_{\text{uv}}$ denotes the introduction of UV masses in all quadratic \textit{denominators}. Applying this formula we find
\begin{align}
    \mathbf{T}_1(\gamma_1)_{ij}&=\frac{[(1+\gamma^0)\gamma^\nu (\slashed{k}+m) \gamma_\nu]_{ij}}{2(k^2-m^2)(k-p^{os})^2}+\frac{[(1-\gamma^0)\gamma^\nu (\slashed{k}+m) \gamma_\nu]_{ij}}{2(k^2-m^2)(k+p^{os})^2} \\
    &+2 k\cdot p\frac{[\gamma^\nu \slashed{k} \gamma_\nu]_{ij}}{(k^2-m_{\text{uv}}^2)^3} 
    -2 k\cdot p^{os}\frac{[\gamma^0\gamma^\nu \slashed{k} \gamma_\nu]_{ij}}{(k^2-m_{\text{uv}}^2)^3}.
\end{align}
with $p^{os}=(m,0,0,0)$. We have, for $\bar{K}(\gamma_1)$:
\begin{equation}
    \bar{K}(\gamma_1)=-\prod_{i=1}^n N(k_i) \left(\mathds{1}-K^{\overline{\text{MS}}}\right)[T(\gamma)-T (T^{os}( \gamma))],
\end{equation}
which yields
\begin{align}
    \begin{split}
     \bar{K}(\gamma_1)_{ij}&=-\frac{2\mathrm{i}(4 \pi)^2 m_{\text{UV}}^2}{(k^2-m_{\text{uv}}^2)^3}\left(\mathds{1}-K^{\overline{\text{MS}}}\right)\left( \frac{\mu^2}{4\pi e^{-\gamma_E}}\right)^{\epsilon} \int  \mathrm{d}^{4-2\epsilon}k' \\
     &\Bigg[2 k'\cdot p\frac{[\gamma^\nu \slashed{k}' \gamma_\nu]_{ij}}{({k'}^2-m_{\text{uv}}^2)^3} 
    -2 k'\cdot p^{os}\frac{[\gamma^0\gamma^\nu \slashed{k}' \gamma_\nu]_{ij}}{({k'}^2-m_{\text{uv}}^2)^3}\Bigg] \\
    &=\frac{2\mathrm{i}(4 \pi)^2 m_{\text{UV}}^2}{(k^2-m_{\text{uv}}^2)^3}\frac{(\slashed{p}-\gamma^0 \slashed{p}^{os})_{ij}}{2}\mathrm{i} \pi^2 \left(-1 - (1 - \epsilon) \ln\left(\frac{\mu^2}{m_{\text{uv}}^2}\right)
    + \frac{\pi^2}{12} \epsilon + \frac{1}{2} \ln\left(\frac{\mu^2}{m_{\text{uv}}^2}\right)^2 \epsilon\right).
    \end{split}
\end{align}

\paragraph{Term $K(K(\gamma_1)* (\Gamma \setminus \gamma_1))$:\\[0.6cm]}

We are now ready to discuss the nested application of $K$. We find
\begin{align}
\begin{split}
    K(K(\gamma_1)* (\Gamma \setminus \gamma_1))&=\mathbf{T}(\mathbf{T}(\gamma_1)* (\Gamma \setminus \gamma_1))+\bar{K}(\mathbf{T}(\gamma_1)* (\Gamma \setminus \gamma_1)) \\
    &+\mathbf{T}(\bar{K}(\gamma_1)* (\Gamma \setminus \gamma_1))+\bar{K}(\bar{K}(\gamma_1)* (\Gamma \setminus \gamma_1))
            \end{split}
\end{align}
Regarding the nested application of the $\mathbf{T}$ operator, we get the following
\begin{align}
    \begin{split}
    \mathbf{T}&(\mathbf{T}(\gamma_1)* (\Gamma \setminus \gamma_1))=\Bigg[\frac{[(\slashed{p}+m) \gamma^\mu (\slashed{p}+m) \gamma^\nu (\slashed{p}+m)]_{ij}}{(p^2-m^2)^3} \\
    &-\frac{[(\slashed{p}+m) \gamma^\mu \slashed{q} \gamma^\nu (\slashed{p}+m)]_{ij}}{(p^2-m^2)^3 }+\frac{2(p\cdot q) [(\slashed{p}+m) \gamma^\mu (\slashed{p}+m) \gamma^\nu (\slashed{p}+m)]_{ij}}{(p^2-m^2)^4} \Bigg] \\
    &*\Bigg[\frac{[(1+\gamma^0)\gamma^\sigma (\slashed{k}+m) \gamma_\sigma]_{ji}}{2(k^2-m^2)(k-p^{os})^2}+\frac{[(1-\gamma^0)\gamma^\sigma (\slashed{k}+m) \gamma_\sigma]_{ji}}{2(k^2-m^2)(k+p^{os})^2}
    +2 k\cdot p\frac{[\gamma^\sigma \slashed{k} \gamma_\sigma]_{ji}}{(k^2-m_{\text{uv}}^2)^3} 
    \\
    &-2 k\cdot p^{os}\frac{[\gamma^0\gamma^\sigma \slashed{k} \gamma_\sigma]_{ji}}{(k^2-m_{\text{uv}}^2)^3}\Bigg]+\Bigg[\frac{[\gamma^\sigma \slashed{k} \gamma_\sigma]_{ij}}{(k^2-m_{\text{uv}}^2)^2}+2(k\cdot p)\frac{[\gamma^\sigma \slashed{k} \gamma_\sigma]_{ij}}{(k^2-m_{\text{uv}}^2)^3}\Bigg]\\
    &*\Bigg[-\frac{2(p\cdot q) [\slashed{p} \gamma^\mu \slashed{q} \gamma^\nu \slashed{p}]_{ji} -q^2 [\slashed{p} \gamma^\mu \slashed{p} \gamma^\nu \slashed{p}]_{ji}}{(p^2-m_{\text{uv}}^2)^4}
    +\frac{4(p\cdot q)^2 [\slashed{p} \gamma^\mu \slashed{p} \gamma^\nu \slashed{p}]_{ji}}{(p^2-m_{\text{uv}}^2)^5 }\Bigg] .
    \end{split}
\end{align}
and similarly as before
\begin{align}
    \begin{split}
    \bar{K}(&\mathbf{T}(\gamma_1)* (\Gamma \setminus \gamma_1))=-\frac{(2\mathrm{i}(4 \pi)^2 m_{\text{UV}}^2)^2}{(p^2-m_{\text{uv}}^2)^3 (k^2-m_{\text{uv}}^2)^3} \left(\mathds{1}-K^{\overline{\text{MS}}}\right)\left( \frac{\mu^2}{4\pi e^{-\gamma_E}}\right)^{2\epsilon} \int \mathrm{d}^{4-2\epsilon}p' \, \mathrm{d}^{4-2\epsilon}k' \\
    &\Bigg[\frac{[\gamma^\sigma \slashed{k}' \gamma_\sigma]_{ji}}{({k'}^2-m_{\text{uv}}^2)^2}+2(k'\cdot p')\frac{[\gamma^\sigma \slashed{k}' \gamma_\sigma]_{ji}}{({k'}^2-m_{\text{uv}}^2)^3}\Bigg]
    *\Bigg[-\frac{2(p'\cdot q) [\slashed{p}' \gamma^\mu \slashed{q} \gamma^\nu \slashed{p}']_{ij} - q^2 [\slashed{p}' \gamma^\mu \slashed{p}' \gamma^\nu \slashed{p}']_{ij} }{({p'}^2-m_{\text{uv}}^2)^4} \\
    &+\frac{4(p'\cdot q)^2 [\slashed{p}' \gamma^\mu \slashed{p}' \gamma^\nu \slashed{p}']_{ij}}{({p'}^2-m_{\text{uv}}^2)^5 }\Bigg] .
    \end{split}
\end{align}
We also have
\begin{align}
    \begin{split}
    \bar{K}(\bar{K}(\gamma_1)&* (\Gamma \setminus \gamma_1))=-\frac{(2\mathrm{i}(4 \pi)^2 m_{\text{UV}}^2)^2}{(p^2-m_{\text{uv}}^2)^3 (k^2-m_{\text{uv}}^2)^3} \left(\mathds{1}-K^{\overline{\text{MS}}}\right)\left( \frac{\mu^2}{4\pi e^{-\gamma_E}}\right)^{\epsilon} \int \mathrm{d}^{4-2\epsilon}p' \,  \\
    &\Bigg[-\frac{2(p'\cdot q) [\slashed{p}' \gamma^\mu \slashed{q} \gamma^\nu \slashed{p}']_{ij} - q^2 [\slashed{p}' \gamma^\mu \slashed{p}' \gamma^\nu \slashed{p}']_{ij} }{({p'}^2-m_{\text{uv}}^2)^4}
    +\frac{4(p'\cdot q)^2 [\slashed{p}' \gamma^\mu \slashed{p}' \gamma^\nu \slashed{p}']_{ij}}{({p'}^2-m_{\text{uv}}^2)^5 }\Bigg] \\
    &* (\mathds{1}-K^{\overline{\text{MS}}}) \left( \frac{\mu^2}{4\pi e^{-\gamma_E}}\right)^{\epsilon}\int \mathrm{d}^{4-2\epsilon}k' \Bigg[2 k'\cdot p'\frac{[\gamma^\sigma \slashed{k}' \gamma_\sigma]_{ji}}{({k'}^2-m_{\text{uv}}^2)^3}\Bigg] \;,
        \end{split}
\end{align}
where we note that the $\epsilon$ contribution of $\bar{K}(\gamma_1)$ will contribute to the final result, as it will hit the pole of $[\Gamma \setminus \gamma_1]$. 
Finally, we have:
\begin{align}
    \begin{split}
    &\mathbf{T}(\bar{K}(\gamma_1)* (\Gamma \setminus \gamma_1))= \frac{2\mathrm{i}(4 \pi)^2 m_{\text{UV}}^2}{(k^2-m_{\text{uv}}^2)^3}\Bigg[\frac{[(\slashed{p}+m) \gamma^\mu (\slashed{p}-\slashed{q}+m) \gamma^\nu (\slashed{p}+m)]_{ij}}{(p^2-m^2)^3} \\
    &+\frac{2(p\cdot q) [(\slashed{p}+m) \gamma^\mu (\slashed{p}+m) \gamma^\nu (\slashed{p}+m)]_{ij}}{(p^2-m^2)^4} \Bigg] \\ &*\left(\mathds{1}-K^{\overline{\text{MS}}}\right) \left( \frac{\mu^2}{4\pi e^{-\gamma_E}}\right)^{\epsilon}\int \mathrm{d}^{4-2\epsilon}k' \Bigg[2 k'\cdot p\frac{[\gamma^\sigma \slashed{k}' \gamma_\sigma]_{ji}}{({k'}^2-m_{\text{uv}}^2)^3} 
    -2 k'\cdot p^{os}\frac{[\gamma^0\gamma^\sigma \slashed{k}' \gamma_\sigma]_{ji}}{({k'}^2-m_{\text{uv}}^2)^3}\Bigg] \\
    &+\Bigg[-\frac{2(p\cdot q) [\slashed{p} \gamma^\mu \slashed{q} \gamma^\nu \slashed{p}]_{ij} -q^2 [\slashed{p} \gamma^\mu \slashed{p} \gamma^\nu \slashed{p}]_{ij} }{(p^2-m_{\text{uv}}^2)^4}
      +\frac{4(p\cdot q)^2 [\slashed{p} \gamma^\mu \slashed{p} \gamma^\nu \slashed{p}]_{ij}}{(p^2-m_{\text{uv}}^2)^5 }\Bigg] \\
    &*\left(\mathds{1}-K^{\overline{\text{MS}}}\right) \left( \frac{\mu^2}{4\pi e^{-\gamma_E}}\right)^{\epsilon}\int \mathrm{d}^{4-2\epsilon}k' \Bigg[2 k'\cdot p\frac{[\gamma^\sigma \slashed{k}' \gamma_\sigma]_{ji}}{({k'}^2-m_{\text{uv}}^2)^3}\Bigg].
           \end{split} 
\end{align}

\paragraph{Term $K(\gamma_2)$:\\[0.6cm]}

Finally, regarding $\gamma_2$, which is logarithmic, we have:
\begin{align}
    K(\gamma_2)* (\Gamma \setminus \gamma_2) &=\frac{[\slashed{p} \gamma^\mu \slashed{p} \gamma^\nu \slashed{p}]_{ij}}{(p^2-m_{\text{uv}}^2)^4}\frac{[\gamma^\sigma \slashed{p} \gamma_\sigma]_{ji}}{(k-p)^2} \\
    &-\frac{g^{\alpha\sigma} 2\mathrm{i}(4 \pi)^2 m_{\text{UV}}^2}{(k-p)^2(p^2-m_{\text{uv}}^2)^3} \int \mathrm{d}^{4-2\epsilon}p'\frac{[\slashed{p}' \gamma^\mu \slashed{p}' \gamma^\nu \slashed{p}']_{ij}[\gamma_\alpha \slashed{p}' \gamma_\sigma]_{ji}}{({p'}^2-m_{\text{uv}}^2)^4}.
\end{align}

\paragraph{Term $K(K(\gamma_2)* (\Gamma \setminus \gamma_2))$:\\[0.6cm]}

Furthermore we have:
\begin{align}
\begin{split}
    \mathbf{T}(K(\gamma_2)* (\Gamma \setminus \gamma_2)) &=\frac{[\slashed{p} \gamma^\mu \slashed{p} \gamma^\nu \slashed{p}]_{ij}}{(p^2-m_{\text{uv}}^2)^4}\frac{[\gamma^\sigma \slashed{p} \gamma_\sigma]_{ji}}{(k-p)^2} \\
    &-\frac{g^{\alpha\sigma} 2\mathrm{i}(4 \pi)^2 m_{\text{UV}}^2}{(k-p)^2(p^2-m_{\text{uv}}^2)^3} \int \mathrm{d}^{4-2\epsilon}p'\frac{[\slashed{p}' \gamma^\mu \slashed{p}' \gamma^\nu \slashed{p}']_{ij}[\gamma_\alpha \slashed{p}' \gamma_\sigma]_{ji}}{({p'}^2-m_{\text{uv}}^2)^4} \;,
\end{split}
\end{align}
and
\begin{align}
     \bar{K}(K(\gamma_2)* (\Gamma \setminus \gamma_2))=0.
\end{align}
This shows that
\begin{equation}
    K(\gamma_2)* (\Gamma\setminus \gamma_2)+K(K(\gamma_2)* (\Gamma \setminus \gamma_2))=0 \,.
\end{equation}

\bibliographystyle{JHEP}
\bibliography{biblio.bib}

\end{document}